\documentclass[11pt, a4paper]{article}
\pdfoutput=1
\usepackage[utf8]{inputenc}
\usepackage[T1]{fontenc}
\usepackage{mlmodern}
\usepackage{textcomp}
\usepackage[american]{babel}
\usepackage{csquotes}

\usepackage{eurosym}
\usepackage{xspace}

\usepackage{graphicx}
\usepackage{subcaption}
\usepackage{float}

\usepackage{makecell}
\usepackage{multirow}
\usepackage{multicol}
\usepackage{enumitem}

\usepackage{marginnote}
\usepackage[multiple]{footmisc}
\usepackage{microtype}

\usepackage[nottoc]{tocbibind}

\usepackage[dvipsnames, svgnames, table]{xcolor}
\selectcolormodel{cmy}

\usepackage[backend=biber, style=numeric-comp]{biblatex}

\usepackage{amsmath, amsfonts, amssymb}
\usepackage{amsthm, thmtools}
\usepackage{mathtools}
\usepackage{mathrsfs}
\usepackage{cancel}
\usepackage{braket}
\usepackage{tensor}
\usepackage{slashed}
\usepackage{siunitx}
\usepackage{listings}
\usepackage{tikz}
\usepackage[compat=1.1.0]{tikz-feynman}

\usepackage{authblk}

\usepackage[
	pdftex, breaklinks=true, linktocpage, colorlinks=true,
	urlcolor=RoyalBlue, linkcolor=RoyalBlue, citecolor=BrickRed
]{hyperref}
\usepackage[all]{hypcap}

\DeclareUnicodeCharacter{00A0}{~}
\DeclareUnicodeCharacter{202F}{~}

\newcommand{\e}[0]{\mathrm{e}}
\newcommand{\I}[0]{\mathrm{i}}

\newcommand{\R}[0]{\mathbb{R}}
\newcommand{\C}[0]{\mathbb{C}}
\newcommand{\dd}[0]{\mathrm{d}}

\newcommand{\pd}[0]{\partial}
\newcommand{\mc}[1]{{\mathcal{#1}}}

\DeclareMathOperator{\tr}{tr}

\newcommand{\abs}[1]{{|#1|}}

\renewcommand{\bra}[2][]{\mathinner{{}_{#1}\hspace{-2pt}\langle #2 |}}

\newcommand{\group}[1]{\mathrm{#1}}
\newcommand{\alg}[1]{\mathfrak{#1}}

\newcommand{\transp}[1]{{#1^t}}

\newcommand{\adj}[1]{{#1^\dagger}}

\newcommand{\email}[1]{\href{mailto:#1}{\nolinkurl{#1}}}
\newcommand{\emailfoot}[1]{\thanks{\email{#1}}}

\newcommand{\doi}[1]{\href{https://dx.doi.org/#1}{\nolinkurl{#1}}}
\newcommand{\urlhttp}[1]{\href{https://#1}{\nolinkurl{#1}}}

\newif\ifdraftcomments
\draftcommentstrue
\newcounter{draftcommentcnt}
\extrafloats{100}
\makeatletter
\newcommand{\draftcomment@margin}[1]{%
	\edef\draftcomment@penalty{\the\@floatpenalty}%
	\let\draftcomment@next\marginpar
	\ifinner\let\draftcomment@next\marginnote\fi
	\ifnum\@floatpenalty<\z@\let\draftcomment@next\marginnote\fi
	\ifnum\floatingpenalty>\z@\let\draftcomment@next\marginnote\fi
	\ifx\@footnotetext\@mpfootnotetext\let\draftcomment@next\marginnote\fi
	\draftcomment@next{#1}%
	\@floatpenalty\draftcomment@penalty\relax
}
\NewDocumentCommand{\draftcomment}{s O{red} +m}{%
	\ifdraftcomments
		\leavevmode
		\IfBooleanF{#1}{%
			\stepcounter{draftcommentcnt}%
			\draftcomment@margin{\textcolor{#2}{$\Leftarrow$ \arabic{draftcommentcnt}}}%
		}%
		{\color{#2}#3}%
	\else
		\@bsphack\@esphack
	\fi
}
\makeatother

\usepackage[heightrounded,
	top=3.5cm, bottom=3cm, left=3.5cm, right=3.5cm,
]{geometry}

\graphicspath{{./images/}}
\numberwithin{equation}{section}

\usepackage[sort&compress, capitalize, english]{cleveref}

\input{arxiv.def}

\hypersetup{
	pdftitle={A cubic closed string field theory},
	pdfauthor={Harold Erbin},
	pdflang={en},
}

\title{A cubic closed string field theory}

\author{Harold Erbin\emailfoot{harold.erbin@gmail.com}}

\tikzset{
	ext/.pic = {\draw[solid] (-2.5pt, -2.5pt) -- (2.5pt, 2.5pt) (-2.5pt, 2.5pt) -- (2.5pt, -2.5pt);},
	diagram/.style = {
		baseline = -0.5ex,
		thick,
		every node/.style = {inner sep=2pt},
	},
	sigma line/.style = {dash pattern=on 2pt off 1.5pt},
}

\newcommand{\IBL}[0]{\ensuremath{\mathrm{IBL}_\infty}}
\newcommand{\Loo}[0]{\ensuremath{L_\infty}}
\newcommand{\Ngh}[0]{\ensuremath{N_{\text{gh}}}}

\DeclareMathOperator{\Hess}{Hess}

\newcommand{\gs}[0]{\ensuremath{g_s}}
\newcommand{\Hs}[0]{\ensuremath{\mathcal{H}}}
\NewDocumentCommand{\Hn}{o m}{\ensuremath{\IfValueTF{#1}{\mathcal{H}_{#1}^{\wedge #2}}{\mathcal{H}^{\wedge #2}}}}
\newcommand{\pair}[3][]{\ensuremath{\braket{#2, #3}_{#1}}}
\newcommand{\vertex}[1]{\ensuremath{\mathcal{V}_{#1}}}
\newcommand{\dvertex}[1]{\ensuremath{\mathcal{W}_{#1}}}
\newcommand{\sprod}[2]{\ensuremath{p_{#1,#2}}}
\newcommand{\wprod}[2]{\ensuremath{w_{#1,#2}}}
\newcommand{\kin}[3]{\ensuremath{\kappa\ifx\relax#3\relax\else^{(#3)}\fi_{#1,#1\ifx\relax#2\relax\else;#2\fi}}}  %

\newcommand{\caction}[1][]{\ensuremath{\widetilde{S}_{#1}}}

\begin{document}

\maketitle

\begin{abstract}
	The action of closed bosonic string field theory is non-polynomial, and for more than thirty years a no-go theorem of Sonoda and Zwiebach (1990) has excluded a cubic formulation with a single string field, a symmetric vertex and the canonical propagator.
	We evade it with a Hubbard--Stratonovich transformation and provide an explicit construction of a cubic closed string field theory at all orders.
	The transformation introduces an infinite tower of auxiliary $k$-string fields $\Sigma_k$, entering at order $\gs^{2 k - 2}$, which turn the vertices with four or more legs into kinetic terms and cubic couplings.
	The action is cubic, at most quadratic in each auxiliary field, and triangular: eliminating the auxiliary fields by Gaussian integrals recovers the original action order by order at tree level, under invertibility and kernel conditions on the auxiliary kinetic operators.
	The original gauge symmetry lifts to the cubic action with no new parameter, and we argue that the auxiliary fields add no physical state at generic momentum, wherever their kinetic operators are invertible.
	On the worldsheet, $\Sigma_2$ is an entangled two-string state: by an operator product expansion, the quartic form reduces to a gluing of two asymmetric cubic vertices dressed by the motion of the local coordinates and defined on states without level matching.
	We expect this picture to generalize to higher orders, defining a gluing of patches carrying $k$ punctures each.
	The string vertices, read as products with several outputs, satisfy an extended form of the genus-zero \IBL{} relations, including operations with no input or no output, exactly when the string products form a cyclic $\Loo$ algebra.
	Finally, we compare our action with the bar--cobar construction of Chiaffrino, Mamade and Zwiebach.
\end{abstract}

\hrule
\pdfbookmark[1]{\contentsname}{toc}
\tableofcontents
\bigskip
\hrule

\newpage

\section{Introduction}
\label{sec:intro}

String field theory (SFT) describes the string by a field $\Psi$ living in the Hilbert space $\mc H$ of the worldsheet conformal field theory (CFT), with an action whose Feynman diagrams reproduce the string amplitudes.
The classical action is a kinetic term plus one interaction for every number of strings $n \ge 3$~\cite{Saadi:1989:ClosedStringField, Kugo:1989:NonpolynomialClosedString, Zwiebach:1993:ClosedStringField}:
\begin{equation}
  \label{eq:intro:action}
  S
    = \frac{1}{2} \, \pair{\Psi}{Q_B \Psi}
      + \sum_{n \ge 3} \frac{\gs^{n-2}}{n!} \, \vertex{n}(\Psi^n),
\end{equation}
where $\gs$ is the string coupling, $\pair{\cdot}{\cdot}$ the BPZ inner product with one insertion of $c_0^-$~\eqref{eq:hs1:pairing}, and $\Psi^n := \Psi^{\otimes n}$.
Each vertex $\vertex{n}$ is the integral of a correlation function of $n$ states over a region $\mc V_{0,n}$ of the moduli space of the $n$-punctured sphere $\mc M_{0,n}$.
The regions are chosen so that the Feynman diagrams built from the vertices and the propagator cover the moduli space of every amplitude exactly once~\cite{Zwiebach:1993:ClosedStringField}.
This covering condition is what the vertices must satisfy, and its algebraic form is a set of relations between the vertices, those of an $\Loo$ algebra for the string products $\ell_n$.
We refer the reader to~\cite{Zwiebach:1993:ClosedStringField-2, deLacroix:2017:ClosedSuperstringField, Erler:2020:FourLecturesClosed, Erbin:2021:StringFieldTheory, Maccaferri:2024:StringFieldTheory, Sen:2024:StringFieldTheory} for reviews.

The action~\eqref{eq:intro:action} is non-polynomial, since it contains an infinite number of terms.
Indeed, the diagrams built from the propagator and the vertices of lower order leave a region in the interior of the moduli space of the $n$-punctured sphere uncovered, which the fundamental vertex $\vertex{n}$ must fill.
Different choices of the vertex regions $\mc V_{0, n}$ give theories related by field redefinitions~\cite{Zwiebach:1992:InterpolatingStringField, Hata:1994:DevelopingCovariantBatalinVilkovisky, Costello:2022:HyperbolicStringVertices}, and the theory is independent of the background around which it is written~\cite{Sen:1994:ProofLocalBackground, Sen:1994:QuantumBackgroundIndependence, Bergman:1995:DilatonTheoremClosed, Belopolsky:1996:WhoChangesString}.
The non-polynomial form is thus the price of a covariant closed SFT, and every computation in the theory, perturbative or not, has to cope with an infinite number of elementary interactions.

For the open bosonic string, a parametrization of the vertices such that the action is cubic was found by Witten~\cite{Witten:1986:NoncommutativeGeometryString}.
It is built from a kinetic term and a single interaction, the star product, which glues the right half of one string to the left half of the other.
In this case, the gluing of cubic vertices covers the whole moduli space, without any gap~\cite{Zwiebach:1991:ProofThatWittens}.

Having a cubic action brings more than an economical notation, as has been amply demonstrated in open SFT.
Witten's theory has been at the core of most of the developments and explicit computations in SFT during the last decades.
The main turning point was the proposal of Sen's conjectures on tachyon condensation~\cite{Sen:1998:TachyonCondensationBrane, Sen:1999:UniversalityTachyonPotential}, which could be tested in Witten's theory by level truncation~\cite{Sen:2000:TachyonCondensationString, Moeller:2000:LevelTruncationTachyon, Gaiotto:2003:ExperimentalStringField} (reviews include~\cite{Ohmori:2001:ReviewTachyonCondensation, Taylor:2004:DBranesTachyonsString, Sen:2005:TachyonDynamicsOpen}).
The analytic solution for the tachyon vacuum~\cite{Schnabl:2006:AnalyticSolutionTachyon, Okawa:2006:CommentsSchnablsAnalytic, Ellwood:2007:ProofVanishingCohomology} rests on the algebra of the star product: a cubic action has a single product, and the solution is built from its algebraic properties.
This approach has led to a wealth of results~\cite{Fuchs:2011:AnalyticalSolutionsOpen, Okawa:2012:AnalyticMethodsOpen, Erler:2022:FourLecturesAnalytic}, and culminated in the derivation of a general solution~\cite{Erler:2014:StringFieldTheory, Erler:2020:StringFieldTheory}.
Other applications, such as the question of locality and causality~\cite{Erler:2004:LocalityCausalityInitial}, could be investigated explicitly only thanks to this structure.

More importantly, Witten's product also underlies open superstring field theory.
While super-SFT is non-polynomial because of odd moduli, it was realized that an action for the Neveu--Schwarz (NS) sector of super-SFT could be built using only Witten's product and picture-changing operator (PCO) insertions~\cite{Berkovits:1995:SuperPoincareInvariantSuperstring}.
Although this action does not have an $A_\infty$ structure, an action with such a structure was later built by different groups~\cite{Iimori:2014:BerkovitsFormulationWitten, Erler:2014:ResolvingWittensSuperstring, Erler:2015:RamondEquationsMotion, Kunitomo:2016:CompleteActionOpen, Konopka:2016:OpenSuperstringField, Erler:2016:CompleteActionOpen} and shown to be equivalent to the Berkovits formulation~\cite{Erler:2015:StructureBerkovitsFormulation, Erler:2015:RelatingBerkovitsSuperstringSmall, Erler:2015:RelatingBerkovitsSuperstringLarge}.

It is thus desirable to find a similar parametrization for the bosonic closed string field theory.

\subsection{Searching for cubic closed string field theory}

\paragraph{Sonoda--Zwiebach no-go theorem.}

Unfortunately, a no-go theorem by Sonoda and Zwiebach shows that no such simple formulation exists for the closed string~\cite{Sonoda:1990:CovariantClosedString}.
Hence, it has been assumed since then that no cubic action exists for the closed string, and one has to deal with the infinite number of vertices.
However, the construction of these vertices is extremely challenging, and only the lowest orders are known in different parametrizations~\cite{Saadi:1989:ClosedStringField,Moeller:2004:ClosedBosonicString,Moeller:2007:ClosedBosonicString-1,Erler:2017:OneLoopTadpole,Firat:2021:HyperbolicThreestringVertex,Erbin:2023:SL2QuarticVertex,Erbin:2024:Characterizing4stringContact,Erbin:toappear:SL2VerticesClosed}.
The construction of the closed superstring takes as input the full non-polynomial tower of bosonic products~\cite{Okawa:2004:HeteroticStringField,Berkovits:2004:WZWlikeActionHeterotic,Erler:2014:NSNSSectorClosed, Erler:2015:RamondEquationsMotion}.
This explains why closed SFT has barely been used for explicit computations (see~\cite{Belopolsky:1995:OffshellClosedString, Belopolsky:1995:EffectiveTachyonicPotential, Okawa:2004:TwistedTachyonCondensation, Yang:2005:ClosedStringTachyon, Moeller:2007:NonperturbativeClosedString, Scheinpflug:2025:ClosedStringTachyon, Mazel:2024:ConformalPerturbationTheory} for the computations which have been done).

The theorem of Sonoda--Zwiebach can be stated as follows~\cite{Sonoda:1990:CovariantClosedString}.
Consider a theory satisfying the following assumptions:
\begin{enumerate}
  \item the theory describes a closed string field built on the matter--ghost conformal field theory, in perturbation theory around a conformal background, with level-matched external states;

  \item the interaction is a cubic vertex which is Hermitian and symmetric under all permutations of its inputs, where symmetric means that the conformal maps exchanging the punctures change the local coordinates at most by phases;

  \item the $4$-point amplitude is obtained by sewing the cubic vertices with the plumbing fixture $w_1 w_2 = q$, where the sewing parameter $q$ ranges over a domain which contains $q = 0$, is the same in the three channels, and is invariant under the symmetries required by covariance and unitarity;

  \item the three channels cover the moduli space $\mc M_{0,4}$ exactly once.
\end{enumerate}
Then, for any choice of the cubic vertex and of the sewing domain, the $4$-point amplitude is not BRST invariant off-shell, so that an elementary quartic interaction is necessary.
The third assumption is more general than assuming the canonical propagator for which $q = \e^{- s + \I \theta}$ and $\abs{q} \le 1$:
\begin{equation}
  \Delta
    := \frac{b_0^+}{L_0^+} \, b_0^- \delta_{L_0^-, 0}
    = b_0^+ b_0^- \int_0^\infty \dd s \int_0^{2\pi} \frac{\dd \theta}{2\pi} \, \e^{- s L_0^+ + \I \theta L_0^-}.
\end{equation}
Adding stubs, $\abs{q} \le \e^{- 2 s_0}$, amounts to a rewriting of the theory with different vertices and the same amplitudes~\cite{Sonoda:1990:ClosedStringField}.
For a more general domain, the states propagating inside the diagrams need not be level matched~\cite[pp.~188--189]{Sonoda:1990:CovariantClosedString}.

With the canonical propagator, the three channels of the closed string analogue of Witten's vertex do not cover $\mc M_{0,4}$: the missing region is the restricted tetrahedron of~\cite{Saadi:1989:ClosedStringField}.
We expect from the general construction of the vertices and explicit examples that such a gap $\mc V_{0,4} \neq \emptyset$ remains for any cubic vertex, even if we allow for stubs.
Sonoda and Zwiebach argue that the gap can always be closed by enlarging the domain of the sewing parameter, and that the difficult task is to make the local coordinates of two channels agree, up to phases, on their common boundary.
This is the condition for the amplitude to be BRST invariant off-shell, which they call off-shell duality.
Covariance and unitarity fix the three regions of $\mc M_{0,4}$ covered by the channels independently of the vertex.
At the point of their common boundary where the four punctures form a square, the condition is satisfied only by Witten's vertex.
However, Witten's vertex satisfies it only at three points of the boundaries~\cite{Sonoda:1990:ClosedStringField,Sonoda:1990:CovariantClosedString}.
Hence, no choice of cubic vertex is compatible with this condition, and an elementary quartic vertex is needed to restore the BRST invariance.
With the canonical propagator, the quartic vertex is integrated over the gap $\mc V_{0,4}$, and its local coordinates match those of the three channels on the common boundaries~\cite{Saadi:1989:ClosedStringField}.
When the gap is instead closed by enlarging the sewing domain, the quartic vertex is a “contact” vertex, localized on the common boundaries of the channels, as in the NS sector of open superstring field theory~\cite{Erler:2014:ResolvingWittensSuperstring}.
Sonoda and Zwiebach conjecture that elementary interactions of all orders are needed.
They also list the cases to which the theorem may not apply~\cite[p.~187]{Sonoda:1990:CovariantClosedString}: theories which do not use closed string fields, or whose local coordinates do not come from a conformal field theory; theories whose propagator has no sewing interpretation; theories with a weaker notion of symmetric vertex.

\paragraph{Evading the no-go theorem.}

A way to evade the no-go theorem by enlarging the field content was proposed in~\cite{Erbin:2018:CubicClosedString}.
The idea is to introduce auxiliary fields using the Hubbard--Stratonovich transformation~\cite{Stratonovich:1957:MethodCalculatingQuantum, Hubbard:1959:CalculationPartitionFunctions} (this is also called the intermediate field formalism).
The theory then contains more than the closed string field: the space of fields is enlarged by an auxiliary field which describes entangled states of two strings.
Its kinetic term does not involve $Q_B$, so that its propagator is not the plumbing fixture of closed string field theory but a different operation on Riemann surfaces (\cref{sec:worldsheet}).
The vertex which couples it to the string field is not symmetric under all permutations of its inputs, since it involves two kinds of fields.
Hence the first three assumptions fail, and these are among the cases listed above to which the theorem may not apply.
Finally, the three channels of the cubic vertex of the original theory are kept as they are, and the contribution of the region of $\mc M_{0,4}$ which they leave uncovered is produced by the exchange of the auxiliary field.
The quartic vertex is not removed: it becomes the kinetic operator of the auxiliary field.

The inspiration comes from writing the quartic term of a quantum field theory (QFT) in momentum space:
\begin{equation}
  S_4
    = \int \dd^d k_1 \cdots \dd^d k_4 \, V_{\alpha_1 \cdots \alpha_4}(k_1, \ldots, k_4) \, \phi_{\alpha_1}(k_1) \cdots \phi_{\alpha_4}(k_4),
\end{equation}
where the indices $\alpha_i$ denote both Lorentz indices and internal labels.
Restricting the fields to scalar fields and taking the indices to be in the fundamental representation of $O(N)$, we obtain a vector model of $N$ scalar fields.
The transformation of such a model to a cubic theory with an auxiliary field $\sigma_{\alpha_1 \alpha_2}(k_1, k_2)$ is described in~\cite[sec.~21.6]{Weinberg:2005:QuantumTheoryFields-2} for fermions (the case of several scalar fields, where the auxiliary field carries a pair of indices, is worked out in \cref{app:scalar-hs}).\footnotemark{}
\footnotetext{%
  The textbook transformation~\cite{Altland:2010:CondensedMatterField,Coleman:2016:IntroductionManyBodyPhysics} is not directly generalizable to the model above.
  Indeed, given $S_4 = \lambda \phi^4 / 2$, we write $\sigma = \lambda \phi^2$ such that $S_4 = - \sigma^2 / (2 \lambda) + \sigma \phi^2$.
  The redefinition $\sigma \to \I \varsigma$ is often performed to change the sign of the kinetic term.
  However, it is not necessarily clear how to take the inverse of a general non-local quartic tensor, and it does not generalize well to SFT.
}%
The trick is to add the term
\begin{equation}
  \begin{aligned}
    \Delta S_4
      = - \int \dd^d k_1 \cdots \dd^d k_4 \,
        &
        V_{\alpha_1 \cdots \alpha_4}(k_1, \ldots, k_4) \,
        \Big(
          \sigma_{\alpha_1 \alpha_2}(k_1, k_2) - \phi_{\alpha_1}(k_1) \phi_{\alpha_2}(k_2)
        \Big)
        \\ &
        \times
        \Big(
          \sigma_{\alpha_3 \alpha_4}(k_3, k_4) - \phi_{\alpha_3}(k_3) \phi_{\alpha_4}(k_4)
        \Big)
  \end{aligned}
\end{equation}
to the action.
The new field $\sigma_{\alpha_1 \alpha_2}(k_1, k_2)$ is auxiliary and lives in two copies of the original state space.
Its equation of motion is algebraic, and solving it gives back the original interaction: setting $\sigma_{\alpha_1 \alpha_2}(k_1, k_2) = \phi_{\alpha_1}(k_1) \phi_{\alpha_2}(k_2)$ gives $\Delta S_4 = 0$, and the auxiliary fields are determined in terms of the physical fields up to the kernel of $V$, which is quotiented out (\cref{sec:hs:quartic:field}).
In terms of Feynman diagrams, the original Feynman vertex $V_{\alpha_1 \cdots \alpha_4}(k_1, \ldots, k_4)$ is reproduced by the $s$-, $t$- and $u$-channels containing two cubic vertices and an intermediate auxiliary field:
\begin{equation}
  \begin{aligned}
    \vcenter{\hbox{\includegraphics[scale=1]{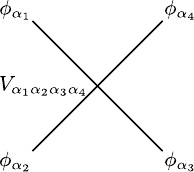}}}
      \qquad
      &
      =
        \qquad
        \vcenter{\hbox{\includegraphics[scale=1]{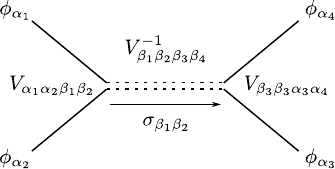}}}
        \\ & \hspace{2cm}
        \medskip
        + \text{$t$-, $u$-channels}
  \end{aligned}
\end{equation}

This can be straightforwardly generalized to SFT~\cite{Erbin:2018:CubicClosedString}.\footnotemark{}
\footnotetext{%
  Studying a toy model of SFT in momentum space is a common practice to generalize QFT techniques (see~\cite{Pius:2016:CutkoskyRulesSuperstring, deLacroix:2017:ClosedSuperstringField, deLacroix:2019:AnalyticityCrossingSymmetry, Erbin:2021:ConstructiveExpansionQuartic}).
}%
Given the quartic vertex $\vertex{4}(\Psi^4)$, we can subtract from the action $\vertex{4}(\Sigma_2 - \Psi^2, \Sigma_2 - \Psi^2)$, where $\Sigma_2$ lives in two copies of the Hilbert space $\Hn{2}$ (the precise notations will be explained in \cref{sec:hs:action}), i.e.\ it is a graded-symmetric tensor with two strands.
It is clear from the form of the term above that it vanishes on the solution $\Sigma_2 = \Psi^2$.
With $S_4$ the quartic term of~\eqref{eq:intro:action}, the quartic part of the action becomes (\cref{sec:hs:quartic})
\begin{equation}
  \label{eq:intro:S4tilde}
  \begin{aligned}
    \caction[4]
      &
      := S_4 - \frac{\gs^2}{4!} \, \vertex{4}(\Sigma_2 - \Psi^2, \Sigma_2 - \Psi^2)
      \\ &
      = - \frac{\gs^2}{4!} \, \vertex{4}(\Sigma_2, \Sigma_2)
        + \frac{2 \gs^2}{4!} \, \vertex{4}(\Sigma_2, \Psi^2),
  \end{aligned}
\end{equation}
a quadratic term for $\Sigma_2$ and a cubic coupling of $\Sigma_2$ to two copies of $\Psi$.
The term quartic in $\Psi$ has disappeared, and the action through this order is cubic in the pair $(\Psi, \Sigma_2)$.
Eliminating $\Sigma_2$ by its equation of motion gives back the quartic term.
In the Feynman rules, the quartic contact interaction of $\Psi$ is replaced by the exchange of $\Sigma_2$ between two cubic vertices.
The exchange diagram reduces to the contact vertex because the propagator of $\Sigma_2$ is the inverse of the form $\vertex{4}$ which appears at its two cubic couplings: the two couplings and the propagator between them collapse to $\vertex{4}(\Psi^4)$.

The action above requires introducing additional structures.
Indeed, in order to vary the action with respect to $\Sigma_2$, we have to introduce a product $p_{2,2}$ with two inputs and two outputs, along with a $2$-state pairing:
\begin{equation}
  \label{eq:intro:p22}
  \delta \vertex{4}(\Sigma_2, \Sigma_2)
    = 2 \, \braket{\delta \Sigma_2, p_{2,2}(\Sigma_2)}_2.
\end{equation}
These products are usually not discussed in SFT, though they naturally sit between the vertices and surface states~\cite{Sen:2015:OffshellAmplitudesSuperstring}.
While this product is completely determined by $\ell_3$ (or equivalently $\vertex{4}$, by~\eqref{eq:hs1:vertex-product}), we can ask whether there is an algebraic structure which puts on an equal footing products $p_{m,n}: \Hn{m} \to \Hn{n}$ (anticipating the higher orders).
The natural candidate~\cite{Erbin:2018:CubicClosedString} is an $\IBL$ homotopy algebra~\cite{Cieliebak:2020:HomologicalAlgebraRelated}, which made a brief appearance in SFT in~\cite{Muenster:2013:QuantumOpenClosedHomotopy, Muenster:2014:HomotopyClassificationBosonic}.
The products $p_{m,n}$ are obtained from the vertices by turning inputs into outputs with the inverse of the odd symplectic form, $p_{2,2}$ being the first.
They satisfy an extended form of the genus-zero $\IBL$ relations, including the operations with no input or no output, if and only if the $\ell_n$ form a cyclic $\Loo$ algebra (\cref{sec:ibl}); whether they satisfy the standard $\IBL$ relations is open.
These products act on $n$-string fields $\Psi_n \in \Hn{n}$ for all $n > 0$, whose spaces are those of the string field $\Psi_1 = \Psi$ and of the auxiliary fields $\Psi_n = \Sigma_n$ introduced below.
The auxiliary fields are however not multi-string fields of an $\IBL$ algebra, since no differential acts on their strands.

There is no inconsistency with the no-go theorem: the amplitude is the one of the original theory, and it is BRST invariant.
We will also discuss the worldsheet interpretation of this transformation in \cref{sec:worldsheet}.
On the worldsheet, the two punctures of a $\Sigma_2$ slot can be enclosed by a circle, and their operator product expansion merges them into a single string state, which is not level matched.
On-shell, we can describe the exchange of $\Sigma_2$ as a gluing of two copies of the cubic vertex of the theory, over the region of the sewing parameter which its Feynman diagrams leave uncovered and with intermediate states which are not level matched.
Each copy is asymmetric, with the merged state at one puncture.
This is a sewing of the kind considered in the Sonoda--Zwiebach theorem~\cite{Sonoda:1990:CovariantClosedString}, with a domain which is not the unit disk, except that the enlarged domain is attached to one channel only for a given pairing of the punctures.
Consistent with the conclusion of the theorem, this gluing with fixed local coordinates is not compatible with the three cubic channels off-shell for the $\group{SL}(2, \C)$ vertex: the local coordinates do not match on their common boundary.
The actual exchange of $\Sigma_2$ reproduces the quartic vertex off-shell by construction.
On the worldsheet, this requires the local coordinates at the four punctures to move with the sewing parameter $q$. Each external state is acted on by a conformal transformation $U_{g_i(q)}$, an exponential of the Virasoro modes $L_{n \ge 0}$ and $\bar L_{n \ge 0}$; the two antighost insertions of the plumbing measure are dressed by $q$-dependent contours of $b$-ghost modes (\cref{sec:worldsheet}).
The propagator of $\Sigma_2$ removes two moduli instead of adding them and is therefore not a sewing in the sense of the theorem.

The main obstacle in~\cite{Erbin:2018:CubicClosedString} was to make the different statements more precise and to extend them to higher orders.
A first attempt was realized in~\cite{Erbin:2024:OpenStringStub}, where we used open bosonic SFT with stubs as a toy model.
This action is obtained by adding stubs ($s_0 \neq 0$) to Witten's theory, such that the action becomes non-polynomial: the action can be made cubic again by introducing a single auxiliary field, which describes the high-energy degrees of freedom of the string field (see also~\cite{Chiaffrino:2022:QFTStubs, Schnabl:2023:OpenStringField, Schnabl:2025:MoreStubsOpen, Stettinger:2025:SliverFrameStubs, Maccaferri:2024:AddingStubsQuantum} for related work).

In retrospect, the fact that it should be possible to find a formulation of closed SFT with quadratic equations of motion could have been expected given the strictification theorems~\cite{Kriz:1995:OperadsAlgebrasModules, Berger:2007:ResolutionColouredOperads}.
These theorems state that any $\Loo$ algebra is quasi-isomorphic to a differential graded Lie algebra ($\mathrm{dgL}$ algebra, or strict $\Loo$ algebra).
This algebra possesses a differential and a single bracket, defined on a larger space built from the tensor powers of the original one.
This result re-emerged recently~\cite{Macrelli:2019:ScatteringAmplitudeRecursion,Borsten:2021:DoubleCopyHomotopy} in the context of describing QFT using $\Loo$~\cite{Hohm:2017:AlgebrasFieldTheory, Jurco:2019:LAlgebrasClassicalField} and made it clear that such a formulation should exist for SFT, though it remained elusive how to describe it explicitly beyond the statements of~\cite{Erbin:2018:CubicClosedString}.
In the present construction, an enlargement of the same kind, though smaller, is obtained by integrating in auxiliary fields.
In the homotopy algebra language, the reverse operation – integrating out – is implemented by homotopy transfer~\cite{Kajiura:2002:HomotopyAlgebraMorphism, Doubek:2019:QuantumAlgebrasHomological, Erbin:2020:ClassicalAlgebraicStructures, Arvanitakis:2022:HomotopyTransferEffective-1, Arvanitakis:2022:HomotopyTransferEffective-2, Maunder:2024:HomotopyTransferLinfinity}.

\paragraph{Construction to higher orders.}

In this paper, we extend the construction from~\cite{Erbin:2018:CubicClosedString} to all orders, introducing an infinite tower of auxiliary fields $\Sigma_n \in \Hn{n}$.
We obtain an action containing only cubic interactions, which provides a strictification of closed string field theory in the sense that its equations of motion are quadratic, not as the Maurer--Cartan action of a $\mathrm{dgL}$ algebra (\cref{sec:ibl:transfer}).
Moreover, the cubic terms are at most quadratic in any given auxiliary field, so that the equation of motion of each auxiliary field is linear in that field.

The higher orders repeat the steps above, with one new feature.
At the quintic order, no new field is needed: we write $\vertex{5}(\Psi^5)$ as the coupling $\vertex{5}(\Sigma_2^2, \Psi)$.
This cubic term can be interpreted as a modification of the kinetic operator of $\Sigma_2$, linear in $\Psi$ seen as a background field, and the classical solution and the gauge transformation of $\Sigma_2$ receive corrections in $\gs$.
Once $\Sigma_2$ propagates and couples through $\Psi \Sigma_2^2$, the Feynman diagrams of the cubic theory contain several $\Sigma_2$ lines glued by this vertex, which does not happen when only the quartic term is decomposed.
Eliminating $\Sigma_2$ therefore produces, besides $\vertex{4}$ and $\vertex{5}$, exchange diagrams of auxiliary fields which contribute from order $\gs^4$ on.
The vertices to be decomposed for $n \ge 6$ are the fundamental vertices minus these exchange contributions: the auxiliary fields of lower orders induce terms of higher order in $\Psi$.

At the sextic order, we need a new $3$-string field $\Sigma_3 \in \Hn{3}$.
As discussed above, we have to decompose a composite vertex made of $\Psi \otimes \Sigma_2$ on top of the vertex $\vertex{6}(\Psi^6)$.
We conjecture that this composite term covers a definite part of the region $\mc V_{0,6}$ of moduli space on which $\vertex{6}$ is defined.
This would extend the geometric picture from usual SFT to auxiliary fields: in the same way that the moduli spaces $\mc M_{0,n}$ are decomposed into regions built from surfaces with and without $\Psi$ propagators, the vertex regions would themselves be decomposed in terms of surfaces with and without $\Sigma_n$ propagators.
We did not manage to prove that the composite vertex corresponds to a definite subregion of $\mc V_{0,6}$, and we leave this question open for future work (see \cref{sec:worldsheet} for the discussion).
Moreover, a naive decomposition of the sextic term would yield the source term $\vertex{6}(\Sigma_3, \Psi^3)$, which is quartic and not cubic.
The solution is to first subtract the exchange contribution of $\Sigma_2$ from $\vertex{6}$, to rewrite the resulting sextic term on $\Psi \otimes \Sigma_2$ in each triple of slots, and then to complete the square in $\Sigma_3 - \Psi \otimes \Sigma_2$. This leaves a quadratic term in $\Sigma_3$ and a cubic coupling of $\Sigma_3$ to $\Sigma_2$ and $\Psi$.
This rewriting induces additional terms in the action, but at higher orders, and they can be decomposed following the same method.
On the solution of the equations of motion, $\Sigma_3 = \Psi \otimes \Sigma_2 + \cdots = \Psi^3 + \cdots$, the dots denoting corrections of higher order in $\gs$.

The $k$-string field $\Sigma_k \in \Hn{k}$ enters at order $\gs^{2 k - 2}$, quadratically in the kinetic term $\mc W_{2k}(\Sigma_k^2)$ and linearly in the cubic interaction $\mc W_{2k}(\Sigma_k, \Sigma_{k-1}, \Psi)$, where $\Sigma_1 := \Psi$ and $\mc W_{2k} = \vertex{2k} + \cdots$.
At the next order, it enters quadratically in $\mc W_{2k+1}(\Sigma_k^2, \Psi)$.
Every vertex $\vertex{n}$ with $n \ge 4$ is thereby replaced by couplings of at most three fields.

We call the vertices which the construction uses at a given order the reduced vertices.
The name follows the conjecture above: if the composite terms cover definite parts of the vertex regions, the reduced vertices are the original vertices restricted to the rest (\cref{sec:worldsheet}).
Their leading-order values, obtained by setting each auxiliary field to $\Sigma_k = \Psi^k$, the leading order of its solution, are the string vertices minus the contributions of the exchange diagrams of the auxiliary fields of lower orders.
The action takes a triangular form: it is quadratic in each auxiliary field separately, and the source of $\Sigma_k$ is linear in $\Sigma_{k-1}$ and $\Psi$.
We can therefore eliminate the auxiliary fields one at a time, starting from the one of highest order, each elimination being a Gaussian integral, exact at tree level.
The result is the original action~\eqref{eq:intro:action} as a formal power series in $\gs$, provided the kinetic operators are invertible on their quotients and the higher vertices vanish on the kernels of these operators (\cref{sec:all:kernel}).
We show these conditions only at the lowest mass level of the quartic vertex.

The gauge symmetry of the original theory lifts to the cubic action, with a transformation of the auxiliary fields determined order by order.
We argue that the auxiliary sector adds no physical state.
The argument is that, around any background on which its dressed kinetic operator is invertible, and to the order to which it has been worked out, the physical spectrum of the cubic theory is that of the original theory at generic momentum (\cref{sec:further}).\footnotemark{}
\footnotetext{%
  The invertibility is established only for the quartic form at the lowest mass level, for a family of vertices with long enough stubs (\cref{sec:hs:quartic:feynman}), and assumed beyond, at generic momentum.
  The null vectors of a kinetic operator are removed by the quotient and need no auxiliary field if the higher vertices vanish on them too (\cref{sec:all:kernel}).
}%

\paragraph{Relation to other work.}

Generalizing the Hubbard--Stratonovich transformation to higher-order interactions $\phi^{2n}$ was done in~\cite{Rivasseau:2010:LoopVertexExpansion, Lionni:2018:NoteIntermediateField} (see also~\cite{Erbin:2021:ConstructiveExpansionQuartic}), as the starting point of the loop vertex expansion.
They also need to introduce a tower of auxiliary fields.
The tower in our paper is simpler because we can reuse fields from lower orders.

Chiaffrino, Mamade and Zwiebach have shown that the equations of motion of any $A_\infty$ field theory can be rewritten as a quadratic equation on an enlarged space of multi-string fields, by the bar--cobar construction, realizing explicitly the strictification theorem~\cite{Chiaffrino:2026:UniversalQuadraticField}.
For $\Loo$ algebras they give the same construction but rely on general results on quasi-isomorphisms, which identify the Maurer--Cartan sets under some assumptions.
They leave open whether these equations derive from an action, and note that simpler constructions, such as a Hubbard--Stratonovich transformation, likely require more knowledge of the theory.
The auxiliary fields of this paper live, up to a quotient and a normalization, in the spaces of their multi-string fields with one output (“type-I”), which they interpret as entangled CFT states; this is the interpretation of the auxiliary field proposed earlier in~\cite{Erbin:2018:CubicClosedString}.
In the sector where only these fields are kept, the two constructions have the same solutions for $\Psi$ order by order, under the assumptions of \cref{sec:all-orders}, with the auxiliary fields agreeing at leading order, $\Sigma_k = \Psi^k$.
The cubic action can be read as a truncation of this sector which replaces their longer fields by products of the ones it keeps.
It is not a consistent truncation of their equations, since their factorization conditions become equations of motion, and their construction has more fields than the cubic action, so that neither is a consistent truncation of the other (\cref{sec:bar-cobar}).

A different route to a closed string field theory determined by its cubic vertex is the hyperbolic one, where the higher vertices are generated recursively from the $3$-string vertex~\cite{Firat:2023:BootstrappingClosedString, Firat:2024:TopologicalRecursionHyperbolic, Kim:2026:RecursivealgebraicSolutionClosed} (see~\cite{Moosavian:2019:HyperbolicGeometryClosed-1,Moosavian:2019:HyperbolicGeometryClosed, Costello:2022:HyperbolicStringVertices, Cho:2020:OpenclosedHyperbolicString, Firat:2023:HyperbolicStringTadpole, Firat:2024:StringVerticesLarge} for the construction of the hyperbolic SFT).
Moreover, Ishibashi constructed closed string field theories in the Fokker--Planck formalism whose Hamiltonian contains only kinetic and $3$-string terms~\cite{Ishibashi:2023:FokkerPlanckFormalismClosed, Ishibashi:2024:StrebelDifferentialsString}.

\subsection{Plan of the paper}

\Cref{sec:hs} carries out the transformation from the quartic to the septic order, where every ingredient of the general construction appears: the auxiliary field and its quotient, the collapse of the exchange diagrams, the dressed kinetic operators, the reduced vertices and the triangular form of the action.
\Cref{sec:all-orders} proves the construction at all orders, under the invertibility of the kinetic operators on their quotients and conditions on their kernels.
Under these conditions, the matching fixes each reduced vertex only through its leading-order value, the auxiliary fields are eliminated exactly from the top of the tower down, and the gauge symmetry lifts to the tower.
\Cref{sec:worldsheet} studies the meaning of the inverse quartic vertex on the worldsheet.
\Cref{sec:ibl} answers the question raised by $p_{2,2}$ and proves the statement above on the $\IBL$ relations of the products with several outputs.
\Cref{sec:bar-cobar} compares the construction, ingredient by ingredient, with the bar--cobar construction of Chiaffrino, Mamade and Zwiebach, which also uses multi-string fields.
\Cref{sec:further} states, without derivations, results on the gauge algebra and the master action, on the Hessian around a background, on level truncation and on a mean-field reading, whose analysis we did not carry to the end.
\Cref{sec:discussion} summarizes the construction and lists the open problems.
The appendices collect the conventions, the exchange sign of the pairing and the reality conditions, a localized form of the action, and the transformation for $N$ scalar fields (\cref{app:conventions,app:reality,app:wrong-sign,app:scalar-hs}).

\section{Hubbard--Stratonovich transformation order by order}
\label{sec:hs}

A Hubbard--Stratonovich transformation~\cite{Stratonovich:1957:MethodCalculatingQuantum,Hubbard:1959:CalculationPartitionFunctions} trades an interaction that is the square of some composite, such as $\lambda f(\phi)^2$, for a Gaussian integral over an auxiliary field coupled linearly to that composite.
For example, the quartic vertex $\lambda \phi^4$ becomes a cubic coupling $\sigma \phi^2$ plus a quadratic term for the auxiliary field, and integrating the auxiliary field out gives back the original interaction exactly (see \cref{app:scalar-hs}).
In closed string field theory the interaction at order $\gs^{n-2}$ is the vertex $\vertex{n}$ evaluated on $n$ copies of the string field.
The natural composites are thus the tensor powers $\Psi^k := \Psi^{\otimes k}$, and the auxiliary fields replacing them are $k$-string fields $\Sigma_k \in \Hn{k}$.
This section carries out the transformation order by order in $\gs$, introducing one auxiliary field at a time.
The purpose is to describe the different ingredients which are needed for rewriting the action at all orders (\cref{sec:all-orders}).

\Cref{sec:hs:action} recalls the action of closed string field theory and fixes the notation.
The quartic order (\cref{sec:hs:quartic}) introduces $\Sigma_2$ and is worked out in full.
At the quintic order (\cref{sec:hs:quintic}), $\Sigma_2$ alone is still sufficient, but its saddle-point solution and its gauge transformation receive corrections.
At the sextic order (\cref{sec:hs:sextic}), a $3$-string field $\Sigma_3$ is introduced, together with the reduced sextic vertex: the sextic vertex minus the exchange diagram induced by $\Sigma_2$.
The septic order (\cref{sec:hs:septic}) makes the general pattern visible.

We treat the quartic order in more detail than its simplicity requires, because every ingredient of the construction already appears there.
We start by describing the equation of motion which identifies the saddle point of the auxiliary field with the tensor power $\Psi^2$, up to the kernel of the quartic vertex, which is quotiented out of the space of the auxiliary field.
We then show in detail how to eliminate the auxiliary field by completing a square.
We describe different properties of the field: its reality condition, its gauge transformation and its spacetime content.
Last come the Feynman rules through which the exchange of the auxiliary field reproduces the quartic vertex.
Each of these points returns, with corrections in $\gs$, at every higher order.

\subsection{The non-polynomial action}
\label{sec:hs:action}

The classical action of closed string field theory~\cite{Zwiebach:1993:ClosedStringField} is a sum of a kinetic term and of one interaction per number of strings:
\begin{equation}
  \label{eq:hs1:action}
  S
    = \frac{1}{2} \, \pair{\Psi}{Q_B \Psi}
      + \sum_{n \ge 3} \frac{\gs^{n-2}}{n!} \, \vertex{n}(\Psi^n),
\end{equation}
where $\gs$ is the string coupling and $\Psi^n := \Psi^{\otimes n}$.

The string field $\Psi \in \Hs$ is an element of the space of states of the matter--ghost CFT satisfying the level-matching conditions
\begin{equation}
  \label{eq:hs1:level-matching}
  b_0^- \ket{\Psi}
    = 0,
  \qquad
  L_0^- \ket{\Psi}
    = 0,
\end{equation}
and carrying ghost number
\begin{equation}
  \label{eq:hs1:ghost-psi}
  \Ngh(\Psi)
    = 2.
\end{equation}
The pairing is the BPZ inner product with one insertion of $c_0^-$:
\begin{equation}
  \label{eq:hs1:pairing}
  \pair{A}{B}
    := \braket{A | c_0^- | B}.
\end{equation}

The maps $\vertex{n} \colon \Hn{n} \to \C$ are the genus-zero string vertices.
They are obtained by integrating the correlation function of the $n$ states over the vertex region $\mathcal V_{0,n} \subset \mc M_{0,n}$ of the moduli space of the $n$-punctured sphere, with the appropriate antighost insertions and local coordinates fixed by a section.
The string products $\ell_n \colon \Hn{n} \to \Hs$ are the maps they define through the pairing:
\begin{equation}
  \label{eq:hs1:vertex-product}
  \vertex{n}(A_1, \ldots, A_n)
    =: \pair{A_1}{\ell_{n-1}(A_2, \ldots, A_n)},
  \qquad
  \ell_1
    := Q_B.
\end{equation}
We follow the conventions of~\cite{Erbin:2021:StringFieldTheory} for the vertices and the products.
They are collected in \cref{app:conventions}, together with the normalization of the sphere correlators, the ghost numbers and the reality condition.

Here $\Hn{n}$ denotes the space of $n$-string states whose components are graded-symmetric under permutations of the strands.
For the Grassmann-even strands of $\Psi^n$ and $\Sigma_n$, it is the $n$-th tensor power of $\Hs$ symmetrized with the Grassmann signs~\cite{Erler:2014:NSNSSectorClosed, Erbin:2020:ClassicalAlgebraicStructures}.
The odd strands of the outputs of the products are treated in \cref{app:conventions}.
The vertices are graded-symmetric in their arguments.
Ghost-number conservation on the sphere requires the insertions to total $\Ngh = 6$, which fixes
\begin{equation}
  \label{eq:hs1:ghost-vertices}
  \Ngh(\vertex{n})
    = 6 - 2n,
  \qquad
  \Ngh(\ell_n)
    = 3 - 2n.
\end{equation}
The vertices and products are extended by multilinearity to tensor arguments, by expanding a tensor on a basis of $\Hn{n}$ built from a basis of $\Hs$.
We write the tensors in the slots they fill, as in $\vertex{4}(\Sigma_2, \Psi^2)$ for $\Sigma_2, \Psi^2 \in \Hn{2}$.

We write the vertices, maps $\Hn{n} \to \C$, with a capital letter, $\vertex{n}$, and the string products with a lowercase one.
A product with $m$ inputs and $n$ outputs, $\Hn{m} \to \Hn{n}$, carries both numbers as subscripts, $\sprod{m}{n}$, except when they equal $0$ or $1$: the products $\ell_n = \sprod{n}{1}$ have one output, the vertices $\vertex{n} = \sprod{n}{0}$ have none, while the surface states $\bra{\mc V_n}$ have no input.
An object that changes from one order in $\gs$ to the next carries the highest power of $\gs$ kept in the action as a parenthesized superscript.
A field living in $\Hn{n}$ carries the index $n$.

The construction of this section rests on reading a vertex with some of its slots as outputs.
The simplest instance is the quartic vertex read as a bilinear form on $2$-string states.
We define the product $\sprod{2}{2} \colon \Hn{2} \to \Hn{2}$ by
\begin{equation}
  \label{eq:hs1:p22}
  \vertex{4}(A_1, A_2, A_3, A_4)
    =: \pair[2]{A_1 \otimes A_2}{\sprod{2}{2}(A_3, A_4)},
  \qquad
  \pair[2]{\cdot}{\cdot}
    := \pair{\cdot}{\cdot}^{\otimes 2},
\end{equation}
where the $2$-string pairing has one $c_0^-$ insertion per strand.
The product $\sprod{2}{2}$ raises the ghost number by two, $\Ngh(\sprod{2}{2}) = 2$.
Since $\vertex{4}$ is symmetric, the product $\sprod{2}{2}$ is symmetric with respect to the $2$-string pairing on even arguments:
\begin{equation}
  \label{eq:hs1:p22-symmetric}
  \pair[2]{A}{\sprod{2}{2}(B)}
    = \pair[2]{\sprod{2}{2}(A)}{B}.
\end{equation}
In the same way, $\ell_3 = \sprod{3}{1}$ is the quartic vertex read with one output.
The general case $\sprod{m}{n}$, its ghost number $3n - 2m$ for $n \ge 1$ (a vertex $\sprod{m}{0}$ keeps the ghost number $6 - 2m$ of~\eqref{eq:hs1:ghost-vertices}), and the transposition identity relating $\sprod{m}{n}$ to $\sprod{n}{m}$ are given in \cref{app:conventions}.

We denote by $S_n$ the term of order $\gs^{n-2}$ in~\eqref{eq:hs1:action}, $S_2$ being the kinetic term, and by
\begin{equation}
  \label{eq:hs1:truncation}
  S
    = \sum_{n \ge 2} S_n,
  \qquad
  S_{\le N}
    := \sum_{n = 2}^N S_n,
\end{equation}
the truncation of the action at order $\gs^{N-2}$.
The construction proceeds truncation by truncation: at each order we look for an action, cubic in the fields, which reproduces $S_{\le N}$ once the auxiliary fields are eliminated.

We write the variation of the action through the pairing, $\delta S = \pair{\delta \Psi}{\mathcal E_\Psi}$, so that the equation of motion of the original theory is $\mathcal E_\Psi = 0$, with
\begin{equation}
  \label{eq:hs1:eom-psi}
  \mathcal E_\Psi
    := Q_B \Psi
      + \sum_{n \ge 2} \frac{\gs^{n-1}}{n!} \, \ell_n(\Psi^n),
\end{equation}
the term of order $\gs^{n-1}$ being the variation of $S_{n+1}$.

The action~\eqref{eq:hs1:action} is invariant, order by order in $\gs$, under the gauge transformation~\cite{Zwiebach:1993:ClosedStringField}
\begin{equation}
  \label{eq:hs1:gauge-psi}
  \delta_\Lambda \Psi
    = Q_B \Lambda
      + \sum_{n \ge 2} \frac{\gs^{n-1}}{(n-1)!} \, \ell_n(\Psi^{n-1}, \Lambda),
\end{equation}
with $\Lambda \in \Hs$ of ghost number one.
This invariance is the content of the $\Loo$ relations satisfied by the products $\ell_n$, which we take as given with the vertices.

The string field satisfies the reality condition
\begin{equation}
  \label{eq:hs1:reality-psi}
  \Psi^\star
    = - \Psi,
\end{equation}
where $\star$ is the star conjugation, the composition of the BPZ and Hermitian conjugations, $A^\star := \adj{(\transp{A})}$, as recalled in \cref{app:conventions}.

\subsection{Quartic order, \texorpdfstring{$O(\gs^2)$}{O(gs**2)}}
\label{sec:hs:quartic}

The quartic order is the first order at which the transformation applies, and we work it out step by step because it is the template that every higher order repeats.
The steps are an auxiliary field with its reality condition and an algebraic equation of motion, the quotient on which it lives, its elimination by a completed square, the gauge transformation it inherits from $\Psi$, its spacetime content, and the Feynman rules through which its exchange reproduces a contact vertex.

\subsubsection{Auxiliary field and equations of motion}
\label{sec:hs:quartic:field}

\paragraph{Auxiliary field.}

The quartic term of the action reads:
\begin{equation}
  S_4
    = \frac{\gs^2}{4!} \, \vertex{4}(\Psi^2, \Psi^2).
\end{equation}
It is the square of the $2$-string state $\Psi^2$ with respect to the bilinear form $\vertex{4}$, in the same way as $\lambda \phi^4$ is the square of $\phi^2$.
A Hubbard--Stratonovich transformation trades a square for a linear coupling to a new field, at the cost of a quadratic term for that field.
We introduce an auxiliary $2$-string field $\Sigma_2 \in \Hn{2}$ and add to the action the square of its fluctuation from $\Psi^2$:
\begin{equation}
  \label{eq:hs4:DeltaS4}
  \Delta S_4
    := - \frac{\gs^2}{4!} \, \vertex{4}(\Sigma_2 - \Psi^2, \Sigma_2 - \Psi^2).
\end{equation}
This term vanishes when $\Sigma_2 = \Psi^2$.
We put a tilde on the cubic action and on its term $\caction[n]$ of order $\gs^{n-2}$.
Expanding the square and using the symmetry of $\vertex{4}$ for the cross term, we find the quartic term of the action:
\begin{equation}
  \label{eq:hs4:S4tilde}
  \caction[4]
    := S_4 + \Delta S_4
    = - \frac{\gs^2}{4!} \, \vertex{4}(\Sigma_2^2)
      + \frac{2 \gs^2}{4!} \, \vertex{4}(\Sigma_2, \Psi^2),
\end{equation}
where $\vertex{4}(\Sigma_2^2) := \vertex{4}(\Sigma_2, \Sigma_2)$.
The term $\vertex{4}(\Psi^4)$ has disappeared: the first term is quadratic in $\Sigma_2$, and the second one is cubic in the fields, with one $\Sigma_2$ and two $\Psi$.
Together with the kinetic term and the cubic vertex of~\eqref{eq:hs1:action}, which are untouched, the action
\begin{equation}
  \label{eq:hs4:action-tilde}
  \caction[\le 4]
    := S_2 + S_3 + \caction[4]
    = S_{\le 4} + \Delta S_4
\end{equation}
is cubic in $(\Psi, \Sigma_2)$.

We could redistribute the powers of $\gs$ in~\eqref{eq:hs4:S4tilde} by absorbing $\gs$ in the auxiliary field, $\Sigma_2 \to \Sigma_2 / \gs$.
The quadratic term then carries no coupling and the mixed term a single power of $\gs$, like the cubic vertex $S_3$, which makes its cubic nature manifest.
We nevertheless keep $\gs$ explicit, so that the solution for $\Sigma_2$ is $\Psi^2$ without a coupling in front and the two actions can be compared term by term in $\gs$.

\paragraph{Reality.}

The string field satisfies the reality condition~\eqref{eq:hs1:reality-psi}, $\Psi^\star = - \Psi$.
The star conjugation acts strand by strand on tensors.
We assume (\cref{app:conventions}) that the vertices are real:
\begin{equation}
  \label{eq:hs4:vertex-real}
  \vertex{n}(A_1, \ldots, A_n)^*
    = (-1)^n \, \vertex{n}(A_1^\star, \ldots, A_n^\star)
  \qquad
  \text{for even arguments}.
\end{equation}
Together with $\Psi^\star = - \Psi$, this property makes the action~\eqref{eq:hs1:action} real.
The solution~\eqref{eq:hs4:saddle} then shows that the natural reality condition on the auxiliary field is:
\begin{equation}
  \label{eq:hs4:reality}
  \Sigma_2^\star
    = \Sigma_2.
\end{equation}
Indeed, $(\Psi \otimes \Psi)^\star = \Psi^\star \otimes \Psi^\star = \Psi \otimes \Psi$ for a string field satisfying its reality condition, and the reality of the vertices makes the cubic action real for real $\Psi$ and $\Sigma_2$.
We derive~\eqref{eq:hs4:reality} and the reality conditions of the higher auxiliary fields in \cref{app:reality}.

\paragraph{Equations of motion.}

We vary the cubic action~\eqref{eq:hs4:action-tilde} with respect to both fields and write each term through the pairing:
\begin{equation}
  \label{eq:hs4:delta-S}
  \begin{aligned}
    \delta \caction[\le 4]
      = \pair{\delta \Psi}{Q_B \Psi}
        &
        + \frac{\gs}{2} \, \pair{\delta \Psi}{\ell_2(\Psi^2)}
        + \frac{\gs^2}{3!} \, \pair{\delta \Psi}{\ell_3(\Psi, \Sigma_2)}
        \\ &
        - \frac{2 \gs^2}{4!} \, \pair[2]{\delta \Sigma_2}{\sprod{2}{2}(\Sigma_2)}
        + \frac{2 \gs^2}{4!} \, \pair[2]{\delta \Sigma_2}{\sprod{2}{2}(\Psi^2)},
  \end{aligned}
\end{equation}
where $\ell_3(\Psi, \Sigma_2)$ is the product $\ell_3$ with $\Sigma_2$ filling two of its three slots.
Since the pairing is non-degenerate, the equations of motion are:
\begin{subequations}
  \label{eq:hs4:eom}
  \begin{align}
    \label{eq:hs4:eom-psi}
    \mathcal E^{(2)}_\Psi
      &
      := Q_B \Psi
        + \frac{\gs}{2} \, \ell_2(\Psi^2)
        + \frac{\gs^2}{3!} \, \ell_3(\Psi, \Sigma_2)
      = 0,
    \\
    \label{eq:hs4:eom-sigma}
    \mathcal E^{(2)}_{\Sigma_2}
      &
      := \sprod{2}{2}(\Psi^2) - \sprod{2}{2}(\Sigma_2)
      = 0.
  \end{align}
\end{subequations}
The equation for $\Sigma_2$ contains no $Q_B$: it is algebraic, and it says that $\Sigma_2$ and $\Psi^2$ have the same image under $\sprod{2}{2}$.
We call $\sprod{2}{2}$ the bare kinetic operator of $\Sigma_2$; in the notation of \cref{sec:all-orders}, $\wprod{2}{2} = \sprod{2}{2}$.
We restrict each strand of $\Sigma_2$ to ghost number two, like the strands of $\Psi^2$: writing $\Hs_g$ for the states of ghost number $g$ and $\Hn[g]{n}$ for the $n$-string states whose strands all have ghost number $g$, $\Sigma_2 \in \Hn[2]{2}$.
An extension to all states of total ghost number four, natural in a theory with a $2$-string gauge parameter, is discussed in \cref{sec:further}.
Here and below, $\sprod{2}{2}$ denotes the block from strands of ghost number two to strands of ghost number three, the only one the variation probes, and every product $\sprod{m}{n}$ acting on the auxiliary fields is read in the same way.

This equation determines $\Sigma_2$ up to the kernel of $\sprod{2}{2}$, which is the kernel of the quartic vertex seen as a bilinear form on $2$-string states:
\begin{equation}
  \label{eq:hs4:kernel}
  N_2 \in \ker \sprod{2}{2}
  \quad \Longleftrightarrow \quad
  \forall A \in \Hn[2]{2}:
  \quad
  \vertex{4}(N_2, A)
    = 0.
\end{equation}
The two statements are equivalent since the $2$-string pairing is non-degenerate.
This kernel can contain entangled $2$-string states, and can then be larger than the space spanned by the symmetrized products $A \otimes B$ with $A$ a single-string state annihilated by $\vertex{4}(A, \cdot, \cdot, \cdot)$.
A kernel direction $N_2$ appears neither in the quadratic term $\vertex{4}(\Sigma_2^2)$ nor in the source term $\vertex{4}(\Sigma_2, \Psi^2)$ of~\eqref{eq:hs4:S4tilde}.
Shifting $\Sigma_2$ by $N_2$ leaves the action invariant, and $N_2$ itself has zero action, so such a direction drops out of every equation and of every amplitude at this order.\footnote{%
  Before the quotient below is taken, this shift is a symmetry of the cubic action, and thus a candidate gauge symmetry of the auxiliary field; the quotient removes it.
}

We therefore take the auxiliary field on the quotient:
\begin{equation}
  \label{eq:hs4:quotient}
  \Sigma_2
    \in \Hn[2]{2} / \ker \sprod{2}{2},
  \qquad
  \Ngh(\Sigma_2)
    = 4.
\end{equation}
On this quotient, $\sprod{2}{2}$ is injective by construction, and the equation of motion~\eqref{eq:hs4:eom-sigma} has the unique solution:
\begin{equation}
  \label{eq:hs4:saddle}
  \Sigma_2^{(2)*}
    := \Psi^2
    = \quad
      \begin{tikzpicture}[diagram]
        \draw (0, 0.1) -- (2, 0.1) pic {ext} node[above right] {$\Psi$};
        \draw (0, -0.1) -- (2, -0.1) pic {ext} node[below right] {$\Psi$};
      \end{tikzpicture}
\end{equation}
We represent the solution graphically by two strands, one for each $\Psi$ of the solution, a description that will be helpful at higher orders.
The value $\Psi^2$ is the leading-order value of $\Sigma_2$.
The asterisk marks the solution, not a conjugation.

Taking the auxiliary field on the quotient~\eqref{eq:hs4:quotient} is consistent with the form of the action: single-string states that never enter the quartic vertex have no quartic term and need no auxiliary field, as the $\phi^4$ model with a degenerate coupling matrix of \cref{app:scalar-hs} illustrates.
The quotient needs no inner product.
The $2$-string pairing provides none: it vanishes identically between two fields whose strands have ghost number two.
Indeed, the pairing of two strands requires ghost numbers adding up to five, whereas the block $\sprod{2}{2}$ raises each strand to ghost number three.
The $2$-string pairing therefore defines no orthogonal complement of the kernel.
Whether $\sprod{2}{2}$ has a kernel, and which properties it has, matters at higher orders, where the composite vertices must be compatible with the quotient (the kernel conditions of \cref{sec:all:kernel}); we come back to the kernel of $\sprod{2}{2}$ with the Feynman rules.

\subsubsection{Elimination and gauge transformation}
\label{sec:hs:quartic:elimination}

\paragraph{Elimination.}

The solution~\eqref{eq:hs4:saddle} shows that $\Sigma_2$ is an auxiliary field: it is fixed entirely by $\Psi$ through an algebraic equation and carries no degree of freedom of its own.
It also shows that $\Delta S_4$ vanishes on the auxiliary solution, since $\Delta S_4$ is the square of $\Sigma_2 - \Psi^2$.
Eliminating $\Sigma_2$ therefore gives back the truncated action.
Substituting the solution into~\eqref{eq:hs4:S4tilde}, we get:
\begin{equation}
  \label{eq:hs4:eliminate}
  \caction[4] \big|_{\Sigma_2 = \Psi^2}
    = \Big( - \frac{1}{4!} + \frac{2}{4!} \Big) \, \gs^2 \, \vertex{4}(\Psi^4)
    = S_4.
\end{equation}
This is the completed square~\eqref{eq:hs4:action-tilde} read backwards, with no extra term.
Since the cubic action is quadratic in $\Sigma_2$, with the quadratic form $\sprod{2}{2}$ independent of $\Psi$, the same holds in the path integral: the Gaussian integral over $\Sigma_2$ is formally exact whatever the signature of the quartic form.
It contributes only a field-independent normalization, so that, formally, all the Green's functions of $\Psi$ computed from $\caction[\le 4]$ coincide with those of $S_{\le 4}$.
From the quintic order on, the kinetic operators of the auxiliary fields depend on $\Psi$, and so do their determinants; the eliminations are exact at tree level, to which we restrict ourselves from then on.

Physically, $\Sigma_2$ is a contact field: its equation of motion is algebraic, and its bare kinetic operator $\sprod{2}{2}$ is a vertex rather than $L_0^+$.
Its inverse therefore has no pole forced at the string masses; its poles, if any, sit at the zeros of $\sprod{2}{2}$.
Two properties of a kinetic operator must be kept apart here: whether it is built from $L_0^+$, and whether it has zeros.
The auxiliary field associated with open-string stubs~\cite{Erbin:2024:OpenStringStub} has a kinetic operator that contains $L_0$ and yet has no zero, since the field lives outside the kernel of $L_0$.
Conversely, an operator that is not built from $L_0^+$ may have zeros: a null vector of $\sprod{2}{2}$ at some total momentum is a $2$-string state with no quadratic term, which would have to be interpreted as an on-shell state of the auxiliary field.
The \enquote{propagator} of $\Sigma_2$ is thus the inverse of a vertex, and the auxiliary field changes the arrangement of the interaction: the $4$-string contact term of closed string field theory is now produced by two cubic couplings joined by a $2$-string line.

\paragraph{Gauge transformation.}

Under the gauge transformation~\eqref{eq:hs1:gauge-psi}, the truncated action is invariant up to the order of the truncation, $\delta_\Lambda S_{\le 4} = O(\gs^3)$, the remainder being canceled by the quintic vertex.
We look for a transformation of $\Sigma_2$ under which the cubic action is invariant to the same order.
Since $\Sigma_2 = \Psi^2$ on the auxiliary solution, the natural candidate is the transformation of $\Psi^2$ itself:
\begin{equation}
  \label{eq:hs4:gauge-sigma}
  \delta_\Lambda \Sigma_2
    = 2 \, \delta_\Lambda \Psi \otimes \Psi.
\end{equation}
We do not symmetrize it, since it only appears inside a symmetric vertex.
The invariance follows from the completed square, by an argument that we use at every higher order.
By~\eqref{eq:hs4:action-tilde}, $\delta \caction[\le 4] = \delta S_{\le 4} + \delta \Delta S_4$ for any variation of the fields, and, for $\delta \Sigma_2$ given by~\eqref{eq:hs4:gauge-sigma} and any $\delta \Psi$,
\begin{equation}
  \label{eq:hs4:delta-DeltaS}
  \delta \Delta S_4
    = - \frac{2 \gs^2}{4!} \, \vertex{4}\big(\delta \Sigma_2 - \delta (\Psi^2), \Sigma_2 - \Psi^2\big)
    = 0,
\end{equation}
since $\delta (\Psi^2) = \delta \Psi \otimes \Psi + \Psi \otimes \delta \Psi$ equals $2 \, \delta \Psi \otimes \Psi$ inside the symmetric vertex.
Hence $\delta \caction[\le 4] = \delta S_{\le 4}$ identically, off-shell, for any variation $\delta \Psi$ and not only for a gauge transformation.
The cubic action with the transformation~\eqref{eq:hs4:gauge-sigma} is thus invariant to the same order $O(\gs^3)$ as the truncated action.
The transformation law~\eqref{eq:hs4:gauge-sigma} is the one under which the fluctuation $\sigma_2^{(2)} := \Sigma_2 - \Psi^2$ is invariant.
It is induced by that of $\Psi$, and no independent parameter enters the matching on the auxiliary solution.
We used the solution~\eqref{eq:hs4:saddle} only through the transformation it suggests.

We do not look for symmetries other than~\eqref{eq:hs4:gauge-sigma} and the transformations proportional to the equations of motion, which every action possesses.
Whether the cubic action can be seen as a gauge-fixed form of a theory with an independent $2$-string parameter $\Lambda_2 \in \Hn{2}$ is a different question, which we take up in \cref{sec:further}.

\subsubsection{Spacetime fields}
\label{sec:hs:quartic:fields}

The string field $\Psi$ expands on a basis $\ket{\phi_\alpha(k)}$ of single-string states of momentum $k$ with coefficients $\psi_\alpha(k)$, the spacetime fields in momentum space.
The auxiliary field expands in the same way on $2$-string states.
Since a general element of $\Hn{2}$ is an entangled state, its coefficient is a function of two momenta:
\begin{equation}
  \label{eq:hs4:spacetime}
  \Sigma_2
    = \sum_{\alpha_1, \alpha_2} \int \frac{\dd^d k_1}{(2\pi)^d} \, \frac{\dd^d k_2}{(2\pi)^d} \,
      \sigma_{\alpha_1 \alpha_2}(k_1, k_2) \,
      \ket{\phi_{\alpha_1}(k_1)} \otimes \ket{\phi_{\alpha_2}(k_2)},
\end{equation}
where $d$ is the spacetime dimension.
The coefficient is symmetric under the joint exchange of the two strands, $(\alpha_1, k_1) \leftrightarrow (\alpha_2, k_2)$.
The components $\sigma_{\alpha_1 \alpha_2}(k_1, k_2)$ are bilocal fields in momentum space, and thus also in position space, where they are functions of two spacetime points.
The $n$-string field introduced at higher orders expands in the same way, with a component $\sigma_{\alpha_1 \ldots \alpha_n}(k_1, \ldots, k_n)$ symmetric, with the Grassmann signs, under the joint exchange of a label and its momentum with another: it is an $n$-local field.

Every vertex conserves only the total momentum of its strands: its single momentum delta function comes from the zero mode of the matter CFT.
The variables adapted to this conservation are the total momentum and the $n - 1$ Jacobi relative momenta,
\begin{equation}
  \label{eq:hs4:jacobi}
  P
    := \sum_{i = 1}^n k_i,
  \qquad
  q_j
    := \sqrt{\frac{j}{j + 1}} \left( k_{j + 1} - \frac{1}{j} \sum_{i \le j} k_i \right),
  \qquad
  j
    = 1, \ldots, n - 1,
\end{equation}
normalized so that
\begin{equation}
  \prod_{i} \dd^d k_i
    = n^{- d / 2} \, \dd^d P \prod_{j} \dd^d q_j,
  \qquad
  \sum_i k_i^2
    = P^2 / n + \sum_j q_j^2.
\end{equation}
At $n = 2$ we use instead the relative momentum $q = - q_1 / \sqrt{2}$:
\begin{equation}
  \label{eq:hs4:momenta}
  P
    := k_1 + k_2,
  \qquad
  q
    := \frac{1}{2} \, (k_1 - k_2),
\end{equation}
for which $\dd^d k_1 \, \dd^d k_2 = \dd^d P \, \dd^d q$.
The quadratic term is then diagonal in $P$:
\begin{equation}
  \label{eq:hs4:diagonal}
  \vertex{4}(\Sigma_2, \Sigma_2')
    = \int \frac{\dd^d P}{(2\pi)^d}
      \int \frac{\dd^d q}{(2\pi)^d} \frac{\dd^d q'}{(2\pi)^d} \;
      \sigma(- P, q') \, V_4(P; q', q) \, \sigma'(P, q),
\end{equation}
where we suppressed the oscillator labels and where $V_4(P; q', q) := V_4\big(- \tfrac{P}{2} + q', - \tfrac{P}{2} - q', \tfrac{P}{2} + q, \tfrac{P}{2} - q\big)$ is the momentum-space quartic vertex on two pairs of states with total momenta $- P$ and $P$.
At fixed total momentum, the quadratic term is thus an integral operator in the relative momentum.
The quadratic term of the $n$-string field, built from $\vertex{2n}$ and exchange diagrams, is diagonal in its total momentum in the same way, an integral operator in the $n - 1$ relative momenta.
By contrast, a cubic coupling of three auxiliary fields is constrained by one delta function in the sum of their total momenta only.
The source term $\vertex{4}(\Sigma_2, \Psi^2)$ is of the same form as~\eqref{eq:hs4:diagonal}, with $\sigma'(P, q)$ replaced by the product $\psi_{\alpha_1}(\tfrac{P}{2} + q) \, \psi_{\alpha_2}(\tfrac{P}{2} - q)$ of two components of $\Psi$.
The component of $\Sigma_2$ at total momentum $P$ is sourced by every pair of components of $\Psi$ whose momenta add up to $P$.

A translation-invariant configuration of $\Sigma_n$, such as a translation-invariant classical solution or a vacuum expectation value, is proportional to $\delta^d(P)$ and depends on the $n - 1$ relative momenta only~\cite{Weinberg:2005:QuantumTheoryFields-2}: it lives on $n$ points modulo an overall translation.
For $n = 2$ it is described by a single momentum, like a local field, though with the non-local quadratic term discussed next.

We define the momentum-space expressions using Sen's generalized Wick rotation~\cite{Pius:2016:CutkoskyRulesSuperstring,deLacroix:2017:ClosedSuperstringField,Erbin:2021:StringFieldTheory,Bautista:2019:QuantumGravityTimelike}.
The theory is defined with Euclidean momenta, and the Lorentzian theory is obtained from it by a Wick rotation afterwards.
We keep the total momentum $P$ as a real label.
The operator $\sprod{2}{2}$ at fixed $P$ acts on the functions of the relative momentum that are square integrable for the $L^2$ product in $q$, combined with the inner product for which the Fock basis of the oscillator labels is orthonormal.
A kernel of $\sprod{2}{2}$ at fixed total momentum is meant in this sense throughout, and its elements, defined by~\eqref{eq:hs4:kernel}, do not depend on the choice of the inner product.
We now restrict each strand to oscillator level at most $L$, as in level truncation.
At level $L$, $V_4(P; q', q)$ is the integral over the compact vertex region of an exponential of a real quadratic form in $(P, q, q')$, the same as at level $0$, times a polynomial in the momenta whose degree grows with the level.
Since the region avoids the collisions of the punctures, $V_4(P; q', q)$ is an entire function of the total momentum and of both relative momenta, rather than a delta function times a polynomial, and the kinetic term of $\Sigma_2$ is not local in the relative coordinate.
The stubs of the vertices~\cite{Sonoda:1990:ClosedStringField} produce a Gaussian decay in $q$ and $q'$ that, when its rate is positive, dominates the polynomial, so that $\sprod{2}{2}$ is a Hilbert--Schmidt, hence compact, operator at every level.
For the vertex family of~\cite{Erbin:toappear:SL2VerticesClosed}, the rate is positive at the sections of the family with stub parameter $\beta > \sqrt{2}$, a condition that does not depend on the level.

At level $0$, the string field reduces to the tachyon, $\Sigma_2$ to its tachyon-pair component $\sigma(P, q)$, even in $q$ by the strand symmetry, and the integral kernel $V_4(P; q', q)$ to the quartic vertex on four tachyons.
For the same sections of the vertex family, this integral kernel is square integrable and trace class, with an absolutely convergent eigenfunction expansion.
At these sections it is also strictly definite on the even functions of $q$ at every real $P^2$, including $P^2 < 0$ after a Wick rotation.
In particular, at level $0$, $\sprod{2}{2}$ is injective at real momenta, with a dense image: it has no null vector at any real $P^2$.
We discuss the form of this kernel and the worldsheet interpretation of the $\Sigma_2$ exchange in \cref{sec:worldsheet}.

\subsubsection{Feynman rules}
\label{sec:hs:quartic:feynman}

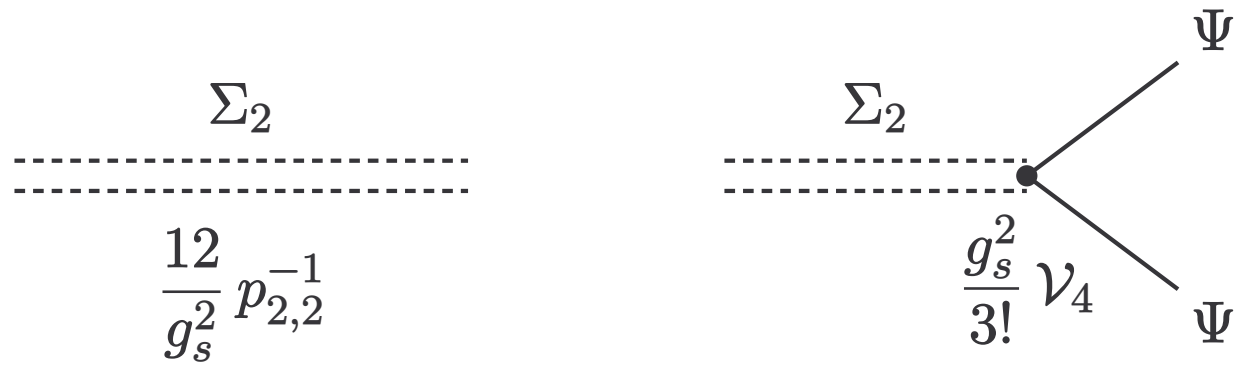
\begin{figure}[htp]
  \centering
  \begin{tikzpicture}[diagram]
    \draw[sigma line] (0, 0.1) -- (3, 0.1) (0, -0.1) -- (3, -0.1);
    \node[above=3pt] at (1.5, 0.1) {$\Sigma_2$};
    \node[below=5pt] at (1.5, -0.1) {$\dfrac{12}{\gs^2} \, \sprod{2}{2}^{-1}$};
  \end{tikzpicture}
  \qquad \qquad
  \begin{tikzpicture}[diagram]
    \draw[sigma line] (0, 0.1) -- (2, 0.1) (0, -0.1) -- (2, -0.1);
    \node[above=3pt] at (1, 0.1) {$\Sigma_2$};
    \draw (2, 0) -- (3, 0.75) node[above right] {$\Psi$};
    \draw (2, 0) -- (3, -0.75) node[below right] {$\Psi$};
    \fill (2, 0) circle (2pt);
    \node[below=5pt] at (2, 0) {$\dfrac{\gs^2}{3!} \, \vertex{4}$};
  \end{tikzpicture}
  \caption{The Feynman rules of the auxiliary sector. Left, the $\Sigma_2$ propagator, a dashed double line with one strand per string of the pair; right, the cubic vertex, a dot joining a $\Sigma_2$ line to two single-string legs $\Psi$.}
  \label{fig:hs4:feynman-rules}
\end{figure}

We read the Feynman rules off the action $\caction[\le 4]$, the path-integral weight being $\e^{\caction}$.
A vertex factor is the derivative of the action with respect to the fields on its legs, and a propagator is minus the inverse of the Hessian of the action.
A tree diagram then gives the corresponding derivative of the action with the internal fields eliminated.
The auxiliary sector of~\eqref{eq:hs4:S4tilde} contributes a propagator and a cubic vertex (\cref{fig:hs4:feynman-rules}).

\paragraph{Assumptions on the propagator.}

The propagator inverts the vertex $\sprod{2}{2}$ on its image, since $\sprod{2}{2}$ is injective on the quotient $\Hn[2]{2} / \ker \sprod{2}{2}$.
By \enquote{invertible on the quotient} we mean injective with a dense image.
At level $L$ and fixed total momentum, $\sprod{2}{2}$ is a finite block of integral operators in the relative momentum, a compact operator, so its inverse on the image is unbounded.
The propagator rests on three assumptions.
First, at each level, $\sprod{2}{2}$ acting on $\Hn[2]{2}$ has no null vector at generic total momentum, the squared total momenta at which it has one forming a discrete set; at level $0$ it has none at real $P^2$, as stated above.
Second, the sources on which the inverse acts, such as the quintic source of \cref{sec:hs:quintic}, lie in the image of $\sprod{2}{2}$ and not only in its closure.
The vanishing of the quintic vertex on the kernel~\eqref{eq:hs5:kernel-condition} gives only the closure, and the image of a compact operator of infinite rank is never closed.
Third, the quotient and the propagator have a limit as the level grows, at generic total momentum.
This is a weak form of the statement that the kernel stabilizes.
The kernels at successive levels are not nested, since the rows added at level $L + 1$ need not annihilate a null vector of level $L$ and can produce null vectors of their own; we assume only the convergence of the quotient and of the propagator.
No finite-level computation decides this statement in either form.

\paragraph{On the first assumption.}

The first assumption does not follow from compactness: zero is never an isolated point of the spectrum of an infinite-rank compact operator when the operator is self-adjoint, as ours is at real momentum.
Moreover, there are entire families of Hilbert--Schmidt operators of the same type with null vectors on a set of momenta accumulating at a finite point.
We recover the finite-dimensional intuition with a Fredholm form.\footnote{%
  A bounded operator is Fredholm when its kernel and its cokernel are finite-dimensional and its range is closed; its index is the difference of the two dimensions.
  For a family of Fredholm operators of index zero depending analytically on a parameter in a connected domain, the analytic Fredholm theorem states that either the operator is non-invertible for every value of the parameter, or it is invertible except on a discrete set of values.
}
We divide $\sprod{2}{2}$ at level $L$ on both sides by the square root of its level-$0$ block at a fixed reference momentum (of its absolute value on the functions odd in the relative momentum, on which we assume the block injective).
We assume that the result is a bounded operator with a bounded inverse at real $P^2$ at level $0$, and a Fredholm operator of index zero depending analytically on $P^2$ near the real axis at every level.
The analytic Fredholm theorem then leaves two possibilities at each level: (i) a null vector at every total momentum, or (ii) null vectors at a discrete set of values of $P^2$ only.
A proof that $\sprod{2}{2}$ has no null vector at a single total momentum would select the second case at that level.
At level $0$ the definiteness stated above does it.
Above level $0$ the form is indefinite near zero momentum (at level $1$, for the vertex family of~\cite{Erbin:toappear:SL2VerticesClosed}, \cref{sec:further}), so the definiteness cannot be used, and we could not establish the second case by other means.
For the same reason, isolated real zeros are not excluded above level $0$, and we state the first assumption at generic momentum.

In the first case the quartic vertex would be blind to a whole spacetime field of $2$-string states.
The quotient removes these states from $\Sigma_2$ and the source $\vertex{4}(\Sigma_2, \Psi^2)$ does not reach them, so the quartic order is unaffected.
The construction then requires the quintic and every higher vertex to vanish on them at every momentum (the kernel conditions of \cref{sec:all:kernel}), imposed on a whole field rather than at isolated momenta.
The composition $\sprod{2}{2}^{-1} \circ \sprod{3}{2}$ needs in addition the second assumption, in either case.
We exclude this case by assumption, which the following argument motivates.
Lengthening the stubs of the vertices adds to the vertex region a strip of surfaces on which a single-string state of every mass is exchanged, in a convergent sum over the masses.
The added length acts as a Laplace variable that separates the intermediate masses, while the stub factor itself, being injective, creates no null vector.
A null vector that persists as the stubs are lengthened must then decouple from the exchange of every mass level separately.
We assume that no $2$-string state decouples from the whole tower of single-string exchanges (at level $0$ the definiteness stated above suggests this); a kernel present at every momentum then cannot be a conservation law of the vertex family.
The argument does not exclude an accidental kernel, present at every momentum without a conservation law behind it, which the assumption excludes.

\paragraph{Collapse of the exchange.}

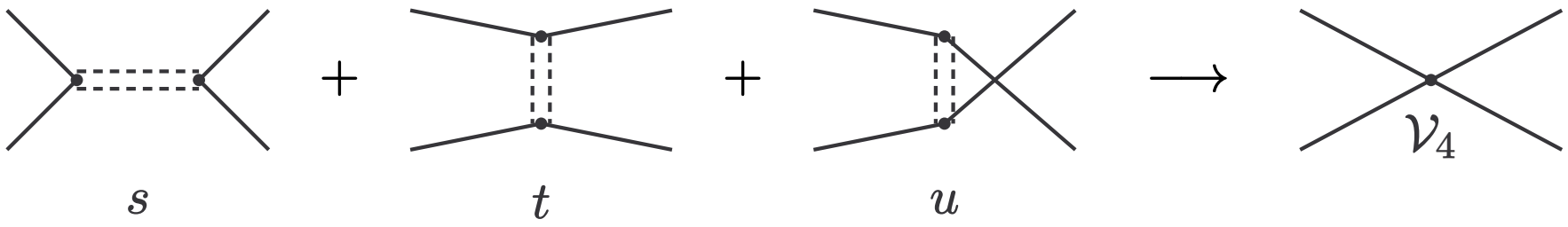
\begin{figure}[htp]
  \centering
  \begin{tikzpicture}[diagram, scale=0.65]
    \draw (0, 0.8) -- (0.8, 0) -- (0, -0.8);
    \draw (3, 0.8) -- (2.2, 0) -- (3, -0.8);
    \draw[sigma line] (0.8, 0.1) -- (2.2, 0.1) (0.8, -0.1) -- (2.2, -0.1);
    \fill (0.8, 0) circle (2pt) (2.2, 0) circle (2pt);
    \node at (1.5, -1.4) {$s$};
  \end{tikzpicture}
  \; $+$ \;
  \begin{tikzpicture}[diagram, scale=0.65]
    \draw (0, 0.8) -- (1.5, 0.5) -- (3, 0.8);
    \draw (0, -0.8) -- (1.5, -0.5) -- (3, -0.8);
    \draw[sigma line] (1.4, 0.5) -- (1.4, -0.5) (1.6, 0.5) -- (1.6, -0.5);
    \fill (1.5, 0.5) circle (2pt) (1.5, -0.5) circle (2pt);
    \node at (1.5, -1.4) {$t$};
  \end{tikzpicture}
  \; $+$ \;
  \begin{tikzpicture}[diagram, scale=0.65]
    \draw (0, 0.8) -- (1.5, 0.5) -- (3, -0.8);
    \draw (0, -0.8) -- (1.5, -0.5) -- (3, 0.8);
    \draw[sigma line] (1.4, 0.5) -- (1.4, -0.5) (1.6, 0.5) -- (1.6, -0.5);
    \fill (1.5, 0.5) circle (2pt) (1.5, -0.5) circle (2pt);
    \node at (1.5, -1.4) {$u$};
  \end{tikzpicture}
  \quad $\longrightarrow$ \quad
  \begin{tikzpicture}[diagram, scale=0.65]
    \draw (0, 0.8) -- (1.5, 0) -- (3, 0.8);
    \draw (0, -0.8) -- (1.5, 0) -- (3, -0.8);
    \fill (1.5, 0) circle (2pt);
    \node[below=5pt] at (1.5, 0) {$\vertex{4}$};
  \end{tikzpicture}
  \caption{Collapse of the $\Sigma_2$ exchange~\eqref{eq:hs4:telescoping}. The three channels $s$, $t$ and $u$, each with two cubic vertices of \cref{fig:hs4:feynman-rules} joined by a $\Sigma_2$ line (dashed double line), add up to the contact term $\gs^2 \, \vertex{4}(\Psi^4)$; the four external legs are the states $\Psi$.}
  \label{fig:hs4:collapse}
\end{figure}

We now consider the four-point function with four identical external states $\Psi$ at order $\gs^2$.
The auxiliary sector contributes three channels, $s$, $t$ and $u$, each with two cubic vertices and one $\Sigma_2$ line (\cref{fig:hs4:collapse}).
Since the external states are identical, the three channels are equal and we get:
\begin{equation}
  \label{eq:hs4:telescoping}
  \begin{aligned}
    3 \times
    \Big( \frac{\gs^2}{3!} \, \vertex{4}(\Psi^2, \cdot) \Big)
    \circ
    \Big( \frac{12}{\gs^2} \, \sprod{2}{2}^{-1} \Big)
    \circ
    \Big( \frac{\gs^2}{3!} \, \vertex{4}(\cdot, \Psi^2) \Big)
      &
      = \gs^2 \, \vertex{4}\Big( \Psi^2, \sprod{2}{2}^{-1} \circ \sprod{2}{2}(\Psi^2) \Big)
      \\ &
      = \gs^2 \, \vertex{4}(\Psi^4),
  \end{aligned}
\end{equation}
where the second vertex is written through the pairing, $\vertex{4}(\cdot, \Psi^2) = \pair[2]{\cdot}{\sprod{2}{2}(\Psi^2)}$, and where $\sprod{2}{2}^{-1} \circ \sprod{2}{2}$ is the identity on the quotient.
We call \emph{collapse} the fact that a $\Sigma_2$ line ending on a source vertex (\cref{fig:hs4:feynman-rules}) reduces to a contact term, $\sprod{2}{2}^{-1} \circ \sprod{2}{2}(\Psi^2) = \Psi^2$.
The exchange~\eqref{eq:hs4:telescoping} is the diagrammatic form of the elimination~\eqref{eq:hs4:eliminate}.

The Feynman rules of the $\Psi$ sector, the propagator $- b_0^+ / L_0^+$ in Siegel gauge and the cubic vertex $\vertex{3}$, are those of closed string field theory.
The $\Sigma_2$ exchange collapses as in~\eqref{eq:hs4:telescoping} for any external states: the two pair states filling the vertices replace $\Psi^2$.
Hence every tree-level Green's function of $\Psi$ computed from the cubic action equals its counterpart in the quartic-truncated theory, for the same vertex family and off-shell.
Besides the invertibility of $\sprod{2}{2}$ on the quotient assumed above, this needs one more assumption.
The composition of vertex, propagator and vertex is a sum over a complete set of intermediate $2$-string states, which runs over the oscillator levels of the two strands and over the relative momentum of the pair.
We assume that this sum converges on the $2$-string states that the external sources produce.

At the quintic order, the quintic vertex couples to $\Sigma_2$ and dresses its kinetic operator with $\Psi$: the field $\Sigma_2$ alone still suffices, but its solution, given by a Neumann series, and its transformation law receive corrections in $\gs$.

\subsection{Quintic order, \texorpdfstring{$O(\gs^3)$}{O(gs**3)}}
\label{sec:hs:quintic}

The quintic order is the first at which the auxiliary field is dressed: the quintic vertex couples to $\Sigma_2$ and adds to its kinetic operator a term linear in $\Psi$, while $\Sigma_2$ alone still suffices.
This dressed operator is the one new ingredient of this order, and we work out what it changes in the steps of the quartic order: the solution, which becomes a Neumann series, the interactions it induces, the gauge law, the kernel condition and the reality of the field.

\subsubsection{Quintic coupling and equations of motion}

\paragraph{Quintic coupling.}

The next term of the action is the quintic term:
\begin{equation}
  \label{eq:hs5:S5}
  S_5
    = \frac{\gs^3}{5!} \, \vertex{5}(\Psi^5),
\end{equation}
where the vertex $\vertex{5}$ has five slots.
The question is how to describe this vertex in terms of auxiliary fields.
We cannot use the same trick as for the quartic order, because it requires an exact square.

Given the form of the saddle-point solution~\eqref{eq:hs4:saddle} for $\Sigma_2$, we can replace $\Psi^4$ by $\Sigma_2^2$ to turn the quintic term into a cubic coupling.
Two auxiliary fields $\Sigma_2$ fill four of the five slots and the string field fills the last one:
\begin{equation}
  \label{eq:hs5:S5tilde}
  \caction[5]
    := \frac{\gs^3}{5!} \, \vertex{5}(\Sigma_2^2, \Psi),
  \qquad
  \caction[\le 5]
    := \caction[\le 4] + \caction[5],
\end{equation}
where $\vertex{5}(\Sigma_2^2, \Psi) := \vertex{5}(\Sigma_2, \Sigma_2, \Psi)$.
At this stage this should be treated as an ansatz: we will now check that eliminating $\Sigma_2$ gives back $S_{\le 5}$.
Note that the partition $5 = 2 + 2 + 1$ of the five strands over three fields drawn from $\{\Sigma_2, \Psi\}$ is the only one, so no other cubic term can be built from the quintic vertex with these two fields.
In principle we could use a $3$-string field to partition $5 = 3 + 1 + 1$, but it is not needed before the sextic order: we want to keep the field content minimal and avoid introducing fields from higher orders into lower orders.
Such a term also appeared when decomposing the quintic term of open SFT with stubs~\cite{Erbin:2024:OpenStringStub}.

We read the vertex with two of its slots as outputs and define the product $\sprod{3}{2} \colon \Hn{3} \to \Hn{2}$ by
\begin{equation}
  \label{eq:hs5:p32}
  \vertex{5}(A_1, A_2, A_3, A_4, A_5)
    =: \pair[2]{A_1 \otimes A_2}{\sprod{3}{2}(A_3, A_4, A_5)},
\end{equation}
as $\sprod{2}{2}$ was defined from the quartic vertex, with the $2$-string pairing on the left.
It has ghost number zero, by the general formula~\eqref{eq:conv:ghost-numbers}.

Since the cubic term $\caction[5]$ contains $\Sigma_2$, it modifies the equation of motion of $\Sigma_2$, whose solution is no longer $\Psi^2$.
The terms of $\caction[\le 5]$ which contain $\Sigma_2$ are:
\begin{equation}
  \label{eq:hs5:F}
  \caction[4] + \caction[5]
    = - \frac{\gs^2}{4!} \, \vertex{4}(\Sigma_2^2)
      + \frac{2 \gs^2}{4!} \, \vertex{4}(\Sigma_2, \Psi^2)
      + \frac{\gs^3}{5!} \, \vertex{5}(\Sigma_2^2, \Psi),
\end{equation}
a quadratic form in $\Sigma_2$ with two terms, and a source term linear in $\Sigma_2$.
The kinetic term and the cubic vertex of $\Psi$ are the same in $\caction[\le 5]$ and in $S_{\le 5}$ and drop out of every comparison.

\paragraph{Equations of motion.}

Varying the cubic action~\eqref{eq:hs5:S5tilde} and writing each term through the pairing, we get:
\begin{equation}
  \label{eq:hs5:delta-S}
\begin{aligned}
    \delta \caction[\le 5]
      &
      = \pair{\delta \Psi}{Q_B \Psi}
        + \frac{\gs}{2} \, \pair{\delta \Psi}{\ell_2(\Psi^2)}
        + \frac{\gs^2}{3!} \, \pair{\delta \Psi}{\ell_3(\Psi, \Sigma_2)}
        + \frac{\gs^3}{5!} \, \pair{\delta \Psi}{\ell_4(\Sigma_2^2)}
        \\ &
        \quad
        - \frac{2 \gs^2}{4!} \, \pair[2]{\delta \Sigma_2}{\sprod{2}{2}(\Sigma_2)}
        + \frac{2 \gs^2}{4!} \, \pair[2]{\delta \Sigma_2}{\sprod{2}{2}(\Psi^2)}
        + \frac{2 \gs^3}{5!} \, \pair[2]{\delta \Sigma_2}{\sprod{3}{2}(\Psi, \Sigma_2)}.
\end{aligned}
\end{equation}
The equations of motion are then:
\begin{subequations}
  \label{eq:hs5:eom}
  \begin{align}
    \label{eq:hs5:eom-psi}
    \mathcal E^{(3)}_\Psi
      &
      := Q_B \Psi
        + \frac{\gs}{2} \, \ell_2(\Psi^2)
        + \frac{\gs^2}{3!} \, \ell_3(\Psi, \Sigma_2)
        + \frac{\gs^3}{5!} \, \ell_4(\Sigma_2^2)
      = 0,
    \\
    \label{eq:hs5:eom-sigma}
    \mathcal E^{(3)}_{\Sigma_2}
      &
      := \sprod{2}{2}(\Psi^2)
        - \sprod{2}{2}(\Sigma_2)
        + \frac{\gs}{5} \, \sprod{3}{2}(\Psi, \Sigma_2)
      = 0.
  \end{align}
\end{subequations}
The equation for $\Sigma_2$ is still algebraic, but it no longer states that $\Sigma_2$ and $\Psi^2$ have the same image under $\sprod{2}{2}$: the quintic vertex adds a term linear in $\Sigma_2$ and in $\Psi$.
We group this term with the quadratic form, since both are linear in $\Sigma_2$, and define the operator:
\begin{equation}
  \label{eq:hs5:dressed}
  \kin{2}{\Psi}{3}
    := \sprod{2}{2} - \frac{\gs}{5} \, \sprod{3}{2}(\Psi, \cdot),
  \qquad
  \kin{2}{\Psi}{3} \colon \Hn{2} \to \Hn{2},
\end{equation}
which we call the \emph{dressed kinetic operator} of $\Sigma_2$.
The quintic vertex, with one slot filled by $\Psi$, dresses the kinetic operator $\sprod{2}{2}$, as interaction insertions dress a propagator.
The superscript records the order $\gs^n$ at which the action is truncated: $\kin{2}{\Psi}{n}$ is the operator of the action truncated at order $\gs^n$, dressed here by the quintic vertex and, from the sextic order on, by the elimination of the higher auxiliary fields, which corrects it at each further order.
The operator of the full tower, written without superscript, is defined in \cref{sec:all:squares}.
The dressed operator is symmetric with respect to the $2$-string pairing, $\pair[2]{A}{\kin{2}{\Psi}{3}(B)} = \pair[2]{\kin{2}{\Psi}{3}(A)}{B}$, since both vertices are symmetric, and it is affine in $\Psi$.
In terms of it, the quadratic part of~\eqref{eq:hs5:F} is a single form:
\begin{equation}
  \label{eq:hs5:F-dressed}
  - \frac{\gs^2}{4!} \, \vertex{4}(\Sigma_2^2)
    + \frac{\gs^3}{5!} \, \vertex{5}(\Sigma_2^2, \Psi)
    = - \frac{\gs^2}{4!} \, \pair[2]{\Sigma_2}{\kin{2}{\Psi}{3}(\Sigma_2)}.
\end{equation}
The equation of motion of $\Sigma_2$ then reads:
\begin{equation}
  \label{eq:hs5:eom-sigma-dressed}
  \mathcal E^{(3)}_{\Sigma_2}
    = \sprod{2}{2}(\Psi^2) - \kin{2}{\Psi}{3}(\Sigma_2)
    = 0.
\end{equation}
We can make an analogy with non-linear sigma models: the quadratic form of $\Sigma_2$ has a $\Psi$-dependent metric, and the field $\Psi$ enters the kinetic term of $\Sigma_2$ as a background.

\subsubsection{Saddle-point solution and induced interactions}

\paragraph{Saddle-point solution.}

On the quotient, where $\sprod{2}{2}$ is injective, the dressed operator is $\sprod{2}{2}$ composed with a perturbation of the identity:
\begin{equation}
  \label{eq:hs5:dressed-factor}
  \kin{2}{\Psi}{3}
    = \sprod{2}{2} \circ \Big( 1 - \frac{\gs}{5} \, \sprod{2}{2}^{-1} \circ \sprod{3}{2}(\Psi, \cdot) \Big),
\end{equation}
provided that $\sprod{3}{2}(\Psi, \cdot)$ descends to the quotient and takes its values in the image of $\sprod{2}{2}$.
The first property is a condition on the quintic vertex, the \emph{kernel condition}~\eqref{eq:hs5:kernel-condition} stated at the end of this subsection, and the second is the assumption on the sources made with the propagator in \cref{sec:hs:quartic}.
For small $\gs$, the inverse of the dressed operator then exists as a Neumann series:
\begin{equation}
  \label{eq:hs5:neumann}
  \big( \kin{2}{\Psi}{3} \big)^{-1}
    = \sum_{n \ge 0} \Big( \frac{\gs}{5} \, \sprod{2}{2}^{-1} \circ \sprod{3}{2}(\Psi, \cdot) \Big)^n \circ \sprod{2}{2}^{-1},
\end{equation}
which converges for $\gs$ small at fixed $\Psi$ when the composite $\sprod{2}{2}^{-1} \circ \sprod{3}{2}(\Psi, \cdot)$ is bounded on the quotient.
Order by order in $\gs$, no invertibility condition appears beyond the one on $\sprod{2}{2}$ assumed in \cref{sec:hs:quartic}.
The statements at finite $\gs$ below assume instead that $\kin{2}{\Psi}{3}$ itself is invertible on the quotient.
The equation of motion~\eqref{eq:hs5:eom-sigma-dressed} then has the unique solution:
\begin{equation}
  \label{eq:hs5:saddle}
  \Sigma_2^{(3)*}
    := \big( \kin{2}{\Psi}{3} \big)^{-1} \circ \sprod{2}{2}(\Psi^2)
    = \Psi^2 + \frac{\gs}{5} \, \big( \kin{2}{\Psi}{3} \big)^{-1} \circ \sprod{3}{2}(\Psi^3),
\end{equation}
where the second form follows from $\sprod{2}{2}(\Psi^2) = \kin{2}{\Psi}{3}(\Psi^2) + \tfrac{\gs}{5} \, \sprod{3}{2}(\Psi, \Psi^2)$, the definition~\eqref{eq:hs5:dressed} applied to $\Psi^2$.
The solution is again fixed entirely by $\Psi$ through an algebraic equation, so that $\Sigma_2$ remains an auxiliary field.
It is no longer $\Psi^2$ but $\Psi^2$ plus corrections at all orders in $\gs$: the leading-order value $\Psi^2$ is unchanged.
We could ask whether the correction of order $\gs$ spoils the matching, but this is not the case by construction.
By stationarity of the quartic-order action at $\Psi^2$, the action is quadratic in any fluctuation around it, and the quintic term, of order $\gs^3$, changes linearly in the correction of order $\gs$, so the correction to the eliminated action starts at order $\gs^4$, as we will see below.

Applying $\sprod{2}{2}^{-1}$ to~\eqref{eq:hs5:eom-sigma} instead gives the solution as a recursion:
\begin{equation}
  \label{eq:hs5:recursion}
\begin{aligned}
    \Sigma_2^{(3)*}
      &
      = \Psi^2 + \frac{\gs}{5} \, \sprod{2}{2}^{-1} \circ \sprod{3}{2}(\Psi, \Sigma_2^{(3)*})
      \\ &
      = \quad
        \begin{tikzpicture}[diagram]
          \draw (0, 0.1) -- (2, 0.1) pic {ext} node[above right] {$\Psi$};
          \draw (0, -0.1) -- (2, -0.1) pic {ext} node[below right] {$\Psi$};
        \end{tikzpicture}
        \quad
        +
        \quad
        \begin{tikzpicture}[diagram]
          \draw[sigma line] (0, 0.1) -- (4, 0.1) pic {ext};
          \draw[sigma line] (0, -0.1) -- (4, -0.1) pic {ext};
          \node[right=3pt] at (4, 0) {$\Sigma_2^{(3)*}$};
          \draw (2, 0.1) -- (2, 1.4) pic {ext} node[above=2pt] {$\Psi$};
          \node[below=3pt] at (0.9, -0.1) {$\sprod{2}{2}^{-1}$};
          \node[below=3pt] at (2, -0.1) {$\sprod{3}{2}$};
        \end{tikzpicture}
\end{aligned}
\end{equation}
Iterating it, we get:
\begin{equation}
  \label{eq:hs5:saddle-series}
\begin{aligned}
    \Sigma_2^{(3)*}
      &
      = \Psi^2
        + \frac{\gs}{5} \, \sprod{2}{2}^{-1} \circ \sprod{3}{2}(\Psi^3)
        + \frac{\gs^2}{25} \, \sprod{2}{2}^{-1} \circ \sprod{3}{2}\Big( \Psi, \sprod{2}{2}^{-1} \circ \sprod{3}{2}(\Psi^3) \Big)
        + O(\gs^3)
      \\ &
      = \quad
        \begin{tikzpicture}[diagram]
          \draw (0, 0.1) -- (2, 0.1) pic {ext} node[above right] {$\Psi$};
          \draw (0, -0.1) -- (2, -0.1) pic {ext} node[below right] {$\Psi$};
        \end{tikzpicture}
        \quad
        +
        \quad
        \begin{tikzpicture}[diagram]
          \draw[sigma line] (0, 0.1) -- (2, 0.1) pic {ext};
          \draw[sigma line] (0, -0.1) -- (2, -0.1) pic {ext};
          \draw (2, 0.1) -- (4, 0.1) pic {ext} node[above right] {$\Psi$};
          \draw (2, -0.1) -- (4, -0.1) pic {ext} node[below right] {$\Psi$};
          \draw (2, 0.1) -- (2, 1.4) pic {ext} node[above=2pt] {$\Psi$};
          \node[below=3pt] at (0.9, -0.1) {$\sprod{2}{2}^{-1}$};
          \node[below=3pt] at (2, -0.1) {$\sprod{3}{2}$};
        \end{tikzpicture}
        \quad
        +
        \quad
        \cdots
\end{aligned}
\end{equation}
which is the Neumann series~\eqref{eq:hs5:neumann} applied to $\sprod{3}{2}(\Psi^3)$ (the numerical factors are not shown in the graphical representations).
Read as Feynman diagrams, the term of order $\gs^n$ is a chain of $n$ quintic vertices joined by $\Sigma_2$ lines, each carrying one external $\Psi$.
The chain ends on the source $\Psi^2$ at one end and on the open $\Sigma_2$ slot at the other.
Up to the factor $12 / \gs^2$, the inverse of the dressed operator is the $\Sigma_2$ propagator in the background $\Psi$.

The field $\Sigma_2$ enters the cubic action through its fluctuation from the solution:
\begin{equation}
  \label{eq:hs5:deviation}
  \sigma_2^{(3)}
    := \Sigma_2 - \Sigma_2^{(3)*},
\end{equation}
which replaces the fluctuation $\sigma_2^{(2)} = \Sigma_2 - \Psi^2$ of the quartic order.
The terms~\eqref{eq:hs5:F} can be written as a completed square in the fluctuation.
By~\eqref{eq:hs5:F-dressed} and the definition~\eqref{eq:hs5:saddle} of the solution (which gives $\sprod{2}{2}(\Psi^2) = \kin{2}{\Psi}{3}(\Sigma_2^{(3)*})$), we get:
\begin{equation}
\label{eq:hs5:square}
\begin{aligned}
  \caction[4] + \caction[5]
    &
    = - \frac{\gs^2}{4!} \Big[ \pair[2]{\Sigma_2}{\kin{2}{\Psi}{3}(\Sigma_2)} - 2 \, \pair[2]{\Sigma_2}{\kin{2}{\Psi}{3}(\Sigma_2^{(3)*})} \Big]
    \\ &
    = - \frac{\gs^2}{4!} \Big[ \pair[2]{\sigma_2^{(3)}}{\kin{2}{\Psi}{3}(\sigma_2^{(3)})} - \pair[2]{\Sigma_2^{(3)*}}{\kin{2}{\Psi}{3}(\Sigma_2^{(3)*})} \Big],
\end{aligned}
\end{equation}
using the symmetry of the dressed operator.
The first term is the quadratic form of the fluctuation and vanishes on the auxiliary solution.
The second term does not depend on $\Sigma_2$ and is the value of the action on the solution.

\paragraph{Induced interactions.}

Expanding the second form of~\eqref{eq:hs5:saddle} in the constant term of~\eqref{eq:hs5:square}, and using $\kin{2}{\Psi}{3} \circ (\kin{2}{\Psi}{3})^{-1} = 1$ together with $\pair[2]{A}{\sprod{3}{2}(\Psi^3)} = \vertex{5}(A, \Psi^2, \Psi)$, we get:
\begin{equation}
  \label{eq:hs5:constant}
  \pair[2]{\Sigma_2^{(3)*}}{\kin{2}{\Psi}{3}(\Sigma_2^{(3)*})}
    = \vertex{4}(\Psi^4)
      + \frac{\gs}{5} \, \vertex{5}(\Psi^5)
      + \frac{\gs^2}{25} \, \vertex{5}\Big( \Psi^3, \big( \kin{2}{\Psi}{3} \big)^{-1} \circ \sprod{3}{2}(\Psi^3) \Big).
\end{equation}
The first two terms come from $\pair[2]{\Psi^2}{\kin{2}{\Psi}{3}(\Psi^2)} = \vertex{4}(\Psi^4) - \tfrac{\gs}{5} \vertex{5}(\Psi^5)$ and from twice the cross term $\tfrac{\gs}{5} \vertex{5}(\Psi^5)$.
Multiplied by $\gs^2 / 4!$, the first two terms are the quartic and quintic terms of $S_{\le 5}$.

Putting~\eqref{eq:hs5:square} and~\eqref{eq:hs5:constant} together, and adding back the terms of $\Psi$ alone, we get:
\begin{equation}
  \label{eq:hs5:identity}
  \caction[\le 5]
    = S_{\le 5}
      - \frac{\gs^2}{4!} \, \pair[2]{\sigma_2^{(3)}}{\kin{2}{\Psi}{3}(\sigma_2^{(3)})}
      + \frac{\gs^4}{600} \, \vertex{5}\Big( \Psi^2, \Psi, \big( \kin{2}{\Psi}{3} \big)^{-1} \circ \sprod{3}{2}(\Psi^3) \Big).
\end{equation}
This is an identity in the fields $(\Psi, \Sigma_2)$, exact in $\gs$, and the counterpart at this order of the completed square~\eqref{eq:hs4:action-tilde}.
The last term starts at order $\gs^4$, where it is of degree six in $\Psi$.
Its leading term is a six-point interaction induced by the exchange of $\Sigma_2$, which the Gaussian elimination produces in addition to the vertex for which the auxiliary field was introduced.

\begin{figure}[htp]
  \centering
  \begin{tikzpicture}[diagram]
    \foreach \y in {0.6, 0, -0.6} {
        \draw (0, \y) -- (1, 0);
        \draw (3, 0) -- (4, \y);
      }
    \draw[sigma line] (1, 0.1) -- (3, 0.1) (1, -0.1) -- (3, -0.1);
    \fill (1, 0) circle (2pt) (3, 0) circle (2pt);
    \node[below=5pt] at (1, 0) {$\vertex{5}$};
    \node[below=5pt] at (3, 0) {$\vertex{5}$};
    \node[above=3pt] at (2, 0.1) {$\sprod{2}{2}^{-1}$};
  \end{tikzpicture}
  \qquad \qquad
  \begin{tikzpicture}[diagram]
    \foreach \y in {0.6, 0, -0.6} {
        \draw (0, \y) -- (1, 0);
        \draw (5, 0) -- (6, \y);
      }
    \draw[sigma line] (1, 0.1) -- (5, 0.1) (1, -0.1) -- (5, -0.1);
    \draw (3, 0) -- (3, 0.9);
    \fill (1, 0) circle (2pt) (3, 0) circle (2pt) (5, 0) circle (2pt);
    \node[below=5pt] at (1, 0) {$\vertex{5}$};
    \node[below=5pt] at (3, 0) {$\vertex{5}$};
    \node[below=5pt] at (5, 0) {$\vertex{5}$};
    \node[above=3pt] at (2, 0.1) {$\sprod{2}{2}^{-1}$};
    \node[above=3pt] at (4, 0.1) {$\sprod{2}{2}^{-1}$};
  \end{tikzpicture}
  \caption{The interaction induced by the exchange of $\Sigma_2$, the last term of~\eqref{eq:hs5:identity}. Left, its leading term at order $\gs^4$, the composite six-point vertex~\eqref{eq:hs5:exchange-vertex}: two quintic vertices joined by a $\Sigma_2$ line (dashed double line) carrying $\sprod{2}{2}^{-1}$. Right, the next term of the Neumann series~\eqref{eq:hs5:neumann} of the dressed inverse, with one more quintic vertex on the line and one more external $\Psi$; it is of order $\gs^5$ and has seven legs. Numerical factors are not shown.}
  \label{fig:hs5:induced-sextic}
\end{figure}
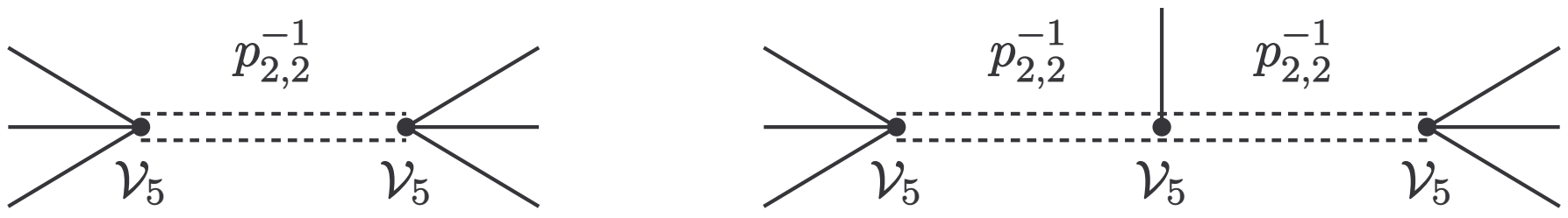

On the auxiliary solution, $\sigma_2^{(3)} = 0$ and the elimination of $\Sigma_2$ gives:
\begin{equation}
  \label{eq:hs5:eliminate}
  \caction[\le 5] \big|_{\Sigma_2 = \Sigma_2^{(3)*}}
    = S_{\le 5}
      + \frac{\gs^4}{600} \, \vertex{5 \cdot 5}(\Psi^6)
      + O(\gs^5),
\end{equation}
where we expanded the dressed inverse with~\eqref{eq:hs5:neumann} and introduced the composite vertex
\begin{equation}
  \label{eq:hs5:exchange-vertex}
\begin{aligned}
    \vertex{5 \cdot 5}(A_1, \ldots, A_6)
      &
      := \vertex{5}\Big( A_1, A_2, A_3, \sprod{2}{2}^{-1} \circ \sprod{3}{2}(A_4, A_5, A_6) \Big)
      \\ &
      = \pair[2]{\sprod{3}{2}(A_1, A_2, A_3)}{\sprod{2}{2}^{-1} \circ \sprod{3}{2}(A_4, A_5, A_6)},
\end{aligned}
\end{equation}
a composite six-point vertex obtained from two quintic vertices joined by a $\Sigma_2$ line (\cref{fig:hs5:induced-sextic}).
It is a six-linear form, graded-symmetric within each triple of arguments and under the exchange of the two triples, since $\sprod{2}{2}^{-1}$ is symmetric with respect to the $2$-string pairing as $\sprod{2}{2}$ is.
We read it also as a trilinear form on $2$-string states: $\vertex{5 \cdot 5}(B_1, B_2, B_3)$ with $B_i \in \Hn{2}$ places $B_1$ in the first two slots, $B_3$ in the fourth and fifth, and the two strands of $B_2$ in the third and sixth slots.
The result is averaged over the three choices of the split argument, so that the form is symmetric in $(B_1, B_2, B_3)$.
This symmetrized form is the one meant whenever $\vertex{5 \cdot 5}$ is evaluated on general $2$-string states, in particular in the reduced sextic vertex of \cref{sec:hs:sextic}.
At the leading-order value $\Sigma_2 = \Psi^2$, this reading gives $\vertex{5 \cdot 5}(\Psi^2, \Psi^2, \Psi^2) = \vertex{5 \cdot 5}(\Psi^6)$.

Once the quintic vertex is included, the auxiliary field thus induces interactions at all orders.
A $\Sigma_2$ line collapses to a contact term when one of its ends is a source vertex, as in~\eqref{eq:hs4:telescoping}.
The line of $\vertex{5 \cdot 5}$ joins two quintic vertices and in general does not collapse.
\Cref{sec:worldsheet} gives the interpretation of this exchange in terms of moduli space.

This also means that the approach of using $\Sigma_2$ directly in the quintic term works only because SFT contains an infinite number of higher-order terms.
Indeed, any term induced at higher orders can be repackaged with the higher-order elementary vertices, as the next two orders show and \cref{sec:all-orders} proves at all orders, under the assumptions of this order.
In normal QFT, we would not be able to use the same technique without also introducing an infinite tower of auxiliary fields.
Another way to deal with odd powers was suggested in~\cite{Rivasseau:2010:LoopVertexExpansion,Lionni:2019:IntermediateFieldRepresentation} and consists in using the identities $2 a^{n+1} = (a^n + a)^2 - a^{2n} - a^2$ or $4 a^{n+1} = (a^n + a)^2 - (a^n - a)^2$.

\paragraph{Five-point diagrams.}

The identity~\eqref{eq:hs5:identity} guarantees that every tree amplitude of the cubic action agrees with that of $S_{\le 5}$ through order $\gs^3$, under the assumptions on the exchange made in \cref{sec:hs:quartic} and the two assumptions stated at the end of this subsection.
We show how the new diagram of the five-point amplitude, the first amplitude to receive one, reduces to the quintic vertex.
The quintic coupling~\eqref{eq:hs5:S5tilde} adds to the Feynman rules of the quartic order (\cref{fig:hs4:feynman-rules}) a vertex with two $\Sigma_2$ legs and one $\Psi$ leg:
\begin{equation}
  \label{eq:hs5:feynman-rule}
  \frac{2 \gs^3}{5!} \, \vertex{5}(\Sigma_2, \Sigma_2, \Psi).
\end{equation}
\begin{figure}[htp]
  \centering
  \begin{tikzpicture}[diagram]
    \draw[sigma line] (0, 0.1) -- (2.4, 0.1) (0, -0.1) -- (2.4, -0.1);
    \draw (1.2, 0) -- (1.2, 0.8) node[above=2pt] {$\Psi$};
    \fill (1.2, 0) circle (2pt);
    \node[left=3pt] at (0, 0) {$\Sigma_2$};
    \node[right=3pt] at (2.4, 0) {$\Sigma_2$};
    \node[below=5pt] at (1.2, 0) {$\vertex{5}$};
  \end{tikzpicture}
  \qquad
  \begin{tikzpicture}[diagram]
    \foreach \y in {0.45, -0.45} {
        \draw (0, \y) -- (0.8, 0);
        \draw (2.4, 0) -- (3.2, \y);
      }
    \draw[sigma line] (0.8, 0.1) -- (2.4, 0.1) (0.8, -0.1) -- (2.4, -0.1);
    \draw (1.6, 0) -- (1.6, 0.8);
    \fill (0.8, 0) circle (2pt) (1.6, 0) circle (2pt) (2.4, 0) circle (2pt);
    \node[below=5pt] at (1.6, 0) {$\vertex{5}$};
    \node[above=3pt] at (1.2, 0.1) {$\Sigma_2$};
    \node[above=3pt] at (2, 0.1) {$\Sigma_2$};
  \end{tikzpicture}
  \quad $\longrightarrow$ \quad
  \begin{tikzpicture}[diagram]
    \foreach \y in {0.5, 0, -0.5} {
        \draw (0, \y) -- (0.8, 0);
      }
    \foreach \y in {0.5, -0.5} {
        \draw (0.8, 0) -- (1.6, \y);
      }
    \fill (0.8, 0) circle (2pt);
    \node[below=5pt] at (0.8, 0) {$\vertex{5}$};
  \end{tikzpicture}
  \caption{Left, the vertex~\eqref{eq:hs5:feynman-rule} of the quintic coupling, with two $\Sigma_2$ legs (dashed double lines) and one $\Psi$ leg (single line); right, the five-point exchange diagram, two source vertices joined by $\Sigma_2$ lines to the quintic vertex, and its collapse to the contact vertex $\vertex{5}$. Every single line is a $\Psi$ leg.}
  \label{fig:hs5:five-point}
\end{figure}
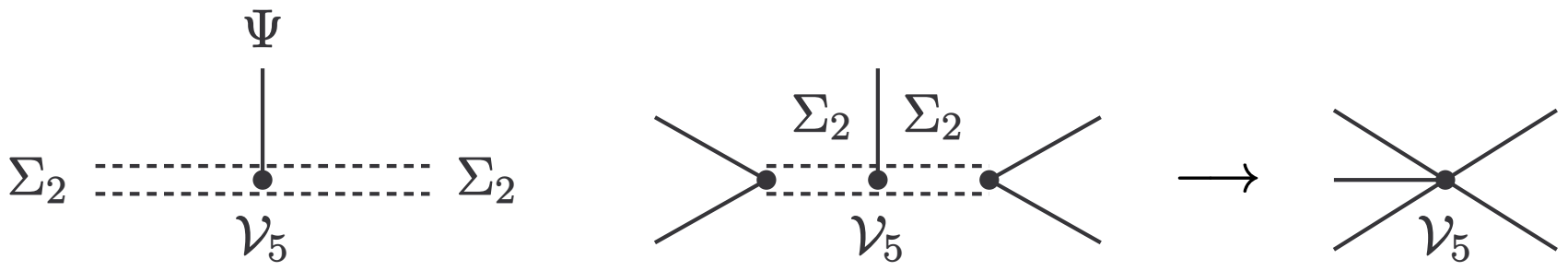

The new diagram joins two source vertices by $\Sigma_2$ lines to the quintic vertex, whose $\Psi$ leg carries the fifth external state (\cref{fig:hs5:five-point}).
Both lines collapse at once, since the propagated source is the pair state, and the diagram evaluates $\vertex{5}(\Psi^2, \Psi^2, \Psi)$, up to an element of $\ker \sprod{2}{2}$ on which the quintic vertex vanishes by the kernel condition~\eqref{eq:hs5:kernel-condition}.
With five distinct external states $A_1, \ldots, A_5$ there are $15$ such channels, five choices of the state on the $\Psi$ leg times three pairings of the other four.
Together they compensate for the factor $1 / 15$ left by the normalizations of the two source vertices, the two propagators and the quintic vertex, so that the sum is $\gs^3 \, \vertex{5}(A_1, \ldots, A_5)$, the five-point vertex of $S_{\le 5}$.
A quintic vertex joined to a cubic vertex by its $\Psi$ leg, or to a second quintic vertex by a $\Sigma_2$ line, has six external legs and is of order $\gs^4$; the second is the diagram behind the induced term of~\eqref{eq:hs5:eliminate}.

\subsubsection{Gauge transformation}

The transformation law~\eqref{eq:hs4:gauge-sigma} of $\Sigma_2$ at quartic order does not leave the action invariant at order $\gs^3$.
Since the solution~\eqref{eq:hs5:saddle} contains new terms which depend on $\Psi$, it transforms under~\eqref{eq:hs1:gauge-psi}, and the transformation of $\Sigma_2$ must be modified to compensate.
We generalize the pattern of the quartic order, where the transformation law~\eqref{eq:hs4:gauge-sigma} left the fluctuation invariant, and require the quadratic form $\pair[2]{\sigma_2^{(3)}}{\kin{2}{\Psi}{3}(\sigma_2^{(3)})}$ to be invariant.
We impose the invariance of this form because it is the only term of the identity~\eqref{eq:hs5:identity} which depends on $\Sigma_2$.
If it is invariant, the variation of the cubic action equals that of $S_{\le 5}$ up to the variation of the last term, which is of order $\gs^4$, the order of the truncation.
Its variation has two pieces, from the fluctuation and from the $\Psi$ inside the dressed operator.
By~\eqref{eq:hs5:dressed} and the symmetry of the dressed operator, we get:
\begin{equation}
  \label{eq:hs5:form-variation}
  \delta_\Lambda \pair[2]{\sigma_2^{(3)}}{\kin{2}{\Psi}{3}(\sigma_2^{(3)})}
    = 2 \, \pair[2]{\delta_\Lambda \sigma_2^{(3)}}{\kin{2}{\Psi}{3}(\sigma_2^{(3)})}
      - \frac{\gs}{5} \, \vertex{5}(\sigma_2^{(3)}, \sigma_2^{(3)}, \delta_\Lambda \Psi).
\end{equation}
This variation vanishes if the fluctuation transforms as:
\begin{equation}
  \label{eq:hs5:gauge-deviation}
  \delta_\Lambda \sigma_2^{(3)}
    = \frac{\gs}{10} \, \big( \kin{2}{\Psi}{3} \big)^{-1} \circ \sprod{3}{2}(\delta_\Lambda \Psi, \sigma_2^{(3)}).
\end{equation}
The fluctuation transforms homogeneously, by a $\Psi$-dependent linear rotation which vanishes on the auxiliary solution.

Translating back to $\Sigma_2 = \Sigma_2^{(3)*} + \sigma_2^{(3)}$, the auxiliary field transforms as its own solution, plus the rotation of the fluctuation:
\begin{equation}
  \label{eq:hs5:gauge-sigma}
  \delta_\Lambda \Sigma_2
    = \delta_\Lambda \Sigma_2^{(3)*}
      + \frac{\gs}{10} \, \big( \kin{2}{\Psi}{3} \big)^{-1} \circ \sprod{3}{2}(\delta_\Lambda \Psi, \sigma_2^{(3)}),
\end{equation}
where $\delta_\Lambda \Sigma_2^{(3)*}$ is the variation of~\eqref{eq:hs5:saddle} induced by $\delta_\Lambda \Psi$.
The transformation law of the quartic order is the order-$\gs^0$ truncation of $\delta_\Lambda \Sigma_2 = \delta_\Lambda \Sigma_2^{(3)*}$.
Through order $\gs$, the transformation law reads:
\begin{equation}
  \label{eq:hs5:gauge-sigma-g1}
\begin{aligned}
    \delta_\Lambda \Sigma_2
      = 2 \, \delta_\Lambda \Psi \otimes \Psi
        &
        + \frac{3 \gs}{5} \, \sprod{2}{2}^{-1} \circ \sprod{3}{2}(\delta_\Lambda \Psi, \Psi^2)
        \\ &
        + \frac{\gs}{10} \, \sprod{2}{2}^{-1} \circ \sprod{3}{2}(\delta_\Lambda \Psi, \Sigma_2 - \Psi^2)
        + O(\gs^2),
    \end{aligned}
\end{equation}
the middle term being the variation of $\tfrac{\gs}{5} \sprod{2}{2}^{-1} \circ \sprod{3}{2}(\Psi^3)$, cubic in $\Psi$.
The last term vanishes on the auxiliary solution but cannot be dropped off-shell: it compensates the variation of the dressed operator.

With~\eqref{eq:hs5:gauge-sigma} the quadratic form is exactly invariant.
Varying the identity~\eqref{eq:hs5:identity} gives:
\begin{equation}
  \label{eq:hs5:delta-S-equal}
  \delta_\Lambda \caction[\le 5] - \delta_\Lambda S_{\le 5}
    = \frac{\gs^4}{600} \, \delta_\Lambda \vertex{5}\Big( \Psi^2, \Psi, \big( \kin{2}{\Psi}{3} \big)^{-1} \circ \sprod{3}{2}(\Psi^3) \Big)
    = O(\gs^4),
\end{equation}
identically, off-shell, for any variation $\delta_\Lambda \Psi$ and exactly in $\gs$.
Since $\delta_\Lambda S_{\le 5} = O(\gs^4)$, the cubic action is invariant up to terms of order $\gs^4$, the order of the truncation it should match.
The nature of the rotation term, the commutator of two gauge transformations and the Batalin--Vilkovisky structure of the cubic action are summarized in \cref{sec:further}.

\subsubsection{Kernel condition and reality}

\paragraph{Kernel condition.}

The auxiliary field lives on the quotient~\eqref{eq:hs4:quotient}, and the quintic coupling must be a function on it.
The two quartic terms of~\eqref{eq:hs5:F} are invariant under $\Sigma_2 \to \Sigma_2 + N_2$ for $N_2 \in \ker \sprod{2}{2}$, by~\eqref{eq:hs4:kernel}, while the quintic term shifts by
\begin{equation}
  \vertex{5}((\Sigma_2 + N_2)^2, \Psi) - \vertex{5}(\Sigma_2^2, \Psi)
    = 2 \, \vertex{5}(N_2, \Sigma_2, \Psi) + \vertex{5}(N_2^2, \Psi).
\end{equation}
Since $\Sigma_2$ is arbitrary, the term linear in $\Sigma_2$ must vanish by itself, and the quadratic term, half its value at $\Sigma_2 = N_2$, vanishes with it.
The cubic action therefore descends to the quotient if and only if the quintic vertex vanishes on the kernel:
\begin{equation}
  \label{eq:hs5:kernel-condition}
  \forall N_2 \in \ker \sprod{2}{2}, \;
  \forall A \in \Hn[2]{2}, \;
  \forall B \in \Hs_2:
  \quad
  \vertex{5}(N_2, A, B)
    = 0,
\end{equation}
that is, $\sprod{3}{2}(A, B)$ pairs to zero with every kernel element.
The condition makes $\sprod{3}{2}(\Psi, \cdot)$ independent of the representative of its argument, so that it descends to the quotient, and with it the dressed operator~\eqref{eq:hs5:dressed-factor} and its inverse.
Since $\vertex{5}$ is symmetric, the condition is equivalent by polarization to $\vertex{5}(N_2, \Psi^2, \Psi) = 0$ for all $\Psi$, which is the only evaluation of the quintic vertex in the matching at order $\gs^3$.

Unless said otherwise, we make two assumptions from this order on.
First, the sources on which $\sprod{2}{2}^{-1}$ acts, here the values of $\sprod{3}{2}$, lie in the image of $\sprod{2}{2}$, as in \cref{sec:hs:quartic}.
Second, the quintic vertex vanishes on the kernel of the quartic vertex,~\eqref{eq:hs5:kernel-condition}.
The first is needed even when $\sprod{2}{2}$ has no kernel: the quotient makes $\sprod{2}{2}$ injective, and the assumption on the sources makes it invertible on them.
Whether the string vertices satisfy the kernel condition is not settled: at level $0$ the quartic form has no kernel at real momenta for the stub parameters of \cref{sec:hs:quartic}, so that the condition holds there trivially, and above level $0$ no null vector is known.
When the condition fails, the action does not descend and one must choose a complement of the kernel.
The eliminated action then differs from $S_{\le 5}$ at order $\gs^3$ by a five-point term which depends on the complement and, for a generic vertex, vanishes for none of them, and the construction breaks down.
The kernel conditions at every order, and the multiplier which may restore the matching, are discussed in \cref{sec:all:kernel}.
The sextic vertex is subject to an analogous condition at the next order.

\paragraph{Reality.}

The quintic order is the first where the solution and the gauge law carry an inverse, so that reality is trickier to prove.
The two maps from which they are built, $\sprod{2}{2}^{-1} \circ \sprod{3}{2}(\Psi, \cdot)$ and $(\kin{2}{\Psi}{3})^{-1} \circ \sprod{2}{2}$, commute with $\star$ for real $\Psi$.
Two relations characterize them: $\vertex{4}\big( A, \sprod{2}{2}^{-1} \circ \sprod{3}{2}(\Psi, C) \big) = \vertex{5}(A, \Psi, C)$, and $\pair[2]{A}{\kin{2}{\Psi}{3}(B')} = \vertex{4}(A, B)$ with $B' := (\kin{2}{\Psi}{3})^{-1} \circ \sprod{2}{2}(B)$.
The result follows by conjugating them, using the reality~\eqref{eq:hs4:vertex-real} of the quartic and quintic vertices, $\Psi^\star = - \Psi$ and the kernel condition; \cref{app:reality} gives the argument at every order.
Since $\Psi^2$ is star-even for $\Psi^\star = - \Psi$ and all coefficients are real rationals times $\gs$, the solution~\eqref{eq:hs5:saddle} satisfies $\Sigma_2^\star = \Sigma_2$ for real $\gs$.
The transformation law~\eqref{eq:hs5:gauge-sigma} maps the real locus of $(\Psi, \Sigma_2)$ to itself for a real variation $\delta_\Lambda \Psi$, which a gauge parameter with $\Lambda^\star = - \Lambda$ produces.

At the sextic order, the composite vertex $\vertex{5 \cdot 5}$ induced here is the first to be subtracted from an elementary vertex, which defines the reduced sextic vertex, and the sextic vertex itself calls for the $3$-string field $\Sigma_3$.

\subsection{Sextic order, \texorpdfstring{$O(\gs^4)$}{O(gs**4)}}
\label{sec:hs:sextic}

The sextic order adds two ingredients: the exchange term induced by $\Sigma_2$ at the quintic order must be subtracted from the sextic vertex, which defines the first reduced vertex, and the sextic vertex calls for a $3$-string field $\Sigma_3$, whose kinetic operator is built from this reduced vertex.
The completed square, the quotient, the elimination and the gauge law follow the two previous orders.

\subsubsection{Reduced vertex and \texorpdfstring{$3$}{3}-string field}
\label{sec:hs:sextic:vertex}

\paragraph{Reduced vertex.}

The target at this order is:
\begin{equation}
  S_{\le 6}
    = S_{\le 5} + \frac{\gs^4}{6!} \, \vertex{6}(\Psi^6).
\end{equation}
Eliminating $\Sigma_2$ from the cubic action $\caction[\le 5]$ of the quintic order produces, beyond $S_{\le 5}$, the six-point term $\tfrac{\gs^4}{600} \, \vertex{5 \cdot 5}(\Psi^6)$ of~\eqref{eq:hs5:eliminate}, the exchange of $\Sigma_2$ between two quintic vertices.
The sextic order must thus reproduce on the auxiliary solution the difference:
\begin{equation}
  \label{eq:hs6:remainder}
  S_{\le 6} - \caction[\le 5] \big|_{\Sigma_2 = \Sigma_2^{(3)*}}
    = \frac{\gs^4}{6!} \Big[ \vertex{6}(\Psi^6) - \frac{6}{5} \, \vertex{5 \cdot 5}(\Psi^6) \Big]
      + O(\gs^5),
\end{equation}
that is, produce the sextic vertex and cancel the induced exchange term at once.

A $\Sigma_2$ line collapses to a contact term when one of its ends is the source vertex $\sprod{2}{2}(\Psi^2)$, since $\sprod{2}{2}^{-1} \circ \sprod{2}{2}$ is the identity on the quotient; for generic vertices, in which the exchange term is not a linear combination of the elementary vertices, it collapses only then.
The line of $\vertex{5 \cdot 5}$ joins two quintic vertices, so it does not collapse, and this is why $\vertex{4}$ and $\vertex{5}$ needed no reduction while $\vertex{6}$ does.
For the same reason the elementary vertices alone cannot cancel the exchange term, which contains $\sprod{2}{2}^{-1}$. The subtraction must itself be an exchange term, produced either by the reduced vertex we now define or by a second $2$-string field whose quadratic form has the sign opposite to that of $\Sigma_2$. This keeps every vertex elementary at the price of one more field (\cref{app:wrong-sign}).
Whether the sign of this field has a consequence beyond tree level, and what its gauge transformation is, are open questions stated in \cref{app:wrong-sign}.

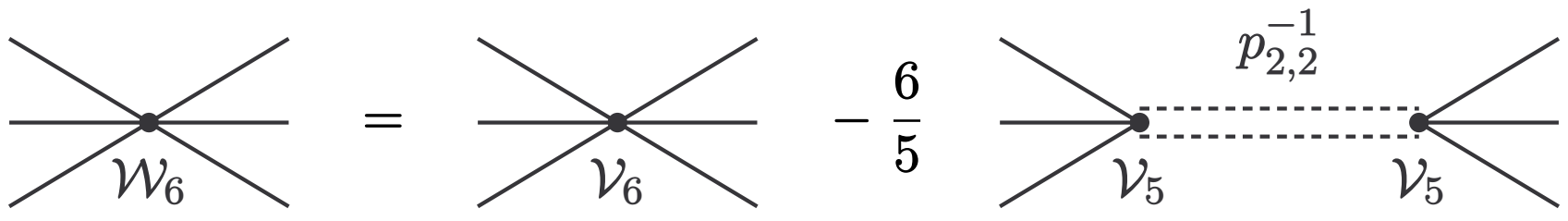
\begin{figure}[htp]
  \centering
  \begin{tikzpicture}[diagram]
    \foreach \y in {0.6, 0, -0.6} {
      \draw (0, \y) -- (1, 0) -- (2, \y);
    }
    \fill (1, 0) circle (2pt);
    \node[below=5pt] at (1, 0) {$\dvertex{6}$};
  \end{tikzpicture}
  \quad $=$ \quad
  \begin{tikzpicture}[diagram]
    \foreach \y in {0.6, 0, -0.6} {
      \draw (0, \y) -- (1, 0) -- (2, \y);
    }
    \fill (1, 0) circle (2pt);
    \node[below=5pt] at (1, 0) {$\vertex{6}$};
  \end{tikzpicture}
  \quad $- \; \dfrac{6}{5}$ \quad
  \begin{tikzpicture}[diagram]
    \foreach \y in {0.6, 0, -0.6} {
      \draw (0, \y) -- (1, 0);
      \draw (3, 0) -- (4, \y);
    }
    \draw[sigma line] (1, 0.1) -- (3, 0.1) (1, -0.1) -- (3, -0.1);
    \fill (1, 0) circle (2pt) (3, 0) circle (2pt);
    \node[below=5pt] at (1, 0) {$\vertex{5}$};
    \node[below=5pt] at (3, 0) {$\vertex{5}$};
    \node[above=3pt] at (2, 0.1) {$\sprod{2}{2}^{-1}$};
  \end{tikzpicture}
  \caption{The reduced sextic vertex~\eqref{eq:hs6:dressed-vertex}: the sextic vertex minus $\tfrac{6}{5}$ times the composite vertex, the exchange of $\Sigma_2$ between two quintic vertices. The six legs are the arguments and the dashed double line is the $\Sigma_2$ line, which carries $\sprod{2}{2}^{-1}$.}
  \label{fig:hs6:dressed-vertex}
\end{figure}

We define the \emph{reduced vertex} $\dvertex{6}$ as the sextic vertex minus $\tfrac{6}{5}$ times the composite vertex~\eqref{eq:hs5:exchange-vertex}; at the two lower orders nothing is subtracted, $\dvertex{4} = \vertex{4}$ and $\dvertex{5} = \vertex{5}$:
\begin{equation}
  \label{eq:hs6:dressed-vertex}
  \dvertex{6}(A_1, \ldots, A_6)
    := \vertex{6}(A_1, \ldots, A_6) - \frac{6}{5} \, \vertex{5 \cdot 5}(A_1, \ldots, A_6)
\end{equation}
In \cref{fig:hs6:dressed-vertex}, the six legs are the arguments and the dashed double line is the $\Sigma_2$ line, which carries $\sprod{2}{2}^{-1}$.
The reduced vertex is a six-linear form, graded-symmetric within each triple of arguments and under the exchange of the two triples, which is the symmetry of its exchange part, but not under the permutations mixing the two triples.
Read as a form in three $2$-string arguments, $\dvertex{6}(B_1, B_2, B_3)$ with $B_i \in \Hn{2}$, it is symmetrized in them, as $\vertex{5 \cdot 5}$ is in \cref{sec:hs:quintic}.
On arguments built from $\Psi$ alone, all readings agree, $\dvertex{6}(\Psi^3, \Psi^3) = \dvertex{6}(\Psi^2, \Psi^2, \Psi^2) = \dvertex{6}(\Psi^6)$, and the term to add on the auxiliary solution is the sextic term with $\dvertex{6}(\Psi^6)$ in place of $\vertex{6}(\Psi^6)$.

\paragraph{The $3$-string field.}

The matching constrains the sextic term only through its leading-order value, a sextic polynomial in $\Psi$, so we have several options to write it with auxiliary fields. Any form in $(\Psi, \Sigma_2)$ with the value $\dvertex{6}(\Psi^6)$ on $\Sigma_2 = \Psi^2$ reproduces the target at this order, and two forms differing by one that vanishes there are equivalent at this truncation.
We keep the one that extends to all orders as a completed square.
The trick of the quartic order does not apply directly with a $3$-string field $\Sigma_3 \in \Hn{3}$, because the source term $\vertex{6}(\Sigma_3, \Psi^3)$ would be quartic.
Writing $\vertex{6}(\Sigma_2^3)$ with $\Sigma_2$ alone does not generalize either, since $\vertex{8}(\Sigma_2^4)$ is quartic.\footnotemark{}
\footnotetext{%
  Such a term appeared in~\cite{Erbin:2024:OpenStringStub}, where it was sufficient, together with the terms linear and quadratic in the auxiliary field, to reproduce the complete action with a single auxiliary field.
  This is specific to the structure of the interactions induced by the stubs.
}%
The way out is to replace $\Psi^2$ by $\Sigma_2$ inside the tensor power, since $\Sigma_2 = \Psi^2 + O(\gs)$ guarantees that only $\dvertex{6}(\Psi^6)$ is produced at this order; the composite vertex, written $\vertex{5 \cdot 5}(\Sigma_2^2, \Psi^2)$, suggests this form.

We write the sextic term as $\dvertex{6}(\Psi \otimes \Sigma_2, \Psi \otimes \Sigma_2)$, with the $3$-string state $\Psi \otimes \Sigma_2$ in each triple of slots, and add the square of the fluctuation $\Sigma_3 - \Psi \otimes \Sigma_2$:
\begin{equation}
  \label{eq:hs6:DeltaS6}
  \begin{gathered}
    \Delta S_6
      := - \frac{\gs^4}{6!} \, \dvertex{6}(\Sigma_3 - \Psi \otimes \Sigma_2, \Sigma_3 - \Psi \otimes \Sigma_2),
    \\
    \Sigma_3 \in \Hn{3},
    \quad
    \Ngh(\Sigma_3)
      = 6.
  \end{gathered}
\end{equation}
We restrict each strand of $\Sigma_3$ to ghost number two, like the strands of $\Psi^3$: $\Sigma_3 \in \Hn[2]{3}$.
Here and below $\Psi \otimes \Sigma_2$ stands for its graded symmetrization.
The term $\Delta S_6$ vanishes when $\Sigma_3 = \Psi \otimes \Sigma_2$.
Expanding the square gives the sextic term of the cubic action:
\begin{equation}
  \label{eq:hs6:S6tilde}
  \begin{aligned}
    \caction[6]
      &
      := \frac{\gs^4}{6!} \, \dvertex{6}(\Psi \otimes \Sigma_2, \Psi \otimes \Sigma_2) + \Delta S_6
      \\ &
      = - \frac{\gs^4}{6!} \, \dvertex{6}(\Sigma_3^2)
        + \frac{2 \gs^4}{6!} \, \dvertex{6}(\Psi \otimes \Sigma_2, \Sigma_3),
  \end{aligned}
\end{equation}
a quadratic term for $\Sigma_3$ and a cubic coupling of $\Psi$, $\Sigma_2$ and $\Sigma_3$, so that the action through this order,
\begin{equation}
  \label{eq:hs6:action-tilde}
  \caction[\le 6]
    := \caction[\le 5] + \caction[6],
\end{equation}
is cubic in $(\Psi, \Sigma_2, \Sigma_3)$.
In terms of the elementary vertices of closed string field theory, we get:
\begin{equation}
  \label{eq:hs6:S6tilde-explicit}
  \begin{aligned}
    \caction[6]
      =
        &
        - \frac{\gs^4}{6!} \, \vertex{6}(\Sigma_3^2)
        + \frac{\gs^4}{600} \, \vertex{5}\Big( \Sigma_3, \sprod{2}{2}^{-1} \circ \sprod{3}{2}(\Sigma_3) \Big)
      \\ &
      + \frac{2 \gs^4}{6!} \, \vertex{6}(\Psi, \Sigma_2, \Sigma_3)
        - \frac{2 \gs^4}{600} \, \vertex{5}\Big( \Sigma_3, \sprod{2}{2}^{-1} \circ \sprod{3}{2}(\Psi, \Sigma_2) \Big).
  \end{aligned}
\end{equation}
The second and fourth terms, the kinetic term of $\Sigma_3$ and its coupling to $(\Psi, \Sigma_2)$, contain the inverse of a vertex, $\sprod{2}{2}^{-1}$, and are non-local: this is the price of canceling an exchange diagram with a contact term.

\paragraph{Kinetic operator.}

Since the sextic vertex is now read with three of its slots as outputs, we define the product $\sprod{3}{3} \colon \Hn{3} \to \Hn{3}$ by:
\begin{equation}
  \label{eq:hs6:p33}
  \vertex{6}(A_1, \ldots, A_6)
    =: \pair[3]{A_1 \otimes A_2 \otimes A_3}{\sprod{3}{3}(A_4, A_5, A_6)},
\end{equation}
with the $3$-string pairing on the left, as $\sprod{2}{2}$ and $\sprod{3}{2}$ were defined from $\vertex{4}$ and $\vertex{5}$; it has ghost number three and is symmetric with respect to the $3$-string pairing since $\vertex{6}$ is graded-symmetric.
The $3$-string field enters $\caction[\le 6]$ only through $\dvertex{6}$, which we write through the same pairing:
\begin{equation}
  \label{eq:hs6:kinetic}
  \begin{aligned}
    \dvertex{6}(A, B)
      &
      =: \pair[3]{A}{\wprod{3}{3}(B)},
    \\
    \wprod{3}{3}
      &
      := \sprod{3}{3} - \frac{6}{5} \, \sprod{2}{3} \circ \sprod{2}{2}^{-1} \circ \sprod{3}{2},
    \qquad
    \wprod{3}{3} \colon \Hn{3} \to \Hn{3},
  \end{aligned}
\end{equation}
for $A, B \in \Hn{3}$, where $\sprod{2}{3} \colon \Hn{2} \to \Hn{3}$ is the transpose of $\sprod{3}{2}$,~\eqref{eq:conv:transpose}.
The operator $\wprod{3}{3}$ is the bare kinetic operator of $\Sigma_3$.
It is symmetric with respect to the $3$-string pairing, as $\sprod{3}{3}$ and $\sprod{2}{2}^{-1}$ are, and, like $\sprod{2}{2}$, it depends neither on the fields nor on the order at which the tower is truncated.
Whereas $\sprod{2}{2}$ is an elementary vertex, $\wprod{3}{3}$ contains a diagram: its second term is two quintic vertices joined by a $\Sigma_2$ line, read with three strands in and three strands out.
We can make an analogy with linear algebra: $\wprod{3}{3}$ has the algebraic form of a Schur complement,\footnote{%
  For a block matrix with blocks $A$, $B$, $B^T$ and $D$, with $D$ invertible, the Schur complement of $D$ is $A - B D^{-1} B^T$.
  Here $A = \sprod{3}{3}$, $B = \sprod{2}{3}$ and $D = \tfrac{5}{6} \, \sprod{2}{2}$.
  Eliminating $\Sigma_2$ itself adds the exchange term instead~\eqref{eq:hs5:eliminate}.
} the part of the sextic interaction left for $\Sigma_3$ to describe once $\Sigma_2$ has produced its share.
The completed square~\eqref{eq:hs6:DeltaS6} is built on this form.

\subsubsection{Elimination of the auxiliary fields}
\label{sec:hs:sextic:eom}

Varying~\eqref{eq:hs6:S6tilde} with respect to the three fields, we get:
\begin{equation}
  \label{eq:hs6:delta-S}
  \begin{aligned}
    \delta \caction[6]
      &
      = \frac{2 \gs^4}{6!} \, \dvertex{6}(\delta \Psi \otimes \Sigma_2, \Sigma_3)
        + \frac{2 \gs^4}{6!} \, \dvertex{6}(\Psi \otimes \delta \Sigma_2, \Sigma_3)
      \\ &
      \quad
      - \frac{2 \gs^4}{6!} \, \pair[3]{\delta \Sigma_3}{\wprod{3}{3}(\Sigma_3)}
        + \frac{2 \gs^4}{6!} \, \pair[3]{\delta \Sigma_3}{\wprod{3}{3}(\Psi \otimes \Sigma_2)}.
  \end{aligned}
\end{equation}
The equations of motion of $\caction[\le 6]$ are then:
\begin{subequations}
  \label{eq:hs6:eom}
  \begin{align}
    \label{eq:hs6:eom-psi}
    \mathcal E^{(4)}_\Psi
      &
      := \mathcal E^{(3)}_\Psi
        + \frac{2 \gs^4}{6!} \, \dvertex{6}(\, \cdot \, \otimes \Sigma_2, \Sigma_3)
      = 0,
    \\
    \label{eq:hs6:eom-sigma2}
    \mathcal E^{(4)}_{\Sigma_2}
      &
      := \mathcal E^{(3)}_{\Sigma_2}
        + \frac{\gs^2}{30} \, \dvertex{6}(\Psi \otimes \cdot \,, \Sigma_3)
      = 0,
    \\
    \label{eq:hs6:eom-sigma3}
    \mathcal E^{(4)}_{\Sigma_3}
      &
      := \wprod{3}{3}(\Psi \otimes \Sigma_2) - \wprod{3}{3}(\Sigma_3)
      = 0,
  \end{align}
\end{subequations}
where $\mathcal E^{(3)}_\Psi$ and $\mathcal E^{(3)}_{\Sigma_2}$ are the equations~\eqref{eq:hs5:eom} of the quintic order, and where $\dvertex{6}(\, \cdot \, \otimes \Sigma_2, \Sigma_3)$ denotes the state paired with the open slot.
Inserting~\eqref{eq:hs5:eom}, the equations of $\Psi$ and $\Sigma_2$ read in full:
\begin{subequations}
  \begin{align}
    \mathcal E^{(4)}_\Psi
      &
      = Q_B \Psi
        + \frac{\gs}{2} \, \ell_2(\Psi^2)
        + \frac{\gs^2}{3!} \, \ell_3(\Psi, \Sigma_2)
      \nonumber
      \\ &
      \quad
      + \frac{\gs^3}{5!} \, \ell_4(\Sigma_2^2)
        + \frac{2 \gs^4}{6!} \, \dvertex{6}(\, \cdot \, \otimes \Sigma_2, \Sigma_3)
      = 0,
    \\
    \mathcal E^{(4)}_{\Sigma_2}
      &
      = \sprod{2}{2}(\Psi^2)
        - \sprod{2}{2}(\Sigma_2)
        + \frac{\gs}{5} \, \sprod{3}{2}(\Psi, \Sigma_2)
        + \frac{\gs^2}{30} \, \dvertex{6}(\Psi \otimes \cdot \,, \Sigma_3)
      = 0.
  \end{align}
\end{subequations}
The composite vertices are not completely symmetric, so a symbol $\wprod{m}{n}$ with $m \neq n$ would need a convention on which slots the inputs fill; since these expressions are rarely needed, we write them out.

The equation for $\Sigma_3$ is algebraic and linear, with the source $\wprod{3}{3}(\Psi \otimes \Sigma_2)$ and the field-independent operator $\wprod{3}{3}$.
As for the kernel of $\sprod{2}{2}$ at the quartic order, a kernel direction $N_3 \in \ker \wprod{3}{3}$ appears neither in the quadratic term nor in the coupling of~\eqref{eq:hs6:S6tilde}, by the symmetry of $\wprod{3}{3}$: it has zero action and couples to nothing at this order.
We therefore take the $3$-string field on the quotient:
\begin{equation}
  \label{eq:hs6:quotient}
  \Sigma_3
    \in \Hn[2]{3} / \ker \wprod{3}{3},
\end{equation}
on which $\wprod{3}{3}$ is injective by construction, and the equation of motion~\eqref{eq:hs6:eom-sigma3} has the unique solution:
\begin{equation}
  \label{eq:hs6:saddle-sigma3}
  \Sigma_3^{(4)*}
    = \Psi \otimes \Sigma_2,
\end{equation}
for any $\Sigma_2$, and $\Psi \otimes \Sigma_2^{(4)*}$ on the joint solution.
The $3$-string field is thus an auxiliary field, fixed by $(\Psi, \Sigma_2)$ through an algebraic equation, with the leading-order value $\Psi^3$ once $\Sigma_2$ is at its saddle.
It satisfies the reality condition $\Sigma_3^\star = - \Sigma_3$, the sign of three star-odd strands; the kernel of $\wprod{3}{3}$ is stable under $\star$, so that the condition is meaningful on the quotient, and the solution is real for real $\Psi$ and $\Sigma_2$ (\cref{app:reality}).

Since $\Delta S_6$ is the square of $\Sigma_3 - \Psi \otimes \Sigma_2$, it vanishes on the solution, and eliminating $\Sigma_3$ gives back the sextic term we started from, exactly in $\gs$ and for any $\Sigma_2$, as in~\eqref{eq:hs4:eliminate}:
\begin{equation}
  \label{eq:hs6:eliminate-sigma3}
  \caction[\le 6] \big|_{\Sigma_3 = \Sigma_3^{(4)*}}
    = \caction[\le 5] + \frac{\gs^4}{6!} \, \dvertex{6}(\Psi \otimes \Sigma_2, \Psi \otimes \Sigma_2).
\end{equation}
This action is quadratic in $\Sigma_2$, and its quadratic part defines the dressed kinetic operator of $\Sigma_2$ at this order:
\begin{equation}
  \label{eq:hs6:dressed-2}
  \pair[2]{A}{\kin{2}{\Psi}{4}(B)}
    := \pair[2]{A}{\kin{2}{\Psi}{3}(B)}
      - \frac{\gs^2}{30} \, \dvertex{6}(\Psi \otimes A, \Psi \otimes B),
\end{equation}
for $A, B \in \Hn{2}$, so that the terms of~\eqref{eq:hs6:eliminate-sigma3} quadratic in $\Sigma_2$ take the form~\eqref{eq:hs5:F-dressed} with $\kin{2}{\Psi}{4}$ in place of $\kin{2}{\Psi}{3}$.
The correction is the exchange of $\Sigma_3$ between two of the cubic couplings $\dvertex{6}(\Psi \otimes \Sigma_2, \Sigma_3)$, collapsed to a contact term by the elimination of $\Sigma_3$.
The operator is symmetric with respect to the $2$-string pairing and depends on $\Psi$ through both terms, and the solution and the fluctuation are defined as at the quintic order:
\begin{equation}
  \label{eq:hs6:saddle-sigma2}
  \Sigma_2^{(4)*}
    := \big( \kin{2}{\Psi}{4} \big)^{-1} \circ \sprod{2}{2}(\Psi^2),
  \qquad
  \sigma_2^{(4)}
    := \Sigma_2 - \Sigma_2^{(4)*},
\end{equation}
the inverse being a Neumann series in $\gs$ around $\sprod{2}{2}^{-1}$ with one more kind of insertion, the term of order $\gs^2$ of~\eqref{eq:hs6:dressed-2}.

The elimination of $\Sigma_2$ then proceeds as at the quintic order.
The sextic term is quadratic in $\Sigma_2$ and of order $\gs^4$, so it adds a correction of order $\gs^2$ to the dressed operator and none to the source, which remains $\sprod{2}{2}(\Psi^2)$, and $\Sigma_2^{(4)*} = \Sigma_2^{(3)*} + O(\gs^2)$.
By stationarity of $\caction[\le 5]$ at $\Sigma_2^{(3)*}$, this shift changes $\caction[\le 5]$ at order $\gs^2 \times (\gs^2)^2 = \gs^6$, and the sextic term at the same order, so that at order $\gs^4$ only the leading-order value $\dvertex{6}(\Psi^6)$ of the sextic term enters.
With~\eqref{eq:hs5:eliminate} and~\eqref{eq:hs6:remainder}, we get:
\begin{equation}
  \label{eq:hs6:eliminate}
  \begin{aligned}
    \caction[\le 6] \big|_{\Sigma_3 = \Sigma_3^{(4)*}, \, \Sigma_2 = \Sigma_2^{(4)*}}
      &
      = S_{\le 5}
        + \frac{\gs^4}{600} \, \vertex{5 \cdot 5}(\Psi^6)
        + \frac{\gs^4}{6!} \Big[ \vertex{6}(\Psi^6) - \frac{6}{5} \, \vertex{5 \cdot 5}(\Psi^6) \Big]
        + O(\gs^5)
      \\ &
      = S_{\le 6} + O(\gs^5).
  \end{aligned}
\end{equation}
The exchange term produced by $\Sigma_2$ at the quintic order and the exchange part of the reduced vertex cancel, and the sextic vertex is reproduced: the elimination of $\Sigma_2$ matches the target through order $\gs^4$.
The remainder of order $\gs^5$ is the next induced interaction, built from the reduced vertex and from chains of quintic vertices on the $\Sigma_2$ line, which the septic order absorbs in its turn.

\subsubsection{Kernel conditions}
\label{sec:hs:sextic:kernel}

Two conditions make the quotients of this order consistent, besides the invertibility of $\sprod{2}{2}$ on its quotient and the kernel condition~\eqref{eq:hs5:kernel-condition}, which the exchange part of $\wprod{3}{3}$ needs.
First, $\wprod{3}{3}$ must be invertible on the quotient~\eqref{eq:hs6:quotient}, so that $\Sigma_3$ has a propagator and is integrated out by a Gaussian integral.
Unlike the dressed operator of $\Sigma_2$, $\wprod{3}{3}$ carries no power of $\gs$: its two terms are of the same order.
Its invertibility is therefore a condition on the sextic vertex and the exchange diagram together, which no expansion in $\gs$ reduces to the condition on $\sprod{2}{2}$; for the string vertices it can be checked level by level.
We assume it as we did for $\sprod{2}{2}$ in \cref{sec:hs:quartic,sec:hs:quintic}: kernels in the square-integrable sense at fixed total momentum, and sources of $(\wprod{3}{3})^{-1}$ and of its dressed forms lying in the image of $\wprod{3}{3}$. We also assume that the quotient and the inverse have a limit as the level grows, at generic total momentum.

Second, the cubic action must be a function on the quotients.
For $\Sigma_3$ nothing is required at this order, since the field enters only through $\wprod{3}{3}$, whose kernel is quotiented; from the septic order on, where $\Sigma_3$ enters the septic vertex, the kernel directions must be null vectors of every vertex in which $\Sigma_3$ appears, and we state this condition there.
For $\Sigma_2$, the sextic term adds one coupling, $\dvertex{6}(\Psi \otimes \Sigma_2, \Sigma_3)$, to the quintic coupling, which descends to the quotient under the kernel condition~\eqref{eq:hs5:kernel-condition}.
Under $\Sigma_2 \to \Sigma_2 + N_2$ with $N_2 \in \ker \sprod{2}{2}$, its exchange part shifts by a term containing $\sprod{3}{2}(\Psi, N_2)$, which vanishes by the same condition, and its contact part by $\vertex{6}(\Psi, N_2, \Sigma_3)$.
The cubic action therefore descends to the quotient if and only if the sextic vertex vanishes on the kernel as well:
\begin{equation}
  \label{eq:hs6:kernel-condition}
  \forall N_2 \in \ker \sprod{2}{2}, \;
  \forall A \in \Hn[2]{3}, \;
  \forall B \in \Hs_2:
  \quad
  \vertex{6}(N_2, A, B)
    = 0.
\end{equation}
The quotient is thus consistent through this order when the quartic form has no kernel or when the quintic and sextic vertices vanish on it, in particular at generic momentum if the quartic form has no kernel there.
As for the quintic vertex, whether the string vertices satisfy the condition is not settled.
As at the quartic order, the quotient needs no inner product and no choice: the kernel directions have zero action and couple to nothing.
When the condition fails, the action does not descend and the eliminated action depends on the chosen complement of the kernel, as at the quintic order: the construction breaks down and would need refinement.
The conditions at every order and the multiplier which may restore the matching are discussed in \cref{sec:all:kernel}.

\subsubsection{Gauge transformation}
\label{sec:hs:sextic:gauge}

Under the gauge transformation~\eqref{eq:hs1:gauge-psi} the truncated action is invariant up to the order of the truncation, $\delta_\Lambda S_{\le 6} = O(\gs^5)$, which the cubic action must reproduce.
The two auxiliary fields enter $\caction[\le 6]$ through two completed squares, and we read off their transformation laws by demanding that each square be invariant, as at the quintic order.

The $3$-string field enters only through $\Delta S_6$, the square of the fluctuation:
\begin{equation}
  \label{eq:hs6:deviation-3}
  \sigma_3^{(4)}
    := \Sigma_3 - \Sigma_3^{(4)*}
    = \Sigma_3 - \Psi \otimes \Sigma_2,
  \qquad
  \Delta S_6
    = - \frac{\gs^4}{6!} \, \pair[3]{\sigma_3^{(4)}}{\wprod{3}{3}(\sigma_3^{(4)})},
\end{equation}
with the field-independent kinetic operator~\eqref{eq:hs6:kinetic}, so that its variation has no term from the operator:
\begin{equation}
  \label{eq:hs6:square-variation-3}
  \delta_\Lambda \Delta S_6
    = - \frac{2 \gs^4}{6!} \, \pair[3]{\delta_\Lambda \sigma_3^{(4)}}{\wprod{3}{3}(\sigma_3^{(4)})}.
\end{equation}
It vanishes for all $\sigma_3^{(4)}$ when $\delta_\Lambda \sigma_3^{(4)}$ lies in the kernel of $\wprod{3}{3}$, that is, when it vanishes on the quotient~\eqref{eq:hs6:quotient}.
We take the fluctuation of the $3$-string field inert, so that $\Sigma_3$ transforms as its solution:
\begin{equation}
  \label{eq:hs6:gauge-sigma3}
  \delta_\Lambda \sigma_3^{(4)}
    = 0,
  \qquad
  \delta_\Lambda \Sigma_3
    = \delta_\Lambda \Psi \otimes \Sigma_2 + \Psi \otimes \delta_\Lambda \Sigma_2.
\end{equation}
There is no rotation term: the rotation at the quintic order compensated the variation of the dressed operator, which depends on $\Psi$, and $\wprod{3}{3}$ does not.
This is the quartic mechanism one level up, with $\Psi \otimes \Sigma_2$ in place of $\Psi^2$.

Once $\Sigma_3$ is eliminated, completing the square in the action~\eqref{eq:hs6:eliminate-sigma3} as in~\eqref{eq:hs5:square} shows that it is quadratic in the fluctuation $\sigma_2^{(4)}$ with the operator $\kin{2}{\Psi}{4}$.
We demand that this quadratic form be invariant.
Its variation has a term from the fluctuation and one from the $\Psi$ inside the operator, and it vanishes, by the symmetry of the operator, for:
\begin{equation}
  \label{eq:hs6:gauge-sigma2}
  \delta_\Lambda \sigma_2^{(4)}
    = - \frac{1}{2} \, \big( \kin{2}{\Psi}{4} \big)^{-1} \circ \big( \delta_\Lambda \kin{2}{\Psi}{4} \big) (\sigma_2^{(4)}),
  \qquad
  \delta_\Lambda \Sigma_2
    = \delta_\Lambda \Sigma_2^{(4)*} + \delta_\Lambda \sigma_2^{(4)},
\end{equation}
where $\delta_\Lambda \kin{2}{\Psi}{4}$ is the variation of the operator~\eqref{eq:hs6:dressed-2} through its argument $\Psi$.
At order $\gs$ the variation of the operator is $- \tfrac{\gs}{5} \, \sprod{3}{2}(\delta_\Lambda \Psi, \cdot)$, and~\eqref{eq:hs6:gauge-sigma2} is the transformation law~\eqref{eq:hs5:gauge-deviation} of the quintic order; the term of order $\gs^2$ is the variation of the sextic correction.
The law has the structure of the quintic one: the auxiliary field transforms as its solution, plus a rotation of the fluctuation which is linear in it, carries no parameter of its own and vanishes on the auxiliary solution.
Both laws are unique up to the trivial symmetries proportional to the equations of motion (\cref{sec:further}).

With~\eqref{eq:hs6:gauge-sigma3} and~\eqref{eq:hs6:gauge-sigma2}, both squares are exactly invariant.
The off-shell identity behind~\eqref{eq:hs6:eliminate}, the analog of~\eqref{eq:hs5:identity}, reads
\begin{equation}
  \caction[\le 6]
    = \caction[\le 6] \big|_{\Sigma_3 = \Sigma_3^{(4)*}, \, \Sigma_2 = \Sigma_2^{(4)*}}
      - \frac{\gs^2}{4!} \, \pair[2]{\sigma_2^{(4)}}{\kin{2}{\Psi}{4}(\sigma_2^{(4)})}
      - \frac{\gs^4}{6!} \, \pair[3]{\sigma_3^{(4)}}{\wprod{3}{3}(\sigma_3^{(4)})},
\end{equation}
where the first term is $S_{\le 6} + O(\gs^5)$, and gives:
\begin{equation}
  \label{eq:hs6:delta-S-equal}
  \delta_\Lambda \caction[\le 6] - \delta_\Lambda S_{\le 6}
    = O(\gs^5),
\end{equation}
identically, for any $\delta_\Lambda \Psi$: the cubic action is invariant to the order of the truncation it reproduces, as at the two previous orders, and the algebra of these transformations is that of closed string field theory through this order (\cref{sec:further}).

The septic order displays the triangular form of this action, in which the auxiliary fields are eliminated from the top of the tower down by Gaussian integrals, exact at tree level, and dresses $\wprod{3}{3}$ by $\Psi$ for the first time, as the quintic vertex dressed $\sprod{2}{2}$.

\subsection{Septic order, \texorpdfstring{$O(\gs^5)$}{O(gs**5)}}
\label{sec:hs:septic}

The septic vertex $\vertex{7}$ is the first one that cannot be written with $\Sigma_2$ alone, and the septic order adds one ingredient: the septic vertex couples to the $3$-string field and dresses its kinetic operator $\wprod{3}{3}$ by $\Psi$, as the quintic vertex dressed $\sprod{2}{2}$.
We first display the form the action has acquired, which lets us eliminate the two auxiliary fields by Gaussian integrals, exact at tree level, and write the interactions they induce in closed form; the septic coupling then cancels those of order $\gs^5$.

\subsubsection{Triangular form and elimination of the auxiliary fields}
\label{sec:hs:septic:structure}

\paragraph{Triangular form.}

The terms of $\caction[\le 6]$ which contain an auxiliary field are, by~\eqref{eq:hs5:F} and~\eqref{eq:hs6:S6tilde}:
\begin{equation}
  \label{eq:hs7:action-6}
  \begin{aligned}
    \caction[\le 6] - S_{\le 3}
      &
      = - \frac{\gs^2}{4!} \, \vertex{4}(\Sigma_2^2)
        + \frac{2 \gs^2}{4!} \, \vertex{4}(\Sigma_2, \Psi^2)
        + \frac{\gs^3}{5!} \, \vertex{5}(\Sigma_2^2, \Psi)
      \\ &
      \quad
      - \frac{\gs^4}{6!} \, \dvertex{6}(\Sigma_3^2)
        + \frac{2 \gs^4}{6!} \, \dvertex{6}(\Sigma_3, \Psi, \Sigma_2),
  \end{aligned}
\end{equation}
where $S_{\le 3}$ is the kinetic term and the cubic vertex of $\Psi$, common to the cubic action and to the target.
The action is quadratic in each auxiliary field separately, in $\Sigma_3$ through the field-independent term $\dvertex{6}(\Sigma_3^2)$, and in $\Sigma_2$ through $\vertex{4}(\Sigma_2^2)$ dressed by $\vertex{5}(\Sigma_2^2, \Psi)$, the quadratic form of the dressed operator $\kin{2}{\Psi}{3}$ of~\eqref{eq:hs5:dressed}.
The source of $\Sigma_3$ is linear in $\Sigma_2$ and $\Psi$, and the source of $\Sigma_2$ involves $\Psi$ only, apart from the coupling to $\Sigma_3$.
We call this the \emph{triangular form} of the action.
It allows us to eliminate the auxiliary fields one at a time, in decreasing order. Once $\Sigma_3$ is eliminated, the action is again quadratic in $\Sigma_2$ with a source that depends on $\Psi$ only, so that each elimination is a Gaussian integral, exact at tree level. The value of the action on the solution then comes out in closed form rather than order by order.
This form is what makes the construction extend to all orders in \cref{sec:all-orders}.

\paragraph{Eliminating the $3$-string field.}

We eliminated $\Sigma_3$ in~\eqref{eq:hs6:eliminate-sigma3}: the value of the $\Sigma_3$ terms is the sextic term $\dvertex{6}(\Psi \otimes \Sigma_2, \Psi \otimes \Sigma_2)$, quadratic in $\Sigma_2$ since the source is linear in it.
The action left for $\Sigma_2$ is quadratic too, since the quintic vertex has only two $\Sigma_2$ slots, and we expand it in the fluctuation $\sigma_2^{(2)} = \Sigma_2 - \Psi^2$ around the quartic-order solution, defined in \cref{sec:hs:quartic:elimination}:
\begin{equation}
  \label{eq:hs7:sigma2-expansion}
  \caction[\le 6] \big|_{\Sigma_3 = \Sigma_3^{(4)*}}
    = S_{\le 6}
      - \frac{\gs^4}{600} \, \vertex{5 \cdot 5}(\Psi^6)
      - \frac{\gs^2}{4!} \, \pair[2]{\sigma_2^{(2)}}{\kin{2}{\Psi}{4}(\sigma_2^{(2)})}
      + \frac{\gs^3}{60} \, \pair[2]{\beta_\Psi}{\sigma_2^{(2)}}.
\end{equation}
Here the second term is the sextic deficit, $- \tfrac{6}{5} \vertex{5 \cdot 5}(\Psi^6)$ inside $\dvertex{6}(\Psi^6)$, which the exchange term of the quintic order compensates as in~\eqref{eq:hs6:eliminate}.
The operator $\kin{2}{\Psi}{4}$ is the dressed operator~\eqref{eq:hs6:dressed-2}, the quartic form dressed by the quintic vertex and by the exchange of $\Sigma_3$ between two sextic couplings.
The source of the fluctuation is the $2$-string state $\beta_\Psi \in \Hn{2}$ defined by
\begin{equation}
  \label{eq:hs7:source-2}
  \begin{gathered}
    \pair[2]{\beta_\Psi}{A_2}
      := \vertex{5}(A_2, \Psi^2, \Psi) + \frac{\gs}{6} \, \dvertex{6}(\Psi^3, \Psi \otimes A_2),
    \\
    \beta_\Psi
      = \sprod{3}{2}(\Psi^3) + O(\gs).
  \end{gathered}
\end{equation}
Completing the square, we get the value of $\caction[\le 6]$ on the joint solution as a single expression, exact in $\gs$:
\begin{equation}
  \label{eq:hs7:closed-form}
  \caction[\le 6] \big|_{\Sigma_3 = \Sigma_3^{(4)*}, \, \Sigma_2 = \Sigma_2^{(4)*}} - S_{\le 6}
    = \frac{\gs^4}{600} \Big[ \pair[2]{\beta_\Psi}{\big( \kin{2}{\Psi}{4} \big)^{-1}(\beta_\Psi)} - \vertex{5 \cdot 5}(\Psi^6) \Big],
\end{equation}
where the last term is the sextic deficit.
The right-hand side is one function of $\gs$: it collects the whole sequence of interactions induced by the two auxiliary fields, and it is as explicit as the inverse of the dressed operator, a Neumann series for small $\gs$ as in \cref{sec:hs:quintic}.
At finite $\gs$ it holds when $\wprod{3}{3}$ and $\kin{2}{\Psi}{4}$ are invertible on their quotients.

\paragraph{Induced interactions at order $\gs^5$.}

We expand~\eqref{eq:hs7:closed-form} to first order in $\gs$ beyond the sextic order, with the inverse of the dressed operator
\begin{equation}
  \big( \kin{2}{\Psi}{4} \big)^{-1}
    = \sprod{2}{2}^{-1}
      + \frac{\gs}{5} \, \sprod{2}{2}^{-1} \circ \sprod{3}{2}(\Psi, \cdot) \circ \sprod{2}{2}^{-1}
      + O(\gs^2),
\end{equation}
in which the $\Sigma_3$ exchange enters only at order $\gs^6$.
We get:
\begin{equation}
  \label{eq:hs7:induced}
  \caction[\le 6] \big|_{\Sigma_3 = \Sigma_3^{(4)*}, \, \Sigma_2 = \Sigma_2^{(4)*}}
    = S_{\le 6}
      + \frac{\gs^5}{600} \Big[ \frac{1}{3} \, \vertex{6 \cdot 5}(\Psi^7) - \frac{1}{5} \, \vertex{5 \cdot 5 \cdot 5}(\Psi^7) \Big]
      + O(\gs^6),
\end{equation}
with the two seven-point composite vertices
\begin{subequations}
  \label{eq:hs7:structures}
  \begin{align}
    \label{eq:hs7:S1}
    \vertex{5 \cdot 5 \cdot 5}(\Psi^7)
      &
      := \vertex{5}\Big( \sprod{2}{2}^{-1} \circ \sprod{3}{2}(\Psi^3), \, \sprod{2}{2}^{-1} \circ \sprod{3}{2}(\Psi^3), \, \Psi \Big),
    \\
    \label{eq:hs7:S2}
    \vertex{6 \cdot 5}(\Psi^7)
      &
      := \vertex{6}\Big( \Psi^4, \, \sprod{2}{2}^{-1} \circ \sprod{3}{2}(\Psi^3) \Big),
  \end{align}
\end{subequations}
named as $\vertex{5 \cdot 5}$ was in~\eqref{eq:hs5:exchange-vertex}; only their values on $\Psi^7$ enter at this order.
Both are exchange diagrams of $\Sigma_2$ between interaction vertices, and neither collapses, by the criterion of \cref{sec:hs:sextic:vertex}.
The first is a chain of three quintic vertices joined by two $\Sigma_2$ lines, the two end vertices carrying three external $\Psi$ each and the middle one a single $\Psi$: it is the next term of the Neumann series~\eqref{eq:hs5:saddle-series} inserted in the exchange term of~\eqref{eq:hs5:identity}, which gives $+ \tfrac{\gs^5}{3000} \vertex{5 \cdot 5 \cdot 5}$.
The second is a sextic vertex with four external $\Psi$ joined by a $\Sigma_2$ line to a quintic vertex with three.
It comes from the sextic term $\dvertex{6}(\Psi \otimes \Sigma_2, \Psi \otimes \Sigma_2)$ evaluated on the quintic-order solution, one $\Sigma_2$ slot receiving the correction $\tfrac{\gs}{5} X_2$ of~\eqref{eq:hs5:saddle-series}, with $X_2 := \sprod{2}{2}^{-1} \circ \sprod{3}{2}(\Psi^3)$; the cross term is
\begin{equation}
  \frac{2 \gs^5}{5 \cdot 6!} \, \dvertex{6}(\Psi^3, \Psi \otimes X_2)
    = \frac{\gs^5}{1800} \Big[ \vertex{6 \cdot 5}(\Psi^7) - \frac{6}{5} \, \vertex{5 \cdot 5 \cdot 5}(\Psi^7) \Big],
\end{equation}
where the second term is the exchange part of the reduced sextic vertex, which gives the chain once more, and the two contributions add up to~\eqref{eq:hs7:induced}.
No other term enters at order $\gs^5$: the correction of order $\gs^2$ to the solution for $\Sigma_2$ from the sextic term contributes at order $\gs^6$ only, by stationarity of $\caction[\le 5]$ at its own solution, as at the sextic order.
In both diagrams each triple of external legs is attached to one and the same vertex, a property we use to build the reduced vertex.

\subsubsection{Septic coupling}
\label{sec:hs:septic:dressed}

\paragraph{Reduced septic vertex.}

The septic order must reproduce $\vertex{7}$ and cancel the induced term of~\eqref{eq:hs7:induced} at the same time:
\begin{equation}
  S_{\le 7}
    = S_{\le 6} + \frac{\gs^5}{7!} \, \vertex{7}(\Psi^7).
\end{equation}
A term added at order $\gs^5$ contributes to the eliminated action at that order through its leading-order value, which must be of degree seven in $\Psi$, and it must be cubic in the fields: the partitions of seven strands over three fields drawn from $\{\Psi, \Sigma_2, \Sigma_3\}$ are $7 = 3 + 3 + 1$ and $7 = 3 + 2 + 2$.
We take the first, in which the $3$-string field plays the role $\Sigma_2$ played in the quintic coupling:
\begin{equation}
  \label{eq:hs7:S7tilde}
  \caction[7]
    := \frac{\gs^5}{7!} \, \dvertex{7}(\Sigma_3, \Sigma_3, \Psi),
  \qquad
  \caction[\le 7]
    := \caction[\le 6] + \caction[7],
\end{equation}
with a trilinear form $\dvertex{7}$ on $\Hn{3} \times \Hn{3} \times \Hs$, symmetric in its two $3$-string arguments, whose leading-order value the matching determines and which we define in~\eqref{eq:hs7:dressed-vertex} below.

Since the matching constrains the septic term through its leading-order value only, the second partition also reproduces the target at this order: a coupling $\vertex{7}(\Sigma_3, \Sigma_2, \Sigma_2)$, reduced by the same two diagrams, has the same leading-order value, and so does any combination of the two with coefficients adding up to one.
We keep the first for three reasons.
In the first partition, the three strands of each $\Sigma_3$ slot enter one constituent vertex, whereas the second splits the two strands of a $\Sigma_2$ slot between two vertices. Each exchange part of the reduced vertex is therefore one of the induced diagrams~\eqref{eq:hs7:structures} with its triples of legs read as $\Sigma_3$ slots.
The first partition also preserves the triangular form, whereas $\vertex{7}(\Sigma_3, \Sigma_2, \Sigma_2)$ makes the source of $\Sigma_3$ quadratic in $\Sigma_2$ and the action left after eliminating $\Sigma_3$ quartic in $\Sigma_2$, and loses the closed form~\eqref{eq:hs7:closed-form} at the next order.
Finally, $\Sigma_3$ comes with an even and an odd coupling, $\dvertex{6}$ and $\dvertex{7}$, as $\Sigma_2$ came with $\vertex{4}$ and $\vertex{5}$. Each auxiliary field is introduced through a completed square at the even order $\gs^{2 k - 2}$, and the odd vertex dresses its kinetic operator at the next order, the pattern of \cref{sec:all-orders}. The septic order introduces no new field.

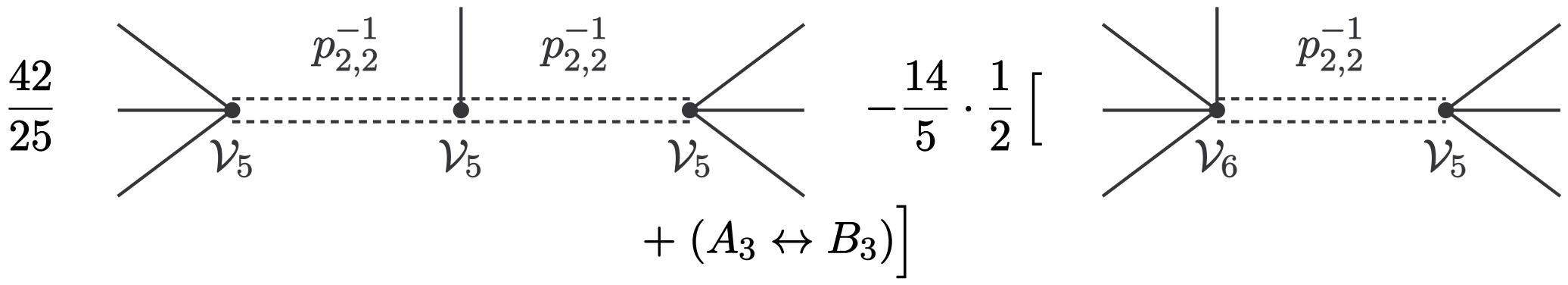
\begin{figure}[htp]
  \centering
  $\dfrac{42}{25}$ \quad
  \begin{tikzpicture}[diagram]
    \foreach \y in {0.75, 0, -0.75} {
        \draw (0, \y) -- (1, 0);
        \draw (5, 0) -- (6, \y);
      }
    \draw[sigma line] (1, 0.1) -- (5, 0.1) (1, -0.1) -- (5, -0.1);
    \draw (3, 0) -- (3, 0.9);
    \fill (1, 0) circle (2pt) (3, 0) circle (2pt) (5, 0) circle (2pt);
    \node[below=5pt] at (1, 0) {$\vertex{5}$};
    \node[below=5pt] at (3, 0) {$\vertex{5}$};
    \node[below=5pt] at (5, 0) {$\vertex{5}$};
    \node[above=3pt] at (2, 0.1) {$\sprod{2}{2}^{-1}$};
    \node[above=3pt] at (4, 0.1) {$\sprod{2}{2}^{-1}$};
  \end{tikzpicture}
  \quad $- \dfrac{14}{5} \cdot \dfrac{1}{2} \, \Big[$ \quad
        \begin{tikzpicture}[diagram]
          \foreach \y in {0.75, 0, -0.75} {
              \draw (0, \y) -- (1, 0);
              \draw (3, 0) -- (4, \y);
            }
          \draw (1, 0) -- (1, 0.9);
          \draw[sigma line] (1, 0.1) -- (3, 0.1) (1, -0.1) -- (3, -0.1);
          \fill (1, 0) circle (2pt) (3, 0) circle (2pt);
          \node[below=5pt] at (1, 0) {$\vertex{6}$};
          \node[below=5pt] at (3, 0) {$\vertex{5}$};
          \node[above=3pt] at (2, 0.1) {$\sprod{2}{2}^{-1}$};
        \end{tikzpicture}
        \quad $+ \; (A_3 \leftrightarrow B_3) \Big]$
  \caption{The two exchange terms of the reduced septic vertex~\eqref{eq:hs7:dressed-vertex}. The three legs on the left are the strands of $A_3$, the three on the right those of $B_3$, the upper leg is $C_1$, and the dashed double line is a $\Sigma_2$ line carrying $\sprod{2}{2}^{-1}$. In both terms, the three strands of a $\Sigma_3$ argument enter one and the same vertex.}
  \label{fig:hs7:dressed-vertex}
\end{figure}

We define the reduced septic vertex, for $A_3, B_3 \in \Hn{3}$ and $C_1 \in \Hs$:
\begin{equation}
  \label{eq:hs7:dressed-vertex}
  \begin{aligned}
    \dvertex{7}(A_3, B_3, C_1)
      &
      := \vertex{7}(A_3, B_3, C_1)
        + \frac{42}{25} \, \vertex{5}\Big( \sprod{2}{2}^{-1} \circ \sprod{3}{2}(A_3), \, \sprod{2}{2}^{-1} \circ \sprod{3}{2}(B_3), \, C_1 \Big)
      \\ &
      \quad
      - \frac{14}{5} \cdot \frac{1}{2} \Big[
        \vertex{6}\Big( A_3, C_1, \sprod{2}{2}^{-1} \circ \sprod{3}{2}(B_3) \Big)
        + \vertex{6}\Big( B_3, C_1, \sprod{2}{2}^{-1} \circ \sprod{3}{2}(A_3) \Big)
      \Big],
  \end{aligned}
\end{equation}
where $\sprod{3}{2}(A_3)$ fills the three input slots of $\sprod{3}{2}$ with the $3$-string state $A_3$, and where $\vertex{6}(A_3, C_1, D_2)$ fills three slots with $A_3$, one with $C_1$ and two with $D_2 \in \Hn{2}$.
The second and third terms are the two diagrams of~\eqref{eq:hs7:structures} with each triple of legs read as a $\Sigma_3$ slot, the third symmetrized (\cref{fig:hs7:dressed-vertex}).
At leading order,
\begin{equation}
  \label{eq:hs7:diagonal}
  \dvertex{7}(\Psi^3, \Psi^3, \Psi)
    = \vertex{7}(\Psi^7)
      + \frac{42}{25} \, \vertex{5 \cdot 5 \cdot 5}(\Psi^7)
      - \frac{14}{5} \, \vertex{6 \cdot 5}(\Psi^7),
\end{equation}
by~\eqref{eq:hs7:structures}.

\paragraph{Elimination and closure.}

The elimination proceeds as before, $\Sigma_3$ first.
The septic coupling is quadratic in $\Sigma_3$ and does not change its source, so the $3$-string sector stays Gaussian, with the dressed kinetic operator:
\begin{equation}
  \label{eq:hs7:dressed-3}
  \pair[3]{A_3}{\kin{3}{\Psi}{5}(B_3)}
    := \pair[3]{A_3}{\wprod{3}{3}(B_3)} - \frac{\gs}{7} \, \dvertex{7}(A_3, B_3, \Psi),
  \qquad
  A_3, B_3 \in \Hn{3}.
\end{equation}
The septic coupling contains no $\Sigma_2$: it does not modify the equation of motion of $\Sigma_2$ directly, and reaches it only through the solution for $\Sigma_3$.
The equation of motion of $\Sigma_3$ and its solution are:
\begin{equation}
  \label{eq:hs7:saddle-3}
  \begin{aligned}
    \mathcal E^{(5)}_{\Sigma_3}
      &
      := \wprod{3}{3}(\Psi \otimes \Sigma_2) - \kin{3}{\Psi}{5}(\Sigma_3)
      = 0,
    \\
    \Sigma_3^{(5)*}
      &
      := \big( \kin{3}{\Psi}{5} \big)^{-1} \circ \wprod{3}{3}(\Psi \otimes \Sigma_2),
  \end{aligned}
\end{equation}
the inverse existing as a Neumann series in $\tfrac{\gs}{7} (\wprod{3}{3})^{-1} \circ \dvertex{7}(\, \cdot \,, \cdot \,, \Psi)$ for small $\gs$.
The solution is now $\Psi \otimes \Sigma_2$ plus a correction of order $\gs$, as was $\Sigma_2^{(3)*}$ at the quintic order.
The value of the $\Sigma_3$ terms on it is the completed square $\tfrac{\gs^4}{6!} \pair[3]{\wprod{3}{3}(\Psi \otimes \Sigma_2)}{\Sigma_3^{(5)*}}$: the first term of the Neumann series gives back the sextic term $\tfrac{\gs^4}{6!} \dvertex{6}(\Psi \otimes \Sigma_2, \Psi \otimes \Sigma_2)$ and the second is the septic coupling at its naive contact value, $\tfrac{\gs^5}{7!} \dvertex{7}(\Psi \otimes \Sigma_2, \Psi \otimes \Sigma_2, \Psi)$, as for the quintic coupling in~\eqref{eq:hs5:constant}.
The action left for $\Sigma_2$ is still  quadratic, since $\Sigma_3^{(5)*}$ is linear in $\Sigma_2$, with the dressed operator
\begin{equation}
  \label{eq:hs7:dressed-2}
  \begin{aligned}
    \pair[2]{A_2}{\kin{2}{\Psi}{5}(B_2)}
      &
      := \vertex{4}(A_2, B_2)
        - \frac{\gs}{5} \, \vertex{5}(A_2, B_2, \Psi)
      \\ &
      \quad
      - \frac{\gs^2}{30} \, \pair[3]{\wprod{3}{3}(\Psi \otimes A_2)}{\big( \kin{3}{\Psi}{5} \big)^{-1} \circ \wprod{3}{3}(\Psi \otimes B_2)},
  \end{aligned}
\end{equation}
the operator $\kin{2}{\Psi}{4}$ of~\eqref{eq:hs6:dressed-2} with the $\Sigma_3$ propagator dressed in its turn.
Its solution $\Sigma_2^{(5)*}$ differs from $\Sigma_2^{(4)*}$ at order $\gs^3$, which by stationarity does not affect order $\gs^5$, where the septic term enters only through its leading-order value.

With the coefficients of~\eqref{eq:hs7:dressed-vertex}, the elimination of the auxiliary fields matches the target through order $\gs^5$:
\begin{equation}
  \label{eq:hs7:eliminate}
  \caction[\le 7] \big|_{\Sigma_3 = \Sigma_3^{(5)*}, \, \Sigma_2 = \Sigma_2^{(5)*}}
    = S_{\le 7} + O(\gs^6).
\end{equation}
The leading-order value~\eqref{eq:hs7:diagonal} is fixed by this single requirement, each exchange part of $\dvertex{7}$ canceling one of the two induced diagrams.
The three coefficients $(1, \tfrac{42}{25}, - \tfrac{14}{5})$ are determined when $\vertex{7}(\Psi^7)$, $\vertex{5 \cdot 5 \cdot 5}(\Psi^7)$ and $\vertex{6 \cdot 5}(\Psi^7)$ are independent functions of $\Psi$.
We assume this for the string vertices (third assumption of \cref{sec:all-orders}, plus mutual independence of the exchange diagrams).
The remainder of order $\gs^6$ is the next induced interaction, which the next level of the tower absorbs in its turn (\cref{sec:all-orders}).
The leading-order value of every reduced vertex is fixed in the same way (see~\eqref{eq:all:diagonal}), as the vertex of the target minus what the previous orders induce:
\begin{equation}
  \label{eq:hs7:deficit}
  \dvertex{n}(\Psi^n)
    = \vertex{n}(\Psi^n) - n! \, \big[ \gs^{n - 2} \big] \, \caction[\le n - 1] \big|_{\text{solution}},
\end{equation}
where the cubic action of the previous order is evaluated on the solution of its auxiliary fields and $[\gs^{n - 2}]$ extracts the coefficient of $\gs^{n - 2}$; the reduced sextic vertex~\eqref{eq:hs6:dressed-vertex} and the leading-order value~\eqref{eq:hs7:diagonal} of the reduced septic vertex are its cases $n = 6$ and $n = 7$.

\paragraph{Dressed kinetic operator of $\Sigma_3$.}

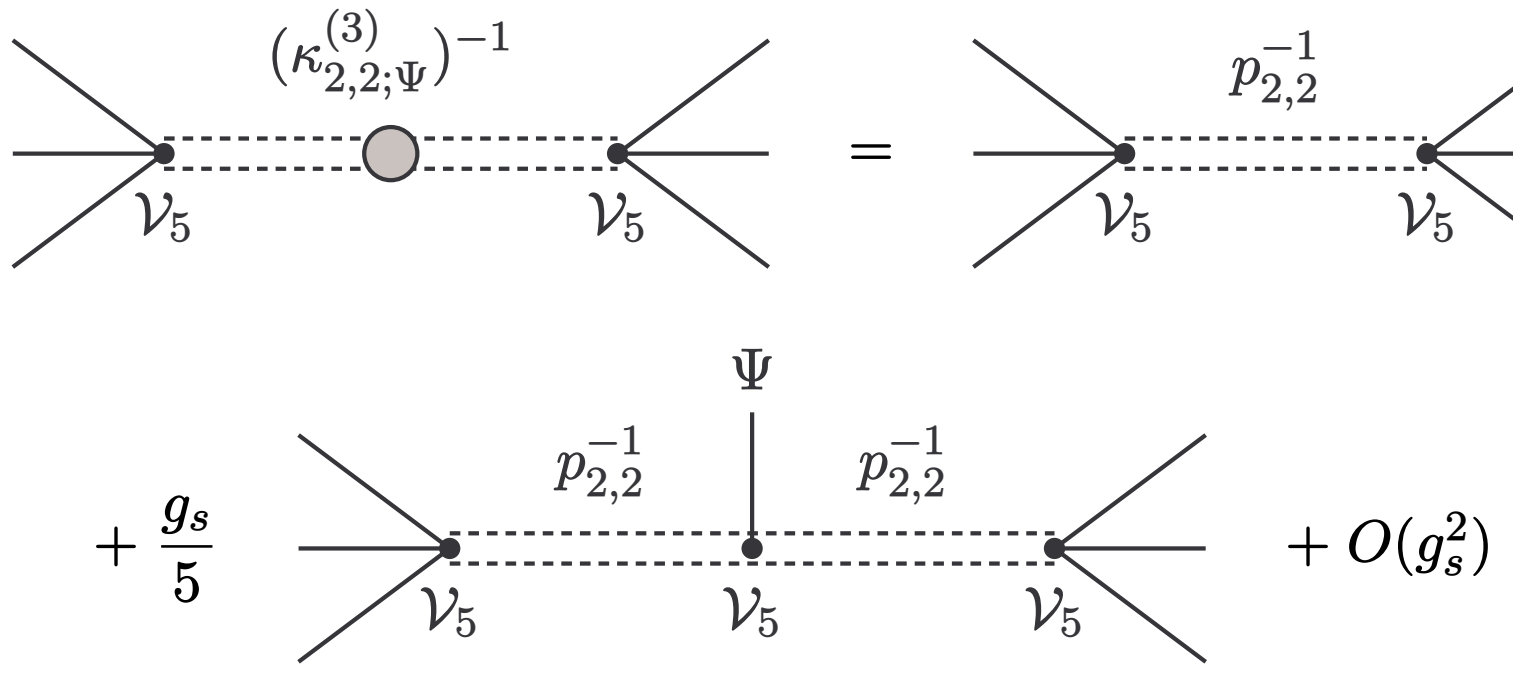
\begin{figure}[htp]
  \centering
  \begin{tikzpicture}[diagram]
    \foreach \y in {0.75, 0, -0.75} {
        \draw (0, \y) -- (1, 0);
        \draw (4, 0) -- (5, \y);
      }
    \draw[sigma line] (1, 0.1) -- (4, 0.1) (1, -0.1) -- (4, -0.1);
    \filldraw[fill=gray!40] (2.5, 0) circle (5pt);
    \fill (1, 0) circle (2pt) (4, 0) circle (2pt);
    \node[below=5pt] at (1, 0) {$\vertex{5}$};
    \node[below=5pt] at (4, 0) {$\vertex{5}$};
    \node[above=5pt] at (2.5, 0.1) {$(\kin{2}{\Psi}{3})^{-1}$};
  \end{tikzpicture}
  \quad $=$ \quad
  \begin{tikzpicture}[diagram]
    \foreach \y in {0.75, 0, -0.75} {
        \draw (0, \y) -- (1, 0);
        \draw (3, 0) -- (4, \y);
      }
    \draw[sigma line] (1, 0.1) -- (3, 0.1) (1, -0.1) -- (3, -0.1);
    \fill (1, 0) circle (2pt) (3, 0) circle (2pt);
    \node[below=5pt] at (1, 0) {$\vertex{5}$};
    \node[below=5pt] at (3, 0) {$\vertex{5}$};
    \node[above=3pt] at (2, 0.1) {$\sprod{2}{2}^{-1}$};
  \end{tikzpicture}

  \bigskip
  $+ \; \dfrac{\gs}{5}$ \quad
  \begin{tikzpicture}[diagram]
    \foreach \y in {0.75, 0, -0.75} {
        \draw (0, \y) -- (1, 0);
        \draw (5, 0) -- (6, \y);
      }
    \draw[sigma line] (1, 0.1) -- (5, 0.1) (1, -0.1) -- (5, -0.1);
    \draw (3, 0) -- (3, 0.9) node[above=2pt] {$\Psi$};
    \fill (1, 0) circle (2pt) (3, 0) circle (2pt) (5, 0) circle (2pt);
    \node[below=5pt] at (1, 0) {$\vertex{5}$};
    \node[below=5pt] at (3, 0) {$\vertex{5}$};
    \node[below=5pt] at (5, 0) {$\vertex{5}$};
    \node[above=3pt] at (2, 0.1) {$\sprod{2}{2}^{-1}$};
    \node[above=3pt] at (4, 0.1) {$\sprod{2}{2}^{-1}$};
  \end{tikzpicture}
  \quad $+ \; O(\gs^2)$
  \caption{Exchange part of the kinetic operator of $\Sigma_3$ with its internal $\Sigma_2$ line dressed, expanded by the Neumann series~\eqref{eq:hs5:neumann}; it enters with the factor $- \tfrac{6}{5}$ of~\eqref{eq:hs6:kinetic}. The three legs on the left and the three on the right are the strands of the two $3$-string arguments, the dashed double line is a $\Sigma_2$ line, and the shaded disk marks the dressed propagator $(\kin{2}{\Psi}{3})^{-1}$. The first term of the expansion is the exchange part of $\wprod{3}{3}$; the second, with one quintic insertion carrying $\Psi$, is the chain term of the reduced septic vertex~\eqref{eq:hs7:dressed-vertex}.}
  \label{fig:hs7:running-operator}
\end{figure}

The chain term of~\eqref{eq:hs7:dressed-vertex} enters $\kin{3}{\Psi}{5}$ with the coefficient $- \tfrac{6}{5} \cdot \tfrac{\gs}{5}$, which is the term of order $\gs$ of the exchange part of $\wprod{3}{3}$ once its bare $\Sigma_2$ propagator $\sprod{2}{2}^{-1}$ is replaced by the dressed one $(\kin{2}{\Psi}{3})^{-1}$, by the Neumann series~\eqref{eq:hs5:neumann}.
The bare operator $\wprod{3}{3}$ is thus the value at $\gs = 0$ of a kinetic operator whose internal $\Sigma_2$ line is dressed (\cref{fig:hs7:running-operator}), and this order tests its first coefficient.
Each order thus dresses the kinetic operators of the lower auxiliary fields by the exchanges of the higher ones, with the propagators inside these exchanges dressed in their turn, as~\eqref{eq:hs7:dressed-2} shows for $\Sigma_2$.

\paragraph{Kernel condition.}

The septic coupling must be a function on the quotient~\eqref{eq:hs6:quotient} on which $\Sigma_3$ lives, as the quintic coupling had to be a function on $\Hn[2]{2} / \ker \sprod{2}{2}$.
The sextic source descends automatically, by the symmetry of $\wprod{3}{3}$, whereas the septic coupling does not.
From this order on we assume, besides the conditions of \cref{sec:hs:quintic,sec:hs:sextic}, that the kernel of the kinetic operator with which each auxiliary field is introduced, $\wprod{3}{3}$ for $\Sigma_3$, consists of joint null vectors of every vertex in which that field appears, which at this order reads:
\begin{equation}
  \label{eq:hs7:kernel-condition}
  \forall N_3 \in \ker \wprod{3}{3}, \;
  \forall A_3 \in \Hn[2]{3}, \;
  \forall C_1 \in \Hs_2:
  \quad
  \dvertex{7}(N_3, A_3, C_1)
    = 0.
\end{equation}
It makes $\dvertex{7}(\, \cdot \,, \cdot \,, \Psi)$, hence $\kin{3}{\Psi}{5}$ and its inverse, descend to the quotient.
When it fails along a kernel direction, that direction has no kinetic term at order $\gs^4$ and receives terms from the septic coupling at order $\gs^5$, and the construction breaks down as described in \cref{sec:all:kernel} and would need refinement.
Whether the string vertices satisfy this condition is not settled.

\subsubsection{Gauge transformation}
\label{sec:hs:septic:gauge}

With the septic coupling, the action decomposes exactly into the target, the induced interactions and two squares:
\begin{equation}
  \label{eq:hs7:identity}
  \begin{aligned}
    \caction[\le 7]
      &
      = S_{\le 7}
        + \big( \text{exchange terms of order } \gs^6 \big)
      \\ &
      \quad
      - \frac{\gs^4}{6!} \, \pair[3]{\sigma_3^{(5)}}{\kin{3}{\Psi}{5}(\sigma_3^{(5)})}
      - \frac{\gs^2}{4!} \, \pair[2]{\sigma_2^{(5)}}{\kin{2}{\Psi}{5}(\sigma_2^{(5)})},
  \end{aligned}
\end{equation}
where the exchange terms are the remainder of~\eqref{eq:hs7:eliminate}, a function of $\Psi$ alone, and where the fluctuations are $\sigma_3^{(5)} := \Sigma_3 - \Sigma_3^{(5)*}$, from the solution~\eqref{eq:hs7:saddle-3} at the given $\Sigma_2$, and $\sigma_2^{(5)} := \Sigma_2 - \Sigma_2^{(5)*}$, from the solution of the $\Sigma_2$ equation once $\Sigma_3$ is eliminated.
Both squares now come with dressed kinetic operators, and we fix the gauge law as at the quintic order, by demanding that each square be exactly invariant; since the action is Gaussian in both fluctuations, the laws are explicit rotations:
\begin{equation}
  \label{eq:hs7:gauge-deviations}
  \begin{aligned}
    \delta_\Lambda \sigma_3^{(5)}
      &
      = - \frac{1}{2} \, \big( \kin{3}{\Psi}{5} \big)^{-1} \circ \big( \delta_\Lambda \kin{3}{\Psi}{5} \big) (\sigma_3^{(5)})
    \\ &
      = \frac{\gs}{14} \, \big( \kin{3}{\Psi}{5} \big)^{-1} \circ \dvertex{7}(\, \cdot \,, \sigma_3^{(5)}, \delta_\Lambda \Psi),
    \\
    \delta_\Lambda \sigma_2^{(5)}
      &
      = - \frac{1}{2} \, \big( \kin{2}{\Psi}{5} \big)^{-1} \circ \big( \delta_\Lambda \kin{2}{\Psi}{5} \big) (\sigma_2^{(5)}),
  \end{aligned}
\end{equation}
where $\dvertex{7}(\, \cdot \,, \sigma_3^{(5)}, \delta_\Lambda \Psi)$ denotes the $3$-string state paired with the open slot, as in~\eqref{eq:hs6:eom}, and where the variation of $\kin{2}{\Psi}{5}$ collects those of the quintic term, of the sextic source $\wprod{3}{3}(\Psi \otimes \cdot)$ and of the inverse of $\kin{3}{\Psi}{5}$ in~\eqref{eq:hs7:dressed-2}.
Each auxiliary field transforms as its solution plus the rotation of its fluctuation, $\delta_\Lambda \Sigma_k = \delta_\Lambda \Sigma_k^{(5)*} + \delta_\Lambda \sigma_k^{(5)}$, where $\delta_\Lambda \Sigma_3^{(5)*}$ includes the variation of~\eqref{eq:hs7:saddle-3} through $\Sigma_2$.
Varying~\eqref{eq:hs7:identity} then gives $\delta_\Lambda \caction[\le 7] = \delta_\Lambda \big[ S_{\le 7} + \big( \text{exchange terms of order } \gs^6 \big) \big]$ identically, for any variation $\delta_\Lambda \Psi$ and exactly in $\gs$; since $\delta_\Lambda S_{\le 7} = O(\gs^6)$, the cubic action is invariant up to order $\gs^6$, the order of the truncation it reproduces.
At the sextic order the fluctuation of $\Sigma_3$ needed no rotation, since its kinetic operator was independent of $\Psi$; from this order on both fluctuations rotate homogeneously, $\delta_\Lambda X \propto X$, and the rotations vanish on the auxiliary solution.
The algebra of the transformations and the Batalin--Vilkovisky structure of the cubic action, through the sextic order, are summarized in \cref{sec:further}.

\paragraph{Reality.}

The septic term is real on real fields, since $\dvertex{7}$ is built from the vertices, $\sprod{3}{2}$ and $\sprod{2}{2}^{-1}$ with real coefficients, and the solution $\Sigma_3^{(5)*}$ and the rotations~\eqref{eq:hs7:gauge-deviations} preserve the reality conditions $\Sigma_3^\star = - \Sigma_3$ and $\Sigma_2^\star = \Sigma_2$, by the argument of \cref{app:reality}.

\Cref{sec:all-orders} states the pattern of these four orders at all orders and proves it.

\section{All-order construction}
\label{sec:all-orders}

The orders worked out in \cref{sec:hs} follow one pattern, and we now show that, under the assumptions below, it continues at every order.
Closed string field theory then admits a cubic action, with a tower of auxiliary $k$-string fields, which reproduces the non-polynomial action as a formal power series in $\gs$.
The cubic action is triangular in the tower: each $\Sigma_k$ enters at order $\gs^{2 k - 2}$ through a completed square on a reduced vertex $\dvertex{2 k}$, with a source linear in the field just below it, and the odd vertex $\dvertex{2 k + 1}$ dresses its kinetic operator.
The matching fixes each reduced vertex only through its leading-order value, its value at $\Sigma_k = \Psi^k$, which is why the proof is short.
We write the action with all the auxiliary fields and its equations of motion, eliminate the fields from the top of the tower down, which puts the action in a sum of completed squares.
We then state and prove the matching which fixes the reduced vertices at every order, recover the original action and its equation of motion, give the gauge transformations, and state the conditions on the kernels of the kinetic operators under which the construction holds.
\Cref{sec:hs} serves as the example: the cases $k = 2, 3$ of every statement below are the quartic to septic orders.

Throughout, we work at tree level and read every identity as a formal power series in $\gs$.
At any truncation of the tower the identities are exact in $\gs$ and, at a fixed mass level with the momenta fixed or discretized, rational.
We do not address the convergence of the series in the number of auxiliary fields; the original action is itself only a formal series in $\gs$, so nothing is lost against it.
Three assumptions carry over from the previous section.
First, the bare kinetic operator $\wprod{k}{k}$ of each auxiliary field, defined below, is invertible on the quotient of $\Hn[2]{k}$ by its kernel.
We also assume, as for $\sprod{2}{2}$ in \cref{sec:hs:quartic}, that the quotient and the inverse have a limit as the mass level of the truncation grows, and that the inverse acts only on sources and dressing terms lying in its image (\cref{sec:hs:sextic:kernel}).
Second, the kernels satisfy the conditions of \cref{sec:all:kernel} (the \emph{kernel conditions}), which make every term of the action a function on the quotients.
Third, the exchange diagrams are not linear combinations of the elementary vertices, which the recursion does not need but which makes the terms subtracted in a reduced vertex a definite set, as for the coefficients of the reduced septic vertex.
The matching uses only the symmetry of the vertices.

\subsection{Cubic action at all orders}
\label{sec:all:action}

We first write the action with every auxiliary field and its equations of motion, and show that it is triangular in the tower.

\paragraph{Tower of auxiliary fields.}

\begin{figure}[htp]
  \centering
  \begin{tikzpicture}[diagram, >=stealth]
    \draw (0, 0) -- (0, 1.2);
    \node at (0, -0.35) {$\Psi$};
    \node at (0, 1.6) {$Q_B, \; \vertex{3}$};
    \node at (0, -0.85) {$\gs^0, \; \gs$};
    \foreach \x in {-0.07, 0.07} {
        \draw[sigma line] ({3.4 + \x}, 0) -- ({3.4 + \x}, 1.2);
      }
    \node at (3.4, -0.35) {$\Sigma_2$};
    \node at (3.4, 1.6) {$\dvertex{4}, \; \dvertex{5}$};
    \node at (3.4, -0.85) {$\gs^2, \; \gs^3$};
    \foreach \x in {-0.14, 0, 0.14} {
        \draw[sigma line] ({6.8 + \x}, 0) -- ({6.8 + \x}, 1.2);
      }
    \node at (6.8, -0.35) {$\Sigma_3$};
    \node at (6.8, 1.6) {$\dvertex{6}, \; \dvertex{7}$};
    \node at (6.8, -0.85) {$\gs^4, \; \gs^5$};
    \foreach \x in {-0.21, -0.07, 0.07, 0.21} {
        \draw[sigma line] ({10.2 + \x}, 0) -- ({10.2 + \x}, 1.2);
      }
    \node at (10.2, -0.35) {$\Sigma_4$};
    \node at (10.2, 1.6) {$\dvertex{8}, \; \dvertex{9}$};
    \node at (10.2, -0.85) {$\gs^6, \; \gs^7$};
    \node at (12.3, 0.6) {$\cdots$};
    \draw[->] (0.55, 0.6) -- (2.85, 0.6) node[midway, above, font=\small] {$\dvertex{4}(\Psi^2, \Sigma_2)$};
    \draw[->] (3.95, 0.6) -- (6.25, 0.6) node[midway, above, font=\small] {$\dvertex{6}(\Psi \otimes \Sigma_2, \Sigma_3)$};
    \draw[->] (7.35, 0.6) -- (9.65, 0.6) node[midway, above, font=\small] {$\dvertex{8}(\Psi \otimes \Sigma_3, \Sigma_4)$};
    \draw[->] (10.75, 0.6) -- (11.9, 0.6);
  \end{tikzpicture}
  \caption{The tower of auxiliary fields. Each level $k$ carries the $k$-string field $\Sigma_k$, drawn as $k$ dashed strands, as the auxiliary lines of the diagrams of \cref{sec:hs}, the pair of reduced vertices $(\dvertex{2 k}, \dvertex{2 k + 1})$ giving its kinetic term and source, and its odd coupling, and the orders $\gs^{2 k - 2}$, $\gs^{2 k - 1}$ at which they enter. The arrows are the sources, each linear in the field of the level below.}
  \label{fig:all:tower}
\end{figure}
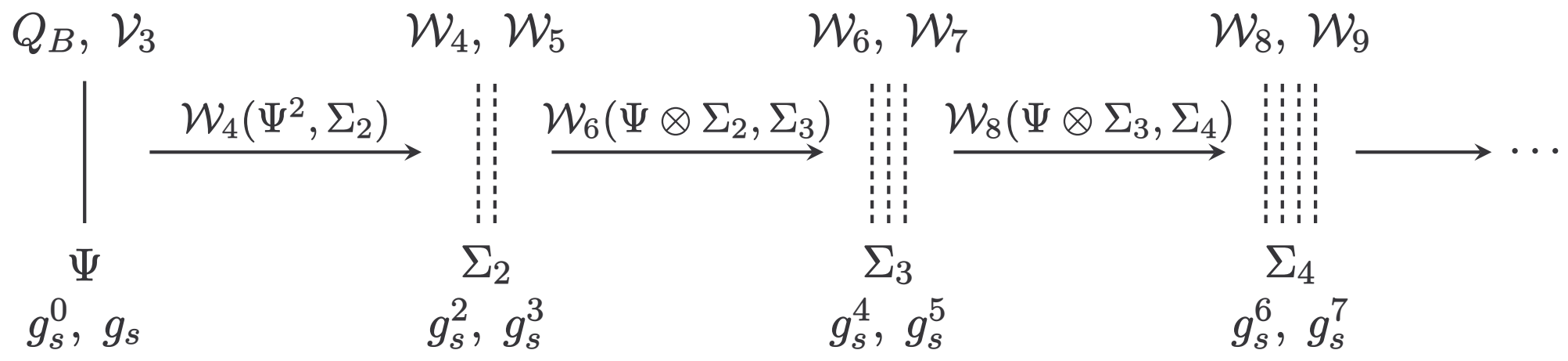

Each even order introduces one auxiliary field, and the odd order which follows couples it to $\Psi$: $\Sigma_2$ at order $\gs^2$, $\Sigma_3$ at order $\gs^4$, and in general
\begin{equation}
  \label{eq:all:counting}
  \Sigma_k \in \Hn[2]{k}
    \;\longleftrightarrow\;
  \Big( \frac{\gs^{2 k - 2}}{(2 k)!} \, \dvertex{2 k}, \; \frac{\gs^{2 k - 1}}{(2 k + 1)!} \, \dvertex{2 k + 1} \Big),
  \qquad
  k \ge 2,
\end{equation}
where $\dvertex{2 k}$ and $\dvertex{2 k + 1}$ are the reduced vertices of the two orders, with $\dvertex{4} = \vertex{4}$ and $\dvertex{5} = \vertex{5}$ (\cref{fig:all:tower}).
The reason is the one met at the sextic and septic orders.
The only quadratic term for a $k$-string field, of ghost number $\Ngh(\Sigma_k) = 2 k$, is $\dvertex{2 k}(\Sigma_k, \Sigma_k)$, which comes with $\gs^{2 k - 2}$, and the even vertex $\vertex{2 k}$ is the first vertex whose $2 k$ slots split as $k + k$.
The odd vertex is then written with the partition $2 k + 1 = k + k + 1$, which keeps the action quadratic in $\Sigma_k$ and dresses its kinetic operator by $\Psi$, as the quintic vertex did for $\Sigma_2$.
We call $k$ the level of $\Sigma_k$ in the tower and keep \enquote{mass level} for the level truncation.\footnote{%
  A cubic action built from the elementary vertices alone could do with fewer fields: splitting $\vertex{n}$ as $n = a + b + c$ over $\Sigma_a$, $\Sigma_b$, $\Sigma_c$ needs $\Sigma_k$ only from $\vertex{3 k - 2}$ on.
  The construction introduces $\Sigma_k$ earlier, at $\vertex{2 k}$, because each field comes with a completed square, which is what makes its elimination exact.
  The action with fewer fields is not triangular in the sense of \cref{sec:all:action}: a vertex $\vertex{n}$ split as $a + b + c$ couples fields of any three levels, and $\vertex{3 k}(\Sigma_k, \Sigma_k, \Sigma_k)$ is cubic in one field, so that the equation of each auxiliary field is non-linear and couples it to levels that are not its neighbors.
  The fields then cannot be eliminated one Gaussian step at a time.
}

The $4$-string field shows the rule at the first order beyond \cref{sec:hs}.
Eliminating $\Sigma_2$ and $\Sigma_3$ from $\caction[\le 7]$ leaves at order $\gs^6$ the exchange diagrams of the two fields between the quintic, sextic and septic vertices, and the leading-order value of the reduced octic vertex is the octic vertex minus these diagrams, by the rule~\eqref{eq:hs7:deficit}, the case $n = 8$ of~\eqref{eq:all:diagonal}.
We read $\dvertex{8}$ as a symmetric bilinear form on $\Hn{4}$, write the octic term as $\dvertex{8}(\Psi \otimes \Sigma_3, \Psi \otimes \Sigma_3)$ and complete its square with a $4$-string field:
\begin{subequations}
  \label{eq:all:S8tilde}
  \begin{align}
    \label{eq:all:DeltaS8}
    \Delta S_8
      &
      := - \frac{\gs^6}{8!} \, \dvertex{8}(\Sigma_4 - \Psi \otimes \Sigma_3, \Sigma_4 - \Psi \otimes \Sigma_3),
    \qquad
    \Sigma_4 \in \Hn[2]{4},
    \\
    \caction[8]
      &
      := \frac{\gs^6}{8!} \, \dvertex{8}(\Psi \otimes \Sigma_3, \Psi \otimes \Sigma_3) + \Delta S_8
      = - \frac{\gs^6}{8!} \, \dvertex{8}(\Sigma_4^2)
        + \frac{2 \gs^6}{8!} \, \dvertex{8}(\Psi \otimes \Sigma_3, \Sigma_4).
  \end{align}
\end{subequations}
The state $\Psi \otimes \Sigma_3$, rather than $\Psi^4$ or $\Sigma_2 \otimes \Sigma_2$, keeps the cross term cubic in the fields and the source of $\Sigma_4$ linear in the field just below it.
Since $\Sigma_3 = \Psi^3 + O(\gs)$ on the solution, only $\dvertex{8}(\Psi^8)$ is produced at this order and the octic order needs nothing more; the odd vertex $\dvertex{9}$ then couples as $\dvertex{9}(\Sigma_4, \Sigma_4, \Psi)$ at order $\gs^7$ and dresses the kinetic operator of $\Sigma_4$, as $\dvertex{7}$ did for $\Sigma_3$.

\paragraph{General form of the action.}

With all the auxiliary fields, the cubic action is:
\begin{equation}
  \label{eq:all:action}
  \begin{aligned}
    \caction
      &
      := S_{\le 3}
        + \sum_{k \ge 2} \Big[
          - \frac{\gs^{2 k - 2}}{(2 k)!} \, \dvertex{2 k}(\Sigma_k, \Sigma_k)
          + \frac{2 \gs^{2 k - 2}}{(2 k)!} \, \dvertex{2 k}(\Psi \otimes \Sigma_{k - 1}, \Sigma_k)
      \\ &
      \qquad\qquad\quad
          + \frac{\gs^{2 k - 1}}{(2 k + 1)!} \, \dvertex{2 k + 1}(\Sigma_k, \Sigma_k, \Psi)
        \Big],
    \qquad
    \Sigma_1
      := \Psi,
  \end{aligned}
\end{equation}
where $S_{\le 3}$ is the kinetic term and the cubic vertex of $\Psi$, untouched by the construction.
Each level contributes three terms: the kinetic term of $\Sigma_k$, its source, linear in $\Sigma_{k - 1}$ and in $\Psi$, and its odd coupling.
The action is cubic in the fields: apart from $S_{\le 3}$, every term contains two auxiliary fields and at most one $\Psi$, except the quartic source, which contains $\Sigma_2$ and two $\Psi$.
It is real on the slice~\eqref{eq:conv:reality-fields} (\cref{app:reality}).
The sum over $k$ is infinite, but each order in $\gs$ receives finitely many terms, two at every even order from $\gs^2$ on and one at every odd order: the coefficient of $\gs^m$ involves $\Psi$, $\Sigma_{\lfloor m / 2 \rfloor + 1}$ and, for even $m$, $\Sigma_{m / 2}$ only.
The leading-order value of a reduced vertex $\dvertex{n}$ is itself a finite polynomial in $\Psi$, the coefficient of one order of the eliminated action, to which finitely many diagrams contribute (\cref{sec:all:reduced-vertices}).
The action is thus well defined as a formal power series in $\gs$, and truncating the series gives the actions of \cref{sec:hs}:
\begin{equation}
  \label{eq:all:truncation}
  \caction[\le 2 K + 1]
    := S_{\le 3} + \sum_{k = 2}^{K} \big[ \ldots \big],
  \qquad
  \caction[\le 2 K]
    := \caction[\le 2 K + 1] - \frac{\gs^{2 K - 1}}{(2 K + 1)!} \, \dvertex{2 K + 1}(\Sigma_K^2, \Psi),
\end{equation}
with the bracket of~\eqref{eq:all:action}: the first is the action with the fields $\Sigma_2, \ldots, \Sigma_K$, the second drops the odd coupling of the top field.

The kinetic term and the source of level $k$ come from one form, $\dvertex{2 k}$, with its $2 k$ slots split as $k + k$ and as $k + (k - 1) + 1$.
We write it through the $k$-string pairing, the $k$-th tensor power~\eqref{eq:conv:pairing-n} of the pairing of \cref{sec:hs:action}, which for the even tensors of this section is the plain product of the pairings of the strands:
\begin{equation}
  \label{eq:all:kinetic}
  \dvertex{2 k}(A_k, B_k)
    =: \pair[k]{A_k}{\wprod{k}{k}(B_k)},
  \qquad
  A_k, B_k \in \Hn{k},
  \qquad
  \wprod{2}{2}
    = \sprod{2}{2},
\end{equation}
which defines the \emph{bare kinetic operator} $\wprod{k}{k} \colon \Hn{k} \to \Hn{k}$ of $\Sigma_k$, symmetric with respect to the pairing; $\wprod{3}{3}$ is the operator of~\eqref{eq:hs6:kinetic}.
The kinetic term and the source then read $\pair[k]{\Sigma_k}{\wprod{k}{k}(\Sigma_k)}$ and $\pair[k]{\Sigma_k}{\wprod{k}{k}(\Psi \otimes \Sigma_{k - 1})}$.
The form $\dvertex{2 k}$ is fixed by the levels below $k$ (\cref{sec:all:reduced-vertices}), so that $\wprod{k}{k}$, its kernel and the quotient $\Hn[2]{k} / \ker \wprod{k}{k}$ on which $\Sigma_k$ lives do not depend on the order at which the tower is truncated.
Every dressing of the operator vanishes at $\Psi = 0$.

\paragraph{Triangular form.}

The action is quadratic in each auxiliary field separately, and the two terms quadratic in $\Sigma_k$ combine into one operator dressed by $\Psi$, as at the quintic and septic orders:
\begin{equation}
  \label{eq:all:dressed-odd}
  \pair[k]{A_k}{\kin{k}{\Psi}{2 k - 1}(B_k)}
    := \pair[k]{A_k}{\wprod{k}{k}(B_k)}
      - \frac{\gs}{2 k + 1} \, \dvertex{2 k + 1}(A_k, B_k, \Psi),
  \qquad
  A_k, B_k \in \Hn{k},
\end{equation}
affine in $\Psi$ and symmetric with respect to the pairing; the label records that it is the operator of $\Sigma_k$ when it is the top field of the tower, before the exchanges of the higher fields dress it.
The action is also quadratic in the tower as a whole, since each source is bilinear in two consecutive fields.
Collecting the fields in $\Sigma := (\Sigma_2, \Sigma_3, \ldots)$, we get:
\begin{equation}
  \label{eq:all:tridiagonal}
  \caction
    = S_{\le 3}
      + \frac{2 \gs^2}{4!} \, \pair[2]{\Sigma_2}{\sprod{2}{2}(\Psi^2)}
      - \sum_{k, l \ge 2} \pair[k]{\Sigma_k}{\mathcal K_{k, l; \Psi}(\Sigma_l)},
\end{equation}
with the blocks
\begin{equation}
  \label{eq:all:blocks}
  \begin{gathered}
    \mathcal K_{k, k; \Psi}
      := \frac{\gs^{2 k - 2}}{(2 k)!} \, \kin{k}{\Psi}{2 k - 1},
    \qquad
    \mathcal K_{k, k - 1; \Psi}
      := - \frac{\gs^{2 k - 2}}{(2 k)!} \, \wprod{k}{k}(\Psi \otimes \, \cdot \,),
    \\
    \mathcal K_{k, l; \Psi}
      = 0
    \quad
    (\abs{k - l} \ge 2),
  \end{gathered}
\end{equation}
and $\mathcal K_{k - 1, k; \Psi}$ the transpose of $\mathcal K_{k, k - 1; \Psi}$ with respect to the pairings,~\eqref{eq:conv:transpose}.
This is the \emph{triangular form} of the action named at the septic order, now for the whole tower: the quadratic form of the tower is block-tridiagonal, each field couples to itself and to its two neighbors only.
The only term linear in the tower is the quartic source of $\Sigma_2$.
Everything below rests on this form.
The equations of the auxiliary fields are linear and are solved, at each truncation, from the top down, one inversion per level (\cref{sec:all:squares} for the full tower).
The elimination of the tower is therefore a chain of Gaussian integrals, each exact at tree level.
The value of the action on the solution is then, at each truncation, one closed expression in $\gs$ rather than a series of diagrams.
Adding a level adds one row and one column to the form, which is why the elimination extends to every order.

\paragraph{Equations of motion.}

We normalize the equations of motion so that each auxiliary equation has coefficient one on its source, as in \cref{sec:hs}:
\begin{equation}
  \label{eq:all:eom-normalization}
  \delta \caction
    = \pair{\delta \Psi}{\mathcal E_\Psi}
      + \sum_{k \ge 2} \frac{2 \gs^{2 k - 2}}{(2 k)!} \, \pair[k]{\delta \Sigma_k}{\mathcal E_{\Sigma_k}}.
\end{equation}
Varying~\eqref{eq:all:action} gives:
\begin{subequations}
  \label{eq:all:eom}
  \begin{align}
    \label{eq:all:eom-psi}
    \mathcal E_\Psi
      &
      = Q_B \Psi
        + \frac{\gs}{2} \, \ell_2(\Psi^2)
        + \frac{\gs^2}{3!} \, \ell_3(\Psi, \Sigma_2)
        + \frac{\gs^3}{5!} \, \ell_4(\Sigma_2^2)
    \notag
    \\ &
    \quad
        + \sum_{k \ge 3} \Big[
          \frac{2 \gs^{2 k - 2}}{(2 k)!} \, \dvertex{2 k}(\, \cdot \, \otimes \Sigma_{k - 1}, \Sigma_k)
          + \frac{\gs^{2 k - 1}}{(2 k + 1)!} \, \dvertex{2 k + 1}(\Sigma_k, \Sigma_k, \, \cdot \,)
        \Big],
    \\
    \label{eq:all:eom-sigma}
    \mathcal E_{\Sigma_k}
      &
      = \wprod{k}{k}(\Psi \otimes \Sigma_{k - 1})
        - \wprod{k}{k}(\Sigma_k)
        + \frac{\gs}{2 k + 1} \, \dvertex{2 k + 1}(\, \cdot \,, \Sigma_k, \Psi)
    \notag
    \\ &
    \quad
        + \frac{\gs^2}{(2 k + 1) (2 k + 2)} \, \dvertex{2 k + 2}(\Psi \otimes \, \cdot \,, \Sigma_{k + 1}),
  \end{align}
\end{subequations}
for $k \ge 2$, where the dot denotes the state paired with the open slot, as in~\eqref{eq:hs6:eom}, and where the terms of level $2$ in the equation of $\Psi$ are written with the products $\ell_3$ and $\ell_4$, as at the quintic order.
The equation of $\Sigma_k$ is row $k$ of the tridiagonal form~\eqref{eq:all:tridiagonal}: it couples $\Sigma_k$ to $\Sigma_{k - 1}$ through its source, to itself through the operator dressed by the odd vertex, and to $\Sigma_{k + 1}$ through the transpose of the next source, which comes with $\gs^2$.
At a truncation $K$ the last term of the equation of $\Sigma_K$ is absent, and in the even truncation the odd term of level $K$ is absent from both equations.
The equation of $\Psi$ is the only non-linear one and the only one which involves every field; once the tower is eliminated, it becomes the equation of motion of the original action (\cref{sec:all:integrating-out}).

\paragraph{Rescaled fields.}

The powers of $\gs$ in~\eqref{eq:all:action} are those of the original action, in which $\vertex{n}$ carries $\gs^{n - 2}$.
Under $\Psi \to t \Psi$, $\Sigma_k \to t^k \Sigma_k$ and $\gs \to \gs / t$, every term of the action scales as $t^2$, so that the fields
\begin{equation}
  \label{eq:all:rescaling}
  \widetilde{\Sigma}_k
    := \gs^{k - 1} \, \Sigma_k,
  \qquad
  k \ge 1,
  \qquad
  \widetilde{\Sigma}_1
    = \Psi,
\end{equation}
all scale as $\Psi$ does.
In terms of them, every kinetic term carries no power of $\gs$ and every cubic term, the cubic vertex of $\Psi$ included, carries one, as for a cubic action with a single coupling constant; this extends the rescaling $\Sigma_2 \to \Sigma_2 / \gs$ of \cref{sec:hs:quartic} to the whole tower.
The sum of the rescaled fields is the total field
\begin{equation}
  \label{eq:all:total-field}
  \boldsymbol{\Psi}
    := \sum_{n \ge 1} \gs^{n - 1} \, \Psi_n
    = \sum_{n \ge 1} \widetilde{\Sigma}_n,
  \qquad
  \Psi_1
    := \Psi,
  \qquad
  \Psi_n
    := \Sigma_n
  \quad
  (n \ge 2),
\end{equation}
an element of the graded sum $\bigoplus_{n \ge 1} \Hn{n}$, each component carrying its own ghost number, in terms of which the kinetic terms form one quadratic form for the pairing taken component by component and the cubic terms one cubic form with a single $\gs$ in front.
\Cref{sec:ibl} reads the cubic action in this way, as an action for one string field with multi-string components.

\subsection{Completed squares and nested operators}
\label{sec:all:squares}

The triangular form lets us eliminate the auxiliary fields from the top of the tower down, one Gaussian step per level, and the result is a sum of completed squares whose coefficients are the dressed kinetic operators.
We write the step at a truncation $K$, where the top field is $\Sigma_K$, and treat the full tower at the end of the subsection.

Only the three terms of level $K$ contain $\Sigma_K$, and its equation is the case $k = K$ of~\eqref{eq:all:eom-sigma} without the last term:
\begin{equation}
  \label{eq:all:eom-top}
  \mathcal E_{\Sigma_K}
    = \wprod{K}{K}(\Psi \otimes \Sigma_{K - 1}) - \kin{K}{\Psi}{2 K - 1}(\Sigma_K)
    = 0,
\end{equation}
whose solution $\Sigma_K^* = (\kin{K}{\Psi}{2 K - 1})^{-1} \circ \wprod{K}{K}(\Psi \otimes \Sigma_{K - 1})$ is linear in $\Sigma_{K - 1}$, the inverse being a Neumann series in $\gs$ around its term of order $\gs^0$, the bare operator $\wprod{K}{K}$, as in~\eqref{eq:hs5:neumann}.
At tree level the Gaussian integral over $\Sigma_K$ reduces to its value at the saddle, its determinant being a loop effect that we drop.
Since the action is quadratic in $\Sigma_K$, substituting the solution back completes a square, and the kinetic term and the source of level $K$ combine into the exchange of $\Sigma_K$ between two of its sources:
\begin{equation}
  \label{eq:all:eliminate-top}
  \begin{aligned}
    \caction[\le 2 K + 1] \big|_{\Sigma_K = \Sigma_K^*}
      &
      = \caction[\le 2 K - 1]
    \\ &
    \quad
      + \frac{\gs^{2 K - 2}}{(2 K)!} \, \pair[K]{\wprod{K}{K}(\Psi \otimes \Sigma_{K - 1})}{\big( \kin{K}{\Psi}{2 K - 1} \big)^{-1} \circ \wprod{K}{K}(\Psi \otimes \Sigma_{K - 1})}.
  \end{aligned}
\end{equation}
The induced term is quadratic in $\Sigma_{K - 1}$, with one $\Psi$ on each side, and has the form of the kinetic term of $\Sigma_{K - 1}$, into which we absorb it.
The action of the levels $2, \ldots, K - 1$ then has the form~\eqref{eq:all:truncation} again, with the operator of its top field dressed by the exchange of the field just eliminated, and the next step eliminates $\Sigma_{K - 1}$ with this operator.
At $K = 3$ this is the elimination of $\Sigma_3$ at the septic order.

Repeating the step down to $\Sigma_2$ builds the \emph{dressed kinetic operators} of the tower, from the top down, as a chain of exchanges with one rung per level.
The operators of \cref{sec:hs} are the truncations of the chain: at a truncation of the action at order $\gs^n$ the chain stops at the top field, with its odd vertex for $n$ odd and with its bare operator for $n$ even.
We write $\kin{k}{\Psi}{n}$ for the operator so obtained, as there.
Without superscript, $\kin{k}{\Psi}{}$ denotes the operator of the full tower, and $\kin{k}{\Psi}{n} = \kin{k}{\Psi}{} + O(\gs^{n - 2 k + 3})$, since each rung of the chain costs $\gs^2$ and the odd vertex of the top field costs $\gs$.
Thus $\kin{2}{\Psi}{5}$ and $\kin{3}{\Psi}{5}$ are the operators of the septic order, and $\kin{K}{\Psi}{2 K - 1}$ of~\eqref{eq:all:dressed-odd} is the top of the chain at the truncation $K$.
The chain reads:
\begin{equation}
  \label{eq:all:nested}
  \begin{aligned}
    \pair[k]{A_k}{\kin{k}{\Psi}{}(B_k)}
      &
      := \pair[k]{A_k}{\wprod{k}{k}(B_k)}
        - \frac{\gs}{2 k + 1} \, \dvertex{2 k + 1}(A_k, B_k, \Psi)
    \\ &
    \quad
      - \frac{\gs^2}{(2 k + 1) (2 k + 2)} \,
    \\ &
    \qquad
      \times \pair[k + 1]{\wprod{k + 1}{k + 1}(\Psi \otimes A_k)}{\kin{k + 1}{\Psi}{}^{-1} \circ \wprod{k + 1}{k + 1}(\Psi \otimes B_k)},
  \end{aligned}
\end{equation}
the bare operator, dressed by the odd vertex at order $\gs$ and by the exchange of $\Sigma_{k + 1}$ at order $\gs^2$, inside which the $\Sigma_{k + 1}$ line carries the dressed operator of $\Sigma_{k + 1}$ in its turn.
With each inverse expanded as a Neumann series, the operator of $\Sigma_2$ is a continued fraction of exchanges whose depth is the number of auxiliary fields.

Every $\Psi$ in these operators comes with one power of $\gs$: the odd vertex enters as $\gs \, \dvertex{2 k + 1}(\, \cdot \,, \cdot \,, \Psi)$, and the source of the next field, linear in $\Psi$, enters the exchange twice, with $\gs^2$.
The dressed operators are therefore functions of $\gs \Psi$ alone, and their expansion starts with the bare operator:
\begin{equation}
  \label{eq:all:scaling}
  \begin{aligned}
    \pair[k]{A_k}{\kin{k}{\Psi}{}(B_k)}
      &
      = \pair[k]{A_k}{\wprod{k}{k}(B_k)}
        - \frac{\gs}{2 k + 1} \, \dvertex{2 k + 1}(A_k, B_k, \Psi)
    \\ &
    \quad
      - \frac{\gs^2}{(2 k + 1) (2 k + 2)} \, \dvertex{2 k + 2}(\Psi \otimes A_k, \Psi \otimes B_k)
        + O(\gs^3 \Psi^3),
  \end{aligned}
\end{equation}
where the exchange term reduces at order $\gs^2$ to the reduced vertex of the next level, because $\wprod{k + 1}{k + 1}^{-1} \circ \wprod{k + 1}{k + 1}$ is the identity on the quotient, as in~\eqref{eq:hs6:dressed-2}.
Under the kernel conditions of \cref{sec:all:kernel}, every dressed operator descends to the quotient of its bare operator, and its inverse exists as a formal power series in $\gs$ because its term of order $\gs^0$ is the bare operator.
The invertibility of the dressed operators thus reduces to that of the bare ones, and the operator with which $\Sigma_k$ is introduced is the value at $\gs = 0$ of its dressed operator at every later order.

The elimination also gives the action off-shell.
We define the fluctuation of each field from its conditional solution, the solution of its own equation at fixed lower fields once the higher ones are eliminated:
\begin{equation}
  \label{eq:all:deviation}
  \sigma_k
    := \Sigma_k - \kin{k}{\Psi}{}^{-1} \circ \wprod{k}{k}(\Psi \otimes \Sigma_{k - 1}),
  \qquad
  k \ge 2,
\end{equation}
so that $\sigma_2 = \Sigma_2 - \kin{2}{\Psi}{}^{-1} \circ \sprod{2}{2}(\Psi^2)$.
Then, off-shell, we have:
\begin{equation}
  \label{eq:all:squares}
  \caction
    = \caction \big|_{\Sigma = \Sigma^*}
      - \sum_{k \ge 2} \frac{\gs^{2 k - 2}}{(2 k)!} \, \pair[k]{\sigma_k}{\kin{k}{\Psi}{}(\sigma_k)},
\end{equation}
where $\Sigma = \Sigma^*$ denotes the joint solution of the auxiliary equations, obtained from the bottom up once the operators are known:
\begin{equation}
  \label{eq:all:saddles}
  \Sigma_k^*
    = \kin{k}{\Psi}{}^{-1} \circ \wprod{k}{k}(\Psi \otimes \Sigma_{k - 1}^*),
  \qquad
  k \ge 2,
  \qquad
  \Sigma_1^*
    := \Psi.
\end{equation}
At $K = 3$ this is the off-shell identity of the septic order.
On the solution the squares vanish and the value of the action is one Gaussian:
\begin{equation}
  \label{eq:all:value}
  \caction \big|_{\Sigma = \Sigma^*}
    = S_{\le 3} + \frac{\gs^2}{4!} \, \pair[2]{\sprod{2}{2}(\Psi^2)}{\kin{2}{\Psi}{}^{-1} \circ \sprod{2}{2}(\Psi^2)}.
\end{equation}
The only source is the quartic one, and every other term of the action sits in the continued fraction $\kin{2}{\Psi}{}$.
At a truncation $K$ the same formulas hold with the chain stopped at $K$, and the value is exact in $\gs$: the whole tower of interactions induced by the auxiliary fields is this one function, as at the septic order,~\eqref{eq:hs7:closed-form}.

For the full tower the chain has no top, and the formulas are read as formal power series.
Since each rung of the chain costs $\gs^2$, the coefficient of $\gs^m$ in the operator of $\Sigma_k$ and in $\Sigma_k^*$ involves the levels up to $k + \lfloor m / 2 \rfloor$ only, and that of the value the levels up to $1 + \lfloor m / 2 \rfloor$.
Every coefficient is therefore unchanged once the tower is long enough, and the operators, solutions and value of the full tower are the limits of those at finite $K$, coefficient by coefficient.

\subsection{Reduced vertices at every order}
\label{sec:all:reduced-vertices}

The orders of \cref{sec:hs} fixed each reduced vertex through its \emph{leading-order value}, its value on the diagonal $\Sigma_k = \Psi^k$, which is the leading order of the solution.
The following statement extends the matching to every order.

\paragraph{Matching theorem.}

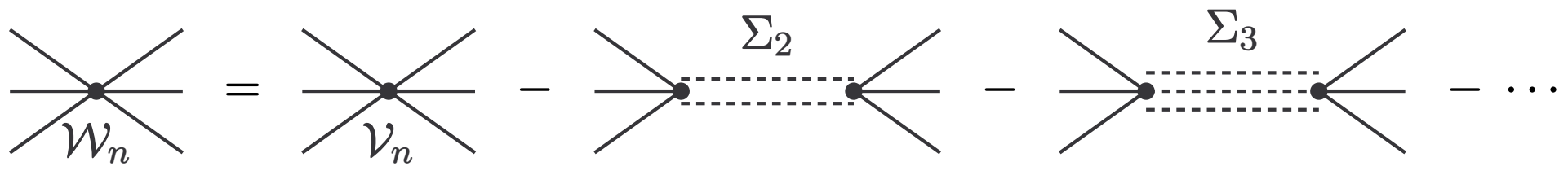
\begin{figure}[htp]
  \centering
  \begin{tikzpicture}[diagram]
    \foreach \y in {0.5, 0, -0.5} {
        \draw (0, \y) -- (0.7, 0) -- (1.4, \y);
      }
    \fill (0.7, 0) circle (2pt);
    \node[below=5pt] at (0.7, 0) {$\dvertex{n}$};
  \end{tikzpicture}
  \enspace $=$ \enspace
  \begin{tikzpicture}[diagram]
    \foreach \y in {0.5, 0, -0.5} {
        \draw (0, \y) -- (0.7, 0) -- (1.4, \y);
      }
    \fill (0.7, 0) circle (2pt);
    \node[below=5pt] at (0.7, 0) {$\vertex{n}$};
  \end{tikzpicture}
  \enspace $-$ \enspace
  \begin{tikzpicture}[diagram]
    \foreach \y in {0.5, 0, -0.5} {
        \draw (0, \y) -- (0.7, 0);
        \draw (2.1, 0) -- (2.8, \y);
      }
    \draw[sigma line] (0.7, 0.1) -- (2.1, 0.1) (0.7, -0.1) -- (2.1, -0.1);
    \fill (0.7, 0) circle (2pt) (2.1, 0) circle (2pt);
    \node[above=3pt] at (1.4, 0.1) {$\Sigma_2$};
  \end{tikzpicture}
  \enspace $-$ \enspace
  \begin{tikzpicture}[diagram]
    \foreach \y in {0.5, 0, -0.5} {
        \draw (0, \y) -- (0.7, 0);
        \draw (2.1, 0) -- (2.8, \y);
      }
    \draw[sigma line] (0.7, 0.15) -- (2.1, 0.15) (0.7, 0) -- (2.1, 0) (0.7, -0.15) -- (2.1, -0.15);
    \fill (0.7, 0) circle (2pt) (2.1, 0) circle (2pt);
    \node[above=3pt] at (1.4, 0.15) {$\Sigma_3$};
  \end{tikzpicture}
  \enspace $- \; \cdots$
  \caption{A reduced vertex $\dvertex{n}$ built from diagrams, drawn schematically: the elementary vertex $\vertex{n}$ minus the exchange diagrams of the lower auxiliary fields, with the reduced vertices at their nodes and with the same $n$ external strands. A dashed double line is a $\Sigma_2$ line, as in the diagrams of \cref{sec:hs}, and a dashed triple line a $\Sigma_3$ line; the sextic and septic cases, drawn with every strand as a leg and with the elementary vertices at the nodes, where the chain term of~\eqref{eq:hs7:diagonal} appears with the opposite sign, are \cref{fig:hs6:dressed-vertex,fig:hs7:dressed-vertex}.}
  \label{fig:all:reduced-vertex}
\end{figure}

Let $\dvertex{4} = \vertex{4}$, $\dvertex{5} = \vertex{5}$ and $\dvertex{m}$, $6 \le m < n$, be symmetric forms satisfying the assumptions of this section.
Then symmetric forms $\dvertex{n}$ exist with the leading-order value
\begin{equation}
  \label{eq:all:diagonal}
  \dvertex{n}(\Psi^n)
    = \vertex{n}(\Psi^n) - n! \, \big[ \gs^{n - 2} \big] \, \caction[\le n - 1] \big|_{\Sigma = \Sigma^*},
  \qquad
  n \ge 6,
\end{equation}
where $[\gs^{n - 2}]$ extracts the coefficient of $\gs^{n - 2}$, and, for each of them satisfying the assumptions that involve it, the cubic action reproduces the original one at this order on the solution of the auxiliary equations.
Hence, for reduced vertices satisfying the assumptions at every level:
\begin{equation}
  \label{eq:all:closure}
  \caction[\le N] \big|_{\Sigma = \Sigma^*}
    = S_{\le N} + O(\gs^{N - 1}),
  \qquad
  N \ge 4,
\end{equation}
and, for the full tower, $\caction |_{\Sigma = \Sigma^*} = S$ as formal power series in $\gs$.
That such vertices exist is part of the assumptions (\cref{sec:all:kernel}).
At each truncation the left-hand side is given exactly by~\eqref{eq:all:value} and differs from $S_{\le N}$ at the order of the first term left out, $\gs^{2 K}$ for $\caction[\le 2 K + 1]$ and $\gs^{2 K - 1}$ for $\caction[\le 2 K]$.
The rule~\eqref{eq:all:diagonal} is the one found at the septic order: the leading-order value of the new vertex is the vertex of the original action minus what the previous orders already produce, and the reduced sextic and septic vertices are its cases $n = 6, 7$.
The reduced vertex itself is any symmetric form with this leading-order value satisfying the required conditions.
One choice is the form built from the diagrams, the elementary vertex minus the exchange diagrams of the lower auxiliary fields, with the reduced vertices at their nodes and with the same external strands (\cref{fig:all:reduced-vertex}); we write it only at the sextic and septic orders, where $\dvertex{6}$ subtracts $\vertex{5 \cdot 5}$ and $\dvertex{7}$ subtracts $\vertex{5 \cdot 5 \cdot 5}$ and the diagram $\dvertex{6} \cdot \vertex{5}$, whose reduced node turns the sign of the chain in~\eqref{eq:hs7:diagonal}.
Another is the polarization of the leading-order value, which exists at every order.

The proof is an induction on $n$, and three facts make it short.
First, the mismatch on the right of~\eqref{eq:all:diagonal} is a homogeneous polynomial of degree $n$ in $\Psi$: a polynomial because finitely many diagrams contribute to it, as noted after~\eqref{eq:all:action}, and homogeneous because of the scaling $\Psi \to t \Psi$, $\Sigma_k \to t^k \Sigma_k$, $\gs \to \gs / t$ noted above.
The coefficient of $\gs^{n - 2}$ in a quantity scaling as $t^2$ has degree $n$.
A homogeneous polynomial of degree $n$ is the leading-order value of a symmetric $n$-linear form, its polarization.
We read the polarization as a form on $\Hn{k} \times \Hn{k}$ for $n = 2 k$ and on $\Hn{k} \times \Hn{k} \times \Hs$ for $n = 2 k + 1$ by grouping its slots, with the same value at $\Psi^n$.
A symmetric form with the required value therefore exists.
Second, the coefficient of $\gs^{n - 2}$ in the eliminated action depends on $\dvertex{n}$ only through its leading-order value, with coefficient $1 / n!$, and on the lower vertices otherwise; this is the collapse lemma below.
By stationarity, the shifts of the lower solutions change the value only at order $\gs^{n}$.
The leading-order value of $\dvertex{n}$ is therefore fixed by the orders below it, and fixing it does not disturb them: the recursion is triangular, and the induction hypothesis that the vertices $\dvertex{m}$, $m < n$, reproduce the original action at the orders below $\gs^{n - 2}$ carries over to $n$.
Third, everything else in $\dvertex{n}$ is free as far as the matching goes.
This freedom feeds forward: the terms of $\dvertex{n}$ which vanish on the diagonal enter the exchange diagrams of the next orders and change the mismatches there, hence the leading-order values of the later vertices, which the recursion fixes in their turn.
At the sextic order, for instance, consider two vertices: the one built from the exchange diagram, and the polarization of its leading-order value, when both satisfy the assumptions.
They reproduce the original action at every order, but differ as forms and can give different leading-order values to $\dvertex{7}$.
Where the bare operators have a kernel, the kernel conditions restrict this freedom (\cref{sec:all:kernel}).

\paragraph{Collapse lemma.}

The lemma behind the second fact is the all-order form of the collapse of \cref{sec:hs:quartic}, where the $\Sigma_2$ exchange between two quartic sources gave back the contact term $\vertex{4}(\Psi^4)$.
It has three parts.
First, the solutions lie on the diagonal at leading order, $\Sigma_k^* = \Psi^k + O(\gs)$.
Since the kinetic term and the source of level $k$ come from the same form, the source at $\Sigma_{k - 1} = \Psi^{k - 1}$ is $\wprod{k}{k}(\Psi^k)$, so that $\wprod{k}{k}^{-1}$ applied to it gives $\Psi^k$ on the quotient; as the dressed operators reduce to the bare ones at $\gs = 0$, the claim follows by induction on $k$.
No symmetry of $\dvertex{2 k}$ beyond the one in its two groups of $k$ slots is used, so that the forms built from the diagrams and the polarized ones qualify equally.
Second, each rung of the chain~\eqref{eq:all:nested} costs $\gs^2$, so that the data of level $k$ first enter the value of the action at order $\gs^2 (\gs^2)^{k - 2} = \gs^{2 k - 2}$, and the odd vertex, which dresses the operator at order $\gs$, one order later.
Third, at its first order each level contributes its leading-order value alone, with the coefficient of the original action: at that order the two sources flanking the level-$k$ exchange are on the diagonal, and
\begin{equation}
  \label{eq:all:collapse}
  \frac{\gs^{2 k - 2}}{(2 k)!} \, \pair[k]{\wprod{k}{k}(\Psi^k)}{\wprod{k}{k}^{-1} \circ \wprod{k}{k}(\Psi^k)}
    = \frac{\gs^{2 k - 2}}{(2 k)!} \, \dvertex{2 k}(\Psi^{2 k}),
\end{equation}
the inverse cancelling against one of the two operators, as $\sprod{2}{2}^{-1}$ did in~\eqref{eq:hs4:telescoping}; the coefficient $1 / (2 k)!$ comes out of the rungs $(2 j)! / (2 j + 2)!$ of the chain, which telescope.
For the odd vertex, expanding the dressed operator to first order in $\gs$ between the same two sources gives $\tfrac{\gs^{2 k - 1}}{(2 k + 1)!} \dvertex{2 k + 1}(\Psi^{2 k + 1})$, as the quintic vertex entered in~\eqref{eq:hs5:constant}.
An exchange diagram of auxiliary fields with all its sources on the diagonal thus collapses to a contact term at every level, whereas the exchanges between interaction vertices, met from the sextic order on, do not collapse and are what the reduced vertices subtract.

\subsection{Recovered action and equations of motion}
\label{sec:all:integrating-out}

Expanding the value~\eqref{eq:all:value} in $\gs$ shows how the original action comes back.
Every term of the expansion is a tree diagram with $\Psi$ lines and auxiliary lines, the sources and the odd vertices at its nodes.
The diagrams whose sources all sit on the diagonal collapse to contact terms by the lemma: at order $\gs^{2 k - 2}$ the $\Sigma_k$ exchange between two diagonal sources gives $\dvertex{2 k}(\Psi^{2 k})$, and at order $\gs^{2 k - 1}$ the odd vertex on the $\Sigma_k$ line gives $\dvertex{2 k + 1}(\Psi^{2 k + 1})$, both with the coefficients of the original action.
The remaining diagrams at order $\gs^{n - 2}$ are the exchanges between interaction vertices, and by~\eqref{eq:all:diagonal} they are the terms by which $\dvertex{n}(\Psi^n)$ differs from $\vertex{n}(\Psi^n)$.
The contact term and the exchange terms therefore add up to $\vertex{n}(\Psi^n)$ at every order:
\begin{equation}
  \label{eq:all:recovered}
  \caction[\le 2 K + 1] \big|_{\Sigma = \Sigma^*}
    = S_{\le 2 K + 1} + O(\gs^{2 K}),
\end{equation}
which is the closure~\eqref{eq:all:closure} at $N = 2 K + 1$, and for the full tower the remainder is absent.
The construction gives back the original action through the two mechanisms of \cref{sec:hs} repeated at every level: the collapse of the diagonal exchanges and the subtraction of the others.

The equation of motion of $\Psi$ follows.
Since the auxiliary fields are stationary on their solution, their variation drops out, and the $\Psi$ equation of the cubic action with the tower eliminated is the equation of motion of the value~\eqref{eq:all:recovered}:
\begin{equation}
  \label{eq:all:eom-recovered}
  \mathcal E_\Psi \big|_{\Sigma = \Sigma^*}
    = Q_B \Psi + \sum_{n = 2}^{2 K} \frac{\gs^{n - 1}}{n!} \, \ell_n(\Psi^n) + O(\gs^{2 K}),
\end{equation}
with the products normalized as in \cref{sec:hs:action}; the remainder of the value is $\gs^{-2}$ times a function of $\gs \Psi$, so that its derivative has the same order.
\Cref{sec:bar-cobar} uses the result in this form.

The solutions correspond in the same way: wherever the dressed operators are invertible, a configuration solves the equations of the cubic action if and only if $\Sigma = \Sigma^*$ and $\Psi$ is a critical point of the value~\eqref{eq:all:value}.
At the quartic order these are exactly the solutions of the original action truncated at the quartic vertex, with $\Sigma_2 = \Psi^2$.
At higher orders they agree with those of the truncated original action as power series in $\gs$, up to the order of the truncation.
Since the action is the sum of a quadratic and a cubic part, its value at any solution is one third of its quadratic part, as for the cubic action of the open string.
When the target space is compact or without boundaries, the on-shell action of closed SFT vanishes~\cite{Erler:2022:ClosedStringField}, so that at a solution of the full theory the quadratic and cubic parts vanish separately,
\begin{equation}
  \frac{1}{2} \langle \Psi, Q_B \Psi \rangle
    = \sum_k \frac{\gs^{2 k - 2}}{(2 k)!} \, \dvertex{2 k}(\Sigma_k, \Sigma_k).
\end{equation}
The deviation from this identity measures the error of a truncation.

The solutions~\eqref{eq:all:saddles} are determined only modulo the kernels of the bare operators, and the statements above do not depend on the representative.
A shift $\Sigma_k \to \Sigma_k + N_k$ with $N_k \in \ker \wprod{k}{k}$ leaves the kinetic term of $\Sigma_k$ and its own source unchanged, by the symmetry of $\wprod{k}{k}$, and changes the action by the shifts of the odd coupling and of the source of $\Sigma_{k + 1}$:
\begin{equation}
  \label{eq:all:kernel-shift}
  \begin{aligned}
    &
    \frac{\gs^{2 k - 1}}{(2 k + 1)!} \Big[
      2 \, \dvertex{2 k + 1}(N_k, \Sigma_k, \Psi)
      + \dvertex{2 k + 1}(N_k, N_k, \Psi)
    \Big]
    \\ &
    \qquad
    + \frac{2 \gs^{2 k}}{(2 k + 2)!} \, \pair[k + 1]{\Sigma_{k + 1}}{\wprod{k + 1}{k + 1}(\Psi \otimes N_k)},
  \end{aligned}
\end{equation}
the last term being absent at the top of a truncated tower.
These are the only terms through which a kernel shift reaches the action and its equations.
We assume that the kernel conditions of \cref{sec:all:kernel} hold, so that these terms vanish: the equations and the value are the same for every representative, and the solution is stationary in every direction, the kernel directions included.

\subsection{Gauge transformations at all orders}
\label{sec:all:gauge}

The gauge transformation~\eqref{eq:hs1:gauge-psi} of the string field extends to the tower by the rule of the quintic and septic orders: each auxiliary field transforms as its solution, plus a rotation of its fluctuation which keeps its square invariant.
The sum of squares~\eqref{eq:all:squares} makes the rule explicit at every level at once: its first term is a function of $\Psi$ alone, and each square is a quadratic form in $\sigma_k$ with the operator $\kin{k}{\Psi}{}$.
Every square is invariant under an arbitrary variation $\delta_\Lambda \Psi$ if:
\begin{equation}
  \label{eq:all:gauge-deviations}
  \delta_\Lambda \sigma_k
    = - \frac{1}{2} \, \kin{k}{\Psi}{}^{-1} \circ \big( \delta_\Lambda \kin{k}{\Psi}{} \big) (\sigma_k),
  \qquad
  k \ge 2,
\end{equation}
and the field itself transforms as:
\begin{equation}
  \label{eq:all:gauge-fields}
  \delta_\Lambda \Sigma_k
    = \delta_\Lambda \Big[ \kin{k}{\Psi}{}^{-1} \circ \wprod{k}{k}(\Psi \otimes \Sigma_{k - 1}) \Big]
      + \delta_\Lambda \sigma_k,
\end{equation}
the first term being the variation of its conditional solution, through $\Psi$ and through $\Sigma_{k - 1}$, which transforms in its turn.
The laws are unique up to trivial symmetries (\cref{sec:further}).
The variation of the dressed operator is read off~\eqref{eq:all:nested}: it collects the variation of the odd coupling, $- \tfrac{\gs}{2 k + 1} \dvertex{2 k + 1}(\, \cdot \,, \, \cdot \,, \delta_\Lambda \Psi)$, those of the two sources flanking the exchange, and that of the inverse of the operator one level up, which contains the same three pieces.
The laws are rotations rather than implicit equations because each square is Gaussian in its fluctuation.
With them, the variation of the cubic action equals the variation of its value on the solution, off-shell, for any $\delta_\Lambda \Psi$:
\begin{equation}
  \label{eq:all:delta-S-equal}
  \delta_\Lambda \caction
    = \delta_\Lambda \Big[ \caction \big|_{\Sigma = \Sigma^*} \Big],
\end{equation}
exact in $\gs$ at a truncation, so that by~\eqref{eq:all:recovered} the action truncated at $K$ is invariant up to order $\gs^{2 K}$, the order of the truncation it reproduces, and the full action is invariant as a formal series.
This last step uses the gauge invariance of the original action, which rests on the relations among the string vertices; the matching itself does not.

The rotation is homogeneous in $\sigma_k$ and vanishes on the solution of the auxiliary equations, and the laws carry no gauge parameter of their own at any level: the auxiliary sector follows the solution and rotates its fluctuations.
Below the top level the dependence of the operator on $\Psi$ comes through the odd coupling and through the exchanges of the higher fields, at the top level through the odd coupling alone.
The rotation vanishes when the operator does not depend on $\Psi$, as for $\Sigma_3$ at the sextic order.
In the Batalin--Vilkovisky formalism, each fluctuation and its antifield form a pair, contractible up to a field redefinition, whose coefficient depends on $\Psi$.
A representative of the transformation law which makes the coefficient constant exists through order $\gs^4$, and we expect it at every order.
The closure of the gauge algebra, this representative, the master action and the question whether the cubic action is the gauge-fixed form of a theory with an independent $2$-string gauge parameter are summarized in \cref{sec:further}.

\subsection{Kernel conditions}
\label{sec:all:kernel}

The auxiliary fields live on quotients, $\Sigma_k$ on $\Hn[2]{k} / \ker \wprod{k}{k}$, and every term of the action must be a function on them.
At the quintic and septic orders this required the odd vertex to vanish on the kernel, and at the sextic order the source of the next field.
Besides its kinetic term, a kernel direction $N_k \in \ker \wprod{k}{k}$ enters three terms of the action: its own source, which pairs to zero with $N_k$ by the symmetry of $\wprod{k}{k}$, the odd coupling, and the source of $\Sigma_{k + 1}$.
The action descends to the quotients if the last two vanish on every kernel direction:
\begin{equation}
  \label{eq:all:kernel-condition}
  \begin{gathered}
    \forall N_k \in \ker \wprod{k}{k}, \;
    \forall A_k \in \Hn[2]{k}, \;
    \forall B \in \Hs_2:
    \\
    \dvertex{2 k + 1}(N_k, A_k, B)
      = 0,
    \qquad
    \wprod{k + 1}{k + 1}(B \otimes N_k)
      = 0,
  \end{gathered}
\end{equation}
for every $k \ge 2$.
The first clause is the condition of the odd orders and the second that of the even ones: at $k = 2$ the second clause is the vanishing of the sextic vertex on $\ker \sprod{2}{2}$,~\eqref{eq:hs6:kernel-condition}, since, for the form built from the diagrams, the exchange part of $\dvertex{6}$ vanishes there by the first clause.
The second clause says that a null direction of one level, tensored with any string field, is a null direction of the next: null directions propagate up the tower.
Every null vector of the kinetic operator of $\Sigma_k$ is a joint null vector of every vertex in which $\Sigma_k$ appears.
A kernel is meant in the square-integrable sense at fixed total momentum, as for $\sprod{2}{2}$ in \cref{sec:hs:quartic}.
Through the second clause, a null vector of $\sprod{2}{2}$ at a single total momentum would give the higher operators null vectors at every total momentum.
For continuous momenta these are plane waves, sharp in one relative momentum, and do not count as kernel directions; for discretized momenta they do.
At $k = 2$ the condition bears on the quintic and sextic vertices themselves; for $k \ge 3$ it bears in part on the choice of the reduced vertices, which are defined up to terms vanishing at leading order.
The choice has a limit: at $k = 2$ a reduced sextic vertex satisfying the second clause exists only if its leading-order value vanishes on the totally symmetric tensors in $\Hn[2]{3} \otimes \ker \sprod{2}{2} \otimes \Hs_2$.
The condition holds trivially when the bare operators have no kernel.
We do not try to settle here whether the string vertices satisfy it; in principle, this can be checked level by level in level truncation.

The conditions are what makes the quotient prescription consistent at every order.
The kernels in~\eqref{eq:all:kernel-condition} are those of the bare operators, while the operators inverted in the elimination are the dressed ones, and the quotient is meaningful only if $\ker \wprod{k}{k}$ stays in the radical of the dressed operator for every $\Psi$, so that every dressed operator descends to the same quotient.
Read order by order in~\eqref{eq:all:scaling}, this is the condition: the first clause makes the term of order $\gs$ vanish on the kernel and the second the exchange term.
Generically the second clause is also necessary: a kernel direction it misses is lifted at order $\gs^2$.
When a clause fails along a kernel direction, that direction has no kinetic term at order $\gs^{2 k - 2}$ and acquires one from the odd coupling at the next order, or from the exchange one order further.
The failure does not show in the dressed operator of the level below at $\gs = 0$, so that the invertibility of the quartic form on its quotient does not certify the higher levels.
The construction as presented then breaks down and must be refined.\footnote{%
  A way out may exist when a condition fails.
  The constraint $\Sigma_k = \Psi \otimes \Sigma_{k - 1}$ can be imposed by a multiplier instead of a kinetic term, either on the whole field, in which case nothing is inverted and no kernel condition arises, but the exchange structure is lost, or on the kernel directions only.
  This hybrid reproduces, at the sextic order, the original action through order $\gs^4$ for every complement of $\ker \sprod{2}{2}$, with a reduced vertex that depends on the complement (shown at first order in the quintic coupling for a scalar model in \cref{app:scalar-hs}).
  The gauge transformations of the hybrid are left for future work.
}

\section{Worldsheet perspective}
\label{sec:worldsheet}

The cubic action is built from three kinds of objects: the string vertices of closed string field theory, the composite vertices such as $\vertex{5 \cdot 5}$, and the inverse vertices $\wprod{k}{k}^{-1}$ that propagate the auxiliary fields.
The first are integrals over the vertex regions $\mathcal V_{0,n}$ of moduli space; the second and the third need an interpretation on the worldsheet, and this section gives it as far as we have one.
We focus on $\Sigma_2$ and the inverse quartic vertex $\sprod{2}{2}^{-1}$, and we examine whether the exchange of $\Sigma_2$ between two vertices can be read as a sewing of Riemann surfaces.

In some cases, where we want to illustrate or check a statement, we consider the $\group{SL}(2, \C)$ vertices of~\cite{Erbin:2023:SL2QuarticVertex,Erbin:toappear:SL2VerticesClosed}, with Möbius local coordinates and a stub parameter $\beta$, and call them collectively the $\group{SL}(2, \C)$ vertex; numbers are quoted at $\beta = 3 / 2$ and $2$.
The coordinate discs of the $\group{SL}(2, \C)$ quartic vertex are non-overlapping only for $\beta \ge 1 + 1 / \sqrt{2}$~\cite{Erbin:toappear:SL2VerticesClosed}: at $\beta = 2$ the section is a string vertex, while at $\beta = 3 / 2$ the numbers are those of a symmetric section that matches the glued coordinates on the seams.
Only the exactness of the rigid gluing and the explicit inverse it gives, and the Sonoda--Zwiebach sewing of \cref{sec:ws:onshell}, hold on-shell; the dressed gluing and the counting hold off-shell.
Throughout we work at tree level and genus zero, on inputs of finite level, with $\Sigma_2$ on the quotient of \cref{sec:hs:quartic}.

\Cref{sec:ws:fields} fixes the dictionary: what a $\Sigma_2$ slot and a $\Sigma_2$ line are on the worldsheet.
\Cref{sec:ws:inverse} collects what constrains the inverse $\sprod{2}{2}^{-1}$, the one element of the Feynman rules without counterpart in closed string field theory, and its expansion in chains of vertices.
\Cref{sec:ws:moduli} shows that the moduli count of every diagram gives the dimension of the moduli space of its external punctures once the line is counted with negative moduli.
\Cref{sec:ws:gluing} gives the mechanism: the quartic form is a gluing of two copies of an asymmetric cubic vertex along a circle, over a region of the plumbing parameter whose two boundaries, the seams, are shared with the Feynman regions of the exchange diagrams.
At the seams the local coordinates of the glued surface must agree with those of the exchanges for the amplitudes to fit together off-shell, and at the outer seam fixed coordinates do not: the coordinates have to move with the plumbing parameter, and this motion dresses the plumbing operator.
\Cref{sec:ws:onshell} shows that on-shell the dressing drops and the gluing becomes a sewing of the kind Sonoda and Zwiebach studied.
\Cref{sec:ws:six-point} turns to the composite vertex at six points and states the covering conjecture.

The dictionary and the counting of \cref{sec:ws:fields,sec:ws:moduli} extend to any $\Sigma_k$ with $k \ge 3$, but not directly the gluing of \cref{sec:ws:gluing}, since its patches then carry moduli.

\subsection{Fields and lines on the worldsheet}
\label{sec:ws:fields}

\begin{figure}[htp]
  \centering
  \includegraphics[width=0.3\linewidth]{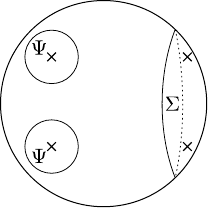}
  \caption{The coupling $\vertex{4}(\Sigma_2, \Psi^2)$ on the worldsheet: the four-punctured sphere of the quartic vertex, with two punctures carrying $\Psi$ and the two punctures of the $\Sigma_2$ slot grouped together.}
  \label{fig:ws:dictionary}
\end{figure}

A vertex $\vertex{n}$ of closed string field theory is the integral over the vertex region $\mathcal V_{0,n}$ of the correlator of $n$ states, one per puncture, with one antighost insertion per real modulus (\cref{sec:hs:action}).
The vertices of the cubic action are the same integrals, with the punctures grouped by the field they belong to.
The coupling $\vertex{4}(\Sigma_2, \Psi^2)$ is the quartic vertex, with its region and its measure, in which two of the four punctures carry the two strands of one field:
\begin{equation}
  \label{eq:ws:slot}
  \vertex{4}(\Sigma_2, \Psi^2)
    = \sum_{r,s} \sigma_{rs} \, \vertex{4}(\phi_r, \phi_s, \Psi, \Psi)
\end{equation}
where we have expanded $\Sigma_2$ on a basis $\{ \phi_r \}$ of $\mc H$:
\begin{equation}
  \label{eq:wq:sigma2-expansion}
  \Sigma_2
    = \sum_{r,s} \sigma_{rs} \, \phi_r \otimes \phi_s.
\end{equation}
A $\Sigma_2$ slot is thus a pair of punctures that belong to one field, and the state on the pair is in general entangled, since the components~\eqref{eq:hs4:spacetime} of $\Sigma_2$ need not factorize.
\Cref{fig:ws:dictionary} shows the quartic vertex with this grouping.

A $\Sigma_2$ line joins a $\Sigma_2$ slot of one vertex to a $\Sigma_2$ slot of another.
By the Feynman rules of \cref{fig:hs4:feynman-rules} it carries the inverse $\sprod{2}{2}^{-1}$ and the $2$-string pairing, which is the single-string pairing strand by strand, and we draw it as two strands, one per puncture of the slot.
The pairing is the contraction that also enters the propagator of $\Psi$; the inverse of the quartic form is the one new element, and the next two subsections describe it.

The strands make one bookkeeping fact visible.
A chain of $k$ quartic vertices joined by $\Sigma_2$ lines is a tree when each line counts as one edge, and its power of $\gs$, its number of moduli and its ghost number are those of a tree.
Resolve each line into its two strands and sew each strand as a pair of punctures: the chain becomes a connected surface of genus $k - 1$, sewn along $2 (k - 1)$ pairs of punctures.
We call it the resolved surface of the diagram.
By the additivity of the Euler characteristic under sewing, a diagram with $L$ loops whose edges $e$ carry $k_e$ strands has resolved genus
\begin{equation}
  \label{eq:ws:genus}
  g_{\mathrm{res}}
    = L + \sum_e (k_e - 1) ,
\end{equation}
so that a $\Sigma_2$ line raises the resolved genus by one and the loop number by zero.
No coupling counts this genus, since the necks that produce it carry no integrated length, and the expansion in $\gs$ organizes the diagrams by $L$.\footnote{%
  In a matrix model the double lines fix the genus of the fat graph and the expansion in $1 / N$ is an expansion in that genus; here the strands fix a genus that no parameter counts.
}

\subsection{The inverse of the quartic form}
\label{sec:ws:inverse}

\begin{figure}[htp]
  \centering
  $\vcenter{\hbox{\includegraphics[scale=0.9]{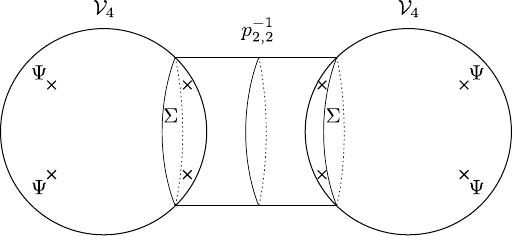}}}
    \quad = \quad
    \vcenter{\hbox{\includegraphics[scale=0.9]{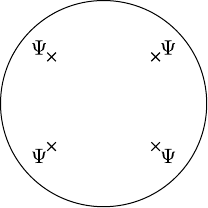}}}$
  \caption{The collapse~\eqref{eq:ws:collapse} on the worldsheet. Left, two four-punctured spheres of the vertex region, each with two $\Psi$ punctures, joined through $\sprod{2}{2}^{-1}$ along the circles enclosing their $\Sigma_2$ slots; right, one four-punctured sphere of the vertex region with the four $\Psi$ punctures.}
  \label{fig:ws:gluing-p22}
\end{figure}

The propagator of the auxiliary field is the inverse of a vertex, and this subsection describes what that means on the worldsheet.
Two facts constrain the answer before any construction.
The first is the ghost number.
We count ghost numbers as those of operators on $\Hn{2}$, with the $2$-string pairing as reference: the quartic form is nonzero on pairs of $2$-string states of total ghost number eight, two less than the ten that the $2$-string pairing~\eqref{eq:conv:pairing-n} requires.
The deficit is the two antighost insertions of the measure on $\mathcal V_{0,4}$, so that $\Ngh(\sprod{2}{2}) = + 2$ and
\begin{equation}
  \label{eq:ws:ghost-obstruction}
  \Ngh\big( \sprod{2}{2}^{-1} \big)
    = - 2 .
\end{equation}
The second is the collapse~\eqref{eq:hs4:telescoping}, which reads, for $A, B \in \Hn{2}$,
\begin{equation}
  \label{eq:ws:collapse}
  \vertex{4}\big( A, \sprod{2}{2}^{-1} \circ \sprod{2}{2}(B) \big)
    = \vertex{4}(A, B),
\end{equation}
a $\Sigma_2$ line between two quartic vertices is one quartic vertex, so that whatever represents the inverse must cancel two antighosts and two moduli of its neighbors.
This is represented geometrically in \cref{fig:ws:gluing-p22}.
In this sense, the inverse of the quartic form cannot be described by a gluing of surfaces with the standard plumbing fixture.

A representation of the inverse that expands in chains of vertices is the Neumann series.
Split the quartic form into its level-diagonal part and the rest,
\begin{equation}
  \label{eq:ws:neumann}
  \sprod{2}{2}
    = D - R,
  \qquad
  \sprod{2}{2}^{-1}
    = \sum_{k \ge 0} (D^{-1} R)^k \, D^{-1},
\end{equation}
with $D$ invertible on the quotient; the series converges when the spectral radius of $D^{-1} R$ is below one, which we assume.
The term of order $k$ is a chain of $k$ level-off-diagonal quartic vertices threaded on $k + 1$ level-diagonal $\Sigma_2$ lines $D^{-1}$: a discrete sum over chains of four-punctured spheres with one algebraic factor $D^{-1}$ per line, rather than an integral over a continuous family of surfaces.
The split is a choice, like a subtraction scheme, and the sum does not depend on it where it converges.
For the $\group{SL}(2, \C)$ vertex on the block of pairs of zero-momentum tachyons and ghost dilatons, the spectral radius is $0.37$ at $\beta = 2$ and $0.81$ at $\beta = 3 / 2$ (the computation is part of the material described in \cref{sec:discussion}).
At other momenta and on the other blocks the convergence is open.
The series in $\gs$ of \cref{sec:hs:quintic}, which dresses $\sprod{2}{2}^{-1}$ with quintic vertices, is a Neumann series of the same kind.

\subsection{Counting moduli}
\label{sec:ws:moduli}

A $\Sigma_2$ line carries no modulus to integrate, and we now show that the moduli count of every diagram makes sense once the line is counted with a negative number of moduli, as the ghost number~\eqref{eq:ws:ghost-obstruction} suggests.
We assign to every vertex the real dimension of its region,
\begin{equation}
  \label{eq:ws:dhat}
  \hat d(\vertex{n})
    := \dim_\R \mathcal V_{0,n}
    = 2 n - 6 ,
\end{equation}
extended to $n \le 2$ by the same formula, and we require the count to be additive under gluing.
Kinetic operators count as the vertices they correspond to, $\hat d(Q_B) = \hat d(\vertex{2}) = - 2$ and $\hat d(\sprod{2}{2}) = \hat d(\vertex{4}) = + 2$, and an inverse counts minus what it inverts.
For the string field this gives $\hat d(Q_B^{-1}) = + 2$, the two real moduli of the plumbing cylinder: its length, integrated with $b_0^+ / L_0^+$, and its twist, integrated with the level-matching insertion $b_0^- \delta_{L_0^-, 0}$.
For the auxiliary field it gives
\begin{equation}
  \label{eq:ws:negative-moduli}
  \hat d\big( \sprod{2}{2}^{-1} \big)
    = - 2 ,
\end{equation}
two moduli removed rather than added.
This is the dimension-counting form of~\eqref{eq:ws:ghost-obstruction}, the two gradings agreeing on $\sprod{2}{2}$ and its inverse, and the sense in which a $\Sigma_2$ line carries negative moduli.
On the resolved surface~\eqref{eq:ws:genus}, $\hat d$ is the dimension of its moduli space, the number of antighost zero modes minus the number of ghost zero modes, minus six per handle, the shift of the ghost-number anomaly.
With~\eqref{eq:ws:negative-moduli} the count gives the dimension of the moduli space at every order:
\begin{equation}
  \label{eq:ws:closure}
  \begin{aligned}
    \hat d\big( \vertex{4} \, \sprod{2}{2}^{-1} \, \vertex{4} \big)
      &
      = 2 - 2 + 2
      = 2
      = \dim_\R \mathcal M_{0,4},
    \\
    \hat d\big( \vertex{4} \, \sprod{2}{2}^{-1} \, \vertex{5} \, \sprod{2}{2}^{-1} \, \vertex{4} \big)
      &
      = 2 - 2 + 4 - 2 + 2
      = 4
      = \dim_\R \mathcal M_{0,5},
    \\
    \hat d\big( \vertex{5} \, \sprod{2}{2}^{-1} \, \vertex{5} \big)
      &
      = 4 - 2 + 4
      = 6
      = \dim_\R \mathcal M_{0,6},
  \end{aligned}
\end{equation}
and for longer chains it reduces to $2 N - 6$ for $N$ external legs: the diagrams of the cubic theory have, formally, the dimension of the moduli space of their external punctures, as the diagrams of closed string field theory do.

The count is necessary for a geometric reading and never sufficient, as two examples show.
The chain $\vertex{4} \, \sprod{2}{2}^{-1} \, \vertex{4}$ has the dimension of $\mathcal M_{0,4}$ and equals $\vertex{4}$ by the collapse~\eqref{eq:ws:collapse}: it covers the single region $\mathcal V_{0,4}$, the line removing the two moduli that the second vertex adds.
Conversely, the double tube, two plumbing cylinders whose lengths and twists are integrated independently, has the same count $\hat d = - 2$, but it gives a torus and not a four-punctured sphere.

At six points the count is the first that no collapse explains.
The composite vertex $\vertex{5 \cdot 5}$ of~\eqref{eq:hs5:exchange-vertex} has the dimension of $\mathcal M_{0,6}$, and its line joins two interaction vertices, so that it does not collapse (\cref{sec:hs:sextic:vertex}).
Whether it covers a region of $\mathcal M_{0,6}$, and covers it once, is an open question, to which we return in \cref{sec:ws:six-point}.

\subsection{The quartic form as a dressed gluing of cubic vertices}
\label{sec:ws:gluing}

We now give the $\Sigma_2$ line a geometric meaning.

As shown in \cref{fig:ws:gluing-p22}, the gluing using $\sprod{2}{2}^{-1}$ is performed by cutting the two four-punctured spheres of the vertex region along a circle that encloses the two punctures of one $\Sigma_2$ slot, and then inserting $\sprod{2}{2}^{-1}$.
The state on the circle is a single string state in $\mc H$, the operator product of the two strands, and the quartic form becomes the pairing of two such states through a plumbing cylinder.
This leads us to consider a generalized gluing of two asymmetric cubic vertices, so that we can use the machinery developed by Sonoda and Zwiebach~\cite{Sonoda:1990:CovariantClosedString}.\footnotemark{}
\footnotetext{%
  Such an asymmetric cubic vertex, denoted by $\vertex{3}'$, also appeared in the proof of background independence~\cite{Sen:1994:ProofLocalBackground,Sen:1994:QuantumBackgroundIndependence}.
}%
As recalled in the introduction, they asked whether a cubic vertex sewn by the plumbing fixture over a domain larger than the unit disc could produce the four-point amplitude by itself.
We take up the construction and show that the quartic vertex is such a sewing of two copies of a cubic vertex, over the vertex region and dressed by the motion of its local coordinates.
We first describe the state that the circle carries, then the construction with fixed local coordinates.
We then state the theorem for a quartic vertex given as a section of local coordinates over $\mathcal V_{0,4}$, whose motion with the plumbing parameter dresses the plumbing operator.
We end with a recipe for an explicit vertex and a remark on the inverse.

\paragraph{OPE for the auxiliary field.}

Consider a single surface contributing to \eqref{eq:ws:slot}, such that the modulus of the four-punctured sphere is fixed to some $z_4$.
Expanding the auxiliary field $\Sigma_2$ as \eqref{eq:wq:sigma2-expansion}, the basis states $\phi_r$ and $\phi_s$ are inserted at the two punctures of the slot, in the local coordinate patch.
We can then use the OPE between both states\footnotemark{}
\footnotetext{%
  If the higher-order vertices are built from the $2$-product, as in Witten's open SFT and in the NS sector of open superstring field theory, we can instead use this product to provide a map $\mc H^{\wedge 2} \to \mc H$.
  This is the case for open SFT with stubs~\cite{Erbin:2024:OpenStringStub}.
}%
\begin{equation}
  \phi_r(z_r) \phi_s(z_s)
    = \sum_t C_{rs}{}^{t}(z_r - z_s) \, \phi_t(z_s)
\end{equation}
to describe them in terms of a single state.
This corresponds to the state on the gluing circle of \cref{fig:ws:gluing-p22}, and we write in a simplified way:
\begin{equation}
  \label{eq:ws:circle-map}
  \ket{\varphi_{2,1}(\phi_r, \phi_s)}
    := f_1 \circ \phi_r(0) \; f_2 \circ \phi_s(0) \, \ket{0},
\end{equation}
with $f_1, f_2$ the local coordinate maps at the two punctures of the slot read in the frame of the circle and the product radially ordered.
Its level expansion is the operator product expansion of the two operators, and it is located at some puncture of the surface, with associated local coordinates $f_\varphi$ induced by $f_1$ and $f_2$.

We call $\varphi_{2,1} \colon \Hn{2} \to \Hs$, extended by linearity, the merging map and $\varphi_{2,1}(\Sigma_2)$ the merged state.
It has ghost number four; its zero mode carries the total momentum $P$ and its level content the relative momentum, so that we expect nothing to be lost in the merging: the map should be injective on inputs of finite level.
It is not level matched, since two punctures inside a circle are not a rotation-invariant configuration.
On a tachyon pair, for instance, nothing makes the $L_0^- = \pm 2$ components vanish; only the even-$L_0^-$ components survive on a graded-symmetric $\Sigma_2$ when the two coordinates are related by $f_2 = - f_1$, as for the vertex~\eqref{eq:ws:half} of the next paragraph.
This means that we have to define a cubic vertex $\vertex{3}'$ that takes two level-matched states of ghost number $2$ and outputs a general (not level-matched) state of ghost number $4$, so that the vertex is not symmetric.

The grouping of punctures by fields of different ghost numbers obtained by OPE suggests a colored operad of moduli spaces, with one color per number of strands, whose construction we leave open.

\paragraph{The asymmetric cubic vertex.}

Consider the asymmetric cubic vertex $\vertex{3}'$ with $\group{SL}(2, \C)$ local coordinates: a three-punctured sphere with two ordinary punctures at $z = 0, \infty$ and a special puncture at $z = 1$, which carries the merged state, with local coordinates
\begin{equation}
  \label{eq:ws:half}
  z
    = \frac{v_1}{\gamma' + \lambda v_1},
  \qquad
  z
    = \frac{\gamma' + \lambda v_2}{v_2},
  \qquad
  z
    = \frac{2 \beta - w}{2 \beta + w},
\end{equation}
where $\beta > 0$ is the stub parameter and $\gamma'$ and $\lambda$ are a scale and a shape parameter of the ordinary coordinates.
In the $w$-plane the ordinary punctures sit at $w = \pm 2 \beta$.
For $\beta < \beta_\circ := (1 + \sqrt{2}) / 2$, the Feynman discs of different channels overlap, so we consider only the cases $\beta \ge \beta_\circ$.

\begin{figure}[htp]
  \centering
  \includegraphics[width=0.55\linewidth]{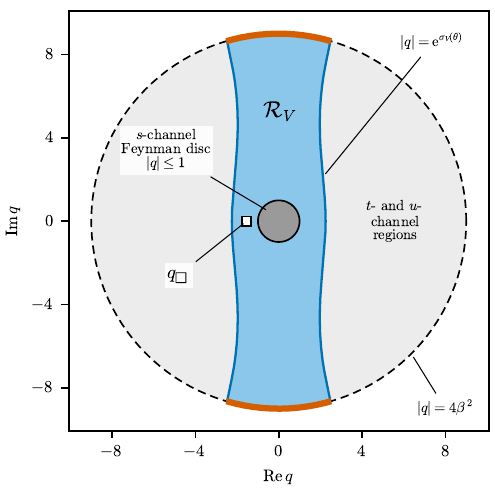}
  \caption{The plumbing plane of the gluing: the $s$-channel disc, the region $\mathcal R_V$ and the outer annulus. At $\beta = 3 / 2$, the unit disc $\abs{q} \le 1$ (dark gray) is the $s$-channel Feynman region and the region $\mathcal R_V$ (blue) lies between $\abs{q} = 1$ and $\abs{q} = \e^{\sigma_V(\theta)}$.
      The outer annulus (light gray), up to the dashed circle $\abs{q} = 4 \beta^2$, covers the $t$- and $u$-channel regions.
      $\mathcal R_V$ reaches that circle on the two thick arcs, and the open square marks the point $q_\square$ of~\eqref{eq:ws:square-point} on the ray $\theta = \pi$.}
  \label{fig:ws:region}
\end{figure}

At $\gamma' = \beta$ and $\lambda = 1 / 2$, $\vertex{3}'$ is the cubic vertex $\vertex{3}$ of the theory.
Glue two copies at their special punctures by the plumbing fixture
\begin{equation}
  \label{eq:ws:plumbing}
  w_1 w_2
    = q,
  \qquad
  q
    = \e^{- s + \I \theta},
\end{equation}
with $s$ the neck length and $\theta$ the twist; for $\abs{q} < 4 \beta^2$ the circle $\abs{w_1} = \sqrt{\abs{q}}$ separates the two punctures of the first copy from those of the second.
The glued sphere has the cross-ratio, in the frame $(0, 1, \infty, Z)$,
\begin{equation}
  \label{eq:ws:cross-ratio}
  Z(q)
    = \frac{(q + 4 \beta^2)^2}{16 \beta^2 \, q} .
\end{equation}
The map is injective on $\abs{q} < 4 \beta^2$.
The unit disc $\abs{q} \le 1$ is the $s$-channel Feynman region, and the pre-image of the vertex region is, as found numerically at $\beta = 3 / 2$, an annulus with an angle-dependent outer radius (\cref{fig:ws:region}),
\begin{equation}
  \label{eq:ws:region}
  \mathcal R_V
    := Z^{-1}(\mathcal V_{0,4})
    = \{\, 1 < \abs{q} < \e^{\sigma_V(\theta)} \,\},
\end{equation}
mapped bijectively up to measure zero onto $\mathcal V_{0,4}$, the outer annulus $\e^{\sigma_V} \le \abs{q} < 4 \beta^2$ covering the $t$- and $u$-channel Feynman regions.

The function $\sigma_V$ is $\pi$-periodic, with, for every $\beta \ge \beta_\circ$,
\begin{equation}
  \sigma_V(\pi)
    = 2 \ln \frac{2 \beta (2 \beta - 1)}{2 \beta + 1},
\end{equation}
and reaches $\ln 4 \beta^2$ on two arcs of the circle $\abs{q} = 4 \beta^2$, which $Z$ maps into the real segment $0 < Z < 1$.

The region depends on $\beta$ alone, not on $(\gamma', \lambda)$, and it is determined by the vertex: the one circle that $\sprod{2}{2}$ prescribes reaches all of $\mathcal V_{0,4}$, with no freedom of splitting the region among channels.
Nothing collides on the two arcs; the singularities of the integrand, the cross-channel collisions, sit at $\theta = 0, \pi$ on $\abs{q} = 4 \beta^2$, outside the closure of $\mathcal R_V$.

We have described only the bosonic part of the gluing.
We next describe what happens if we consider the usual insertion $b_0^+ b_0^-$, and show that it works only on-shell, as expected from~\cite{Sonoda:1990:CovariantClosedString}.
We then explain how to correct the gluing to describe the operation corresponding to $\sprod{2}{2}$ off-shell.

\paragraph{On-shell gluing.}

With the coordinates of the two copies fixed, the only deformations of the glued surface are those of $q$, a dilation and a rotation of the circle, whose antighost insertions are the zero modes in the frame of the circle,
\begin{equation}
  \label{eq:ws:schiffer}
  b(\pd_s)
    = - b_0^+,
  \qquad
  b(\pd_\theta)
    = \I \, b_0^- ,
\end{equation}
and the gluing over $\mathcal R_V$ defines the bilinear form
\begin{equation}
  \label{eq:ws:p-circ}
  \begin{gathered}
    \sprod{2}{2}^{\mathrm{rig}}(\Sigma_2, \Sigma_2')
      := - \I \mathcal N \int_{\mathcal R_V} \dd s \, \dd\theta \;
        \big\langle \varphi_{2,1}(\Sigma_2') \big| \, b_0^+ b_0^- \, \mathcal S(q) \, \big| \varphi_{2,1}(\Sigma_2) \big\rangle,
    \\
    \mathcal S(q)
      := \e^{- s L_0^+} \e^{\I \theta L_0^-},
  \end{gathered}
\end{equation}
in the BPZ inner product, with $\mathcal N$ the normalization of the measure, fixed by the propagator on the unit disc.
We call it the rigid gluing.

At the inner seam $\abs{q} = 1$ the glued surface is the $s$-channel exchange, and a vertex consistent with the exchanges carries the coordinates of the two cubic vertices there.
At the outer seam $\abs{q} = \e^{\sigma_V(\theta)}$ the same surface is the $t$- or $u$-channel exchange at unit plumbing modulus.
The coordinates of the two descriptions must agree up to a phase for the vertex and the exchanges to fit together off-shell with level-matched external states; this is the condition that Sonoda and Zwiebach impose for the BRST invariance of the amplitude~\cite[eq.~(3.49)]{Sonoda:1990:CovariantClosedString}.

The condition fails at first order: at $\theta = \pi$ the logarithms of the mapping radii of the two descriptions differ, at all four punctures, by
\begin{equation}
  \label{eq:ws:seam-mismatch}
  \ln \frac{4 \beta^2 - 1}{4 \beta} ,
\end{equation}
exactly, for every $\beta \ge \beta_\circ$; it vanishes only at $\beta_\circ$, where the region pinches at $\theta = \pi$.
At eleven other seam points at $\beta = 3 / 2$ the mismatch is of the same size.
It does not depend on the coordinates at the ordinary punctures, so that no gluing of fixed copies of~\eqref{eq:ws:half}, over any region reaching the outer seam and for any $(\gamma', \lambda)$, is the quartic form off-shell.

The seam condition thus requires more than fixed coordinates can give, and the off-shell form needs a more general gluing: its local coordinates move with $q$ and join the coordinates of the exchanges at the seams, and the input operators carry the mismatch.
The logarithm of their mapping radius at the outer seam at $\theta = \pi$ is minus~\eqref{eq:ws:seam-mismatch} for the $\group{SL}(2, \C)$ vertex (numerically, at $\beta = 3 / 2$ and $2$).

\paragraph{Off-shell gluing.}

A quartic vertex is a section of local coordinates over the vertex region: at each $Z \in \mathcal V_{0,4}$, four maps $f_i(\cdot \,; Z)$ from the unit disc to the sphere with punctures at $0, 1, \infty, Z$, and its form is the correlator with the antighost insertions of the Schiffer vectors of the tangent vectors~\cite{Zwiebach:1993:ClosedStringField}.
Read in the frame of the glued sphere at $q$, through the inverse of the Möbius map that sends its first three punctures to $0, 1, \infty$, the coordinates become local coordinate maps $F_i(\cdot \,; q)$ at the four punctures, defined near the origin of the disc, which we split as
\begin{equation}
  \label{eq:ws:split}
  F_i(\cdot \,; q)
    = h_i \circ g_i(q),
\end{equation}
with $h_i$ the fixed coordinate of the copy at that puncture and $g_i(q)$ a holomorphic map defined near the origin of the disc and fixing it.
By the transformation law of a vertex operator under a change of local coordinate, $F_i \circ A_i(0) = h_i \circ (U_{g_i} A_i)(0)$, with $U_{g_i}$ the exponential of $L_{n \ge 0}$ and $\bar L_{n \ge 0}$ that represents $g_i$, so that the state on the circle at $q$ is the merged state of the fixed copy on the inputs $U_{g_i} A_i$.
The tangent vectors $\pd_s$ and $\pd_\theta$ of the family act on the neck and on the four coordinates at once, and their antighost insertions are the zero modes~\eqref{eq:ws:schiffer} dressed by a contour in each disc:
\begin{equation}
  \label{eq:ws:dressing}
  \begin{gathered}
    b(\pd_s)
      = - b_0^+ + \sum_{i = 1}^4 B_i[\pd_s g_i],
    \qquad
    b(\pd_\theta)
      = \I \, b_0^- + \sum_{i = 1}^4 B_i[\pd_\theta g_i],
    \\
    B_i[\pd_x g_i]
      := b^{(i)} \Big[ \frac{\pd_x g_i}{g_i'} \Big],
  \end{gathered}
\end{equation}
with $b^{(i)}[v]$ the antighost contour in the frame of the $i$-th input weighted by the vector field $v$, acting on $A_i$ before $U_{g_i}$.
  Here $v = \pd_x g_i / g_i'$, for $x = s, \theta$, is the vector field on the disc of the $i$-th input whose flow generates the variation of $g_i$, $g_i(W + \epsilon \, v(W)) = g_i(W) + \epsilon \, \pd_x g_i(W)$ to first order.
Since $v$ vanishes at the puncture, the contour contains the modes $b_k$ with $k \ge 0$ only.
The quartic form is then the dressed gluing
\begin{multline}
  \label{eq:ws:recipe-plumbing}
  \sprod{2}{2}(A_1 \otimes A_2, A_3 \otimes A_4)
    = \mathcal N \int_{\mathcal R_V} \dd s \, \dd\theta \;
    \\
      \times \big\langle \varphi_{2,1}(U_{g_3} A_3, U_{g_4} A_4) \big| \, b(\pd_s) \, b(\pd_\theta) \, \mathcal S(q) \, \big| \varphi_{2,1}(U_{g_1} A_1, U_{g_2} A_2) \big\rangle ,
\end{multline}
for any fixed coordinates $h_i$ and inputs of finite level, provided the section is piecewise $C^1$ in $Z$; the constant is the $\mathcal N$ of~\eqref{eq:ws:p-circ}, since the form is one $2$-form on $\mathcal M_{0,4}$ and the unit disc gives the $s$-channel exchange.
The inputs of this formula are the section, the transformation law and the Schiffer formalism.
The $q$-derivative of the family $q \mapsto (\mathcal S(q), U_{g_i(q)})$ has one term from the neck and one from each disc, each the stress-tensor insertion of a vector field, so that the sum of these vector fields is a Schiffer vector of the tangent vector.
  The antighost insertion of a Schiffer vector depends only on the tangent vector it represents~\cite{Zwiebach:1993:ClosedStringField}.
For the $\group{SL}(2, \C)$ vertex we assume this regularity: that its section is real-analytic on each tile of $\mathcal V_{0,4}$ and continuous across the arcs where the tiles meet, which a $2$-form integral does not see.
The choice of the fixed coordinates $h_i$ is free: the surface states and the total antighost insertions are fixed by the vertex, and another choice moves a $q$-independent factor into the plumbing operator.
The canonical choice is the coordinates of $\vertex{3}$.
With these coordinates and a section that joins the coordinates of the $s$-channel exchange at the inner seam, the input operators reduce to the identity on level-matched inputs there; the $\group{SL}(2, \C)$ vertex joins them numerically.

This formula gives the form itself; its inverse, which the Feynman rules need, is constrained by~\eqref{eq:ws:ghost-obstruction} and~\eqref{eq:ws:collapse}, is expanded by the Neumann series~\eqref{eq:ws:neumann} where that converges, and is explicit on-shell (\cref{sec:ws:onshell}); what its construction would take off-shell is discussed at the end of this subsection.

\paragraph{Recipe for an explicit vertex.}

For a given quartic vertex, Möbius or not, the dressed gluing is assembled as follows.
\begin{enumerate}
  \setlength{\itemsep}{0.5ex}

  \item \emph{Transport the section.}
    At each $q \in \mathcal R_V$, pull the four coordinates $f_i(\cdot \,; Z(q))$ back to the glued sphere by the inverse of the Möbius map that sends the punctures of the first copy, at $u = \pm 1 / 2 \beta$ in the frame $u = 1 / w_1$ of the circle, and those of the second, at $u = \pm 2 \beta / q$, to $0, 1, \infty, Z(q)$; this gives the maps $F_i$, with nothing chosen.

  \item \emph{Split the coordinates.}
    Choose the fixed coordinates $h_i$ and read off $g_i(q) = h_i^{-1} \circ F_i(\cdot \,; q)$.
    For the $\group{SL}(2, \C)$ vertex the maps $g_i$ are Möbius: relative to the canonical copy, $g_i(W) = \rho W / (1 + \nu W)$ with $\nu = \rho \, (\lambda(q) - 1 / 2) / \beta$, where $\rho(q) = F_i'(0; q) / h_i'(0)$ is the mapping radius of the coordinate of the vertex relative to that of the copy and $\lambda(q)$ is read off the pole of $g_i$.
    One pair $(\lambda, \rho)(q)$ serves the four legs, by the symmetry of the section under relabeling of the punctures, and
    \begin{equation}
      \label{eq:ws:input-operator}
      U_{g_i}
        = \rho^{L_0} \bar\rho^{\bar L_0} \, \e^{- \nu L_1 - \bar\nu \bar L_1} .
    \end{equation}
    For a section that is not Möbius, $U_{g_i}$ is $\rho^{L_0} \bar\rho^{\bar L_0}$ times an exponential of the $L_{n \ge 1}$ and $\bar L_{n \ge 1}$ whose coefficients follow order by order from the Taylor coefficients of $g_i / \rho$~\cite{LeClair:1989:StringFieldTheory-1,Rastelli:2001:TachyonPotentialsStar}.
      On inputs of level at most $L$ only $L_{n \le L}$ act, so that the $(L + 1)$-jet of $g_i$, with its first derivatives in $s$ and $\theta$, is all a numerical section has to supply.

  \item \emph{Dress the antighosts.}
    For the Möbius maps $g_i$ and level-matched inputs the contours $B_i$ of~\eqref{eq:ws:dressing} are
    \begin{equation}
      \label{eq:ws:dressing-moebius}
      B_i[\pd_x g_i]
        = (\pd_x \ln \abs{\rho}) \, b_0^{+ (i)}
          - \frac{\rho}{\beta} \, \pd_x \lambda \; b_1^{(i)}
          - \frac{\bar\rho}{\beta} \, \pd_x \bar\lambda \; \bar b_1^{(i)},
      \qquad
      x
        = s, \theta ,
    \end{equation}
    with coefficients real-analytic on each tile under the regularity assumed above, and not smooth across the arcs where tiles meet; on inputs of level at most $L$ only the modes $b_k$ with $k \le L$ act.

  \item \emph{Evaluate.}
    Insert everything in~\eqref{eq:ws:recipe-plumbing}.
    On inputs $A_i = c_1 \bar c_1 V_i$ with $V_i$ a matter primary of any momentum, every $b_{k \ge 0}$ and every $L_{n \ge 1}$ annihilates $A_i$, so that the dressing drops and the input operators reduce to weights:
    \begin{multline}
      \label{eq:ws:recipe-primaries}
      \sprod{2}{2}(A_1 \otimes A_2, A_3 \otimes A_4)
        = - \I \mathcal N \int_{\mathcal R_V} \dd s \, \dd\theta \; \prod_{i = 1}^4 \abs{\rho_i(q)}^{\Delta_i}
      \\
          \times \big\langle \varphi_{2,1}(A_3, A_4) \big| \, b_0^+ b_0^- \, \mathcal S(q) \, \big| \varphi_{2,1}(A_1, A_2) \big\rangle ,
    \end{multline}
    with $\Delta_i$ the eigenvalue of $L_0^+$ on $A_i$, $\alpha' k_i^2 / 2 - 2$ for a tachyon, and $\rho_i = F_i'(0; q) / h_i'(0)$ the four mapping radii, equal for the $\group{SL}(2, \C)$ vertex.
\end{enumerate}

\paragraph{Inverse of the dressed gluing.}

The dressed gluing constructs the form; the Feynman rules need its inverse, and we end with what its construction would take.
On-shell the inverse is explicit sector by sector, up to a projector onto the merged states (\cref{sec:ws:onshell}): the pair of ghost zero modes dual to the two antighosts, times the inverse of a bosonic operator, the transform of the region.
Off-shell the antighosts are the dressed insertions $b(\pd_s)$ and $b(\pd_\theta)$ of~\eqref{eq:ws:dressing}, and one may look for a factorization of the same shape,
\begin{equation}
  \label{eq:ws:inverse-factorized}
  \sprod{2}{2}^{-1}
    = C_s C_\theta \, G,
\end{equation}
with $C_s$ and $C_\theta$ ghost operators dual to the two dressed antighosts and $G$ the inverse of the bosonic part $\mathcal P$ of the form.
On the primary states of~\eqref{eq:ws:recipe-primaries} the antighosts are $b_0^+ b_0^-$ for every $q$ and factor out, and $\mathcal P$ is the remaining integral, of ghost number zero, between merged states.
On general inputs the coefficients of $b_0^{+ (i)}$, $b_1^{(i)}$, $\bar b_1^{(i)}$ in~\eqref{eq:ws:dressing-moebius} depend on $q$, and whether a single pair of dual operators can be extracted is open.
If it can, $G$ can be represented as for the string field.
The propagator of the string field $\Psi$ in Siegel gauge contains the operator $1 / L_0^+$, which can be written as a Laplace transform,
\begin{equation}
  \label{eq:ws:propagator-psi}
  \frac{1}{L_0^+}
    = \int_0^\infty \dd s \; \e^{- s L_0^+} ,
\end{equation}
which has a worldsheet reading: $\e^{- s L_0^+}$ is a cylinder of length $s$, and the integral over $s$ is the integral over the length of the plumbing cylinder, with $b_0^+$ the antighost of this modulus.
Because of the tachyon, the spectrum of $L_0^+$ is not positive and the right-hand side does not converge, strictly speaking; the left-hand side is well defined and serves as a definition of the right-hand side wherever the latter is not~\cite{Sen:2015:OffshellAmplitudesSuperstring,deLacroix:2017:ClosedSuperstringField,Erbin:2021:StringFieldTheory}.
In the same spirit we would write
\begin{equation}
  \label{eq:ws:propagator-sigma}
  G
    = \int_0^\infty \dd t \; \e^{- t \mathcal P}
\end{equation}
on the quotient, the left-hand side defining the right-hand side where the integral does not converge.
Unlike $s$ in~\eqref{eq:ws:propagator-psi}, $t$ is not a modulus of any surface.
Expanded in powers of $t$, $\e^{- t \mathcal P}$ is a sum over chains of $k$ copies of $\mathcal P$ contracted through the pairing, with weight $(- t)^k / k!$, and the integral over $t$ does not commute with the expansion, since each $\int_0^\infty t^k \, \dd t$ diverges.
This representation therefore does not expand the inverse term by term in chains of vertices, as the Neumann series~\eqref{eq:ws:neumann} does, whereas the integral over $s$ is part of the integral over the moduli of the sewn surface.
The off-shell Veneziano amplitude in Schnabl gauge~\cite{Rastelli:2008:OffshellVenezianoAmplitude}, where the propagator is also an integral of an operator exponential with its ghost insertion pulled out, may offer an analogy; we leave this direction for later work.

\subsection{On-shell exactness and the Sonoda--Zwiebach sewing}
\label{sec:ws:onshell}

On four on-shell primaries the dressing is invisible: by the last step of the recipe the input operators reduce to the weights $\abs{\rho_i}^{\Delta_i}$, which equal one at $\Delta_i = 0$, so that the quartic form is the rigid gluing~\eqref{eq:ws:p-circ}.
This holds for every choice of the fixed coordinates, since the merged states of two choices differ by operators $U_g$ that act trivially on such inputs.
A second argument, which uses nothing about the dressing, gives the same: the integrand of the rigid gluing and the pull-back by $Z$ of the integrand of $\vertex{4}$ on these states, which on-shell does not depend on the local coordinates, are two real-analytic $2$-forms on the connected set $0 < \abs{q} < 4 \beta^2$, where the four punctures stay distinct.
  On the unit disc they agree by the standard factorization of the $s$-channel exchange, which fixes $\mathcal N$; hence they agree everywhere, on $\mathcal R_V$ in particular.
In words, the plumbing measure $\dd s \, \dd\theta \, b_0^+ b_0^-$ is the canonical measure of $\mathcal M_{0,4}$ on the reversed range $s < 0$ as on the Feynman disc.

On-shell the rigid gluing is the quartic form, and its inverse is explicit sector by sector.
Pushed to $\Hs$ through the merging map, it is the pair of ghost zero modes $c_0^- c_0^+$ dual to the two antighosts~\eqref{eq:ws:schiffer}, which realizes the count~\eqref{eq:ws:ghost-obstruction}, times the reciprocal of the transform of the region (Laplace in the neck length and Fourier in the twist), in each sector of $(L_0^+, L_0^-)$.
This holds up to a projector onto the merged states, which is invisible between them.
Off-shell the merged states and the antighosts~\eqref{eq:ws:dressing} depend on $q$ and contain the modes $b_1^{(i)}$, $\bar b_1^{(i)}$.
The form is then not given by a single merging map and such an operator, and a closed form of the inverse is not known.

\paragraph{The Sonoda--Zwiebach sewing.}

At the parameters of the theory the two copies are $\vertex{3}$, which is the symmetric Hermitian $\group{SL}(2, \C)$ vertex of Sonoda and Zwiebach with parameter $a = 2 \beta$~\cite[sect.~2.3.1, eq.~(2.18)]{Sonoda:1990:CovariantClosedString}.
On-shell and at finite level, the $s$-channel exchange and the quartic form together are then their sewing of that vertex over a domain that is not a disc, the integral~\eqref{eq:ws:p-circ} taken over $\{\abs{q} \le 1\} \cup \mathcal R_V$ instead of $\mathcal R_V$, with the $t$- and $u$-channel exchanges keeping their unit discs.
Over $\mathcal R_V$ the range of $\theta$ at fixed $s < 0$ is not the full circle, so that the integral over the twist does not project onto $L_0^- = 0$: the propagating states are the non-level-matched components of the merged states.
Sonoda and Zwiebach allow for such states~\cite[pp.~188--189]{Sonoda:1990:CovariantClosedString}, since their proof examines only amplitudes with level-matched external states and holds as long as the set of off-shell states contains the level-matched ones,\footnote{%
  Different sets of on-shell and off-shell states are considered in different instances~\cite{Pius:2016:CutkoskyRulesSuperstring,Bautista:2019:QuantumGravityTimelike,Bautista:2020:BRSTCohomologyTimelike}.
}
and the sewing over $\{\abs{q} \le 1\} \cup \mathcal R_V$ realizes the enlargement they contemplate.
Constructions of closed string field theory without level matching have been considered in~\cite{Okawa:2022:ClosedStringField,Erbin:2023:ClosedStringTheory}.

Their theorem~\cite[sect.~4]{Sonoda:1990:CovariantClosedString} states that no such vertex, sewn by the plumbing fixture over a domain common to the three channels so that they cover $\mathcal M_{0,4}$ once (a hypothesis the construction here does not meet, the extended domain belonging to the $s$-channel alone), has coordinates that agree up to a phase at the seams.
Its proof runs through the point where the four punctures form a square, $Z = - 1$~\cite[sect.~4.3, fig.~8]{Sonoda:1990:CovariantClosedString}, which lies on a seam between two of their channels.
Here that point sits at
\begin{equation}
  \label{eq:ws:square-point}
  q_\square
    = - 4 \beta^2 \, (\sqrt{2} - 1)^2 ,
\end{equation}
on the ray $\theta = \pi$, strictly inside $\mathcal R_V$ for every $\beta > \beta_\circ$, between the $s$-seam $\abs{q} = 1$ and the cross-channel seam at $\e^{\sigma_V(\pi)}$.
At $\beta = \beta_\circ$ it sits on both seams at once, where the region pinches and the three Feynman regions touch, their configuration, and below $\beta_\circ$ it lies inside the Feynman discs, which then overlap.
Its images $2$, at $- q_\square$, and $\tfrac{1}{2}$, at the midpoints $q = \pm 4 \I \beta^2$ of the two arcs where the region folds, are also interior points of the vertex region, by its $S_3$ covariance.
Continued off-shell with fixed coordinates, the sewing fails the seam condition at a seam their framework does not have, so that their proof neither covers nor contradicts the construction; the quartic form is the dressed sewing, whose coordinates satisfy the seam condition.
We read this as follows: the theory evades the theorem through the quartic vertex, written as the exchange of $\Sigma_2$.
The field $\Sigma_2$ is a state of two closed strings and not a closed string field, so that the coupling of $\Sigma_2$ to two string fields is not a symmetric vertex of three closed string fields.
The enlarged set of propagating states is a consequence of the same construction, not what lets the theory evade the theorem.

\subsection{Composite vertices and the covering conjecture}
\label{sec:ws:six-point}

We have introduced reduced vertices $\dvertex{n}$, built from the fundamental vertex $\vertex{n}$ by subtracting the composite vertices, the exchange diagrams of the auxiliary fields between lower reduced vertices, as in~\eqref{eq:all:diagonal}.
For example, at the lowest order, the reduced sextic vertex~\eqref{eq:hs6:dressed-vertex} is the sextic vertex minus the composite vertex~\eqref{eq:hs5:exchange-vertex}, $\dvertex{6} = \vertex{6} - \tfrac{6}{5} \vertex{5 \cdot 5}$.
At seven points the composite vertices are~\eqref{eq:hs7:structures} and the reduced vertex is~\eqref{eq:hs7:dressed-vertex}.

The minus sign of the sextic example is general: by~\eqref{eq:all:diagonal}, $\dvertex{n}(\Psi^n)$ is $\vertex{n}(\Psi^n)$ minus every tree diagram of the cubic action of the previous order, $\caction[\le n - 1]$, at order $\gs^{n - 2}$, that is, the exchanges of the auxiliary fields between the lower reduced vertices.
Each diagram comes with a positive coefficient, since the kinetic term of every auxiliary field carries a minus sign and its odd couplings a plus sign, so that the elimination produces every exchange with a positive weight.
The exchange diagrams, with the reduced vertices at their nodes, are thus always subtracted, which supports the reading of each of them as a region removed from $\mathcal V_{0, n}$; when the diagrams are rewritten with the string vertices at their nodes, the sign depends on the diagram, as the chain term $+ \tfrac{42}{25} \vertex{5 \cdot 5 \cdot 5}$ of~\eqref{eq:hs7:diagonal} shows, because a diagram nested in a reduced vertex is counted once more with the opposite sign.

The natural question is whether the reduced vertex is the integral over a sub-region of $\mathcal V_{0,n}$.
In that case, the sum of all composite vertices would be the complementary sub-region, so that together they reproduce $\mathcal V_{0,n}$.
The count of \cref{sec:ws:moduli} gives each composite vertex the dimension of the moduli space of its external punctures, and the statement below should be provable from the gluing and the local coordinates alone, that is, from how the $\Sigma_2$ line between two vertex regions acts on their sections of local coordinates, with no reference to the states inserted at the punctures.
We conjecture that $\dvertex{n}$ is the integral of the standard measure over a sub-region of $\mathcal V_{0,n}$, though we did not manage to prove it.
Another possibility is that the composite vertex covers the whole region with a constant weight, a construction similar to the generalized sections of~\cite{Sen:2015:OffshellAmplitudesSuperstring,deLacroix:2017:ClosedSuperstringField}.

The motivation is the structure of closed string field theory itself: the construction of the vertices for $\Psi$ starts from $\mathcal V_{0,3} = \mathcal M_{0,3}$, and $\mathcal M_{0,4}$ is decomposed into the Feynman regions and $\mathcal V_{0,4}$.
Here $\mathcal V_{0,4}$ plays for $\Sigma_2$ the role that $\mathcal M_{0,3}$ plays for $\Psi$, and we may expect the higher vertex regions to decompose in the same way.

\section{\IBL{} perspective}
\label{sec:ibl}

The construction of \cref{sec:hs} reads the string vertices with several of their slots as outputs.
Operations with several inputs and several outputs are the defining data of an \IBL{} algebra, the homotopy version of an involutive Lie bialgebra: a bracket and a cobracket with their higher homotopies.
Involutive means that the bracket composed with the cobracket vanishes~\cite{Cieliebak:2020:HomologicalAlgebraRelated, Markl:2017:MVFormalismIBL}.
In string field theory, this structure appeared earlier in~\cite{Muenster:2013:QuantumOpenClosedHomotopy, Muenster:2014:HomotopyClassificationBosonic}.

In this section, we write the cubic action in that language and study which relations the products read from its vertices satisfy.
After studying the lowest product $\sprod{2}{2}$ in detail to gain some intuition, we generalize to higher orders.

From the products $\ell_n$ alone, we build a whole family $p^{\Omega}_{m, n}$ by raising some of their slots with the inverse of the odd symplectic form; on inputs of even parity it coincides with the family $\sprod{m}{n}$ obtained with the inverse pairing.
We show that this family satisfies the genus-zero \IBL{} relations if and only if the $\ell_n$ form a cyclic $\Loo$ algebra, provided the vertices, read with no output or with no input, are counted among the operations.

We then explain what the homotopy transfer theorem for \IBL{} algebras says about the physical states, and why the elimination of $\Sigma_2$ is a homotopy transfer of a different kind.

In this section, we focus on the products built from the fundamental vertices $\vertex{n}$: from the sextic order on, the cubic action is built on reduced vertices $\dvertex{n}$ that differ from the fundamental ones (\cref{sec:all-orders}), and we have not studied the \IBL{} relations of the products read from them.
Note that we write everything with physical ghost numbers and without suspension (\cref{app:conventions}).

\subsection{Operations with several outputs}
\label{sec:ibl:operations}

An $\Loo$ algebra is a set of operations $\ell_n$ with $n$ inputs and one output, the string products of the original theory (\cref{sec:hs:action}).
The cubic action contains instead the product $\sprod{2}{2} \colon \Hn{2} \to \Hn{2}$ of~\eqref{eq:hs1:p22}, with two inputs and two outputs, as the kinetic operator of $\Sigma_2$.
At higher orders the construction uses $\sprod{3}{2}$, $\sprod{3}{3}$ and the products built in the same way from the reduced vertices.
In an \IBL{} algebra, the single-output operations form an $\Loo$ algebra, the operation with one input and two outputs is a cobracket, and the operations with more outputs are their higher homotopies.

An \IBL{} structure also acts on multi-string states.
Each operation $\sprod{m}{n}$ is extended to the symmetric algebra $\bigoplus_{n \ge 1} \Hn{n}$ by acting on every subset of $m$ strands of a multi-string state and leaving the other strands untouched, and the master equation is an identity between operators on that algebra~\cite{Cieliebak:2020:HomologicalAlgebraRelated, Muenster:2013:QuantumOpenClosedHomotopy}.
The field of the theory is accordingly a sum of components with any number of strands~\cite[sec.~9]{Cieliebak:2020:HomologicalAlgebraRelated}.
This is the field content of the cubic action: one component per number of strands, $\Psi$ with one strand and the auxiliary field $\Sigma_n$ with $n$ strands.
We use the total field
\begin{equation}
  \boldsymbol{\Psi}
    = \sum_{n \ge 1} \gs^{n-1} \, \Psi_n
\end{equation}
of~\eqref{eq:all:total-field}, with $\Psi_1 = \Psi$ and $\Psi_n = \Sigma_n$ for $n \ge 2$, in terms of which every cubic term of the action carries one power of $\gs$ and every kinetic term none.

\subsection{Cubic action on the symmetric algebra}
\label{sec:ibl:symmetric-algebra}

Each $n$-string space carries the pairing $\pair[n]{\cdot}{\cdot}$, with one $c_0^-$ per strand, which is graded-symmetric for every $n$ and non-degenerate.
The products $\sprod{m}{n}$ are defined from the vertices by reading $n$ slots through it, as in~\eqref{eq:conv:sprod}.
A vertex $\vertex{k}$ is nonzero only when its arguments total ghost number $2 k$ (\cref{app:conventions}), and each pairing saturates at five.
Hence an output strand of a product sits at ghost number three and an input strand at two, and we get:
\begin{equation}
  \label{eq:ibl:ghost-numbers}
  \Ngh(\sprod{m}{n})
    = 3 n - 2 m,
\end{equation}
where in particular $\Ngh(\sprod{2}{2}) = 2$.
This grading differs from the one of~\cite{Cieliebak:2020:HomologicalAlgebraRelated}, in which the degree of an operation does not depend on its number of outputs: the only strand degree affine in the ghost number that reproduces their degrees is $3 - \Ngh$, and it requires $d = \tfrac{1}{2}$, a value their definition excludes.
At this stage, the question of whether these products satisfy the \IBL{} relations is still open.

The cubic action is a function on the symmetric algebra.
At quartic order, with the components $\Psi_1 = \Psi$ and $\Psi_2 = \Sigma_2$ of~\eqref{eq:all:total-field}, the action~\eqref{eq:hs4:action-tilde} reads:
\begin{equation}
  \label{eq:ibl:action-components}
  \begin{aligned}
    \caction[\le 4]
      &
      = \frac{1}{2} \, \pair{\Psi_1}{Q_B \Psi_1}
        - \frac{1}{4!} \, \pair[2]{\gs \Psi_2}{\sprod{2}{2}(\gs \Psi_2)}
      \\ &
      \quad
      + \frac{\gs}{3!} \, \pair{\Psi_1}{\ell_2(\Psi_1^2)}
        + \frac{2 \gs}{4!} \, \pair[2]{\gs \Psi_2}{\sprod{2}{2}(\Psi_1^2)},
  \end{aligned}
\end{equation}
using the definition of $\sprod{2}{2}$ (\cref{sec:hs:action}).
Every term is the pairing, on some $\Hn{n}$, of one component $\gs^{n-1} \Psi_n$ of $\boldsymbol{\Psi}$ with a product applied to other components.
This is the form of an action built from the operations $\sprod{m}{n}$ and the multi-string pairings, whose direct sum is the pairing on $\bigoplus_{n \ge 1} \Hn{n}$.

Ghost-number conservation fixes the kinetic operator of each component.
Since the $n$-string pairing saturates at ghost number $5 n$, a quadratic term $\pair[n]{\Psi_n}{K \Psi_n}$ for the component $\Psi_n$, of ghost number $2 n$, requires $\Ngh(K) = n$, which is the ghost number of $\sprod{n}{n}$.
In an \IBL{} algebra, by contrast, the differential acts on every strand of a multi-string state~\cite[Rem.~2.2]{Cieliebak:2020:HomologicalAlgebraRelated}, as $1 \otimes \cdots \otimes Q_B \otimes \cdots \otimes 1$.
Such an operator has ghost number one and cannot be the kinetic operator of $\Sigma_n$ with the field content above.
A kinetic term of this form needs a different field content, with a partner field of ghost number five paired with $\Sigma_2$; we come back to it in \cref{sec:further}.
We compare the two structures on the physical states in \cref{sec:ibl:transfer}.

Two facts about the $2$-string component enter the comparison of \cref{sec:ibl:transfer}.
It is auxiliary: its equation of motion, $\sprod{2}{2}(\Sigma_2) = \sprod{2}{2}(\Psi^2)$, fixes $\Sigma_2 = \Psi^2$ at this order on the quotient~\eqref{eq:hs4:quotient}, where $\sprod{2}{2}$ is injective.
Its quadratic form is the quartic vertex itself, $\pair[2]{\Sigma_2}{\sprod{2}{2}(\Sigma_2')} = \vertex{4}(\Sigma_2, \Sigma_2')$, and $\sprod{2}{2}$ is self-adjoint for the $2$-string pairing because $\vertex{4}$ is symmetric.

\subsection{\IBL{} relations of the raised products}
\label{sec:ibl:general}

Every product $\sprod{m}{n}$ is built from the $\ell_n$ by raising indices, so the $\Loo$ relations of the original theory may already imply the \IBL{} relations.
Let
\begin{equation}
  \label{eq:ibl:inverse-pairing}
  \Delta
    := \pair{\cdot}{\cdot}^{-1}
\end{equation}
be the inverse pairing, an element of $\Hn{2}$ whose two strands have ghost numbers adding up to five.
Feeding one strand of $\Delta$ into an input of $\ell_n$ and keeping the other strand as a second output defines
\begin{equation}
  \label{eq:ibl:raising}
  \sprod{n-1}{2}
    = \Delta \circ \ell_n,
\end{equation}
and iterating produces every $\sprod{m}{n}$ of~\eqref{eq:conv:sprod} from the $\ell_n$ alone.

The raising that defines the antibracket uses instead the inverse of the odd symplectic form $\Omega$ of~\eqref{eq:conv:odd-symplectic}, which differs from $\Delta$ by a sign $-1$ on every slot of the vertex that holds an odd state.
Raising every slot in this way defines the family $p^{\Omega}_{m, n}$:
\begin{equation}
  \label{eq:ibl:raised-family}
  \vertex{m+n}(A_1, \ldots, A_n, B_1, \ldots, B_m)
    =: \Omega_n \bigl( A_1 \otimes \cdots \otimes A_n, p^{\Omega}_{m, n}(B_1, \ldots, B_m) \bigr),
\end{equation}
where $\Omega_n(X, Y) := (-1)^{\abs{X}} \, \pair[n]{X}{Y}$ for $X$ of definite parity, extended by linearity, is the $n$-th tensor power of $\Omega$ with the Koszul sign of~\eqref{eq:conv:pairing-n}.
Here $\abs{X}$ is the parity of the states, not of their coefficients, so that on a string field of unrestricted ghost number the sign acts component by component.
Since the pairing joins states of opposite parity and the vertices are even, this family differs from the family of~\eqref{eq:conv:sprod} by the parity of the inputs:
\begin{equation}
  \label{eq:ibl:raised-vs-sprod}
  p^{\Omega}_{m, n}(B_1, \ldots, B_m)
    = (-1)^{\abs{B_1} + \cdots + \abs{B_m}} \, \sprod{m}{n}(B_1, \ldots, B_m).
\end{equation}
The two families carry the same information, and the choice between them is a sign convention, like the choice of a suspension for an $A_\infty$ algebra; they coincide on the classical string field, of ghost number two, and on the auxiliary fields $\Sigma_k$, of ghost number $2 k$.
We write the relations below for $p^{\Omega}_{m, n}$, and we include in the family the vertices themselves, $p^{\Omega}_{m, 0} = \sprod{m}{0} = \vertex{m}$, read with no output, and the fully raised vertices $p^{\Omega}_{0, n} = \sprod{0}{n} \in \Hn{n}$, with no input.
The whole family is encoded in the original action $S$ of~\eqref{eq:hs1:action} evaluated at the sum of two string fields $\Psi$ and $\Psi'$ of unrestricted ghost number:
\begin{equation}
  \label{eq:ibl:generating}
  \begin{aligned}
    S(\Psi + \Psi')
      &
      = \sum_{\substack{m, n \ge 0 \\ m + n \ge 2}} \frac{\gs^{m+n-2}}{m! \, n!} \, \Omega_n \bigl( \Psi'^n, p^{\Omega}_{m, n}(\Psi^m) \bigr)
    \\
      &
      = \sum_{\substack{m, n \ge 0 \\ m + n \ge 2}} \frac{\gs^{m+n-2}}{m! \, n!} \, \pair[n]{\Psi'^n}{\sprod{m}{n}(\Psi^m)},
  \end{aligned}
\end{equation}
where each family is read through its own form and the two sums agree term by term, each term being $\tfrac{\gs^{m+n-2}}{m! \, n!} \, \vertex{m+n}(\Psi'^n, \Psi^m)$, with $\vertex{2}(A, B) = \pair{A}{Q_B B}$ so that $\sprod{1}{1} = Q_B$.
An $\Loo$ algebra is itself an \IBL{} algebra whose operations with two or more outputs, and those of higher genus, vanish~\cite{Cieliebak:2020:HomologicalAlgebraRelated}, and the question is whether the raised family is one as well.

The $\Loo$ relations of the $\ell_n$ are equivalent to the classical master equation $(S, S) = 0$ for the action $S(\Psi)$~\cite{Zwiebach:1993:ClosedStringField}, with the bracket on the string field defined by the inverse of $\Omega$.
Applied to $S(\Psi + \Psi')$, with the bracket contracting a derivative with respect to $\Psi$ against a derivative with respect to $\Psi'$, the same equation reads:
\begin{equation}
  \label{eq:ibl:polarized-master}
  \frac{\pd^r S(\Psi + \Psi')}{\pd \Psi^a} \, (\Omega^{-1})^{ab} \, \frac{\pd^l S(\Psi + \Psi')}{\pd \Psi'^b}
    = 0,
\end{equation}
where $\Psi = \Psi^a e_a$ and $\Psi' = \Psi'^a e_a$ in a basis $e_a$ of $\Hs$.
In this subsection $\Psi$ and $\Psi'$ are string fields of unrestricted ghost number, whose coefficients carry the Grassmann parity that makes each of them even; on fields of ghost number two alone the contraction with $\Omega^{-1}$ would vanish identically, and~\eqref{eq:ibl:polarized-master} would hold for any products.
Products and forms are evaluated on them with the coefficients moved to the left, each coefficient passing only the basis states to its left, so that $\Omega$ is symmetric on two odd fields.
Since $S(\Psi + \Psi')$ depends on the sum only, \eqref{eq:ibl:polarized-master} is the master equation at the point $\Psi + \Psi'$, and it holds if and only if the $\ell_n$ form a cyclic $\Loo$ algebra.
Sorting~\eqref{eq:ibl:polarized-master} by the degrees $(m, n)$ in $(\Psi, \Psi')$ gives one relation per pair $(m, n)$.
In each of its terms the derivative with respect to $\Psi'$ frees an output strand of one product, and $\Omega^{-1}$ carries it into an input of another product, freed by the derivative with respect to $\Psi$.
The other outputs of both products stay paired with $\Psi'$, while their other inputs are filled with $\Psi$.
Read through the first form of~\eqref{eq:ibl:generating}, $\Omega^{-1}$ inverts $\Omega$, so that the freed strand enters the second product as the output of the first, and each relation is, up to a sign common to all $(m, n)$, a sum of compositions of two products $p^{\Omega}_{m, n}$.
The relation at $(m, n)$ states that the sum over all ways of composing two products of the family through one contracted strand, with $m$ inputs and $n$ outputs in total, each ordered configuration counted once, vanishes.
These are the genus-zero \IBL{} relations in the form of~\cite[sec.~8]{Cieliebak:2020:HomologicalAlgebraRelated}, where operations with no input or no output are allowed, written in the generating-function language of their section~7.
With $\Delta$ in place of $\Omega^{-1}$, a configuration and its mirror image, in which the two products exchange their roles, would enter with opposite signs, and every relation would hold identically.

Equation~\eqref{eq:ibl:polarized-master} can therefore be read as follows: \emph{the family $p^{\Omega}_{m, n}$ obtained from the vertices by raising every index with $\Omega^{-1}$, with $m + n \ge 2$, satisfies the genus-zero \IBL{} relations at every $(m, n)$, in the generating-function form above, if and only if the products $\ell_n$ form a cyclic $\Loo$ algebra.}
Two conventions meet in this statement: the relations of the family are evaluated with the coefficient rule fixed above, and the $\Loo$ relations of the $\ell_n$ are those of the standard Koszul convention for odd maps.
The proof, in both directions, is the observation that~\eqref{eq:ibl:polarized-master} is the master equation at the point $\Psi + \Psi'$.
The argument is algebraic, and for the string field it holds in the same formal sense as the master equation itself, since the inverse pairing $\Delta$ is a sum over a basis of $\Hs$.
The relations contain nothing beyond the master equation of the original theory, since the family is its Taylor expansion about a second field.

The relations just obtained include the operations with no input or no output, the vertices $p^{\Omega}_{m, 0}$ and the fully raised vertices $p^{\Omega}_{0, n}$.
The definition of an \IBL{} algebra in~\cite[Def.~2.3]{Cieliebak:2020:HomologicalAlgebraRelated} excludes them: every operation has at least one input and one output, and the relations are written among these operations only.
Their section~8 admits the extended relations above.
Their degree bookkeeping, with an integer $d$, corresponds here to $d = \tfrac{1}{2}$, with one suspension sign per slot when the slots of an operation are permuted.
In the conventions of~\cite{Muenster:2013:QuantumOpenClosedHomotopy,Muenster:2014:HomotopyClassificationBosonic}, where the string field has degree zero and the only operation with several outputs is the inverse symplectic form, every operation has degree one and $d = 0$; the value $d = \tfrac{1}{2}$ reflects our reading of the vertices with several outputs, not a property of closed string field theory.
We use their relations only as identities between operations, which this regrading does not affect, and their Theorem~6.1 below only as a comparison.
For $\min(m, n) = 1$, read in the same generating-function form, the two sets of relations coincide, and both hold.
At $(m, 1)$ they are the $\Loo$ relations, and at $(1, 2)$ they state that $p^{\Omega}_{1, 1}$ and the cobracket $p^{\Omega}_{1, 2}$, equal to $Q_B$ and $\sprod{1}{2}$ on inputs of even parity, are compatible.
At $(2, 2)$ they do not coincide.

The relation~\eqref{eq:ibl:polarized-master} at degree $(2, 2)$ is the relation $[Q_B, \ell_3] + \ell_2 \circ \ell_2 = 0$ of the $\ell_n$, written with two fields.
It involves four configurations, up to the exchange of the two products:
\begin{equation}
  \label{eq:ibl:configurations-22}
  \begin{aligned}
    T_1
      &
      = \tfrac{1}{2} \, \Omega(Q_B \Psi', \ell_3(\Psi, \Psi, \Psi')),
    &
    T_2
      &
      = \tfrac{1}{2} \, \Omega(\ell_3(\Psi, \Psi', \Psi'), Q_B \Psi),
    \\
    T_3
      &
      = \tfrac{1}{4} \, \Omega(\ell_2(\Psi', \Psi'), \ell_2(\Psi, \Psi)),
    &
    T_4
      &
      = \Omega(\ell_2(\Psi, \Psi'), \ell_2(\Psi, \Psi')),
  \end{aligned}
\end{equation}
where the numerical factors are those of the Taylor expansion of $S(\Psi + \Psi')$.
The part of degree $(2, 2)$ of~\eqref{eq:ibl:polarized-master}, divided by $2 \gs^2$, is:
\begin{equation}
  \label{eq:ibl:polarized-22}
  T_1 + T_2 + T_3 + \tfrac{1}{2} \, T_4
    = 0.
\end{equation}
The four terms are the configurations in which an output of $\sprod{2}{2}$ enters $Q_B$, an output of $Q_B$ enters $\sprod{2}{2}$, an output of $\ell_2$ enters $\sprod{1}{2}$ and an output of $\sprod{1}{2}$ enters $\ell_2$.
In each, the joined strand is raised with $\Omega^{-1}$, and the term is written as the pairing of the two products joined by the contracted strand.
In~\eqref{eq:ibl:polarized-master} each of $T_1$, $T_2$, $T_3$ also appears with the roles of the two products exchanged, an output of $\sprod{0}{2}$ entering $\ell_3$, of $\sprod{1}{3}$ entering $\sprod{2}{0}$ and of $\sprod{0}{3}$ entering $\sprod{3}{0}$, with the same value, whereas $T_4$ is its own mirror image and appears once.
After this division, $T_4$ therefore enters~\eqref{eq:ibl:polarized-22} with half the weight of $T_1$, $T_2$, $T_3$.
The same contraction restricted to the operations with at least one input and one output contains none of the mirror terms, and the relation it imposes at degree $(2, 2)$ gives the four configurations the same weight:
\begin{equation}
  \label{eq:ibl:io-22}
  T_1 + T_2 + T_3 + T_4
    = 0.
\end{equation}
The difference of the left-hand sides of~\eqref{eq:ibl:io-22} and~\eqref{eq:ibl:polarized-22}, $\tfrac{1}{2} \, T_4$, does not vanish for a general cyclic $\Loo$ algebra: the cyclic Lie algebra $\alg{sl}_2 \oplus \alg{sl}_2^*$, placed at ghost numbers one and four, is a counterexample.
The two sets differ at the level of the operations, on inputs of free parity.
Take $[e, f] = h$, $[h, e] = 2 e$, the coadjoint action on $\alg{sl}_2^*$ and $\pair{x}{\chi} = \chi(x)$.
One then finds for the polarized form $T_4(e, f; e^*, e) = -2$ (slots of $\Psi$ first, then $\Psi'$), while~\eqref{eq:ibl:polarized-22} holds, as the theorem requires.
Whether the residual vanishes for the closed string vertices themselves is left for a future study.

\subsection{Homotopy transfer}
\label{sec:ibl:transfer}

We now turn to what the transfer theorem says about the physical states.
Theorem~6.1 of~\cite{Cieliebak:2020:HomologicalAlgebraRelated} states that every \IBL{} algebra over a field of characteristic zero is homotopy equivalent to an \IBL{} structure on its homology with respect to the differential $\sprod{1}{1}$.
The transferred operations are given by a recursion of the type of the homological perturbation lemma, built from a splitting of the space and a contracting homotopy.
For $\sprod{1}{1} = Q_B$, the transferred structure lives on the symmetric algebra of the one-particle states $H(Q_B)$, the Fock space of the free theory, and every other operation becomes an operation on it.
In this sense only $\sprod{1}{1} = Q_B$ determines the physical states.
In an \IBL{} algebra, however, the differential acts on each strand of the multi-string components, so that their cohomology is $\mathrm{Sym}^k H(Q_B)$, read as the $k$-particle states, whereas in the cubic action as written $\Sigma_2$ is auxiliary.
This comparison is with the cubic action as written; whether a parent theory with a $2$-string gauge parameter, of which the cubic action would be a gauge-fixed form, represents states in $\Sigma_2$ is the open question of \cref{sec:further}.
The operations $\sprod{k}{k}$ enter as perturbations of this differential, weighted by powers of the parameter $\hbar$ of~\cite{Cieliebak:2020:HomologicalAlgebraRelated}, of degree $2 d$, which would be odd for the value $d = \tfrac{1}{2}$ found above.

Three operations on homotopy algebras should be kept apart.
A strictification replaces a theory by a quasi-isomorphic one on a larger space with simpler operations.
For $\Loo$ algebras it is a classical result~\cite{Loday:2012:AlgebraicOperads}, realized for closed string field theory by the bar--cobar construction of \cref{sec:bar-cobar}.
The cubic action is one in the sense that its equations of motion are quadratic, and a quasi-isomorphic one on the classical complex at quartic order, as shown below.
The elimination of an auxiliary field goes the other way: it is a homotopy transfer along a contraction whose kernel, the pair formed by the field and its equation of motion, is acyclic, and it inverts the strictification up to homotopy.
The minimal model is a further transfer, onto the cohomology $H(Q_B)$, in which every off-shell mode is integrated out with the homotopy as propagator, and its operations are the tree-level on-shell amplitudes, as for the $A_\infty$ algebras of open strings~\cite{Kajiura:2007:NoncommutativeHomotopyAlgebras}.
The canonical model of Theorem~6.1 is of the third kind, and it is reached from the original theory as well as from any of its strictifications.

The elimination of $\Sigma_2$ is of the second kind.
In the cubic action, whose coupling $\gs$ carries no ghost number, the kinetic operator of $\Sigma_2$ has ghost number two: it is $\sprod{2}{2}$ at quartic order and the dressed operator~\eqref{eq:hs5:dressed} beyond.
On the quotient on which it is injective, the pair formed by $\Sigma_2$ and its equation of motion carries no cohomology.
On the classical complex, the fields and their equations of motion, with $\Psi$ at ghost numbers two and three and $\Sigma_2$ at four and six, integrating $\Sigma_2$ out is a homotopy transfer of $\Loo$ algebras along the contraction of that pair.
The contracting homotopy is the inverse of $\sprod{2}{2}$ on the quotient.
The transfer map is $\Psi \mapsto (\Psi, \Psi^2)$, the solution of the auxiliary equation, with no higher component, so that the map is not strict.
The transferred brackets are those of the original theory truncated at the quartic order: the three-bracket is the $\Sigma_2$-exchange tree between two quartic sources, which collapses to the contact vertex as in \cref{sec:hs:quartic}, and no higher bracket is generated at this order.
Under the assumptions on the quartic form stated at the head of \cref{sec:all-orders}, the map is a cyclic $\Loo$ quasi-isomorphism from the quartic-truncated theory to the cubic one, which controls the solutions and their linearization.
Off the classical complex, $\Psi \mapsto (\Psi, \Psi^2)$ is not a morphism of $\Loo$ algebras, because the kinetic operator of $\Sigma_2$ has ghost number two and not one: its linearization lands at ghost number five, where no $2$-string state of the classical complex lies.
The transfer therefore says nothing by itself about the gauge algebra or the BRST spectrum.
Around a classical solution, the spectrum of the cubic theory is nevertheless that of the original theory, by the Batalin--Vilkovisky argument summarized in \cref{sec:further}, to the order to which it has been worked out.

At quartic order the cubic theory and the original one thus have the same physical states, as Theorem~6.1 would suggest, but by a different mechanism: $\Sigma_2$ is not a multi-string direction of an \IBL{} algebra, since no differential acts on its strands, and it carries no state.
For the higher auxiliary fields we expect the same, since their kinetic operators have ghost number $n$ by the same counting, but their dressed kinetic operators and the transfer of the whole tower are left for a future study.
The \IBL{} formalism also comes with a tower of gauge parameters $\Lambda_n \in \Hn{n}$, one per number of strands~\cite[app.~A.2]{Muenster:2014:HomotopyClassificationBosonic}, whereas the cubic action has the single parameter $\Lambda$.
Whether the cubic action is the gauge-fixed form of a theory with an independent $2$-string parameter is discussed in \cref{sec:further}.
The bar--cobar construction of the next section strictifies closed string field theory on another space of multi-string fields, and we compare the two field by field.

\section{Comparison with the bar–cobar construction}
\label{sec:bar-cobar}

Chiaffrino, Mamade and Zwiebach have shown that the equations of motion of any $A_\infty$ field theory can be rewritten as a quadratic equation on an enlarged space of multi-string fields, and they leave open whether this equation derives from an action~\cite{Chiaffrino:2026:UniversalQuadraticField}.
The auxiliary fields of the cubic action, introduced as multi-string fields in~\cite{Erbin:2018:CubicClosedString}, live, up to a quotient and a normalization, in the spaces of their type-I multi-string fields.
We compare the two constructions ingredient by ingredient: we recall their construction in our notations, we state their results on the fields, the equations and the gauge symmetry, and we map them onto the cubic action.
For $\Loo$ algebras they give the construction and rely on general results on quasi-isomorphisms, which identify the gauge classes of solutions under some assumptions.
  We transcribe their $A_\infty$ analysis of the solutions and of the gauge transformations to the symmetric coalgebra, which brings the binomial factors and the normalization $1 / r!$ below.

\subsection{The bar–cobar construction}
\label{sec:barcobar:construction}

The products $\ell_n$ of closed string field theory form an $\Loo$ algebra.
The bar construction packages all of them into a single coderivation $B$ of the symmetric tensor coalgebra built on the state space, and the cobar construction then builds the free Lie algebra on this coalgebra.
The result is a differential graded Lie algebra, quasi-isomorphic to the original $\Loo$ algebra~\cite{Chiaffrino:2026:UniversalQuadraticField}.
Its Maurer--Cartan equation is quadratic because a differential graded Lie algebra has no product beyond the bracket, and the price is the size of the field space.

The field $\Phi$ of the enlarged theory has components labeled by a \emph{factor number} $f$ and by a \emph{length}, the number of strands.
The components of factor number one, which they call type I, are one multi-string field per length:
\begin{equation}
  \label{eq:barcobar:type-I}
  \Phi
    = \Phi_1 + \Phi_2 + \cdots,
  \qquad
  \Phi_1
    = \sum_{r \ge 1} \Phi_{1, r},
  \qquad
  \Phi_{1, r}
    \in \Hn{r},
  \qquad
  \Ngh(\Phi_{1, r})
    = 2 r,
\end{equation}
and the length-one component $\Phi_{1,1}$ is the string field itself; a type-I field of length $r$ and total ghost number $N$ has degree $2 r - 1 - N$ in their homological grading.
The components of factor number $f \ge 2$, their type II, are independent fields valued in the formal products $\Hn{r_1} \otimes' \cdots \otimes' \Hn{r_f}$ generated by the free Lie algebra of the cobar step.
The equation of motion reads:
\begin{equation}
  \label{eq:barcobar:mc}
  (B + \Delta_c) \Phi + \Phi \otimes' \Phi
    = 0,
\end{equation}
where $\Phi \otimes' \Phi = \tfrac{1}{2} [\Phi, \Phi]$ for a field of their degree $-1$.
All the data of the theory sit in $B$: on a type-I field of length $r$ it applies the products $\gs^{n-1} \ell_n$ to $n$ of the strands, for every $n \ge 1$ and every choice of the strands, so that it lowers the length by $n - 1$ and preserves the factor number.
The other two operations are universal and raise the factor number by one: the coproduct $\Delta_c$ splits a type-I field of length $n$ into the pairs of lengths $(i, n - i)$, summing over the ways of distributing the strands, and $\otimes'$ juxtaposes two fields.

\subsection{Canonical solutions and gauge symmetry}
\label{sec:barcobar:canonical}

Since $B$ preserves the factor number and the two other operations raise it by one, an ansatz with type-I components only, $\Phi = \Phi_1$, reduces~\eqref{eq:barcobar:mc} to two equations:
\begin{equation}
  \label{eq:barcobar:canonical-eqs}
  B \Phi_1
    = 0,
  \qquad
  \Delta_c \Phi_1 + \Phi_1 \otimes' \Phi_1
    = 0,
\end{equation}
of factor numbers one and two, all higher factor numbers vanishing identically.
The second equation is a factorization condition, one for every length $n$ and every split $1 \le i \le n - 1$:
\begin{equation}
  \label{eq:barcobar:factorization}
  \binom{n}{i} \, \Phi_{1, n}
    = \Phi_{1, i} \otimes \Phi_{1, n - i},
\end{equation}
as tensors with $n$ strands, the right-hand side being required to be symmetric under all permutations of its strands.
Its unique solution is the exponential of the string field:
\begin{equation}
  \label{eq:barcobar:canonical}
  \Phi_{1, r}
    = \frac{\Psi^r}{r!}.
\end{equation}
On this solution, the component of $B \Phi_1$ with $L$ output strands is $\mathcal E_\Psi \otimes \Psi^{L-1} / (L-1)!$, with $\mathcal E_\Psi$ the equation of motion~\eqref{eq:hs1:eom-psi} of the original theory, so that the first equation of~\eqref{eq:barcobar:canonical-eqs} reduces to $\mathcal E_\Psi = 0$.
They call these solutions \emph{canonical} and prove, for $A_\infty$ algebras, that every solution of~\eqref{eq:barcobar:mc} is gauge equivalent to a canonical one~\cite[sec.~6]{Chiaffrino:2026:UniversalQuadraticField}.
  The gauge classes of solutions of the two theories are in one-to-one correspondence by the quasi-isomorphism, under the assumptions above~\cite[secs.~3.4 and~3.5]{Chiaffrino:2026:UniversalQuadraticField}.
We call canonical sector the type-I fields of the form~\eqref{eq:barcobar:canonical} for an arbitrary, off-shell $\Psi$.

Their gauge transformation,
\begin{equation}
  \label{eq:barcobar:gauge}
  \delta \Phi
    = (B + \Delta_c) \hat\Lambda + [\Phi, \hat\Lambda],
\end{equation}
has one parameter $\hat\Lambda_{f, r}$ for every factor number $f$ and every length $r$, and in particular an independent $2$-string parameter $\hat\Lambda_{1, 2} \in \Hn{2}$ of ghost number three.
Among the parameters of factor number one, those that keep the canonical sector free of type-II components are
\begin{equation}
  \label{eq:barcobar:canonical-parameters}
  \hat\Lambda_{1, r}
    = \frac{1}{(r - 1)!} \, \Lambda \otimes \Psi^{r - 1},
\end{equation}
symmetrized over the strands, with a single-string parameter $\Lambda \in \Hs$ of ghost number one, and on them the transformation generates the gauge transformations of the original theory~\cite[sec.~6.3]{Chiaffrino:2026:UniversalQuadraticField} (transcribed to the symmetric coalgebra).
An independent $\hat\Lambda_{1, 2}$ generates type-II components and leaves the canonical sector.

Their construction gives equations but no action: an action requires a non-degenerate cyclic bilinear form of degree three, which they argue the enlarged space does not admit, and they leave open whether an action exists with another pairing~\cite[sec.~8]{Chiaffrino:2026:UniversalQuadraticField}.

\subsection{Map to the cubic action}
\label{sec:barcobar:map}

The type-I component of length $r$ lives in $\Hn{r}$ with ghost number $2 r$, the space and the ghost number of the auxiliary field $\Sigma_r$.
With $\Sigma_r := r! \, \Phi_{1, r}$, the canonical solution reads $\Sigma_r = \Psi^r$ and the factorization conditions read $\Sigma_n = \Sigma_i \otimes \Sigma_{n - i}$ for every split.
Two differences remain at the level of the fields: $\Sigma_2$ has ghost number two on each strand, not only four in total, and lives on the quotient by $\ker \sprod{2}{2}$ of~\eqref{eq:hs4:quotient}, where their field lives on the whole $2$-string space.

The equations differ more.
In their construction, $\Sigma_k = \Psi^k$ holds exactly at every order on the canonical solutions and $B$ acts on the strands by the elementary products.
In the cubic action, the equation of $\Sigma_2$ reads $\sprod{2}{2}(\Sigma_2 - \Psi^2) = O(\gs)$, so that $\Sigma_2 = \Psi^2$ holds modulo $\ker \sprod{2}{2}$ and at leading order only, the mismatch being absorbed order by order by the reduced vertices of \cref{sec:all-orders}.
The cubic action can be read as a truncation of their type-I sector that keeps, at order $\gs^{n-2}$, the fields of length $k \le n / 2$, replaces the longer ones by the products of shorter fields, $\Psi$ among them, that the factorization conditions prescribe. The truncation also drops the components of $B \Phi_1 = 0$ with several output strands.
It is not a consistent truncation of their equations: fields are substituted rather than set to zero and the factorization conditions become equations of motion through the quadratic forms of the auxiliary fields. The solution for $\Sigma_2$ differs from $\Psi^2$ at order $\gs$, while the original action is recovered order by order.
Conversely, their construction has type-II fields and the whole tower at every order, so that neither construction is a consistent truncation of the other.
The two constructions share their solutions: in the canonical gauge $\Phi_{f \ge 2} = 0$, their equations and ours have the same solutions for $\Psi$, order by order in $\gs$ and modulo the kernels of the quadratic form and of the dressed operators of the auxiliary fields, under the assumptions stated at the head of \cref{sec:all-orders}. The auxiliary fields agree with $\Sigma_k = \Psi^k$ at leading order and modulo these kernels.

The cubic action does not supply the form of uniform degree that their space lacks; it pairs each auxiliary field with itself, through forms of a different kind.
In their grading, the $r$-string pairing $\pair[r]{\cdot}{\cdot}$ has the length-dependent degree $- r - 2$, so that none of these pairings can serve as a cyclic form on the whole tower.
The cubic action pairs $\Sigma_2$ with itself through the quartic vertex, a Grassmann-even symmetric form supplied by the theory. This form needs no odd symplectic structure because $\Sigma_2$ has no BRST kinetic term, and it is degenerate on $\ker \sprod{2}{2}$, the reason for the quotient. The higher auxiliary fields are paired by the reduced even vertices in the same way. A formulation that requires an odd symplectic form of uniform degree on the multi-string space meets their obstruction unchanged.

The gauge symmetries can be compared on the canonical sector.
On it and on the parameters~\eqref{eq:barcobar:canonical-parameters}, their transformation reads on $\Sigma_2$:
\begin{equation}
  \label{eq:barcobar:gauge-sigma}
  \delta_\Lambda \Sigma_2
    = 2 \, \delta_\Lambda \Psi \otimes \Psi
      + 2 \, \mathcal E_\Psi \otimes \Lambda,
\end{equation}
where the products are understood inside a symmetric vertex as in \cref{sec:hs:quartic}.\footnote{%
  The relative sign of the second term is fixed for commuting fields only; it plays no role below, since the term vanishes on-shell.
}
The first term is the composite law of \cref{sec:hs:quartic}.
The second term is proportional to the equation of motion of the original theory, so that the two laws agree on-shell.
It is forced by the form~\eqref{eq:barcobar:gauge} of their transformation, which preserves the diagonal $\Sigma_2 = \Psi^2$ on-shell only, whereas the composite law preserves it off-shell.
The second term has strands of ghost numbers three and one, outside the space of $\Sigma_2$.
By their theorem, the type-II components of any solution can be gauged away, after which only the single-string parameters remain.
At the quartic order, the gauge transformation of $\Sigma_2$ in the cubic action is then on-shell the residue of theirs in the canonical gauge.
Whether the cubic action is the gauge-fixed form of a theory with an independent $2$-string parameter is discussed in \cref{sec:further}. In the classes we have examined, every such theory is a trivial enlargement in which the parameter shifts $\Sigma_2$ along null directions of the quadratic form; whether their $\hat\Lambda_{1, 2}$ falls into one of these classes is left open.

The comparison is classical, the quantum theory on their enlarged space being left open in their paper.
We carry out our transformation for the bosonic closed string only; whether the kernel conditions and the reality conditions carry over to the superstring is left for future work.

\section{Further results}
\label{sec:further}

We collect in this section results that this paper does not develop.
We state them with their scope and without their derivations: we think they are correct, but we did not carry their analysis to the end, and we leave their completion to future work.
The derivations are in the complete write-up in the repository of this paper.\footnote{\url{https://github.com/HaroldErbin/paper-cubic-sft}}

\paragraph{Gauge algebra.}

The transformation law of an auxiliary field is fixed by the invariance of its completed square only up to transformations antisymmetric with respect to the quadratic form.
These are proportional to its own equation of motion: they are the trivial gauge symmetries that every action possesses~\cite[ch.~3]{Henneaux:1994:QuantizationGaugeSystems}.
The rotation of the fluctuation in the quintic law~\eqref{eq:hs5:gauge-sigma} compensates the variation of the dressed kinetic operator $\kin{2}{\Psi}{3}$ with the same parameter $\Lambda$; it is neither a trivial symmetry nor a symmetry on its own.

The commutator of two transformations in the $\Psi$ sector is that of closed string field theory, open from order $\gs^2$ on~\cite{Zwiebach:1993:ClosedStringField}.
In the $\Sigma_2$ sector the commutator is a gauge transformation up to terms proportional to both equations of motion for the law~\eqref{eq:hs5:gauge-sigma}.
In the representative under which the rescaled fluctuation
\begin{equation}
  \begin{gathered}
    \label{eq:further:flat-variable}
    \hat\sigma_2^{(3)}
      := m_\Psi^{1/2} \, \sigma_2^{(3)},
    \qquad
    m_\Psi
      := \sprod{2}{2}^{-1} \circ \kin{2}{\Psi}{3},
    \\
    \pair[2]{\sigma_2^{(3)}}{\kin{2}{\Psi}{3}(\sigma_2^{(3)})}
      = \vertex{4}(\hat\sigma_2^{(3)}, \hat\sigma_2^{(3)}),
  \end{gathered}
\end{equation}
is inert, $\delta_\Lambda \hat\sigma_2^{(3)} = 0$, it is one up to terms proportional to the $\Psi$ equation alone; the square root is a formal power series in $\gs$.
We constructed this flat representative through order $\gs^4$, for $\Sigma_2$ at the quintic order and for $\Sigma_2$ and $\Sigma_3$ at the sextic order.
There it reaches the string vertices under the assumption that the exchange diagrams are not linear combinations of the elementary vertices (\cref{sec:all-orders}).
We expect one flat variable per auxiliary field at every order, but beyond the sextic order the kinetic operator of $\Sigma_3$ is dressed by the background in turn, and its flat variable remains to be constructed.

\paragraph{Master action.}

In the Batalin--Vilkovisky formalism, and under the assumptions of \cref{sec:all-orders}, the inert fluctuation and its antifield form a contractible pair with a constant coefficient:
\begin{equation}
  s \hat\sigma_2^{(3)}
    = 0,
  \qquad
  s \hat\sigma_2^{(3)+}
    = - \frac{\gs^2}{12} \, \sprod{2}{2}(\hat\sigma_2^{(3)}).
\end{equation}
The master action of the cubic theory is the master action of the eliminated theory plus this pair, plus a second one at the sextic order.
Its antibracket with itself equals that of the eliminated theory and vanishes up to terms of order $\gs^{N-1}$ for the action truncated at $\vertex{N}$, the order to which the cubic action is gauge invariant.
The master action is proper for $\gs \neq 0$ at generic momentum.

At the quartic order the master action can also be written, up to contractible pairs, as in~\cite{Zwiebach:1993:ClosedStringField}, with string fields of unrestricted ghost number.
The fields are $\Psi \in \Hs$ and $\Sigma_2 \in \Hn{2} / \ker \sprod{2}{2}$, with the kernel taken at every ghost number, and their coefficients make both fields even.
The antifields of $\Psi$ are its components along the states of ghost number at least three.
Those of $\Sigma_2$ form an independent field $\Sigma_2^+$ of total (worldsheet plus spacetime) ghost number five, the ghost number of the partner field of the conjecture below; the antifield of $\Sigma_n$ has total ghost number $3 n - 1$.
We take the antibracket canonical in $\Psi$ and in the pair $(\Sigma_2 - \Psi^2, \Sigma_2^+)$, so that the fluctuation $\Sigma_2 - \Psi^2$, rather than $\Sigma_2$, has vanishing bracket with $\Psi$.
The cubic action itself then satisfies the master equation to the same order and is proper under the same conditions: $\Sigma_2^+$ does not appear in it, and it remains cubic in the ghosts and antifields.
If $\Sigma_2$ itself is taken to have vanishing bracket with $\Psi$ instead, the master equation holds only up to terms proportional to the equation of motion of $\Sigma_2$.

Around a solution $(\Psi^*, \Sigma_2^*)$ at which the dressed kinetic operator is invertible on the quotient, the linearization of the shift to the fluctuation block-diagonalizes the linearized BRST operator of the cubic theory.
The blocks are the linearized BRST operator of the eliminated theory and the contractible pair.
The first block is the shifted operator $Q_{\Psi^*} = Q_B + \gs \, \ell_2(\Psi^*, \cdot) + \dots$ of the original theory~\cite{Zwiebach:1993:ClosedStringField} up to the order of the truncation.
The cohomologies therefore agree in every ghost number: to the same orders, the spectrum of fluctuations of the cubic theory around a classical solution is that of the original theory.

The similarity itself uses only the shift and holds wherever the dressed operator is invertible, also for a background of order $1 / \gs$.
There its first block is the operator of the eliminated action.
From the quintic order on, this operator differs from the truncated original one by terms of the size of the first vertex dropped.

Whether the cubic action is the gauge-fixed form of a theory with an independent $2$-string parameter $\Lambda_2 \in \Hn{2}$ is open.
We examined the classes where $\Lambda_2$ acts on $\Sigma_2$ alone through a field-independent operator, while $\Psi$ keeps its law and the enlarged quadratic form reduces to the quartic one.
In these classes every such theory is a trivial enlargement: $\Lambda_2$ shifts $\Sigma_2$ along null directions of the enlarged quadratic form.
The classes left open are a $\Lambda_2$ acting on $\Psi$, a field-dependent operator on $\Sigma_2$, the quartic vertex off level matching, and a symmetry appearing at order $\gs^3$.

We conjecture that the cubic action is the gauge-fixed form of a parent theory with a $2$-string parameter $\Lambda_2 \in \Hn{2}$.
In this parent, as the \IBL{} gauge law predicts, $\Lambda_2$ acts on $\Psi$, at leading order by $\gs \, \ell_2(\Lambda_2)$, and on $\Sigma_2$, extended to all of $\Hn{2}$ at ghost number four, by $Q_B$ on each strand.
The parent also has a partner field of ghost number five, paired with $\Sigma_2$ through $Q_B$ on each strand.
The gauge chosen here represents every cohomology class in $\Psi$; a gauge fixed on $\Psi$ would represent in $\Sigma_2$ the classes, if any, that the orbits of $\Lambda_2$ reach, with no new state if the parent is proper.
With a compensating field of ghost number three in place of the partner, such a parent exists trivially.
It is the cubic action evaluated on fields shifted by the compensator, $\Sigma_2$ by the level-matched part of the image of the compensator under $Q_B$ on each strand.
The conjecture is that one exists which is not of this form.

\paragraph{Hessian around a background.}

The second variation of the cubic action around a background $(\Psi^*, \Sigma_2^*)$ is affine in the background, whereas the Hessian of the original quartic action is quadratic in $\Psi^*$.
Through the quintic order its blocks are built from $Q_B$ and the vertices $\vertex{3}$ to $\vertex{5}$, each contracted once with one background field, the auxiliary block being the dressed kinetic operator at the background
\begin{equation}
  H_{\Sigma_2 \Sigma_2}
    = - \frac{\gs^2}{12} \, \pair[2]{\cdot}{\kin{2}{\Psi^*}{3}(\cdot)}.
\end{equation}
Wherever this block is invertible, completing the square in the auxiliary fluctuation shows that the joint Hessian is congruent to a block-diagonal form:
\begin{equation}
  \label{eq:further:congruence}
  \Hess \caction
    = L^{\mathsf T} \, \big( H_{\mathrm{eff}} \oplus H_{\Sigma_2 \Sigma_2} \big) \, L,
  \qquad
  H_{\mathrm{eff}}
    := H_{\Psi \Psi}
      - H_{\Sigma_2 \Psi}^{\mathsf T} \, H_{\Sigma_2 \Sigma_2}^{-1} \, H_{\Sigma_2 \Psi},
\end{equation}
with $L$ unipotent, so that the numbers of positive, negative and zero directions add (Sylvester's law) and the kernels correspond.
On the auxiliary shell the Schur complement $H_{\mathrm{eff}}$ is the Hessian of the eliminated action, which agrees with that of the original truncated action through the matched order.
At the quartic order the split can be read on the vertex: $H_{\Psi \Psi}$ holds $\tfrac{\gs^2}{6} \, \vertex{4}(\Psi^{*2}, \psi, \psi)$, one third of the quartic term of $\pair{\psi}{Q_{\Psi^*} \psi}$, the direct channel of the exchange.
The exchange of $\sigma_2$ supplies the other two thirds.
Freezing $\Sigma_2$ at its classical value is therefore not a consistent truncation of the fluctuations.
At the sextic order, with the sextic term written with $\Sigma_2$ alone, the auxiliary equation is quadratic and its two sheets meet at a fold where $H_{\Sigma_2 \Sigma_2}$ degenerates.
The Hessian of the eliminated action diverges there while $\Hess \caction$ stays finite, so that the fold is an artifact of the elimination.
The price of the cubic form is the auxiliary block: an indefinite algebraic sector with no $L_0^+$ in its quadratic term, whose negative directions are rotated in the path integral and subtracted from every count.
Whether it carries a mass shell at nonzero momentum is open.

We consider the truncation to the tachyon $T$ and the ghost dilaton $D$ at level $(1, 1)$ and zero momentum, with the $\mathrm{SL}(2, \C)$ quartic vertex~\cite{Erbin:toappear:SL2VerticesClosed} at both values of the family parameter studied there.
The quartic form restricted to the three pairs $T \otimes T$, $T \otimes D$, $D \otimes D$ is indefinite.
On the potential side (\cref{app:conventions}) the auxiliary block has one negative direction, which the Gaussian integral over $\Sigma_2$ rotates to the imaginary axis (as is common in Euclidean signature for the Hubbard--Stratonovich transformation).
Counting the negative directions of the $\Psi \Psi$ block alone, with $\Sigma_2$ frozen, gives the wrong count, two negative directions instead of one, at the critical points of largest amplitude.
At one value of the family parameter it also gives two instead of one at the point where both fields are positive.
Freezing the fluctuation $\sigma_2^{(2)} = \Sigma_2 - \Psi^2$ instead gives the right count.
This is the distinction between a mean field and the random-phase approximation in a Hubbard--Stratonovich treatment~\cite{Negele:1988:QuantumManyParticleSystems}.
The stability of the theory is not decidable from zero-momentum data: in string field theory the potential is obtained after integrating out many fields and levels, and the lowest order at the lowest level does not settle it.

\paragraph{Level truncation.}

A level truncation of the cubic theory restricts $\Psi$ to a finite-dimensional space $\mathcal U$ of levels and $\Sigma_2$ to its symmetric square $\mathcal U^{\wedge 2}$, the smallest auxiliary space containing $\Psi^2$.
We assume the quartic form non-degenerate on $\mathcal U^{\wedge 2}$ and evaluate the vertices as in the untruncated theory.
Eliminating $\Sigma_2$ inside the truncation gives back the level-truncated original action exactly at the quartic order and through $O(\gs^3)$ at the quintic order.
At the sextic order, for the matching to hold through $O(\gs^4)$, the reduced vertex has to be built with the inverse of $\vertex{4}$ on $\mathcal U^{\wedge 2}$, $\dvertex{6}^{[\mathcal U]} = \vertex{6} - \tfrac{6}{5} \, \vertex{5 \cdot 5}^{[\mathcal U]}$.
This truncated action is then not the restriction of the full one: along $\mathcal U^{\wedge 2}$ itself, its solutions satisfy the auxiliary equation of the full cubic action only up to terms of relative order $\gs^2$.
The mismatch is a question of how the auxiliary fields should be truncated: $\mathcal U^{\wedge 2}$ is the smallest choice, and other choices are possible.
One option is to truncate $\Sigma_n$ at the level $L$ of $\Psi$, the level of a component of $\Sigma_n$ being the sum of the levels of its $n$ strands.
The mismatch may also be a repackaging of the ordinary level-truncation error; we have not examined this.
The residual of a truncated solution in the other equations of the full action follows from the universality rule~\cite{Sen:1999:UniversalityTachyonPotential,Yang:2005:ClosedStringTachyon}: at zero momentum a genus-zero vertex vanishes when exactly one of its legs has its matter part outside the identity Virasoro module.
Take for $\mathcal U$ a space of zero-momentum universal states in Siegel gauge, such as the tachyon and the ghost dilaton, and a solution of the truncated equations on the branch connected to $\Sigma_2 = \Psi^2$.
It satisfies the $\Psi$ equation of the full cubic action along every non-universal direction identically, at the quartic, quintic and sextic orders.
Along the universal states outside $\mathcal U$ it has the ordinary level-truncation error.
It satisfies the auxiliary equation along every direction at the quartic order.
From the quintic order on, the auxiliary equation has in general a residual along some pairs outside $\mathcal U^{\wedge 2}$, at relative order $\gs$.
At the sextic order it has in general also one along the pairs of a state of $\mathcal U$ and a non-universal state, at relative order $\gs^2$.
These orders are Taylor orders in $\gs$ at fixed $\Psi$ and carry no accuracy statement at a tachyon vacuum, where $\gs \Psi = O(1)$.
The same-level universal states outside $\mathcal U$ and the limit of infinite level are not covered.

\paragraph{Mean-field reading.}

In many-body theory the Hubbard--Stratonovich field is also the mean field.
A quartic interaction is written as the exchange of an auxiliary field coupled to a pair, the fermions are integrated out at fixed auxiliary field, and the saddle point of the resulting effective action is the gap equation.
It sets the auxiliary field equal to the pair correlator in its own background, $\Delta \propto \langle \psi \psi \rangle_\Delta$ for the pairing field of BCS theory~\cite{Negele:1988:QuantumManyParticleSystems}.
The same steps are available here at the quartic order.
With the weight $\e^{S}$ of \cref{app:conventions}, integrating $\Psi$ at fixed $\Sigma_2$ gives the stationarity condition of the auxiliary field, its one-loop effective potential at $\Psi = 0$ and the kinetic operator of $\Psi$ deformed by the background $\Sigma_2$:
\begin{equation}
  \label{eq:further:gap}
  \begin{gathered}
    \sprod{2}{2}\big( \Sigma_2 - \langle \Psi^2 \rangle_{\Sigma_2} \big)
      = 0,
    \qquad
    V[\Sigma_2]
      = \frac{\gs^2}{24} \, \vertex{4}(\Sigma_2^2)
        + \frac{\hbar}{2} \, \tr \log \big( - Q_{\Sigma_2} \big),
    \\
    Q_{\Sigma_2}
      := Q_B + \frac{\gs^2}{6} \, \ell_3(\Sigma_2, \cdot),
  \end{gathered}
\end{equation}
where the potential is computed in the Gaussian approximation, in Siegel gauge, with the cubic vertex dropped and the trace taken over the fields of ghost number two.
The gap equation that follows, $\Sigma_2 = \hbar \, G_{\Sigma_2}$ on $\Hn[2]{2} / \ker \sprod{2}{2}$ with $G_{\Sigma_2}$ the propagator of $Q_{\Sigma_2}$ read as a $2$-string state, is a self-consistent propagator dressed by the auxiliary field, as in BCS theory.
The analogy stops there.
The dressing $\tfrac{\hbar \gs^2}{6} \, \ell_3(G_{\Sigma_2}, \cdot)$ is one third of the one-loop self-energy of the quartic theory, the direct channel.
The other two thirds are the diagrams with an internal $\Sigma_2$ line, which the mean field drops.
We know of no small parameter of the many-body kind, a Fermi surface or a large $N$, that suppresses them; the expansion available here is perturbation theory in $\gs$, or in $\hbar$.
The cubic vertex, moreover, makes the slice $\Psi = 0$ non-stationary at one loop.
At one loop the effective potential of the string field involves the loop vertices of the original theory, which the classical construction does not contain.
Without them the approximation is not controlled, and we do not pursue it here.

Two of the questions these results leave open, the Batalin--Vilkovisky structure of the auxiliary fields and the $2$-string gauge parameter, are among the directions listed in \cref{sec:discussion}.

\section{Discussion and outlook}
\label{sec:discussion}

We have written the classical action of closed bosonic string field theory in a cubic form.
A Hubbard--Stratonovich transformation trades the quartic vertex, the square of $\Psi^2$ with respect to the quadratic form $\vertex{4}$, for an auxiliary $2$-string field $\Sigma_2$ with a quadratic term and a cubic coupling to $\Psi^2$.
Repeating the step order by order, we introduced a tower of $k$-string fields $\Sigma_k$, each entering at order $\gs^{2 k - 2}$ through a completed square on a reduced vertex $\dvertex{2 k}$.
Its leading-order value is the string vertex minus the exchange diagrams induced by the lower auxiliary fields, and the odd vertex $\dvertex{2 k + 1}$ dresses the kinetic operator of $\Sigma_k$.
The action is triangular in the tower, so that the auxiliary fields are eliminated one at a time from the top by Gaussian integrals, exact at tree level. It reproduces the non-polynomial action as a formal power series in $\gs$, under the invertibility of the bare kinetic operators on their quotients and the kernel conditions stated in \cref{sec:all-orders}.
The matching fixes each reduced vertex only through its leading-order value, which is what makes the proof of their existence at every order short.
The gauge symmetry of the original theory lifts to the cubic action with no new parameter: each auxiliary field transforms as its solution, plus a rotation of its fluctuation.
The construction lies outside the hypotheses of the no-go theorem of Sonoda and Zwiebach~\cite{Sonoda:1990:CovariantClosedString}, and the quartic and higher vertices of the original theory remain, in the kinetic operators of the auxiliary fields.

On the worldsheet, $\Sigma_2$ is a state of two entangled closed strings.
The inverse of the quartic form can be read as a more general gluing involving several punctures at the same time.
Using an OPE to merge the punctures, one finds that the quartic form itself is a gluing of two copies of an asymmetric cubic vertex along a circle, dressed by the motion of the local coordinates. On-shell, this gluing is a sewing of the kind studied by Sonoda and Zwiebach~\cite{Sonoda:1990:CovariantClosedString}, over a domain outside the hypotheses of their theorem.
In the language of homotopy algebras, the products with several outputs read from the string vertices satisfy an extended form of the genus-zero \IBL{} relations, with the operations with no input or no output included, if and only if the string products form a cyclic $\Loo$ algebra.
The elimination of $\Sigma_2$ is a homotopy transfer on the classical complex at the quartic order.
Our auxiliary fields live, up to a quotient and a normalization, in the spaces of the type-I fields of the bar--cobar construction of Chiaffrino, Mamade and Zwiebach~\cite{Chiaffrino:2026:UniversalQuadraticField}, but neither construction is a consistent truncation of the other.
The cubic action does not supply the form of uniform degree that their space lacks.
Finally, we stated in \cref{sec:further}, without their derivations, results we think correct but whose analysis we did not carry to the end: the gauge algebra and its flat representative, the master action and the spectrum around a classical solution, the congruence of the Hessian, the rule for a level truncation, and the mean-field reading.
We have worked at tree level; the quantum action, which involves the $\Psi$-dependent determinants of the auxiliary kinetic operators and the loop vertices, is left for future work.

Many aspects of this construction deserve to be developed further:
\begin{enumerate}
  \item Establish for the string vertices the invertibility of the auxiliary kinetic operators and the kernel conditions of \cref{sec:all-orders}.

  \item Understand the Batalin--Vilkovisky (BV) structure of the action beyond the quartic order (\cref{sec:further}).
    We expect the cubic action on fields of unrestricted ghost number to remain its own master action, with an antifield of total ghost number $3 n - 1$ for $\Sigma_n$.
    An antibracket which is canonical in the variables in which the action is cubic is not known.

  \item Determine whether the current action is a gauge-fixed version of a more general cubic theory, which also includes gauge parameters $\Lambda_n \in \mc H^{\wedge n}$ (a conjecture on such a parent theory is stated in \cref{sec:further}), and whether this action has an \IBL{} structure.
    A parent with a compensating field exists trivially, so that the question is whether the partner field can be the antifield of $\Sigma_2$.
    The products raised from the reduced vertices satisfy the \IBL{} relations only if the reduced vertices form a cyclic $\Loo$ algebra by themselves.

  \item Improve the worldsheet description of the different operators, their OPE reductions and their inverses.
    The ghost insertions of the dressed gluing factorize on primary states, and whether they do in general decides if the inverse has the Laplace form of \cref{sec:ws:gluing}; for $k \ge 3$ the patches carry moduli of their own.

  \item Understand whether the reduced vertices $\dvertex{n}$ correspond to a subregion of the vertex region $\mathcal V_{0,n}$, i.e., some sort of fundamental vertex from the point of view of $\Sigma_n$ (see \cref{sec:ws:six-point}).
    The BRST variation of the reduced sextic vertex gives a first test.
    The exchange of $\Sigma_2$ collapses on part of the faces of the vertex region where a cubic vertex splits off, and the conjecture is a condition on the remainder, which contains $\sprod{2}{2}^{-1}$.

  \item Explore classical level-truncated and analytic solutions.
    The coupling scales out of the cubic action, with $\gs \Psi$ and $\gs^k \Sigma_k$ as variables, so that a truncation of the tower is not controlled by a small parameter and has to be tested numerically.

  \item Determine whether the cubic form helps beyond perturbation theory.
    With multi-string fields $\Sigma_k \in \Hn{k}$, the cubic action resembles a tensor model~\cite{Gurau:2016:RandomTensors}, and one may ask whether background independence and the construction of classical solutions become more tractable than in the non-polynomial theory.

  \item Extend the construction to open--closed string field theory and to superstring field theories, in which the vertices carry picture-changing operators and the auxiliary fields would have to carry pictures as well; the stubbed open SFT of~\cite{Erbin:2024:OpenStringStub} is a first step.

\end{enumerate}

\paragraph{Acknowledgments.}

I am grateful to the numerous people with whom I have discussed this project over the past years, and in particular Ashoke Sen, Atakan Hilmi Fırat, Barton Zwiebach, Carlo Maccaferri, Manki Kim, Martin Schnabl, Subhroneel Chakrabarti, Ted Erler and Yuji Okawa.
I would also like to thank Manki Kim, Riccardo Finotello, Sid Mishra-Sharma and Sven Krippendorf for sparking my interest in agentic workflows for research.

I would like to thank Anthropic and Sid Mishra-Sharma for providing me with a Claude subscription that made this project possible.

\paragraph{AI disclosure.}

I worked on this project for several years (in particular, during my postdocs at LMU and MIT), during which I obtained the main ingredients of the construction, but I did not manage to obtain a complete cubic action.
These ideas have been presented at several conferences and talks, such as~\cite{Erbin:2018:CubicClosedString}.

With the recent progress in agentic large language models (LLMs), I decided to attempt to finish the construction.
Using the materials I had collected, I carried the project forward with Claude.
Most computations were carried out with Fable 5 and 5.1, while simpler derivations used Opus 4.6, 4.8, 5 and 5.5.
Mechanical tasks (such as paper fetching and summarizing) used Sonnet 4.6 and 5, and Opus 4.8 to 5.5.
The paper has been written using several models: Fable 5.1, Opus 5.5 and Sonnet 5.5.
An independent verification of the manuscript was performed with Qwen 3.8 Max.
The final check was performed with Fable 5.2.

All the derivations (including the SymPy code and finite-dimensional models) performed by the agents can be found at \url{https://github.com/HaroldErbin/paper-cubic-sft}.
They are provided in two formats: the original Markdown files (for agents) and a LaTeX write-up (for humans, though the style is poor).
The repository contains more than this paper, since I have included only what I have been able to review with a high degree of confidence.
Many algebraic identities of the construction (the matching, the gauge invariance, the master equation, the Hessian congruence, the homotopy-transfer reading, etc.) were checked in finite-dimensional models, with random symmetric vertices in exact arithmetic or with cyclic $\Loo$ algebras obtained by homotopy transfer from a matrix Chern--Simons theory, and the dressed gluing of the quartic form was tested numerically against the $\group{SL}(2, \C)$ vertex; the code is in the repository.

This paper was written outside academia, alongside a full-time job in industry, and with the help of language models, as described above.
I did not have the time to prepare the text as I would have liked, and the writing suffers.
I chose to share the results now rather than wait indefinitely for the time to revise it in full.

\appendix

\section{Conventions and normalization}
\label{app:conventions}

This appendix collects the conventions of the paper, following~\cite{Erbin:2021:StringFieldTheory}.

\paragraph{State space.}

The zero modes of the ghosts and of the Virasoro generators are combined into
\begin{equation}
  \label{eq:conv:pm-operators}
  L_0^\pm
    := L_0 \pm \bar L_0,
  \qquad
  b_0^\pm
    := b_0 \pm \bar b_0,
  \qquad
  c_0^\pm
    := \frac{1}{2} \, (c_0 \pm \bar c_0).
\end{equation}
$\Hs$ denotes the space of states of the matter--ghost CFT, at every ghost number, that satisfy the level-matching conditions~\eqref{eq:hs1:level-matching}.
We include all ghost numbers so that the string products map into $\Hs$.
Ghost numbers are physical: the $\group{SL}(2, \C)$ vacuum has ghost number zero, $c$ counts $+1$ and $b$ counts $-1$, so that the tachyon $c_1 \bar c_1 \ket{0}$ has ghost number two.
The Grassmann parity of a state is its ghost number modulo two, $\abs{A} := \Ngh(A) \bmod 2$.

The $n$-string space $\Hn{n}$ is the space of tensors in $\Hs^{\otimes n}$ whose components are graded-symmetric under permutations of the strands.
We write $\Hs_g$ for the subspace of ghost number $g$ and $\Hn[g]{n}$ for the $n$-string states whose strands all have ghost number $g$.
Given a basis $\{ \phi_r \}$ of $\Hs$, a state $\Sigma_n \in \Hn{n}$ reads
\begin{equation}
  \label{eq:conv:exp-sigma-n}
  \Sigma_n
    = \sum_{\{ r_i \}} \sigma_{r_1 \cdots r_n} \phi_{r_1} \cdots \phi_{r_n},
\end{equation}
and we write $\Psi^n := \Psi^{\otimes n} \in \Hn{n}$.
The parity in which the symmetry is graded depends on the role of the strands.
For the arguments of the vertices and products, the fields $\Psi^n$ and $\Sigma_n$ and the gauge parameters, it is the parity $\abs{\phi_{r_i}}$ of the basis state in the strand $i$ of~\eqref{eq:conv:exp-sigma-n}, and $\Hn{n}$ is the symmetrized tensor product of $n$ copies of $\Hs$~\cite{Erler:2014:NSNSSectorClosed,Erbin:2020:ClassicalAlgebraicStructures}.
Even strands then have symmetric components and odd strands antisymmetric ones, so that a product is antisymmetric in two gauge parameters.
For the strands that the $n$-string pairing~\eqref{eq:conv:pairing-n} sets against a field, those of the outputs of the products, it is the shifted parity $\abs{\phi_{r_i}} + 1$.
This is the dual of the former space under the pairing: with the symmetry graded by the strand parity, the outputs at ghost number three would have antisymmetric components and would pair to zero with $\Sigma_n$.
For a tensor with one odd and one even strand, the two parities give the same sign and we need not specify the role.

The fields of the paper all have commuting coefficients, so that the parity of a field is that of its states.
The string field $\Psi \in \Hs$ has ghost number $2$ and is even, and the gauge parameter $\Lambda \in \Hs$ has ghost number $1$ and is odd (\cref{sec:hs:action}).
The auxiliary field $\Sigma_k$, which replaces the tensor power $\Psi^{\otimes k}$, lives on the quotient of $\Hn[2]{k}$ by the kernel of its quadratic operator, $\sprod{2}{2}$ for $k = 2$ and $\wprod{3}{3}$ for $k = 3$.
It has ghost number $2 k$, two on each strand, and is even (\cref{sec:hs:quartic,sec:hs:sextic}), and the outputs of the products are read on strands of ghost number three.
The partner field $\Upsilon_2 \in \Hn[2]{2} / \ker \sprod{2}{2}$ of \cref{app:wrong-sign} has ghost number $4$ and is even.
In \cref{app:wrong-sign}, $\Sigma_3$ lives instead on $\Hn[2]{3} / \ker \sprod{3}{3}$.
Three letters should not be confused: $d$ is the spacetime dimension, $D$ the ghost dilaton and $\hat d$ the formal dimension of \cref{sec:worldsheet}.

\paragraph{Sphere normalization.}

We normalize the correlation functions on the sphere, with the momentum-conservation factor $(2\pi)^d \delta^{(d)}(\sum k)$ stripped off, as
\begin{equation}
  \label{eq:conv:sphere}
  \braket{c_{-1} \bar c_{-1} c_0^- c_0^+ c_1 \bar c_1}
    = - \frac{1}{2},
  \qquad
  \braket{c_{-1} \bar c_{-1} c_0 \bar c_0 c_1 \bar c_1}
    = - 1.
\end{equation}
The value is real, with no factor $\I$.
A sphere correlator is nonzero only when the ghost numbers of its insertions add up to six, the ghost-number anomaly, and all the ghost-number formulas below follow from this selection rule.

\paragraph{Pairing.}

The pairing~\eqref{eq:hs1:pairing} is $\pair{A}{B} = \braket{A | c_0^- | B}$, with $\bra{A} = \transp{A}$ the BPZ conjugate of $\ket{A}$.
It is nonzero only when $\Ngh(A) + \Ngh(B) = 5$, since the insertion of $c_0^-$ supplies the sixth unit; hence it pairs states of opposite parity, $\abs{A} + \abs{B} = 1$.
BPZ conjugation acts on the modes as
\begin{equation}
  \label{eq:conv:bpz-modes}
  \transp{c_n}
    = - c_{-n},
  \qquad
  \transp{b_n}
    = b_{-n},
  \qquad
  \transp{(c_0^\pm)}
    = - c_0^\pm,
\end{equation}
and in the same way for $\bar c_n$ and $\bar b_n$.
These are the rules for the inversion $I^+(z) = 1/z$, the standard choice for the closed string~\cite[footnote p.~13]{Zwiebach:1993:ClosedStringField}.
On a monomial of creation modes acting on the vacuum, BPZ conjugation conjugates each mode in place, without a reordering sign:
\begin{equation}
  \label{eq:conv:bpz-word}
  \transp{(O_1 \cdots O_k \ket{0})}
    = \bra{0} \transp{O_1} \cdots \transp{O_k}.
\end{equation}
With these rules and the normalization~\eqref{eq:conv:sphere}, the pairing is graded-symmetric in the shifted parity:
\begin{equation}
  \label{eq:conv:pairing-symmetry}
  \pair{B}{A}
    = (-1)^{(1 + \abs{A}) (1 + \abs{B})} \, \pair{A}{B}.
\end{equation}
Since the pairing is nonzero only for $\abs{A} + \abs{B} = 1$, the sign is $+1$ on its support: the pairing is symmetric.
This is the rearrangement rule of~\cite{Erbin:2021:StringFieldTheory} restricted to the support; it is derived in \cref{app:reality}.
The BRST operator is graded self-adjoint for the pairing, $\pair{Q_B A}{B} = (-1)^{\abs{A}} \, \pair{A}{Q_B B}$, which is the symmetry of $\vertex{2}(A_1, A_2) = \pair{A_1}{Q_B A_2}$.

\paragraph{Multi-string pairing.}

On $n$-string states the pairing is the tensor power of~\eqref{eq:hs1:pairing}, one $c_0^-$ per strand, with the Koszul sign of a tensor product of bilinear forms,
\begin{equation}
  \label{eq:conv:pairing-n}
  \pair[n]{A_1 \otimes \cdots \otimes A_n}{B_1 \otimes \cdots \otimes B_n}
    := (-1)^{\sum_{i < j} \abs{B_i} \abs{A_j}} \,
      \pair{A_1}{B_1} \cdots \pair{A_n}{B_n},
\end{equation}
extended bilinearly; for $n = 1$ it is the pairing itself, written without index.
When the strands of the first tensor are even, as for $\Psi^n$ and $\Sigma_n$, the sign is $+1$ and the pairing is the plain product of the pairings of the strands, each $c_0^-$ sitting inside the pairing of its own strand.
The multi-string pairing is nonzero only when the ghost numbers of the two tensors add up to $5n$.
For every $n$, it is graded-symmetric in the total parity:
\begin{equation}
  \label{eq:conv:pairing-n-symmetry}
  \pair[n]{Y}{X}
    = (-1)^{\abs{X} \abs{Y}} \, \pair[n]{X}{Y},
  \qquad
  X, Y \in \Hn{n}.
\end{equation}
Whenever one of the two tensors is even, as for every pair met in this paper, the sign is $+1$ and the pairing is symmetric.

The Batalin--Vilkovisky formalism uses instead the pairing twisted by the parity of its first argument, the odd symplectic form
\begin{equation}
  \label{eq:conv:odd-symplectic}
  \Omega(A, B)
    := (-1)^{\abs{A}} \, \pair{A}{B},
  \qquad
  \Omega(B, A)
    = - (-1)^{\abs{A} \abs{B}} \, \Omega(A, B),
\end{equation}
which is odd and graded-antisymmetric on every pair of states in its support: the pairing itself is symmetric, and the twist $(-1)^{\abs{A}}$ changes sign under the exchange of arguments of opposite parity.

\paragraph{String vertices and products.}

The vertices $\vertex{n} \colon \Hn{n} \to \C$ are the genus-zero string vertices of \cref{sec:hs:action}, graded-symmetric multilinear forms.
Their arguments, and those of the products, are graded-symmetric in the parity of the strands for every parity of the arguments, and the outputs of the multi-string products are read in the shifted parity.
The selection rule fixes the ghost number carried by every map.
For a product it is the ghost number of the output minus the sum of those of the inputs; for a vertex it is that of its antighost insertions, so that $\vertex{n}(A_1, \ldots, A_n)$ is nonzero only when $\sum_i \Ngh(A_i) = 2n$.
The products with several outputs, $\sprod{m}{n} \colon \Hn{m} \to \Hn{n}$, whose output strands have ghost number three, are defined by splitting the slots of a vertex into $n$ outputs and $m$ inputs and reading the result through the $n$-string pairing,
\begin{equation}
  \label{eq:conv:sprod}
  \vertex{m+n}(A_1, \ldots, A_n, B_1, \ldots, B_m)
    =: \pair[n]{A_1 \otimes \cdots \otimes A_n}{\sprod{m}{n}(B_1, \ldots, B_m)},
\end{equation}
so that $\sprod{n}{1} = \ell_n$ and, for instance, $\vertex{4}(A_1, A_2, A_3, A_4) = \pair[2]{A_1 \otimes A_2}{\sprod{2}{2}(A_3, A_4)}$.
Each output raises the ghost number by three and each input lowers it by two,
\begin{equation}
  \label{eq:conv:ghost-numbers}
  \Ngh(\sprod{m}{n})
    = 3n - 2m,
\end{equation}
so that $\Ngh(\sprod{2}{2}) = 2$, $\Ngh(\sprod{3}{2}) = 0$ and $\Ngh(\sprod{3}{3}) = 3$, and $\Ngh(\ell_n) = 3 - 2n$ of~\eqref{eq:hs1:ghost-vertices} is the case of one output; the rule holds for $n \ge 1$, and a vertex $\vertex{m} = \sprod{m}{0}$ carries $6 - 2m$.
Since $\vertex{m+n}$ is graded-symmetric, reading the same vertex with inputs and outputs exchanged gives the transposition identity
\begin{equation}
  \label{eq:conv:transpose}
  \pair[n]{X}{\sprod{m}{n}(Y)}
    = \pair[m]{\sprod{n}{m}(X)}{Y},
  \qquad
  X \in \Hn{n},
  \quad
  Y \in \Hn{m},
\end{equation}
for even arguments, the case of $\Psi$ and the auxiliary fields (odd arguments bring a sign).
$\sprod{n}{m}$ is thus the transpose of $\sprod{m}{n}$ with respect to the multi-string pairing.
We write compositions with $\circ$.

\paragraph{Normalization of the action.}

The action of closed string field theory, its equation of motion and its gauge transformation are those of \cref{sec:hs:action}, with the vertices normalized as in~\eqref{eq:hs1:vertex-product} and $\gs$ the string coupling.
The Euclidean path integral is weighted by $\e^{S}$, that is, $S = - S_E$ with $S_E$ the Euclidean action.
The potential is the action at zero momentum with the sign reversed, $V := - S \big\vert_{k = 0}$, so that the tachyon sits at a maximum of $V$.
A vertex quoted on the potential side carries the sign opposite to that of $\vertex{n}$, and the numbers of positive and negative directions of the Hessians of $S$ and $V$ are exchanged.

\paragraph{Reality.}

We impose reality with the star conjugation of~\cite{Erbin:2021:StringFieldTheory}, the composition of the BPZ and Hermitian conjugations:
\begin{equation}
  \label{eq:conv:star}
  A^\star
    := \adj{(\transp{A})}
    = \transp{(\adj{A})}.
\end{equation}
On the modes, Hermitian conjugation is $\adj{(c_n)} = c_{-n}$ and $\adj{(b_n)} = b_{-n}$, and it reverses the order of a product of modes; the two conjugations commute on every mode, since each maps $c_n$, $b_n$ and $\alpha_n$ to a real multiple of the mode of opposite index.
Star conjugation is antilinear and squares to one on every state, even or odd: on a monomial of modes acting on a momentum state it returns the same monomial, up to a sign, on the opposite momentum.
On a tensor it acts strand by strand, without a reordering sign.

The pairing is equivariant up to a minus sign,
\begin{equation}
  \label{eq:conv:pairing-equivariance}
  \pair{A^\star}{B^\star}
    = - \, \pair{A}{B}^*,
\end{equation}
for all $A$ and $B$.
The reason is that the vacuum functional is anti-Hermitian, $\braket{W}^* = - \braket{\adj{W}}$ for every monomial of modes $W$~\cite[eq.~(2.37)]{Zwiebach:1993:ClosedStringField}; the value~\eqref{eq:conv:sphere} fixes this sign on $W_0 = c_{-1} \bar c_{-1} c_0 \bar c_0 c_1 \bar c_1$, whose Hermitian conjugate is $- W_0$.
Since star conjugation acts strand by strand and preserves the parity of every strand, the multi-string pairing is equivariant as well, $\pair[n]{X^\star}{Y^\star} = (-1)^n \, \pair[n]{X}{Y}^*$.

The vertices are real in the sense that
\begin{equation}
  \label{eq:conv:vertex-reality}
  \vertex{n}(A_1, \ldots, A_n)^*
    = (-1)^{n + \sum_{i < j} \abs{A_i} \abs{A_j}} \, \vertex{n}(A_1^\star, \ldots, A_n^\star).
\end{equation}
On even arguments the sign is $(-1)^n$, and this property makes the action real.
For $n = 2$ it follows from~\eqref{eq:conv:pairing-equivariance} and the star conjugation of $Q_B$,~\eqref{eq:rea:QB-star}, in \cref{app:reality}.
For $n \ge 3$, we take it as a standing assumption on the vertices.
It is expected to follow from the invariance of the vertex regions and measures under complex conjugation of the moduli together with the Hermiticity conditions of~\cite[sec.~5.1 and eq.~(7.98)]{Zwiebach:1993:ClosedStringField}; for the $\group{SL}(2, \C)$ vertices of~\cite{Erbin:toappear:SL2VerticesClosed} it has been checked for the quartic vertex.

On the string field, the reality condition is that of~\cite[eq.~(3.6)]{Zwiebach:1993:ClosedStringField}, $\adj{(\ket{\Psi})} = - \bra{\Psi}$, that is, $\Psi^\star = - \Psi$, since the BPZ bra of $A^\star$ is the Hermitian bra of $A$.
The fields of the paper are subject to
\begin{equation}
  \label{eq:conv:reality-fields}
  \Psi^\star
    = - \Psi,
  \qquad
  \Sigma_k^\star
    = (-1)^k \, \Sigma_k,
  \qquad
  \Lambda^\star
    = - \Lambda.
\end{equation}
The condition on $\Lambda$ is the one under which every gauge transformation preserves that of $\Psi$~\cite[eq.~(3.18)]{Zwiebach:1993:ClosedStringField}, and the condition on $\Sigma_k$ follows from the reality of $\Psi$ and of the vertices once the auxiliary fields are introduced; both are derived in \cref{app:reality}.

\section{Exchange sign of the pairing and reality of the auxiliary fields}
\label{app:reality}

This appendix derives the exchange sign~\eqref{eq:conv:pairing-symmetry} of the pairing, with its consequences for the multi-string pairings.
It then derives the reality of the kinetic vertex and, from that of the higher vertices, the reality of the products, of the fields and of the solution of the auxiliary equations.
We use throughout the normalization~\eqref{eq:conv:sphere} and the BPZ rules~\eqref{eq:conv:bpz-modes}.

\subsection{Exchange sign of the pairing}

The pairing $\pair{A}{B} = \braket{A | c_0^- | B}$ differs from the plain BPZ product $\braket{A | B}$, which is graded-symmetric in the parity of its arguments, by the odd insertion $c_0^-$, and its exchange sign comes from moving the insertion from one argument to the other.
Moving an operator $O$ from the ket to the bra gives the sign of passing $O$ through $A$: $\braket{A | O B} = (-1)^{\abs{O} \abs{A}} \, \braket{\transp{O} A | B}$.
Applied to $O = c_0^-$, with $\transp{(c_0^-)} = - c_0^-$ from~\eqref{eq:conv:bpz-modes}, we get:
\begin{equation}
  \label{eq:rea:exchange-general}
  \begin{aligned}
    \pair{A}{B}
      &
      = \braket{A | c_0^- B}
      = (-1)^{\abs{A} (1 + \abs{B})} \, \braket{c_0^- B | A}
    \\
      &
      = - (-1)^{\abs{A} (1 + \abs{B})} \, (-1)^{\abs{B}} \, \pair{B}{A}
      = (-1)^{(1 + \abs{A}) (1 + \abs{B})} \, \pair{B}{A}.
  \end{aligned}
\end{equation}
The first sign comes from exchanging $A$ with the state $c_0^- B$, of parity $\abs{B} + 1$, in the plain product; in the second step, we move $c_0^-$ into the bra, which costs the sign $(-1)^{\abs{B}}$ of passing the insertion through $B$ and the minus sign of its BPZ conjugate.
This is~\eqref{eq:conv:pairing-symmetry}; the in-place rule~\eqref{eq:conv:bpz-word} brings no further sign.

The $n$-string pairing~\eqref{eq:conv:pairing-n} inherits the symmetry strand by strand: for $X = x_1 \otimes \cdots \otimes x_n$ and $Y = y_1 \otimes \cdots \otimes y_n$, exchanging the two tensors exchanges the roles of the $x_i$ and the $y_j$ in the Koszul sign of the definition, and the arguments of every strand pairing,
\begin{equation}
  \label{eq:rea:exchange-n}
  \begin{aligned}
    \pair[n]{Y}{X}
      &
      = (-1)^{\sum_{i < j} \abs{x_i} \abs{y_j}} \, \prod_i \pair{y_i}{x_i}
      \\ &
      = (-1)^{\sum_{i < j} \abs{x_i} \abs{y_j} + \sum_i \abs{x_i} \abs{y_i} + \sum_{i < j} \abs{y_i} \abs{x_j}} \, \pair[n]{X}{Y}
      \\ &
      = (-1)^{\abs{X} \abs{Y}} \, \pair[n]{X}{Y},
  \end{aligned}
\end{equation}
since the three sums cover every pair $(i, j)$ once; this is~\eqref{eq:conv:pairing-n-symmetry}.

\subsection{Reality conditions}

We first extend the star~\eqref{eq:conv:star} to bras and operators.
The BPZ bra of $A^\star$ is the Hermitian bra of $A$, and conversely: $\bra{A^\star} = \adj{A}$ and $\adj{(A^\star)} = \bra{A}$.
On operators the star is the same composition as~\eqref{eq:conv:star}, $O^\star := \transp{(\adj{O})}$, and it distributes over the action on a state with a Koszul sign, $(O A)^\star = (-1)^{\abs{O} \abs{A}} \, O^\star A^\star$, since $O^\star$ is written to the left of $A^\star$ while Hermitian conjugation reverses the order.

\paragraph{Reality of the vertices.}

The BRST operator is Hermitian and odd under BPZ conjugation, so that $Q_B^\star = - Q_B$.
With the Koszul rule above and $\abs{Q_B} = 1$, it commutes with star conjugation up to a sign fixed by the parity of its argument:
\begin{equation}
  \label{eq:rea:QB-star}
  (Q_B A)^\star
    = (-1)^{\abs{A} + 1} \, Q_B A^\star.
\end{equation}
For $n = 2$, the reality~\eqref{eq:conv:vertex-reality} then follows from the equivariance~\eqref{eq:conv:pairing-equivariance} of the pairing:
\begin{equation}
  \label{eq:rea:V2-star}
  \begin{aligned}
    \vertex{2}(A^\star, B^\star)
      &
      = \pair{A^\star}{Q_B B^\star}
      = (-1)^{\abs{B} + 1} \, \pair{A^\star}{(Q_B B)^\star}
    \\
      &
      = (-1)^{\abs{B}} \, \pair{A}{Q_B B}^*
      = (-1)^{\abs{A} \abs{B}} \, \vertex{2}(A, B)^*,
  \end{aligned}
\end{equation}
since on the support $\abs{A} = \abs{B}$, so that $\abs{A} \abs{B} = \abs{B}$ modulo two.
We obtain the reality of the products by stripping the first argument of~\eqref{eq:hs1:vertex-product} with~\eqref{eq:conv:pairing-equivariance}, since the pairing is non-degenerate on the level-matched space.
In $\vertex{m + 1}(A, B_1, \ldots, B_m) = \pair{A}{\ell_m(B_1, \ldots, B_m)}$, the output has parity $1 + \sum_i \abs{B_i}$, so that $\abs{A} = \sum_i \abs{B_i}$ modulo two on the support of the pairing, and the rule for arguments of any parity reads
\begin{equation}
  \label{eq:rea:product-star}
  \ell_m(B_1^\star, \ldots, B_m^\star)
    = (-1)^{m + \sum_i \abs{B_i} + \sum_{i < j} \abs{B_i} \abs{B_j}} \, \ell_m(B_1, \ldots, B_m)^\star.
\end{equation}
On even arguments the sign is $(-1)^m$, and with one odd argument $(-1)^{m + 1}$.

\paragraph{Reality of the gauge parameter.}

The condition $\Lambda^\star = - \Lambda$ is the one under which every gauge transformation~\eqref{eq:hs1:gauge-psi} preserves $\Psi^\star = - \Psi$: for the free term, \eqref{eq:rea:QB-star} with $\abs{\Lambda} = 1$ gives $(Q_B \Lambda)^\star = Q_B \Lambda^\star$, which fixes $\Lambda^\star = - \Lambda$ up to the kernel of $Q_B$; for the interaction terms, the product rule~\eqref{eq:rea:product-star} with one odd argument gives the same condition.

\paragraph{Reality of the auxiliary fields.}

Every term of the cubic action is a real multiple of a vertex, or of a composition of vertices and inverses of lower kinetic operators such as $\dvertex{6}$, on a tensor whose $N$ strands all have ghost number two.
On such a tensor, the reality of the vertices and of the inverses (shown below) carries the sign $(-1)^N$.
On the slice $\Psi^\star = - \Psi$, $\Sigma_k^\star = (-1)^k \, \Sigma_k$, the strand signs multiply to $(-1)^N$ as well:
\begin{equation}
  \label{eq:rea:cubic-real}
  \begin{aligned}
    \vertex{N}(\Sigma_k, \Sigma_\ell, \Psi^n)^*
      &
      = (-1)^N \, \vertex{N}\big( \Sigma_k^\star, \Sigma_\ell^\star, (\Psi^\star)^n \big)
    \\
      &
      = (-1)^{N + k + \ell + n} \, \vertex{N}(\Sigma_k, \Sigma_\ell, \Psi^n)
      = \vertex{N}(\Sigma_k, \Sigma_\ell, \Psi^n),
  \end{aligned}
\end{equation}
with $N = k + \ell + n$.
The cubic action is thus real on the slice~\eqref{eq:conv:reality-fields}: each strand of ghost number two gives a factor $-1$ in the reality of the vertices, compensated on the slice by the sign of $\Psi^\star = - \Psi$.
The field replacing $\Psi^k$ must therefore carry the sign $(-1)^k$ of $(\Psi^\star)^k$.
The source of $\Sigma_k$ is linear in $\Sigma_k$, with a real coefficient once the lower fields are fixed, so that the sign is determined on every direction the source reaches.
For generic vertices, the sources reach every direction and the sign is fixed on the whole field.
For the vertices of a given conformal field theory the source map need not be surjective; the directions of $\Sigma_k$ it misses are absent from the dynamics, and on them the condition~\eqref{eq:conv:reality-fields} is a choice.
Read on strands, the conditions follow one rule: $\Psi$ and $\Lambda$ are star-odd, and a tensor is star-odd if and only if it has an odd number of star-odd strands.

\paragraph{Reality of the solution.}

The auxiliary fields are eliminated by inverting their dressed kinetic operators on the sources.
For real $\Psi$ and real lower auxiliary fields, a map built from the vertices and the inverses of lower operators, with real coefficients, commutes with the star conjugation up to a sign, by the reality of the vertices~\eqref{eq:conv:vertex-reality}.
For the dressed kinetic operator of $\Sigma_k$, the rule reads $\kin{k}{\Psi}{}(B^\star) = (-1)^k \, \kin{k}{\Psi}{}(B)^\star$ on the real slice~\eqref{eq:conv:reality-fields}.
Its kernel is then star-stable, so that the condition on the auxiliary field is meaningful on the quotient, and its inverse on the quotient obeys the same rule.
The solution therefore satisfies~\eqref{eq:conv:reality-fields} level by level, as a formal series in $\gs$ and wherever the dressed operators are invertible; where one of them is degenerate and the source lies in its range, the class of the solution in the quotient is real.
\Cref{sec:hs:quintic} works this out at the first order where an inverse appears, with the kernel of $\sprod{2}{2}$ quotiented.

\section{Localized form of the action}
\label{app:wrong-sign}

Eliminating $\Sigma_2$ from the quintic-order action produces at order $\gs^4$ the exchange term $\vertex{5 \cdot 5}(\Psi^6)$ of~\eqref{eq:hs5:eliminate}, which contains the inverse vertex $\sprod{2}{2}^{-1}$ and, for generic vertices, has no counterpart among the elementary vertices.
In the body of the paper, the inverse vertex is absorbed in the reduced vertex $\dvertex{6}$ and in the kinetic operator of $\Sigma_3$.
In this appendix, we keep every vertex elementary instead, at the price of a second $2$-string field, and we count the fields this costs at the higher orders.
We work at tree level, with the auxiliary fields at their classical values, and assume that $\sprod{2}{2}$, $\sprod{3}{3}$ and $\sprod{4}{4}$ are invertible on their quotients and that the vertices vanish on their kernels (\cref{sec:all:kernel}).
The results carry over to the string vertices provided the exchange terms are not linear combinations of the elementary vertices, which we assume.

\subsection{Inverse operators in field theory}

In field theory, inverse operators arise in effective actions, and we know how to make them local.
By the Gaussian identity, a field $X$ entering the action through a quadratic form $K$ and a linear form $J$ has the classical value $X^*$ and, at tree level, integrates out to a non-local term:
\begin{equation}
  \label{eq:wsp:gaussian}
  - K(X, X) + 2 J(X),
  \qquad
  X^*
    = K^{-1} J,
  \qquad
  - K(X^*, X^*) + 2 J(X^*)
    = J K^{-1} J.
\end{equation}
Read backwards, the identity writes a non-local term $J K^{-1} J$ as a local action with one more field.
An explicit example is two-dimensional gravity~\cite{Polyakov:1981:QuantumGeometryBosonic}.
The Coulomb gas action and its value on the solution $\phi^* = - \triangle^{-1} R$ are:
\begin{equation}
  \label{eq:wsp:coulomb}
  S
    = \int \dd^2 x \, \sqrt{g} \, \left(
      \frac{1}{2} \, \phi \triangle \phi
      + R \phi
    \right),
  \qquad
  S[\phi^*]
    = - \frac{1}{2} \int \dd^2 x \, \sqrt{g} \, R \triangle^{-1} R,
\end{equation}
where $R[g]$ is the Ricci scalar of the metric $g$ and $\triangle$ its Laplacian.
The second expression is the non-local induced gravitational action.

\subsection{Partner field at order \texorpdfstring{$\gs^4$}{gs4}}

We apply the same procedure to the exchange vertex $\vertex{5 \cdot 5}$.
There are two choices: decompose it before introducing $\Sigma_3$, or after.
We start with the first, which needs a single new field, and come back to the second at the end of this subsection.

We add a $2$-string field $\Upsilon_2 \in \Hn[2]{2} / \ker \sprod{2}{2}$, with $\Ngh(\Upsilon_2) = 4$ and the parity of $\Sigma_2$, since it fills two slots of the quartic vertex.
We take $X = \Upsilon_2$ in~\eqref{eq:wsp:gaussian}, with $K(\Upsilon_2, \Upsilon_2) := - \frac{\gs^2}{4!} \, \vertex{4}(\Upsilon_2^2)$ and $J(\Upsilon_2) := \frac{\gs^3}{5!} \, \vertex{5}(\Upsilon_2, \Sigma_2, \Psi)$.
The action of the partner field is then:
\begin{equation}
  \label{eq:wsp:partner-action}
  S_{\Upsilon}
    := \frac{\gs^2}{4!} \, \vertex{4}(\Upsilon_2^2)
      + \frac{2 \gs^3}{5!} \, \vertex{5}(\Upsilon_2, \Sigma_2, \Psi).
\end{equation}
With $\vertex{4}(\Upsilon_2, \Upsilon_2) = \pair[2]{\Upsilon_2}{\sprod{2}{2}(\Upsilon_2)}$ and $\vertex{5}(\Upsilon_2, \Sigma_2, \Psi) = \pair[2]{\Upsilon_2}{\sprod{3}{2}(\Sigma_2, \Psi)}$, the equation of motion of $\Upsilon_2$, normalized as the one of $\Sigma_2$ in \cref{sec:hs:quartic}, and its solution read:
\begin{equation}
  \label{eq:wsp:saddle}
  \mathcal{E}^{(4)}_{\Upsilon_2}
    := \frac{\gs}{5} \, \sprod{3}{2}(\Sigma_2, \Psi)
      + \sprod{2}{2}(\Upsilon_2)
    = 0,
  \qquad
  \Upsilon_2^{(4)*}
    = - \frac{\gs}{5} \, \sprod{2}{2}^{-1} \circ \sprod{3}{2}(\Sigma_2, \Psi).
\end{equation}
There is no source $\vertex{4}(\Upsilon_2, \Psi^2)$, so the classical value of $\Upsilon_2$ starts at order $\gs$, and inserting it gives $J K^{-1} J$:
\begin{equation}
  \label{eq:wsp:eliminated}
  S_{\Upsilon}(\Upsilon_2^{(4)*})
    = - \frac{\gs^4}{600} \, \vertex{5}\big(\Sigma_2, \Psi, \sprod{2}{2}^{-1} \circ \sprod{3}{2}(\Sigma_2, \Psi)\big)
    = - \frac{\gs^4}{600} \, \vertex{5 \cdot 5}(\Psi^6) + O(\gs^5),
\end{equation}
where the second equality uses the classical value $\Sigma_2 = \Psi^2 + O(\gs)$ of~\eqref{eq:hs5:saddle}.
This is the exchange term of~\eqref{eq:hs5:eliminate} with the opposite sign: in the Feynman rules of \cref{sec:hs:quartic}, an $\Upsilon_2$ line carries the propagator $- 12 \gs^{-2} \, \sprod{2}{2}^{-1}$, opposite in sign to the one of a $\Sigma_2$ line.

The sextic contact term is carried by the $3$-string field $\Sigma_3 \in \Hn[2]{3} / \ker \sprod{3}{3}$, introduced as in \cref{sec:hs:sextic} but through the elementary vertex, since the exchange term that the reduced vertex subtracted is now canceled by $\Upsilon_2$.
The action through order $\gs^4$ is then:
\begin{equation}
  \label{eq:wsp:action}
  \begin{aligned}
    \caction[\le 6]^{\Upsilon}
      &
      = \frac{1}{2} \, \pair{\Psi}{Q_B \Psi}
        + \frac{\gs}{3!} \, \vertex{3}(\Psi^3)
        - \frac{\gs^2}{4!} \, \vertex{4}(\Sigma_2^2)
        + \frac{2 \gs^2}{4!} \, \vertex{4}(\Sigma_2, \Psi^2)
        + \frac{\gs^3}{5!} \, \vertex{5}(\Sigma_2^2, \Psi)
      \\ &
      \qquad
        + \frac{\gs^2}{4!} \, \vertex{4}(\Upsilon_2^2)
        + \frac{2 \gs^3}{5!} \, \vertex{5}(\Upsilon_2, \Sigma_2, \Psi)
        - \frac{\gs^4}{6!} \, \vertex{6}(\Sigma_3^2)
        + \frac{2 \gs^4}{6!} \, \vertex{6}(\Sigma_3, \Sigma_2, \Psi).
  \end{aligned}
\end{equation}
At this order, the sextic term could also be written as $\vertex{6}(\Sigma_2^3)$ without $\Sigma_3$, but $\Sigma_3$ is needed from order $\gs^5$ on, since no cubic coupling of $(\Psi, \Sigma_2, \Upsilon_2)$ fills seven strands, and we introduce it at once.
The action is cubic in $(\Psi, \Sigma_2, \Upsilon_2, \Sigma_3)$ and every vertex is elementary.
With the auxiliary fields at their classical values, it reproduces $S_{\le 6}$ through order $\gs^4$: $\Sigma_3 = \Psi \otimes \Sigma_2$ exactly, the exchange term is canceled by~\eqref{eq:wsp:eliminated}, and the sextic vertex comes from the square of $\Sigma_3$.

The partner field satisfies the reality condition of $\Sigma_2$, whose strands it shares, $\Upsilon_2^\star = \Upsilon_2$ (\cref{app:reality}).
Its quadratic form has the sign opposite to that of $\Sigma_2$: at tree level, where the elimination is the evaluation at the classical point, the sign has no consequence, and whether it has one beyond tree level is open.
The gauge transformation of $\Upsilon_2$ and the completed-square form of its sector are not derived here.

In the second choice, $\Sigma_3$ is introduced first with the reduced vertex, and the sextic action~\eqref{eq:hs6:S6tilde-explicit} contains two non-local terms, its second and fourth, in the kinetic term of $\Sigma_3$ and in its coupling to $\Psi \otimes \Sigma_2$.
The first one has the form $J K^{-1} J$ and is made local by a field $\Xi_2$ whose quadratic form has the sign of that of $\Sigma_2$, with the action and the classical value:
\begin{equation}
  \label{eq:wsp:xi}
  - \frac{\gs^4}{600} \, \vertex{4}(\Xi_2^2)
    + \frac{2 \gs^4}{600} \, \vertex{5}(\Xi_2, \Sigma_3),
  \qquad
  \Xi_2^*
    = \sprod{2}{2}^{-1} \circ \sprod{3}{2}(\Sigma_3).
\end{equation}
The second one, bilinear in $\Sigma_3$ and $\Psi \otimes \Sigma_2$, is not made local by the same field: coupling $\Xi_2$ to $\Sigma_3 - \Psi \otimes \Sigma_2$ would produce, after elimination, the term $\vertex{5}\big(\Psi, \Sigma_2, \sprod{2}{2}^{-1} \circ \sprod{3}{2}(\Psi, \Sigma_2)\big)$, which the completed square had removed.

This choice therefore needs more new fields than the first one.
The intermediate-field representations of~\cite{Rivasseau:2010:LoopVertexExpansion, Lionni:2018:NoteIntermediateField} likewise need several auxiliary fields to decompose an interaction of degree $2 n$.
This choice also leaves $\Sigma_3$ with a kinetic term built on a diagram.
Decomposing the exchange term before introducing $\Sigma_3$ gives $\Sigma_3$ the kinetic term $\vertex{6}$ alone, and, as we find below by an explicit computation, keeps every vertex elementary through order $\gs^8$.

\subsection{Substitution rule and higher orders}

The couplings of $\Upsilon_2$ in~\eqref{eq:wsp:action} follow a rule.
Let $S[\Psi, \Sigma_2, \Sigma_3, \ldots]$ be a cubic action with elementary vertices and without a partner, such as~\eqref{eq:wsp:action} without its $\Upsilon_2$ terms.
Replace $\Sigma_2$ by $\Sigma_2 + \Upsilon_2$ in every coupling with a $\Sigma_2$ leg, keep the quadratic and source terms of $\Sigma_2$ unchanged, and add $+ \vertex{4}(\Upsilon_2^2)$ beside $- \vertex{4}(\Sigma_2^2)$, without any source $\vertex{4}(\Upsilon_2, \Psi^2)$.
By bilinearity of $\vertex{4}$, the result is:
\begin{equation}
  \label{eq:wsp:rule}
  \caction^{\Upsilon}[\Psi, \Sigma_2, \Upsilon_2, \Sigma_3, \ldots]
    = S[\Psi, \Sigma_2 + \Upsilon_2, \Sigma_3, \ldots]
      + \frac{2 \gs^2}{4!} \, \vertex{4}(\Upsilon_2, \Sigma_2 + \Upsilon_2 - \Psi^2).
\end{equation}
At order $\gs^4$ this is~\eqref{eq:wsp:action}, where the quintic coupling $\vertex{5}(\Sigma_2^2, \Psi)$ gives the coefficient $2$ of~\eqref{eq:wsp:partner-action}.
At order $\gs^5$ it adds the couplings $\vertex{5}(\Upsilon_2^2, \Psi)$ and $\vertex{6}(\Upsilon_2, \Sigma_3, \Psi)$, with the coefficients of $\vertex{5}(\Sigma_2^2, \Psi)$ and $\vertex{6}(\Sigma_3, \Sigma_2, \Psi)$ in~\eqref{eq:wsp:action}; they contribute first at that order since the classical value of $\Upsilon_2$ starts at order $\gs$.

In the variables $\Sigma_2 + \Upsilon_2$ and $\Upsilon_2$, the partner field enters linearly: it is a Lagrange multiplier for the constraint $\Sigma_2 + \Upsilon_2 = \Psi^2$, the multiplier form of the constraint mentioned in \cref{sec:all:kernel}.
Its equation of motion sets $\Sigma_2 + \Upsilon_2 = \Psi^2$ on the quotient, the equation of $\Sigma_2 + \Upsilon_2$ fixes $\Upsilon_2$, and the linear term vanishes on the solution, so that the eliminated action is $S$ at $\Sigma_2 = \Psi^2$ with $\Sigma_3, \Sigma_4, \ldots$ at their classical values.
Every exchange term with a $\sprod{2}{2}^{-1}$ line therefore cancels, at every order and for every choice of the couplings of $\Sigma_3, \Sigma_4, \ldots$.

At order $\gs^5$, for instance, the lower orders generate two such terms, three quintic vertices joined by two $\Sigma_2$ lines and a sextic and a quintic vertex joined by one:
\begin{equation}
  \label{eq:wsp:exchange-5}
    \vertex{5}\big(\sprod{2}{2}^{-1} \circ \sprod{3}{2}(\Psi^3), \sprod{2}{2}^{-1} \circ \sprod{3}{2}(\Psi^3), \Psi\big),
    \qquad
    \vertex{6}\big(\sprod{2}{2}^{-1} \circ \sprod{3}{2}(\Psi^3), \Psi^4\big).
\end{equation}
The two couplings of the rule cancel them, one structure each, and the contact term $\vertex{7}(\Psi^7)$ is supplied by a septic coupling of $\Sigma_3$:
\begin{equation}
  \label{eq:wsp:septic}
  \frac{\gs^5}{7!} \, \big[
    a_7 \, \vertex{7}(\Sigma_3^2, \Psi)
    + b_7 \, \vertex{7}(\Sigma_3, \Sigma_2^2)
  \big],
  \qquad
  a_7 + b_7
    = 1.
\end{equation}

The partners needed at the higher orders are decided by the action the rule leaves, that of $\Sigma_3, \Sigma_4, \ldots$ at $\Sigma_2 = \Psi^2$, whose exchange terms carry $\sprod{3}{3}^{-1}$ or $\sprod{4}{4}^{-1}$ lines and are reached by no coupling of $\Upsilon_2$.
At order $\gs^6$ there is one such term, the exchange of $\Sigma_3$ between two septic couplings, with a coefficient proportional to $(2 a_7 + b_7)^2$.
With $a_7 = 1$, a partner $\Upsilon_3$ of $\Sigma_3$ is needed from this order on, with the quadratic term $\vertex{6}(\Upsilon_3^2)$ and the source $2 \, \vertex{7}(\Upsilon_3, \Sigma_3, \Psi)$, the construction of~\eqref{eq:wsp:rule} one strand up.
Suppose $\Sigma_4$ is introduced with the source $\vertex{8}(\Sigma_4, \Sigma_3, \Psi)$, and the field with the most strands is again squared at the next order, in $\vertex{9}(\Sigma_4^2, \Psi)$.
Then an $\Upsilon_4$ follows at order $\gs^8$, and the partners form a tower, which is the multiplier form one strand at a time.

With $(a_7, b_7) = (- 1, 2)$ instead, a completed square on $\Sigma_2 = \Psi^2$, the term is absent.
Through order $\gs^8$, we find that the remaining exchange terms vanish for couplings whose terms linear in $\Sigma_3 - \Psi^3$ and $\Sigma_4 - \Psi^4$ at the classical point do not contribute before that order.
Such couplings are this square at order $\gs^5$, the coupling
\begin{equation}
  \label{eq:wsp:sigma4}
  - \vertex{8}(\Sigma_4^2)
    + 2 \, \vertex{8}(\Sigma_4, \Sigma_2^2)
\end{equation}
at order $\gs^6$, which brings in $\Sigma_4$ at that order as in \cref{sec:all-orders}, and analogous choices at orders $\gs^7$ and $\gs^8$, each of which leaves a parameter free through order $\gs^8$.
At orders $\gs^7$ and $\gs^8$ these are $\lambda \big[ \vertex{9}(\Sigma_4^2, \Psi) - 2 \, \vertex{9}(\Sigma_4, \Sigma_3, \Sigma_2) + \vertex{9}(\Sigma_3^3) \big] + \vertex{9}(\Sigma_3^3)$ and $\rho \, \vertex{10}(\Sigma_4^2, \Sigma_2) + (1 - \rho) \, \vertex{10}(\Sigma_4, \Sigma_3^2)$, with $\lambda$ and $\rho$ the free parameters.

The fields are then $\Psi$, $\Sigma_2$, $\Sigma_3$, $\Sigma_4$ and the single partner $\Upsilon_2$, and every vertex is elementary.
The couplings of $\Upsilon_2$ that complete the matching are those of the rule, up to splits between couplings that produce the same exchange terms.
Whether one partner suffices at every order is not settled: the rule removes every $\sprod{2}{2}^{-1}$ line once and for all, and what remains is decided order by order by the partition of the couplings of $\Sigma_3, \Sigma_4, \ldots$.

\section{Hubbard--Stratonovich transformation for \texorpdfstring{$N$}{N} scalar fields}
\label{app:scalar-hs}

In this appendix, we perform the Hubbard--Stratonovich transformation for a theory of several scalar fields with a quartic interaction, the model of \cref{sec:intro} with constant couplings.
We see on this example how the auxiliary fields are organized when there are several physical fields: the auxiliary field carries a pair of indices, as $\Sigma_2$ carries two strings.
We also see, in a setting where the linear algebra is explicit, what happens when the quadratic form of the auxiliary field has a kernel.

\paragraph{Action of the scalar fields.}

Consider $N$ scalar fields $\phi_i$, $i = 1, \ldots, N$, with the action:
\begin{equation}
  \label{eq:shs:action}
  S
    = \int \dd^d x \left[
      - \frac{1}{2} \, \partial_\mu \phi_i \, \partial^\mu \phi_i
      - \frac{1}{2} \, M_{ij} \phi_i \phi_j
      + \frac{1}{3!} \, W_{ijk} \phi_i \phi_j \phi_k
      + \frac{1}{4!} \, V_{ijk\ell} \phi_i \phi_j \phi_k \phi_\ell
    \right],
\end{equation}
where repeated indices are summed over.
The mass matrix $M_{ij}$ and the coupling tensors $W_{ijk}$ and $V_{ijk\ell}$ are constant and completely symmetric.
We write $S_4$ for the quartic term.

\paragraph{Auxiliary fields.}

The quartic term is the square of the products $\phi_i \phi_j$ with respect to the tensor $V_{ijk\ell}$.
We introduce one auxiliary field for each such product, that is $N (N + 1) / 2$ fields $\sigma_{ij} = \sigma_{ji}$, and add to the action the term~\cite{Stratonovich:1957:MethodCalculatingQuantum, Hubbard:1959:CalculationPartitionFunctions, Weinberg:2005:QuantumTheoryFields-2}:
\begin{equation}
  \label{eq:shs:DeltaS4}
  \Delta S_4
    := - \frac{1}{4!} \int \dd^d x \, V_{ijk\ell} \, \big(
      \sigma_{ij} - \phi_i \phi_j
    \big) \big(
      \sigma_{k\ell} - \phi_k \phi_\ell
    \big),
\end{equation}
which vanishes when $\sigma_{ij} = \phi_i \phi_j$.
We expand the square and use $V_{ijk\ell} = V_{k\ell ij}$ for the cross term, so that the quartic term of the action becomes:
\begin{equation}
  \label{eq:shs:S4tilde}
  \caction[4]
    := S_4 + \Delta S_4
    = \int \dd^d x \left[
      - \frac{1}{4!} \, V_{ijk\ell} \, \sigma_{ij} \sigma_{k\ell}
      + \frac{1}{12} \, V_{ijk\ell} \, \sigma_{ij} \phi_k \phi_\ell
    \right].
\end{equation}
The kinetic, mass and cubic terms are untouched, so that the action
\begin{equation}
  \label{eq:shs:action-tilde}
  \begin{aligned}
    \caction
      := S + \Delta S_4
      = \int \dd^d x \bigg[
        &
        - \frac{1}{2} \, \partial_\mu \phi_i \, \partial^\mu \phi_i
        - \frac{1}{2} \, M_{ij} \phi_i \phi_j
        + \frac{1}{3!} \, W_{ijk} \phi_i \phi_j \phi_k
        \\ &
        + \frac{1}{12} \, V_{ijk\ell} \, \sigma_{ij} \phi_k \phi_\ell
        - \frac{1}{4!} \, V_{ijk\ell} \, \sigma_{ij} \sigma_{k\ell}
      \bigg]
  \end{aligned}
\end{equation}
is cubic in $(\phi_i, \sigma_{ij})$.
For $N = 1$, \eqref{eq:shs:S4tilde} is the transformation of the introduction at constant coupling, with the normalization $1 / 4!$ of the vertices.

The fields $\sigma_{ij}$ have no kinetic term, and their equation of motion is algebraic:
\begin{equation}
  \label{eq:shs:eom-sigma}
  V_{ijk\ell} \, \big( \sigma_{k\ell} - \phi_k \phi_\ell \big)
    = 0.
\end{equation}
When $V_{ijk\ell}$ is invertible as a matrix acting on the symmetric pairs of indices $(ij)$ and $(k\ell)$ (the matrix $V_{IJ}$ defined below), the solution is:
\begin{equation}
  \label{eq:shs:solution}
  \sigma_{ij}
    = \phi_i \phi_j,
\end{equation}
on which $\Delta S_4$ vanishes and $\caction$ reduces to the original action $S$.

\paragraph{Pair-index basis.}

Let $\lambda_{I;ij}$, $I = 1, \ldots, N (N + 1) / 2$, be an orthonormal and complete basis of the real symmetric $N \times N$ matrices:
\begin{equation}
  \label{eq:shs:basis}
  \lambda_{I;ij} \lambda_{J;ij}
    = \delta_{IJ},
  \qquad
  \lambda_{I;ij} \lambda_{I;k\ell}
    = \frac{1}{2} \, \big( \delta_{ik} \delta_{j\ell} + \delta_{i\ell} \delta_{jk} \big),
\end{equation}
and define:
\begin{equation}
  \label{eq:shs:pair-index}
  \sigma_I
    := \lambda_{I;ij} \, \sigma_{ij},
  \qquad
  V_{IJ}
    := \lambda_{I;ij} \lambda_{J;k\ell} \, V_{ijk\ell},
  \qquad
  V_{I;k\ell}
    := \lambda_{I;ij} \, V_{ijk\ell}.
\end{equation}
Since $\sigma_{ij}$ and $V_{ijk\ell}$ are symmetric in each pair of indices, the completeness relation gives $\sigma_{ij} = \lambda_{I;ij} \sigma_I$ and $V_{I;k\ell} = V_{IJ} \lambda_{J;k\ell}$, and the quartic term~\eqref{eq:shs:S4tilde} becomes:
\begin{equation}
  \label{eq:shs:S4tilde-pair}
  \caction[4]
    = \int \dd^d x \left[
      - \frac{1}{4!} \, V_{IJ} \, \sigma_I \sigma_J
      + \frac{1}{12} \, V_{I;k\ell} \, \sigma_I \phi_k \phi_\ell
    \right].
\end{equation}
The matrix $V_{IJ}$ is real and symmetric, so an orthogonal change of the basis $\lambda_{I;ij}$ diagonalizes it, and we assume for now that all its eigenvalues are nonzero.
Otherwise, the fields $\sigma_I$ along the eigenvectors with zero eigenvalue appear neither in the quadratic term nor in the coupling to $\phi_i$, since $V_{I;k\ell} = V_{IJ} \lambda_{J;k\ell}$: they do not enter the action, and we leave them out.

\paragraph{Four-point amplitude.}

We read the Feynman rules off $\caction$ as in \cref{sec:hs:quartic}: a vertex factor is the derivative of the action with respect to the fields on its legs, and a propagator is minus the inverse of the Hessian.
The auxiliary sector~\eqref{eq:shs:S4tilde-pair} contributes the propagator of $\sigma_I$ and the vertex coupling $\sigma_I$ to $\phi_k \phi_\ell$, respectively:
\begin{equation}
  \label{eq:shs:feynman-rules}
  12 \, (V^{-1})_{IJ},
  \qquad
  \frac{1}{3!} \, V_{I;k\ell}.
\end{equation}
The exchange of $\sigma$ between two such vertices, with the external fields $\phi_i, \phi_j$ on one vertex and $\phi_k, \phi_\ell$ on the other, gives:
\begin{equation}
  \label{eq:shs:channel}
  \frac{1}{3!} \, V_{I;ij} \;
  12 \, (V^{-1})_{IJ} \;
  \frac{1}{3!} \, V_{J;k\ell}
    = \frac{1}{3} \, \lambda_{K;ij} V_{KI} (V^{-1})_{IJ} V_{JL} \lambda_{L;k\ell}
    = \frac{1}{3} \, V_{ijk\ell},
\end{equation}
where the first equality uses $V_{I;ij} = V_{IK} \lambda_{K;ij}$ and the second $V_{ijk\ell} = \lambda_{K;ij} V_{KL} \lambda_{L;k\ell}$, which follows from the completeness relation.
Summing over the three ways of distributing the four external fields on the two vertices and using the complete symmetry of $V_{ijk\ell}$, the four-point amplitude is:
\begin{equation}
  \label{eq:shs:A4}
  A_{ijk\ell}
    = \frac{1}{3} \, \big( V_{ijk\ell} + V_{ikj\ell} + V_{i\ell jk} \big) + \cdots
    = V_{ijk\ell} + \cdots,
\end{equation}
where the dots are the $s$-, $t$- and $u$-channel exchanges of $\phi$ built from the cubic coupling $W_{ijk}$, which the transformation leaves untouched.
The exchange of $\sigma$ thus reproduces the contact vertex $V_{ijk\ell}$ of the original action~\eqref{eq:shs:action}.

\paragraph{Relation to the string field.}

We obtain the construction of \cref{sec:hs:quartic} from this one by replacing the products $\phi_i \phi_j$ by the $2$-string state $\Psi^2$, the tensor $V_{ijk\ell}$ by $\gs^2 \vertex{4}$ (so that $V_{IJ}$ becomes $\gs^2 \sprod{2}{2}$), the auxiliary field $\sigma_{ij}$ by $\Sigma_2 \in \Hn{2}$, and the propagator $12 \, (V^{-1})_{IJ}$ by $12 \gs^{-2} \, \sprod{2}{2}^{-1}$.

\paragraph{Degenerate coupling matrix.}

To couple the directions of zero eigenvalue of $V_{IJ}$, left out above, we now add a quintic term to the action~\eqref{eq:shs:action}, with a constant and completely symmetric coupling $U_{ijk\ell m}$:
\begin{equation}
  \label{eq:shs:S5}
  S_5
    = \frac{1}{5!} \int \dd^d x \, U_{ijk\ell m} \, \phi_i \phi_j \phi_k \phi_\ell \phi_m.
\end{equation}
As at the quintic order of the string construction, we replace two pairs of fields by the auxiliary field, which turns the quintic term into a cubic coupling:
\begin{equation}
  \label{eq:shs:S5tilde}
  \caction[5]
    := \frac{1}{5!} \int \dd^d x \, U_{IJm} \, \sigma_I \sigma_J \phi_m,
  \qquad
  U_{IJm}
    := \lambda_{I;ij} \lambda_{J;k\ell} \, U_{ijk\ell m},
\end{equation}
which agrees with $S_5$ at $\sigma_{ij} = \phi_i \phi_j$.
When the matrix $V_{IJ}$ is invertible, eliminating $\sigma_I$ from $\caction + \caction[5]$ gives back $S + S_5$ up to terms of second order in $U_{ijk\ell m}$, the exchange of $\sigma$ between two quintic couplings, whose string counterpart is canceled at the sextic order (\cref{sec:hs:sextic}).

We now suppose that $V_{IJ}$ has null vectors, $V_{IJ} \xi_J = 0$, which span the null space $K$, and we choose a complement $C$ of $K$, $\R^{N (N + 1) / 2} = C \oplus K$.
The orthogonal complement is one choice, which the string construction does not single out, having no inner product on $2$-string states (\cref{sec:hs:quartic}).
We write $\pi_C$ and $\pi_K$ for the projections on $C$ along $K$ and on $K$ along $C$.
The null components of $\sigma_I$ drop out of the quartic terms~\eqref{eq:shs:S4tilde-pair}, as noted above, but they enter the quintic coupling~\eqref{eq:shs:S5tilde}.
Under the shift $\sigma_I \to \sigma_I + \xi_I$, $\caction[5]$ changes by:
\begin{equation}
  \label{eq:shs:null-shift}
  \frac{1}{5!} \int \dd^d x \, \big(
    2 \, U_{IJm} \, \xi_I \sigma_J \phi_m
    + U_{IJm} \, \xi_I \xi_J \phi_m
  \big).
\end{equation}
For this to vanish for all $\sigma_I$, the term linear in $\sigma_J$ must vanish, which then implies that the quadratic one vanishes.
The null components therefore decouple if and only if the quintic coupling annihilates the null vectors, and they can then be left out as before:
\begin{equation}
  \label{eq:shs:kernel-condition}
  \forall \xi \in K, \;
  \forall k, \ell, m:
  \quad
  \xi_I \, U_{I;k\ell m}
    = 0,
  \qquad
  U_{I;k\ell m}
    := \lambda_{I;ij} \, U_{ijk\ell m}.
\end{equation}
This is the kernel condition~\eqref{eq:hs5:kernel-condition} of the string construction, with $\xi$ for $N_2$ and $U_{ijk\ell m}$ for $\vertex{5}$.

When this condition fails, the null components of $\sigma_I$ have no quadratic term in $\caction$, but they have a source, the term of~\eqref{eq:shs:S5tilde} linear in them.
We could keep the null components as fields, but their quadratic form in $\caction[5]$ is proportional to $\phi_m$, so that around $\phi_i = 0$ they have neither a propagator nor a perturbative expansion.
If we leave them out instead, that is, if we take $\sigma_I$ in $C$, we get an action of the form~\eqref{eq:shs:action-tilde} with $V_{IJ}$ restricted to $C$, where it is invertible, and $\sigma_I$ can be eliminated as before.
The eliminated action reproduces the quartic term, but at first order in $U_{ijk\ell m}$ it contains $U_{IJm} \, (\pi_C \Phi)_I (\pi_C \Phi)_J \phi_m$ in place of the quintic term $U_{IJm} \, \Phi_I \Phi_J \phi_m$.
Here $\Phi_I := \lambda_{I;ij} \, \phi_i \phi_j$ is the pair-index form of the products $\phi_i \phi_j$.
The missing terms are those with a null direction in at least one of the two pair indices; they depend on the complement and cannot vanish for all complements.

The construction is repaired by a Lagrange multiplier on the null directions.
We keep all the components of $\sigma_I$, and we add to the action a field $\theta_I$ on the null directions, paired with the null component of $\sigma_I - \Phi_I$:
\begin{equation}
  \label{eq:shs:multiplier}
  \caction + \caction[5]
    + \int \dd^d x \, \theta_I \, \big( \pi_K (\sigma - \Phi) \big)_I.
\end{equation}
Since $\theta_I \Phi_I$ is cubic, every term has at most three fields and the action stays cubic.
The fields are eliminated in three steps.
The equation of motion of $\theta_I$ sets the null components of $\sigma_I$ to those of $\Phi_I$:
\begin{equation}
  \label{eq:shs:multiplier-eom}
  \pi_K (\sigma - \Phi)
    = 0.
\end{equation}
The equation of motion of the null components of $\sigma_I$ fixes $\theta_I$, which appears nowhere else, to minus the derivative of $\caction[5]$ with respect to these components.
Finally, the equation of motion of the components along $C$ is the one of the action with $\sigma_I$ in $C$, with a source shifted by the quintic coupling evaluated on the null component of $\Phi_I$.
Eliminating all the fields gives back $S + S_5$ up to terms of second order in $U_{ijk\ell m}$, for every choice of complement, because the null components supply the part of $\Phi_I$ that the complement was missing.
At second order in $U_{ijk\ell m}$, the exchange of the components along $C$, the scalar counterpart of the reduced sextic vertex, depends on the complement when the kernel condition fails, through its source restricted to $C$.
The multiplier vanishes on the solution if and only if the quintic coupling annihilates the null vectors, in which case it is not needed.
This is how the multiplier replaces the quotient when a vertex does not vanish on the kernel of the quadratic form.
For the string vertices, the multiplier on the kernel directions mentioned in \cref{sec:all:kernel}, which imposes $\Sigma_k = \Psi \otimes \Sigma_{k - 1}$ there, does the same for $\Sigma_2$, leaving a sextic exchange term that depends on the complement.
Whether $\sprod{2}{2}$ has such null vectors is not settled for the string vertices, and none is known (\cref{sec:hs:quintic}).

\printbibliography[heading=bibintoc]

\end{document}